\documentclass[a4paper,12pt, pdftex, twoside]{book}

\usepackage{amsmath} 
\usepackage{amssymb}
\usepackage[ngerman, english]{babel} 
\usepackage{color, colortbl}
\usepackage{lmodern}
\usepackage[T1]{fontenc}
\usepackage{geometry} 
\usepackage{graphicx} 
\usepackage[utf8]{inputenc} 
\usepackage{csquotes}
\usepackage{isotope}
\usepackage{makecell} 
\usepackage{mhchem}
\usepackage{multirow}
\usepackage[section]{placeins} 
\usepackage{pifont}
\usepackage{rotating} 
\usepackage[headsepline,plainheadsepline,nouppercase]{scrpage2} 
\usepackage{setspace}
\usepackage[outercaption]{sidecap}
\sidecaptionvpos{figure}{c}

\usepackage[separate-uncertainty, version-1-compatibility, space-before-unit]{siunitx} 
 \DeclareSIUnit[number-unit-product = {}]\degree{\SIUnitSymbolDegree}
 \DeclareSIUnit\year{yr}
 \DeclareSIUnit\cpkky{cts/(keV\,kg\,y)}

\usepackage{subcaption} 
\usepackage{ wasysym }
\usepackage{xspace} 
\usepackage[bookmarksopenlevel=1,bookmarksopen, pdfstartview=FitH, pdfborder={0 0 0}, unicode]{hyperref} 

\makeindex 

\newcommand{\abs}{\\[0.5mm]}

\begin{document}

\begin{titlepage}
  \vspace*{0.5cm}
   \begin{center}
  \begin{Huge} Dissertation \\[10mm] \parbox{15cm}{\centering  Commissioning of the COBRA demonstrator and investigation of surface events as its main background} \end{Huge}
  \\[2cm]
  \includegraphics[width=5cm]{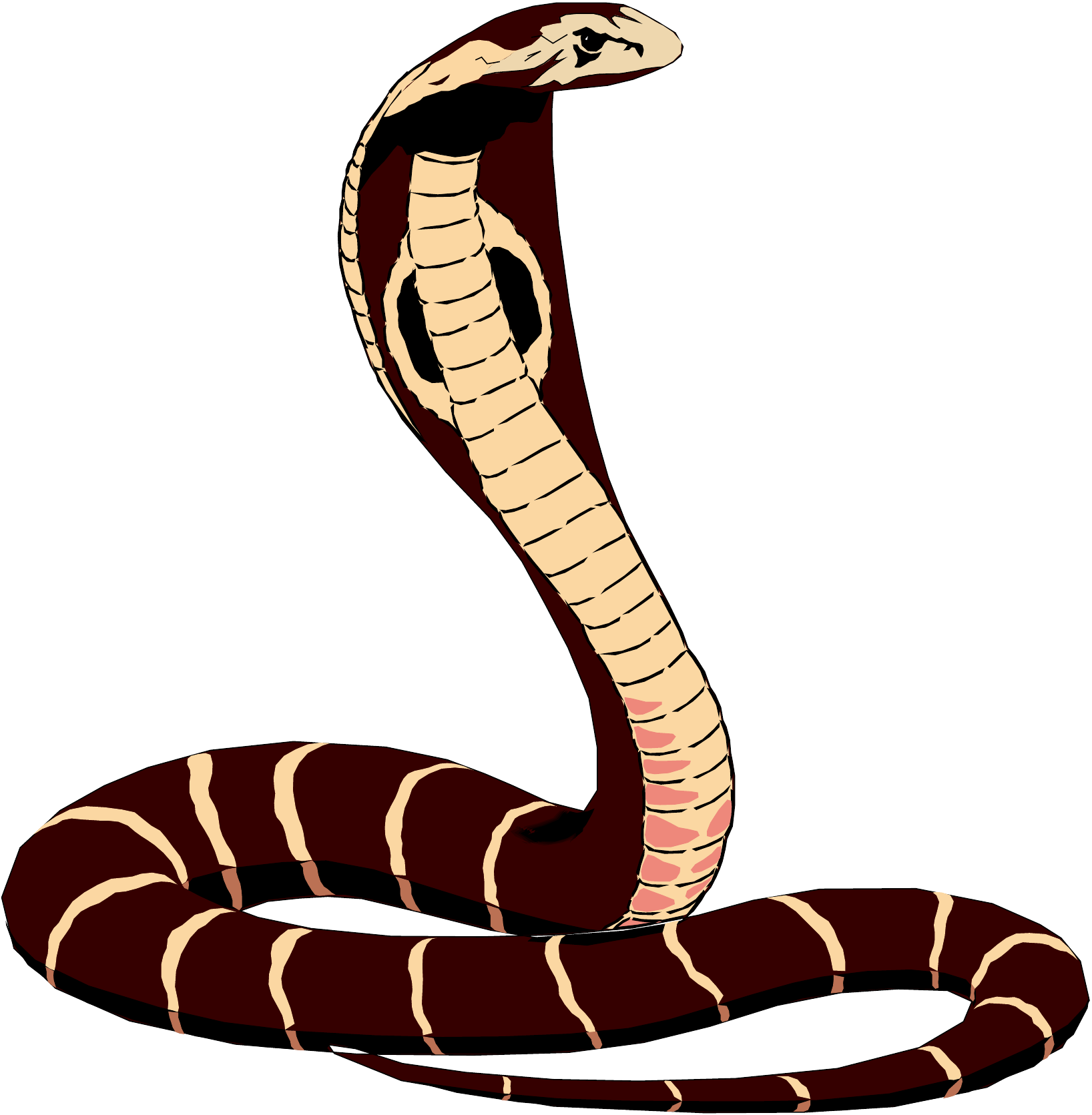}
  \\[1.5cm]
  \begin{Large} vorgelegt von \\[2mm] Jan Tebr\"ugge \abs Lehrstuhl f\"ur Experimentelle Physik IV \abs Fakult\"at f\"ur Physik \\[.5cm] \end{Large}
  \includegraphics[width=8cm]{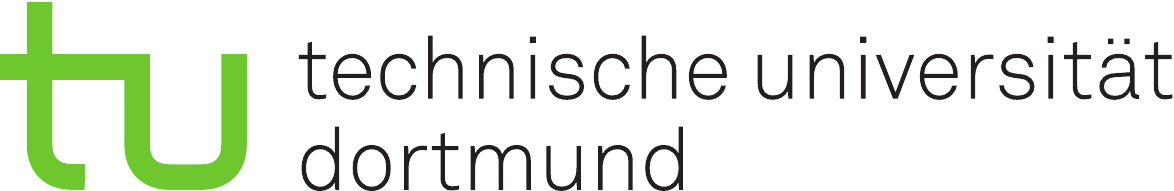}
  \\[.5cm]
  \Large{Dortmund, 03.Mai 2016}

\cleardoublepage

\begin{verbatim}
 
\end{verbatim}

\begin{Huge} {\centering   Commissioning of the COBRA demonstrator and investigation of surface events as its main background} \end{Huge}

\vspace{2cm}

Dissertation zur Erlangung des akademischen Grades eines\\
Doktors der Naturwissenschaften der Fakult\"at Physik an der\\
Technischen Universit\"at Dortmund\\

\vspace{2cm}

vorgelegt von\\
Dipl. Phys. Jan Tebr\"ugge\\
Lehrstuhl f\"ur Experimentelle Physik IV\\
Fakult\"at f\"ur Physik\\
TU Dortmund\\

\vspace{2cm}

Dortmund, 03.Mai 2016\\

\end{center}

Teilergebnisse dieser Arbeit wurden bereits auf Konferenzen und in Publikationen vorgestellt.
Eine Liste davon ist auf Seite \pageref{sec_list_of_publications} zu finden.

\vspace{2cm}

\begin{table}[h!] 
\begin{tabular}{ll}
Gutachter: &Prof. Dr. C. G\"o{\ss}ling, TU Dortmund\\
Zweitgutachter: &Prof. Dr. K. Zuber, TU Dresden\\
Vorsitz der Pr\"ufungskommission: &Prof. Dr. M. Betz, TU Dortmund\\
Vertreterin der wissenschaftlichen Mitarbeiter: &Dr. B. Siegmann, TU Dortmund\\
Termin der m\"undlichen Pr\"ufung: &29. Juni 2016\\
\end{tabular}
\end{table}

\end{titlepage}

\chapter*{Abstract}
%%%%%%%%%%%%%%%%%%%%%%%%%%%%%%%%%%%%%%%%%%%%%%%%%%%%%%%%%%%%%%%%%%%%%%%%%%%%%%%%%%%%%%%%%%%%%%%%%%%%%%%%%%%%%
%\vspace{-2cm}
The COBRA collaboration investigates 0$\nu\beta\beta$-decays (neutrinoless double beta-de-cays).
Therefore, a demonstrator setup using coplanar-grid CdZnTe detectors is operated at the LNGS underground laboratory.\\
In this work, the demonstrator was commissioned and completed, which is discussed extensively.
The demonstrator works reliably and collects low-background physics data.
One result of the analysis of the data is that surface events are the dominating background component.
To better understand and possibly discriminate this background, surface events were studied in detail.
This was done mainly using laboratory measurements. 
For a better interpretation of these measurements, simulations of particle trajectories and ranges were done.
The surface sensitivity tests showed large differences between the individual detectors.
Often, a dead-layer was determined, especially at the surfaces where the non-collecting anode (NCA) is the outermost anode rail.
Due to this, the sensitivity of the surfaces where the collecting anode (CA) is adjacent was typically about a factor of three larger than the NCA sensitivity.
A comparison of the pulse shape analysis methods LSE and A/E was done. Laboratory measurements indicate, that the latter performs better.
Alpha scanning measurements were done to spatially investigate the surface sensitivity. 
Plausible variations were measured. However, no hints were found how to improve the surface event recognition.
The instrumentation of the guard ring, which surrounds the anode structure, was tested and improved the surface event discrimination significantly.
The fraction of surviving alpha events was at a per-mill level.
Furthermore, the electron mobility was determined to \SI{968(28)}{\frac{cm^2}{Vs}}, which is in very good agreement with literature values. 
A variation at the detector edges was found.\\
Important steps for a future large-scale COBRA experiment are discussed briefly, mainly the use of an integrated read-out system.\\
Overall, the results indicate a large potential in background reduction for the \text{COBRA} experiment.

\tableofcontents

\chapter{Introduction}
\label{sec_introduction}
Neutrinos were postulated more than 80 years ago to explain theoretical problems of beta-decay by Wolfgang Pauli, who is quoted as: ``I have done a terrible thing, I have postulated a particle that cannot be detected'' \cite{sutton1992spaceship}.

Although neutrinos are ubiquitous and one of the most abundant particles in the universe, many of their properties are still unknown today.
Determining these properties affects many fields in physics, like particle and nuclear physics.\\
The detection of 0$\nu\beta\beta$-decays (neutrinoless double beta-decays) can help answer some of the open questions in neutrino physics.\\

\begin{SCfigure}[][h!]
 \centering
  \includegraphics[width=0.3\textwidth]{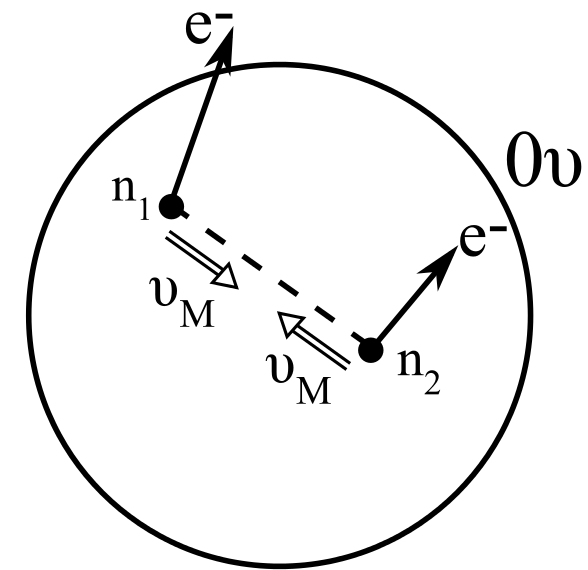}
  \caption[Sketch of double beta-decays]
          {Sketch of\\ 0$\nu\beta\beta$-decay. Adapted\\ from \cite{zuber_04}.}
  \label{fig_sketch_double_beta_decays}
\end{SCfigure}

0$\nu\beta\beta$-decay is a hypothetical radioactive decay which was also proposed about 80 years ago.
During 0$\nu\beta\beta$-decay, two beta-decays happen at the same time, and due to special conditions, no neutrinos are emitted.
This is indicated in Figure \ref{fig_sketch_double_beta_decays}. Two neutrons (n) undergo a beta-decay each.
The two electrons ($e^-$) leave the nucleus (black circle), but the neutrinos ($\nu_{M}$) do not.
All information about neutrinos that could be obtained by detecting this decay rely on measuring these electrons.

0$\nu\beta\beta$-decay is associated with half-lives of more than \SI{E26}{yr}. Consequently, it is a very rare decay, most obvious if the half-life is compared to the age of the universe of about \SI{E10}{yr}.

In order to be able to detect such a rare decay, the following measures are important:\\
Acquire a large number of atoms that can undergo the decay, wait for the decay to happen, and measure it with a high detection efficiency.
One of the challenges is to avoid measuring unimportant signals of other origin like cosmic particles or other radioactive decays, called background.
These background events can happen much more often than the decay under study.

Two issues are principally necessary to reduce the background rate:\\
One is to avoid background using a suitable shielding, and apply only carefully selected materials that emit very little radiation.\\
The other is an active recognition and discrimination of background events in the analysis.\\

In the course of this thesis, the author worked on both of these two measures.
First, the COBRA (\textbf{C}admium Zinc Telluride \textbf{0}$\nu$ double \textbf{B}eta \textbf{R}esearch \textbf{A}pparatus) demonstrator setup was commissioned and completed, which is extensively discussed.
This is an experimental setup in the underground laboratory LNGS (\textbf{L}abora\-tori \textbf{N}azionali del \textbf{G}ran \textbf{S}asso), which shall demonstrate the feasibility of such a concept in the search for 0$\nu\beta\beta$-decays.\\
Second, as events on the detector surfaces were identified as the main background component, 
comprehensive studies were done on surface events. 
These comprise detailed laboratory measurements to obtain information about surface events, where they typically occur, and how they can be minimized and discriminated.
For a better interpretation of these measurements, simulations of particle trajectories and ranges were done.
The goal is to reduce the background further, which is an essential step in the search for 0$\nu\beta\beta$-decays.\\

The following simplified comparison shall give an idea of how rare the 0$\nu\beta\beta$-decay, and how important background reduction is.
One banana contains about \SI{0.5}{g} of potassium (K) \cite{chiquita_banana}.
The radioactivity of \SI{1}{g} K due to the ubiquitous \isotope[40]K is approximately \SI{30}{Bq} \cite{0031-9155-42-2-012}.
Consequently, one typical banana has an activity of \SI{15}{Bq} = \SI{15}{decays\per second} = \SI{5E8}{decays\per year}.
The COBRA demonstrator consists of about \SI{380}{g} of the material CdZnTe.
Natural CdZnTe contains \SI{1.73E23}{atoms\per\kg} of the main isotope of interest \isotope[116]Cd.
The activity of the demonstrator concerning only \isotope[116]Cd with an assumed half-life for the 0$\nu\beta\beta$-decay of \SI{2E26}{yr} can be estimated to \num{2E-4} decays/year.
Even for a future large-scale COBRA experiment with about \SI{500}{kg} of detector material enriched in \isotope[116]Cd to \SI{90}{\%}, the activity of \isotope[116]Cd is only approximately \SI{4}{decays\per year}.\\
This demonstrates that a banana (or similar ``normal'' things) close to the detector would be a serious problem to detect 0$\nu\beta\beta$-decay.

\chapter{Double beta-decay and neutrino physics}
\label{sec_dbd_neutrino_physics}

This chapter provides a short introduction to neutrino physics. 
Then a historical overview shows the important contributions that were achieved by investigating beta-decays and neutrino physics, resulting in general progress in particle physics.

Finally, the latest results in the search for $0\nu\beta\beta$-decays are presented.

%%%%%%%%%%%%%%%%%%%%%%%%%%%%%%%%%%%%%%%%%%%%%%%%%%%%%%%%%%%%%%%%%%%%%%%%%%%%%%%%%%%%%%%%%%%%%%%%%%%%%%%%%%%%%%%%%%%%%%%%%%%%
%%%%%%%%%%%%%%%%%%%%%%%%%%%%%%%%%%%%%%%%%%%%%%%%%%%%%%%%%%%%%%%%%%%%%%%%%%%%%%%%%%%%%%%%%%%%%%%%%%%%%%%%%%%%%%%%%%%%%%%%%%%%
\section{Theory of double beta-decay}
\label{sec_theory}

Atomic nuclei attempt to lower their energy state by conversions of their nucleons. Nuclear beta-decays are a result. 
However, some isotopes exist, where the normal beta-decay is energetically forbidden.
If this is the case, the 2$\nu\beta\beta$-decay (neutrino-accompanied double beta-decay) can be measured. 
In this decay, two (single) beta-decays occur simultaneously.
It is a second order process, and consequently heavily suppressed.
This decay can occur at 35 isotopes \cite{Schwingenheuer:2012zs}.

\begin{figure}
 \centering
  \captionsetup{width=0.9\textwidth} 
  $\vcenter{\hbox{\includegraphics[width=0.29\textwidth]{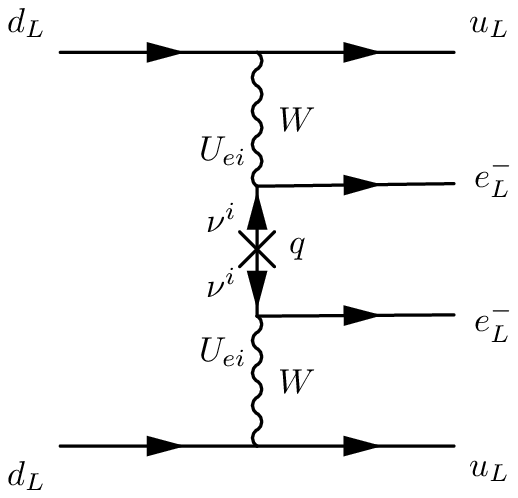}}}$
  $\vcenter{\hbox{\includegraphics[width=0.7\textwidth]{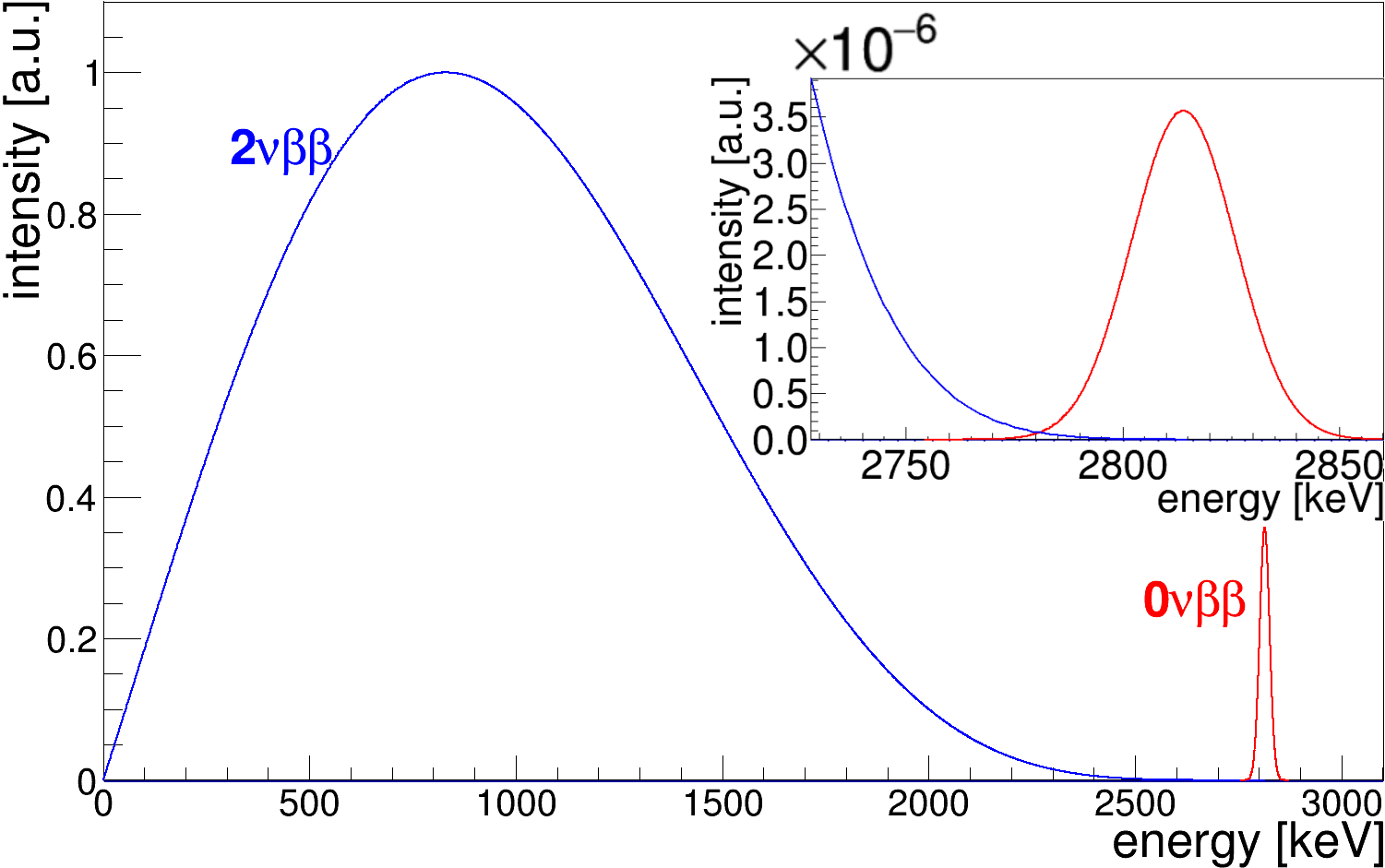}}}$
  \caption[Feynman diagram and energy spectrum of $0\nu\beta\beta$]
          {Left: Feynman diagram of the $0\nu\beta\beta$-decay in the standard interpretation, taken from \cite{Pas:2015eia}. 
           Right: Principal shape of the expected spectra for the 2$\nu\beta\beta$-decay (blue) and $0\nu\beta\beta$-decay (red, scaled up by a factor of \num{100000}) of \isotope[116]Cd. The inset shows the suppression of the $0\nu\beta\beta$-decay correctly scaled. Adapted from \cite{zatschler_diss}.}
  \label{fig_feynman_dbd_standard}
\end{figure}

If neutrinos are their own antiparticles, called Majorana-particles, a $0\nu\beta\beta$-decay is also possible:
The neutrino emitted by one of the nucleons (with proton number Z and mass number A) is absorbed after an helicity-flip by the other. 
Consequently, no neutrinos leave the nucleus, and the total decay energy (Q-value) is transferred to the two electrons ($e^-$) leaving the nucleus. 
\begin{equation}
    (Z,A) \, \rightarrow \, (Z+2,A) + 2 e^- \,.
    \label{eq:0nubb}
\end{equation}
This signature can be used to detect that decay: 
The energy of the two electrons is measured.
As the 2$\nu\beta\beta$-decay occurs, the energy spectrum is continuously distributed. 
If the $0\nu\beta\beta$-decay happens, additionally a line at the Q-value of the decay can be found. 
In an actual experiment, this is difficult to measure, because the background has to be low enough that the rare $0\nu\beta\beta$-decay can be detected, amongst other things. 
A Feynman diagram of the $0\nu\beta\beta$-decay and the principal shapes of the electron spectra are shown in Figure \ref{fig_feynman_dbd_standard}. 
An energy resolution of \SI{1}{\%} FWHM (\textbf{f}ull \textbf{w}idth at \textbf{h}alf \textbf{m}aximum) at \SI{2.8}{MeV} was assumed, and half-lives of \SI{2.88E19}{yr} \cite{Barabash:2015eza} and \SI{2E26}{yr} for the 2$\nu\beta\beta$-decay and $0\nu\beta\beta$-decay. 

The expected half-life of the $0\nu\beta\beta$-decay ($T^{0\nu}_{1/2}$) is directly related to the effective Majorana neutrino mass ($\langle m_{\nu_e} \rangle$) \cite{zuber_04}: 
\begin{equation}
    \left( T^{0\nu}_{1/2} \right)^{-1} = G^{0\nu}(Q,Z) \, \left|M^{0\nu}_{GT} - M^{0\nu}_F \right|^2 \left( \frac{\langle m_{\nu_e} \rangle}{m_e} \right)^2 \,. 
    \label{eq_half_live_neutrino_mass}
\end{equation}
$G^{0\nu}(Q,Z)$ is the phase space integral, which can be calculated exactly. It scales with $Q^5$, so large Q-values result in lower half-lives.
A main source of uncertainty arises from the numerical calculation of the NME (\textbf{n}uclear \textbf{m}atrix \textbf{e}lements) $M^{0\nu}_{\text{Gamow\ Teller}}$ and $M^{0\nu}_{\text{Fermi}}$. 
The values for them differ by more than a factor of two, depending on the underlying model and isotope \cite{Dev:2013vxa}.

The $\langle m_{\nu_e} \rangle$ (effective Majorana neutrino mass) \cite{zuber_04} is given by
\begin{equation}
    \langle m_{\nu_e} \rangle = \left| \sum\limits_i |U_{li}|^2 \cdot \eta_i \cdot m_i \right| \, .
 \label{eq_effective_neutrino_majorana_mass}
\end{equation}
The factor $\eta_i = \pm1$ comes from the CP (\textbf{c}harge conjugation and \textbf{p}arity transforma-tion)-phases, $m_i$ are the corresponding mass eigenvalues. 
The $U_{li}$ are the  elements of the PMNS (\textbf{P}ontecorvo-\textbf{M}aki-\textbf{N}akagava-\textbf{S}akata) matrix, the leptonic mixing matrix (\ref{eq_PMNS_matrix}), analog to the CKM (\textbf{C}abibbo-\textbf{K}obayashi-\textbf{M}askawa) matrix in the quark sector:
\begin{eqnarray}
 \begin{pmatrix} \nu_{e} \\ \nu_{\mu} \\ \nu_{\tau} \end{pmatrix} = 
 \begin{pmatrix}
  	U_{e1} 		& U_{e2} 	& U_{e3} \\
  	U_{\mu1}	& U_{\mu2}	& U_{\mu3} \\
	U_{\tau1}	& U_{\tau2}	& U_{\tau3}
 \end{pmatrix}
 \begin{pmatrix} \nu_1 \\ \nu_2 \\ \nu_3 \end{pmatrix} \, . 
\label{eq_PMNS_matrix}
\end{eqnarray}

One main reason for detecting the $0\nu\beta\beta$-decay apart from neutrino physics is the search for such a lepton number violating process \cite{Pas:2015eia}:\\
The lepton and baryon numbers are only accidentally conserved global conservation laws. Extended theories imply a violation of those quantities.
The baryon number cannot be conserved, as the universe contains more matter than antimatter. Consequently, lepton number violation is somehow expected.

Modern GUT (\textbf{g}rand \textbf{u}nified \textbf{t}heories) and SUSY (\textbf{su}per\textbf{sy}mmetry) models predict the violation of the lepton number conservation at a small degree. 
This quantity might only be conserved at low energies.
Furthermore, GUTs connect the lepton and baryon numbers.

All theories beyond the SM (\textbf{s}tandard \textbf{m}odel of particle physics) which violate the lepton number lead to $0\nu\beta\beta$-decays. 
Nearly all mechanisms generating and suppressing neutrino masses lead to Majorana neutrinos and $0\nu\beta\beta$-decay \cite{Pas:2015eia}.

The so-called see-saw mechanism is a way for neutrinos to acquire mass, which is seen more natural from theorist's point of view.
It postulates that neutrinos have a small Majorana mass \cite{GomezCadenas:2011it}.

Another issue is to clarify the nature of neutrinos as either Dirac- or Majorana-particles.

\begin{SCfigure}[][h!]
 \centering
  \includegraphics[width=0.59\textwidth]{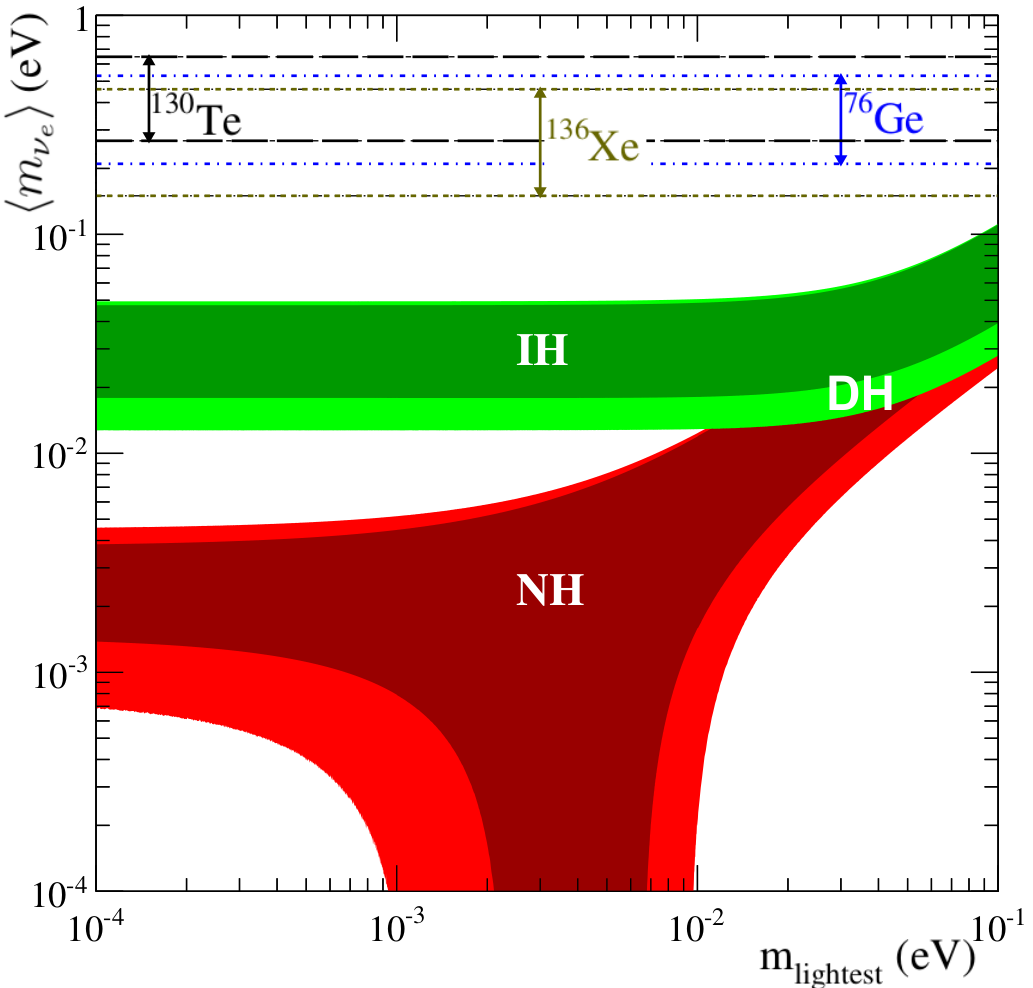}
  \caption[Allowed regions for $\langle m_{\nu_e} \rangle$ in dependence of the lightest neutrino mass]
          {Allowed regions for $\langle m_{\nu_e} \rangle$ in dependence of the lightest neutrino mass. 
          The darker bands are obtained from best-fit neutrino oscillation parameters, the lighter bands incorporate their \SI{3}{\sigma} uncertainty. 
          The spread stems from the uncertainty of the NME. 
          As will be discussed in section \ref{sec_latest_results}, the horizontal constraints arise from experimental results of $0\nu\beta\beta$-decays, the degenerated region more to the right is disfavored by cosmology. Adapted from \cite{Alduino:2016zrl}.}
  \label{fig_neutrino_hierarchy}
\end{SCfigure}

$0\nu\beta\beta$-decays can also determine the absolute scale of the neutrino mass, which is still unknown. 
Neutrino oscillation experiments can only reveal information about the squared mass differences between the neutrinos ($\Delta m_{ij}^2$).
Furthermore, the hierarchy type of neutrino masses can be investigated: NH (\textbf{n}ormal neutrino mass \textbf{h}ierarchy), IH (\textbf{i}nverted neutrino mass \textbf{h}ierarchy) or DH (\textbf{d}egenerated neutrino mass \textbf{h}ierarchy):
In NH, $m_1 < m_2 < m_3$, while this is changed in the inverted case to $m_3 < m_1 < m_2$. The IH splits up due to a cancellation of terms if $m_1 \le \sqrt{\Delta m_{21}^2}$ \cite{Schwingenheuer:2012zs}. The area where IH and NH overlap is called DH, where the lightest neutrino mass is larger
than all squared mass differences. Experimental results pose constraints on certain areas.
See Figure \ref{fig_neutrino_hierarchy} for a graphical representation of the allowed mass regions for $\langle m_{\nu_e} \rangle$.

\newpage
A $0\nu\beta\beta$-decay in the standard interpretation is seen as a long-ranging light neutrino interacting in a SM V-A (\textbf{v}ector minus \textbf{a}xial vector) manner.
Other models include short-range mechanisms by heavy particles, like Left-Right Symmetry, R-parity violating SUSY, leptoquarks or extra dimensions \cite{Pas:2015eia}.

On the other hand, even if $0\nu\beta\beta$-decays are measured, fundamental questions could remain. The possible maximal Majorana mass could be in the range of \SI{e-24}{eV}, the rest arising from SUSY-particles or heavy Majorana neutrinos. 
Consequently, even if $0\nu\beta\beta$-decays are detected, neutrinos could be effective Dirac particles \cite{Schwingenheuer:2012zs}.

%%%%%%%%%%%%%%%%%%%%%%%%%%%%%%%%%%%%%%%%%%%%%%%%%%%%%%%%%%%%%%%%%%%%%%%%%%%%%%%%%%%%%%%%%%%%%%%%%%%%%%%%%%%%%%%%%%%%%%%%%%%%
%%%%%%%%%%%%%%%%%%%%%%%%%%%%%%%%%%%%%%%%%%%%%%%%%%%%%%%%%%%%%%%%%%%%%%%%%%%%%%%%%%%%%%%%%%%%%%%%%%%%%%%%%%%%%%%%%%%%%%%%%%%%
\section[Investigation of beta-decays and particle physics]{Investigation of beta-decays and particle\\physics}
This historical overview shows how progress in particle physics motivated and encouraged the investigation of (double) beta-decays, and how this led to general advances in physics, resulting in the SM and potentially beyond \cite{Barabash:2011mf, Vergados:2012xy}.\\

Pauli postulated the existence of neutrinos in December 1930 \cite{Pauli_1930} to solve the problem of energy and momentum conservation in beta-decays.
This happened in a letter to ''The group of radioactives`` in T\"ubingen.  
Pauli used the name neutron, 14 months before Chadwick's discovery of the particle today known as neutron \cite{Amaldi:1984as}.
The postulate was in contradiction to Bohr, who argued that the energy in beta-decays is only conserved statistically.

In June of 1931, Pauli suggested to decide between the two concepts by experimental proof \cite{Amaldi:1984as}. It should be measured, if the energy of the electrons in beta-decays has a clear upper limit, or shows a Poisson distribution with decreasing intensity. 
The first possibility would prove his neutrino hypothesis.
This method is being invested still today, for example by experiments like KATRIN (\textbf{Ka}rlsruhe \textbf{tri}tium \textbf{N}eutrino Experiment) \cite{KATRIN}.

The name neutrino is an incorrect diminutive of the Italian word neutronino (little neutron) to distinguish it from the neutron. It was promoted internationally by Fermi, although the name itself came from his collaborator Amaldi \cite{Amaldi:1984as}.

A theory of $\beta$-decays as a new type of interaction of four particles (''fermions``) was proposed by Fermi in 1934 \cite{Fermi:1934hr}.

After an idea of Wigner, Goeppert-Mayer derived an expression for the 2$\nu\beta\beta$-decay-rate, and estimated a half-life of \SI{E17}{\year} in 1935 \cite{GoeppertMayer:1935qp}.\\

Majorana stated in 1937, that the theory of $\beta$-decay is unchanged if the neutrino ($\nu$) and antineutrino ($\overline{\nu}$) were indistinguishable \cite{Majorana:1937vz}.
This possibility is referred to as Majorana-particle.
Racah \cite{Racah:1937qq} proposed to test the Majorana hypothesis with neutrinos in a chain reaction
\begin{equation} 
  \begin{split}
              (A,Z) &\rightarrow (A, Z+1) + e^- + \overline{\nu} \\
      \nu + (A',Z') &\rightarrow (A',Z'+1) + e^-\,,
  \end{split}
\end{equation}
where the antineutrino of the first reaction triggers the second reaction as a neutrino. 
This is only possible, if neutrinos are Majorana particles ($\nu \equiv \overline{\nu} \equiv \nu_{\text{Majorana}}$).

The possibility of a $0\nu\beta\beta$-decay was proposed in 1939 by Furry \cite{Furry:1939qr} as a Racah-chain with $(A,Z+1) \equiv (A',Z')$ as two decays occurring simultaneously:
A decay to an intermediate state and virtual antineutrino, which is absorbed by the nucleus as a neutrino in the second decay, as ($\overline{\nu} \equiv \nu \equiv \nu_{\text{Majorana}}$):
\begin{equation}
 (A,Z) \rightarrow (A, Z+2) + 2e^-\,. 
\end{equation}

The chirality suppression of the $0\nu\beta\beta$-decay was not known at that time \cite{Vergados:2012xy}. 
Because of the better phase-space in the Majorana-case, the half-life estimations of $\sim$ \SIrange{E15}{E16}{\year} for the $0\nu\beta\beta$-decay were smaller than $\sim$ \SIrange{E21}{E22}{\year} for the 2$\nu\beta\beta$-decay.
 
In 1949, Fireman claimed the discovery of a 2$\nu\beta\beta$-decay of \isotope[124]Sn in laboratory experiments with $T_{1/2}$ (half-life) = \SIrange[range-phrase = --,range-units = brackets,fixed-exponent = 15,scientific-notation = fixed]{4e15}{9e15}{\year} \cite{Fireman:1949qs}, but disclaimed it later \cite{Fireman:1952qt}.

Geochemical experiments had a much higher sensitivity than counter experiments in laboratories. 
First geochemical 2$\nu\beta\beta$-decay-measurements were done in 1950 for \isotope[130]Te with $T_{1/2}$ = \SI{1.4E21}{\year} \cite{Inghram:1950qv}. These needed some 20 years to become widely accepted \cite{Barabash:2011mf}.

Davis tried to measure the inverse electron capture process in 1955 \cite{Davis:1955bi}:
\begin{equation}
  \nu_e + \isotope[37]Cl \rightarrow \isotope[37]Ar + e^-\,.
  \label{eq_davis_reaction}
\end{equation}
The experiment was done at a reactor producing antineutrinos, while the reaction obviously required neutrinos. 
It is only possible for Majorana neutrinos. 
A zero result was found. This was interpreted as a proof of neutrinos being Dirac particles.
As a consequence, the lepton number conservation was introduced to separate neutrinos from antineutrinos:
$2\nu\beta\beta$-decays are allowed, while $0\nu\beta\beta$-decays are forbidden.

In 1956, Reines and Cowan published the first direct reactor antineutrino measurement using the inverse beta-decay in water \cite{Reines:1956rs, Cowan:1992xc}:
\begin{equation}
  \overline{\nu_e} + p \rightarrow n + e^+\,.
\end{equation}

The (space) parity violation in weak interactions was theoretically formulated by Lee and Yang in 1956 \cite{Lee:1956qn}, contradictory to the common interpretation at that time.
Its experimental verification was done by two crucial experiments shortly later: Wu \cite{Wu:1957my} and Goldhaber \cite{Goldhaber:1958nb}.\\
Wu et al. measured the violation of parity in the angular distribution at the beta-decay of \isotope[60]Co. 
Goldhaber et al. used the EC (\textbf{e}lectron \textbf{c}apture) and subsequent de-excitation of \isotope[152]Eu to determine the neutrino-helicity to \num{-1}.

In 1958, the V-A theory of the weak interaction was introduced, stating that parity is violated maximally. This outdated Fermi's theory.
The V-A theory is realized in the lepton sector by using two-component massless neutrinos, which was proposed in 1957 by Landau, Lee, Yang and Salam.
This idea was already proposed by Weyl in 1929, but was rejected by Pauli in 1933 - because it violated parity \cite{Vergados:2012xy}.
Consequently, neutrinos are considered as left-handed, antineutrinos as right-handed.
Furthermore, the question of neutrinos being Dirac or Majorana particles remained unanswered.

Zero-results for $0\nu\beta\beta$-decay do not mean that neither neutrinos are Dirac-particles, nor the conservation of lepton number.

In 1966, Mateosian and Goldhaber reached a sensitivity in counter experiments above \SI{E20}{\year} for the first time, and introduced the detector = source principle \cite{derMateosian:1966qz}. 
This is widely used today, amongst others also at the COBRA experiment.

An important experiment whose conclusions increased the search for $0\nu\beta\beta$-decays started in the late 1960s. The Homestake Experiment by Davis et al. tried to measure the solar neutrino flux using the Chlorine-Argon reaction
\begin{equation}
  \nu_e + \isotope[37]Cl \rightarrow \isotope[37]Ar + e^-\,.
  \label{eq_davis_reaction_antineutrino}
\end{equation}
The measured flux of solar neutrinos relative to their expected number showed a deficit, only about one third of the expected neutrinos were detected \cite{Bahcall:1976zz, 0004-637X-496-1-505}.
If the experimental setup and its results were considered to be correct, two main explanations are possible:
First, the expected solar neutrino flux and hence the standard solar model could be incorrect. 
The second possibility is the disappearance of neutrinos while traveling to the earth by neutrino oscillation. 
It took until 2002 for finally proving the latter possibility and solving this so-called solar neutrino problem.

In the beginning of the 1980s, the theoretical pioneer work of Kotani et al. \cite{Doi:1980ze, Doi:1985dx} raised new attention to $0\nu\beta\beta$-decay.

In 1982 Schechter and Valle stated in the so called Black-Box theorem, that the observation of $0\nu\beta\beta$-decay ensures that neutrinos have a
Majorana component \cite{Schechter:1981bd}.
This was an important theoretical issue to increase the experimental searches for $0\nu\beta\beta$-decays.

An agreement between experimental and theoretical results for double beta-decay rates was achieved for the first time in 1986. In \cite{Vogel:1986nj}, QRPA (\textbf{Q}uasi-particle \textbf{r}andom \textbf{p}hase \textbf{a}pproximation) nuclear structure methods are used to calculate the NME, which are an important input for Equation \ref{eq_half_live_neutrino_mass}.

The first new 2$\nu\beta\beta$-decay observations in laboratory measurements were done in 1987. A time-projection chamber was used to determine the half-life of \isotope[82]Se to \SI{1.1E20}{\year}  \cite{Elliott:1987kp}.

Important experimental results concerning the solar neutrino problem were published in June 1998: Neutrino flavor oscillation was concluded by Super-Kamiokande (\textbf{Super Kamioka N}ucleon \textbf{d}ecay \textbf{e}xperiment) measuring the disappearance of atmospheric neutrinos \cite{Fukuda:1998mi}.

In 2000, the first measurement of tau-neutrinos were published \cite{Kodama:2000mp}.

The final proof of neutrino oscillation was achieved with SNO (\textbf{S}udbury \textbf{N}eutrino \textbf{O}bservatory) in 2001 and 2002. The flux of solar neutrinos was measured via charged current reactions, the total neutrino flux using neutral current reactions and elastic scattering. The experiment used deuterium (d) in a heavy water detector \cite{Ahmad:2001an, Ahmad:2002ka}:
\begin{align}
 \text{charged current: } \nu_e + d &\rightarrow p + p + e^-\\
 \text{neutral current: } \nu_x + d &\rightarrow \nu_x + p + n\\
 \text{elastic scattering: } \nu_x + e^- &\rightarrow \nu_x + e^-
\end{align}
The neutral current reaction can be triggered by any particle above the energy of \SI{2.2}{MeV} to dissociate the deuterium (but only neutrinos are present in the detector). 
Elastic scattering is sensitive to all neutrino flavors, but the sensitivity to electron neutrinos is larger by a factor of six. 
The SNO results confirmed the total neutrino flux predicted by solar model calculations and the disappearance of electron neutrinos to muon- and tau-neutrinos.
Results of the MSW-effect (\textbf{M}ikheyev-\textbf{S}mirnov-\textbf{W}olfenstein effect) are included. This states, that a stronger oscillation of electron neutrinos happens in the sun due to matter effects. 
The large electron density in the sun affects electron-neutrinos via coherent forward-scattering more than other neutrinos.

Neutrino oscillations have been measured at atmospheric, solar, reactor and accelerator neutrinos, i.a. by Super-Kamiokande \cite{Fukuda:2001nj}, SNO \cite{Ahmad:2002ka} and KamLAND (\textbf{Kam}ioka \textbf{l}iquid scintillator \textbf{an}tineutrino \textbf{d}etector) \cite{Eguchi:2002dm}.
The neutrino flavor oscillation results confirmed theoretical predictions of Pontecorvo \cite{Pontecorvo:1957cp}, and solved the solar neutrino problem of the  Homestake-Experiment.

Furthermore, these results proved without any doubt, that neutrinos do have mass.\\

Concluding, one can state that the theoretical arguments and the proof of neutrino oscillations strongly increased the searches for $0\nu\beta\beta$-decay.%\\

%%%%%%%%%%%%%%%%%%%%%%%%%%%%%%%%%%%%%%%%%%%%%%%%%%%%%%%%%%%%%%%%%%%%%%%%%%%%%%%%%%%%%%%%%%%%%%%%%%%%%%%%%%%%%%%%%%%%%%%%%%%%
%%%%%%%%%%%%%%%%%%%%%%%%%%%%%%%%%%%%%%%%%%%%%%%%%%%%%%%%%%%%%%%%%%%%%%%%%%%%%%%%%%%%%%%%%%%%%%%%%%%%%%%%%%%%%%%%%%%%%%%%%%%%
\section{State of the art at double beta-decay experiments}
\label{sec_0nbb_experiments_state_of_the_art}

The state of the art in experimental results concerning double beta-decays is shown here briefly.\\

$2\nu\beta\beta$-decays were measured directly for eleven isotopes:
\isotope[48]Ca, \isotope[76]Ge, \isotope[82]Se, \isotope[96]Zr, \isotope[100]Mo, \isotope[116]Cd, \isotope[128]Te, \isotope[130]Te, \isotope[136]Xe, \isotope[150]Nd, \isotope[238]U, as well as the $2\nu EC EC$ (neutrino accompanied double electron-capture) in \isotope[130]Ba.
The half-lives are between \SI{7E18}{\year} and \SI{2E24}{\year} \cite{Schwingenheuer:2012zs}.\\

Several experimental concepts for detecting $0\nu\beta\beta$-decay are used:
\begin{itemize}
   \item A setup known from particle-physics detectors comprising a source, tracking devices and a calorimeter: SuperNEMO (\textbf{Super} \textbf{n}eutrino \textbf{E}ttore \textbf{M}ajorana \textbf{o}bservatory).
	 An advantage is that a wide choice of source isotopes is possible, and a low background can be reached due to the topological event reconstruction. However, the efficiency and energy resolution are not that good.  
   \item Large tanks filled with source material surrounded by light detection detectors: SNO+ (\textbf{S}udbury \textbf{N}eutrino \textbf{O}bservatory Plus), KamLAND-Zen (\textbf{Kam}ioka \textbf{L}iquid Scintillator \textbf{An}tineutrino \textbf{D}etector \textbf{ze}ro-\textbf{n}eutrino double-beta decay) or EXO (\textbf{E}n\-riched \textbf{x}enon \textbf{e}xperiment).\\
         These types acquire a high mass, but suffer from a low energy resolution. 
   \item Solid state detector experiments in the source = detector mode like GERDA (\textbf{Ge}rmanium \textbf{d}etector \textbf{a}rray), CUORE (\textbf{C}ryogenic \textbf{u}nderground \textbf{o}bservatory for \textbf{r}are \textbf{e}vents) or COBRA.\\
         Advantages are a high detection efficiency and a good energy resolution. However, it is not that easy to acquire large source masses.
\end{itemize}

Furthermore, several types of neutrino masses have to be distinguished:\\
Classical Kurie-plot experiments like KATRIN measure the incoherent sum
\begin{equation}
 m_\beta = \sqrt{\sum |U_{li}|^2 \, m_i^2} \,. 
\end{equation}

Cosmology experiment are sensitive to the sum of all neutrino masses ($\sum m_i$).

$0\nu\beta\beta$-decay experiments in the standard interpretation measure the $\langle m_{\nu_e} \rangle$, given in Equation \ref{eq_effective_neutrino_majorana_mass} on page 12.%\pageref{eq_effective_neutrino_majorana_mass}.

%%%%%%%%%%%%%%%%%%%%%%%%%%%%%%%%%%%%%%%%%%%%%%%%%%%%%%%%%%%%%%%%%%%%%%%%%%%%%%%%%%%%%%%%%%%%%%%%%%%%%%%%%%%%%%%%%%%%%%%%%%%%
\subsubsection{Latest experimental results}
\label{sec_latest_results}
A controversial claim for an observed $0\nu\beta\beta$-decay was made by Klapdor-Kleingrot\-haus \cite{KlapdorKleingrothaus:2006ff}: A half-life of
$\left(2.23^{+0.44}_{-0.31}\right)\cdot\SI{e25}{\year}$ in \isotope[76]Ge was reported with a statistical significance of about \SI{6}{\sigma}.

This result is excluded with \SI{99}{\%} probability by the successive GERDA experiment \cite{Agostini:2015fga}, which sets a lower half-life limit to 
\begin{equation}
 T_{1/2}^{0\nu} > \SI{2.1E25}{\year} \ \ (\text{\SI{90}{\%} C.L.})
\end{equation}
at a background of \SI{0.01}{\cpkky}.

Similar results were found in \isotope[136]Xe by KamLAND-Zen \cite{Asakura:2014lma} and EXO \cite{Auger:2012ar}:
\begin{equation}
 \begin{split}
  T_{1/2}^{0\nu} > \SI{2.6E25}{\year} \\
  T_{1/2}^{0\nu} > \SI{1.6E25}{\year}
 \end{split}
\end{equation}

The CUORE group (and predecessor) published a limit for \isotope[130]Te \cite{Alduino:2016zrl}:
\begin{equation}
 T_{1/2}^{0\nu} > \SI{4E24}{\year} \ \ (\text{\SI{95}{\%} C.L.})\,.
\end{equation}

These experiments give roughly the same number on the effective Majorana neutrino mass $\langle m_{\nu_e} \rangle$, depending especially on the NME:
\begin{equation}
\langle m_{\nu_e} \rangle  \lesssim \SI{0.3}{eV}
\end{equation}

Measurements of the cosmic microwave background by the Planck Surveyor satellite mission are published in \cite{1367-2630-16-6-065002}, setting a limit on the sum of neutrino masses to
\begin{equation}
  \sum_i m_i < \SI{0.23}{eV} \ \ (\text{\SI{95}{\%} C.L.})\,.
\end{equation}

All these measured limits are compiled graphically in Figure \ref{fig_neutrino_hierarchy} on page 13. %\pageref{fig_neutrino_hierarchy}.\\

New results were published for \isotope[116]Cd \cite{Danevich:2016eot}:\\
A scintillating crystal of \SI{1.162}{kg} of \ce{CdWO4}, enriched in \isotope[116]Cd to \SI{82}{\%}, was used in the source = detector mode.
A half-life for the 2$\nu\beta\beta$-decay is given (\SI{90}{\%} C.L.):
\begin{equation}
   T_{1/2}^{2\nu} = \SI{2.62(14)E19}{yr}\,,
\end{equation}
and limit on the $0\nu\beta\beta$-decay of
\begin{equation}
   T_{1/2}^{0\nu} \geq \SI{1.9E23}{yr}\,.
\end{equation}\\

Concluding, one can state that $2\nu\beta\beta$-decays are measured, while no convincing proof for the existence of $0\nu\beta\beta$-decays has been found so far.

\chapter{CdZnTe detector technology for the COBRA experiment}
\label{sec_COBRA}

This chapter discusses the semiconductor detector material CdZnTe ($\text{\textbf{C}a\textbf{d}mium}_{0.45}-\text{\textbf{Z}i\textbf{n}c}_{0.05}-\text{\textbf{Te}lluride}_{0.5}$). 
First, some general properties of CdZnTe for a semiconductor detector material are discussed. 
Some of the properties necessitate a special electrode configuration, called CPG (\textbf{c}o\textbf{p}lanar-\textbf{g}rid) technology, which is discussed in the following.
Then, the basic measurement and analysis principles for CPG detectors are explained briefly.

%%%%%%%%%%%%%%%%%%%%%%%%%%%%%%%%%%%%%%%%%%%%%%%%%%%%%%%%%%%%%%%%%%%%%%%%%%%%%%%%%%%%%%%%%%%%%%%%%%%%%%%%%%%%%%%%%%%%%%%%%%%%
%%%%%%%%%%%%%%%%%%%%%%%%%%%%%%%%%%%%%%%%%%%%%%%%%%%%%%%%%%%%%%%%%%%%%%%%%%%%%%%%%%%%%%%%%%%%%%%%%%%%%%%%%%%%%%%%%%%%%%%%%%%%
\section{CdZnTe semiconductor detectors}
The COBRA experiment uses CdZnTe semiconductor detectors as source and detector material. It contains nine isotopes, that can undergo double beta-decays in all possible modes, shown in \autoref{tab_double_beta_isotopes}. 
The main focus is on \isotope[116]Cd due to its high Q-value of \SI{2813}{keV}, which is above all prominently occurring natural gamma lines.

\begin{SCtable}
  \centering
  \begin{small}
    \begin{tabular}{|c|c|S|S|}
    \hline
    isotope 	        & decay mode	                                      & {\makecell{natural abun-\\dance $a$ [\%]} } & {Q-value [keV]}  \\ \hline 
    $\isotope[106]{Cd}$ & \makecell{$\beta^+\beta^+,\beta^+/EC$,\\ EC/EC} & \num{1.245(22)} & \num{2775.39(10)} \\
    $\isotope[108]{Cd}$ & $ EC/EC$				      & \num{0.888(11)} & \num{271.8(8)}	  \\
    $\isotope[114]{Cd}$ & $\beta^-\beta^-$			       	      & \num{28.754(81)}& \num{542.5(9)}	  \\
    $\isotope[116]{Cd}$ & $\beta^-\beta^-$				      & \num{7.512(54)} & \num{2813.44(13)} \\ \hline
    $\isotope[64]{Zn}$  & $\beta^+/EC,EC/EC$		      & \num{49.17(75)} & \num{1094.7(7)}   \\
    $\isotope[70]{Zn}$  & $\beta^-\beta^-$				      & \num{0.61(10)}  & \num{997.1(21)}   \\ \hline
    $\isotope[120]{Te}$ & $\beta^+/EC,EC/EC$		      & \num{0.09(1)}   & \num{1730(3)}     \\
    $\isotope[128]{Te}$ & $\beta^-\beta^-$				      & \num{31.74(8)}  & \num{866.5(9)}	  \\        
    $\isotope[130]{Te}$ & $\beta^-\beta^-$				      & \num{34.08(62)} & \num{2527.51(1)}  \\ \hline 
  \end{tabular}
  \end{small}
  \caption[Decay modes, natural abundance and Q-values of double beta-decay isotopes of CdZnTe]
          {Decay modes, natural abundance and Q-values of double beta-decay isotopes of CdZnTe. Values for $a$ taken from \cite{list_isotopic_abundance}, $Q$-values from \cite{1674-1137-36-12-003}.}
  \label{tab_double_beta_isotopes}
\end{SCtable}

CdZnTe is a commercially available room-temperature semiconductor detector with a high Z (proton number) of approximately 49 and a good energy resolution.
The average energy resolution of the COBRA demonstrator is \SI{1.3}{\%} FWHM at \SI{2.6}{MeV}, as discussed in \autoref{sec_cobra_demonstrator}.

The sensitivity of an experiment to detect a half-life can be approximated to
\begin{equation}
 T_{1/2}^{\text{exp}} \sim a\cdot \epsilon \sqrt{\frac{M\cdot t}{\Delta E\cdot B}}\,.
 \label{eq_half_life_sensitivity}
\end{equation}
where $M$ is the source mass and $t$ the measurement live time. The energy resolution $\Delta E$ and the background $B$ are obtained at the ROI (\textbf{r}egion \textbf{o}f \textbf{i}nterest) of the decay under study \cite{zuber_04}. 
These appear only in a square-root dependence, whereas the natural abundance of the isotope under study $a$ and the detection efficiency $\epsilon$ act linearly. This illustrates the advantage of the ``source = detector'' principle, which ensures a high $\epsilon$. \\

The basic properties of CdZnTe are listed in \autoref{tab_properties_CdZnTe}.
\begin{SCtable}
  \centering
      \begin{tabular}{|c|c|c|}
	\hline
        property 				& Cd$_{0.45}$Zn$_{0.05}$Te$_{0.5}$	& Ge    \\ \hline %\hline
	atomic numbers 				& 48, 30, 52 				& 32 	\\ \hline
	average atomic number			& 49.1					& 32	\\ \hline		
	density $\rho$ [g/cm$^3$] 		& 5.78 					& 5.33 	\\ \hline
	band gap $E_g$ [eV]			& 1.57 					& 0.67 	\\ \hline
	pair creation energy $E_{pair}$ [eV] 	& 4.64 					& 2.95 \\ \hline
	resistivity $\rho$ [$\Omega$\,cm] 	& $3\times 10^{10}$ 			& 50 	\\ \hline
	electron mobility $\mu_e$ [cm$^2$/Vs]   & 1000 					& 3900  \\ \hline
	electron lifetime $\tau_e$ [s] 		& $3\times 10^{-6}$ 			& $> 10^{-3}$ \\ \hline
	hole mobility $\mu_h$ [cm$^2$/Vs] 	& 50 - 80 				& 1900 \\ \hline
	hole lifetime $\tau_h$ [s] 		& $10^{-6}$ 				& $10^{-3}$ \\ \hline
	$(\mu \cdot \tau)_e$ [cm$^2$/V] 	& $(3-5) \times 10^{-3}$ 		& $> 1$ \\ \hline
	$(\mu \cdot \tau)_h$ [cm$^2$/V] 	& $5 \times 10^{-5}$ 			& $> 1$ \\ \hline
      \end{tabular}
  \caption[Properties of CdZnTe and Germanium]
          {Properties of CdZnTe and Germanium at room temperature. Important for the principles of COBRA are the higher pair creation energy and resistivity, and especially the low $\mu\tau$ products, in particular for the holes. Taken from \cite{eV_detector_material}.}
  \label{tab_properties_CdZnTe}
\end{SCtable}

The product of mobility and lifetime ($\mu\tau$) of the charge carrier is low, and especially two orders of magnitude lower for the holes than for electrons. Due to this large spread of the two of them, a simple detector with a planar anode and cathode is not possible, the signals would be depth-dependent. As a solution to this, the CPG technology was introduced \cite{cpg94}, which is explained in the following.
A calculation of the $\mu_e$ (electron mobility) using laboratory measurements is done in \autoref{sec_drift_time}.

%%%%%%%%%%%%%%%%%%%%%%%%%%%%%%%%%%%%%%%%%%%%%%%%%%%%%%%%%%%%%%%%%%%%%%%%%%%%%%%%%%%%%%%%%%%%%%%%%%%%%%%%%%%%%%%%%%%%%%%%%%%%
%%%%%%%%%%%%%%%%%%%%%%%%%%%%%%%%%%%%%%%%%%%%%%%%%%%%%%%%%%%%%%%%%%%%%%%%%%%%%%%%%%%%%%%%%%%%%%%%%%%%%%%%%%%%%%%%%%%%%%%%%%%%
\section{Coplanar-grid principles}
\label{sec_cpg_principles}
\begin{SCfigure}
 \centering
  \includegraphics[width=0.5\textwidth]{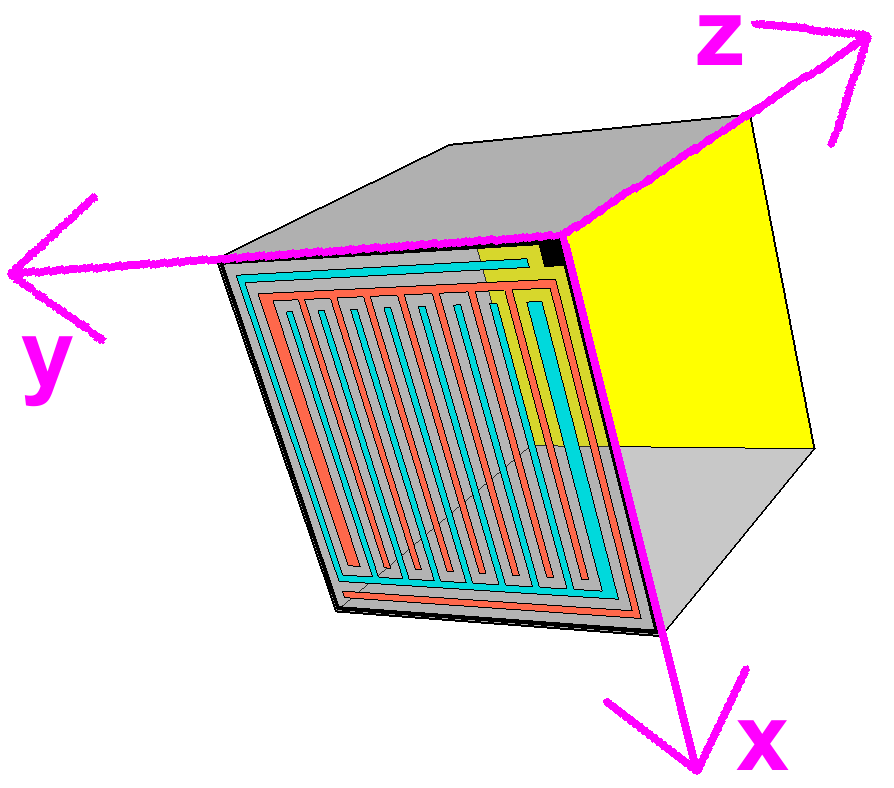}
  \caption[Geometry of a CPG Detector]
	  {Geometry of a standard COBRA CPG detector. The two anodes are marked in red and blue, surrounded by a GR (black). The cathode is the yellow area on the opposite side of the anodes. The magenta arrows indicate the detector axes. Adapted from \cite{Fritts:2014qdf}}
  \label{fig_cpg_detector_geometry}
\end{SCfigure}

The different $\mu\tau$ product of electrons and holes in CdZnTe is the reason that a simple readout of electrons and holes at the anode and cathode is not useful.
The resulting signal would be depending on the interaction depth of the incident particle, as especially the holes are not collected completely, due to their lower mobility and shorter lifetime. 
A technique to compensate for this is to focus only on detecting electrons in a single-polarity charge detection \cite{cpg94}, called CPG.
In a CPG, there are two interleaved comb-shaped anode structures, which are set on a slightly different electric potential. 
These are surrounded by a GR (\textbf{g}uard \textbf{r}ing), which increases the performance of the detector due to a more balanced weighting potential $\varphi_0$ (explained at \autoref{eq_Shockly-Ramo}).
At most applications, the GR is on a floating potential as it is not connected to any electronics. Laboratory measurements using an instrumented GR are discussed in \autoref{sec_guard_ring}.
The cathode is metalized completely as a planar electrode. 
It is set on a negative voltage, called BV (\textbf{b}ias \textbf{v}oltage), of typically \SI{-1}{kV}. The anode, that eventually collects the electrons is called CA (\textbf{c}ollecting \textbf{a}node),
the other NCA (\textbf{n}on-\textbf{c}ollecting \textbf{a}node). Their biases are zero and typically \SI{-50}{V}.
A  sketch of a CPG detector is shown in \autoref{fig_cpg_detector_geometry}.\\

The Shockley-Ramo theorem \cite{1938JAP.....9..635S, 1686997} and its application to (single-polarity charge sensing) semiconductor detectors \cite{He2001250} is an easy method to calculate the detector properties.
It connects a moving charge cloud $q(x)$, generated by an energy deposition of an incident particle in a detector at a position $x$, to the induced charges measured at the electrodes $Q_E$.
The resulting output signal is calculated by means of a weighting potential $\varphi_0$. 
Applying the Shockley-Ramo theorem relies on the following conditions: the selected electrode is set to unity potential, all others to zero. Furthermore, no space charges shall be in the considered detector volume. 
An advantage is that only one weighting field has to be calculated, independent of the moving charges $q(x)$. 

The Shockley-Ramo theorem states, that the measured charge is the product of the moving charge cloud $q(x)$ and the weighting potential $\varphi_0$:
\begin{equation}
 Q_E = -q(x) \cdot \varphi_0(x)
 \label{eq_Shockly-Ramo}
\end{equation}

\begin{figure}
 \centering
  \includegraphics[width=0.99\textwidth]{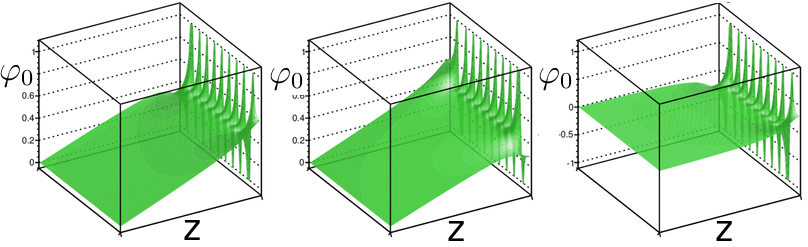}
  \captionsetup{width=0.9\textwidth} 
  \caption[Weighting potentials in a CPG detector]
	  {From left to right: Weighting potential for CA and NCA anodes, and the difference between them. The cathode is left at each plot, the anodes right. Adapted from \cite{Fritts:2014qdf}.}
  \label{fig_weighing_potentials}
\end{figure}

The calculated weighting potentials for the standard COBRA CPG detectors ($1\times1\times1\,$cm$^3$ in volume) are shown in \autoref{fig_weighing_potentials}.\\
The weighting potentials are the same for NCA and CA throughout the bulk of the detector.
Discrepancies arise only in the vicinity of the anodes. 
Consequently, the difference weighting potential is zero throughout most of the detector, except in the near-anode region.
The charge induction occurs only there, as the induced charge is proportional to the  weighting potential $\varphi_0$ (\autoref{eq_Shockly-Ramo}). 
The charge induction is now independent of the interaction depth and the  hole-component.
The deposited energy $ E_{\text{deposited}}$ can hence be calculated as the difference of the amplitudes $A$ of the CA and NCA signals.
The introduction of a weighting factor $w$ to correct for electron trapping greatly improves the detector performance:
\begin{equation}
 E_{\text{deposited}} \propto A_{CA} - w \cdot A_{NCA}
 \label{eq_energy_calculation}
\end{equation}

Using the two independent anode signals, one can calculate the interaction depth in z-direction, i.e. the location of the event parallel to the anodes and the cathode.
A simple formula uses the ratio of the amplitudes of the anode signals $A$ \cite{HeDepth}:
\begin{equation}
 \text{z} = \frac{\text{cathode}}{\text{energy}} = \frac{ A_{CA} + A_{NCA} }{ A_{CA} - w \cdot A_{NCA} }\,.
 \label{eq_cal_ipos_z}
\end{equation}
A more complex model including electron trapping correction (tc) was derived in \cite{Fritts:2012xq}:
\begin{equation}
 \text{z}_{\text{tc}} = \lambda \ln\left(1 + \frac{1}{\lambda}\frac{ A_{CA} + A_{NCA} }{ A_{CA} - A_{NCA} }\right), \text{with } \lambda = \frac{1+w}{1-w}\,.
 \label{eq_cal_ipos_ztc}
\end{equation}
The z-values are given in relative detector units, where $0$ is at the anodes, and $1$ at the cathode, as shown in \autoref{fig_cpg_detector_geometry}.
In the following, the trapping corrected model $z_{tc}$ (interaction depth including trapping correction) is used, unless not stated otherwise.\\

\begin{figure}
 \centering
  \includegraphics[width=0.325\textwidth]{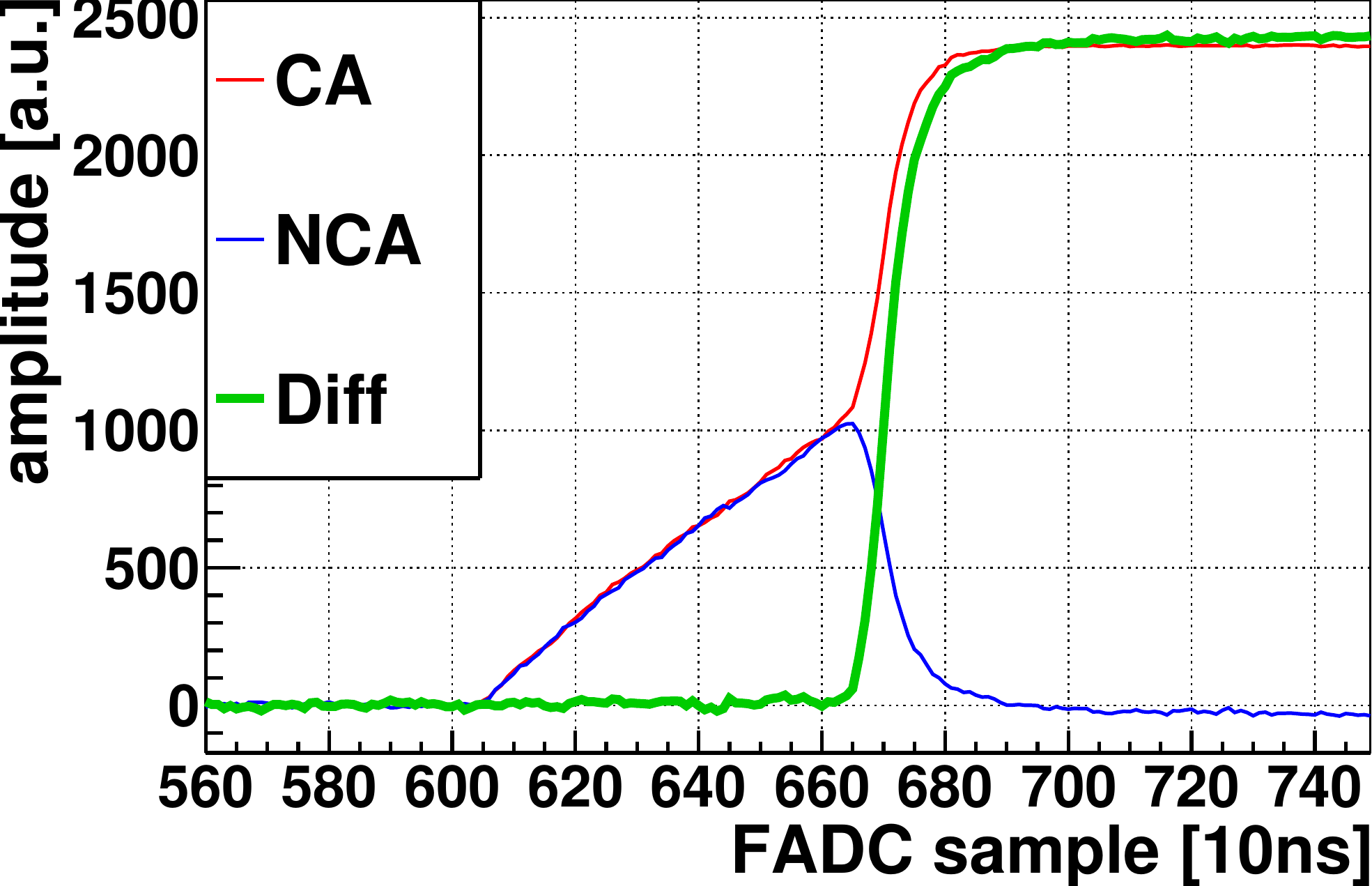} 
  \includegraphics[width=0.325\textwidth]{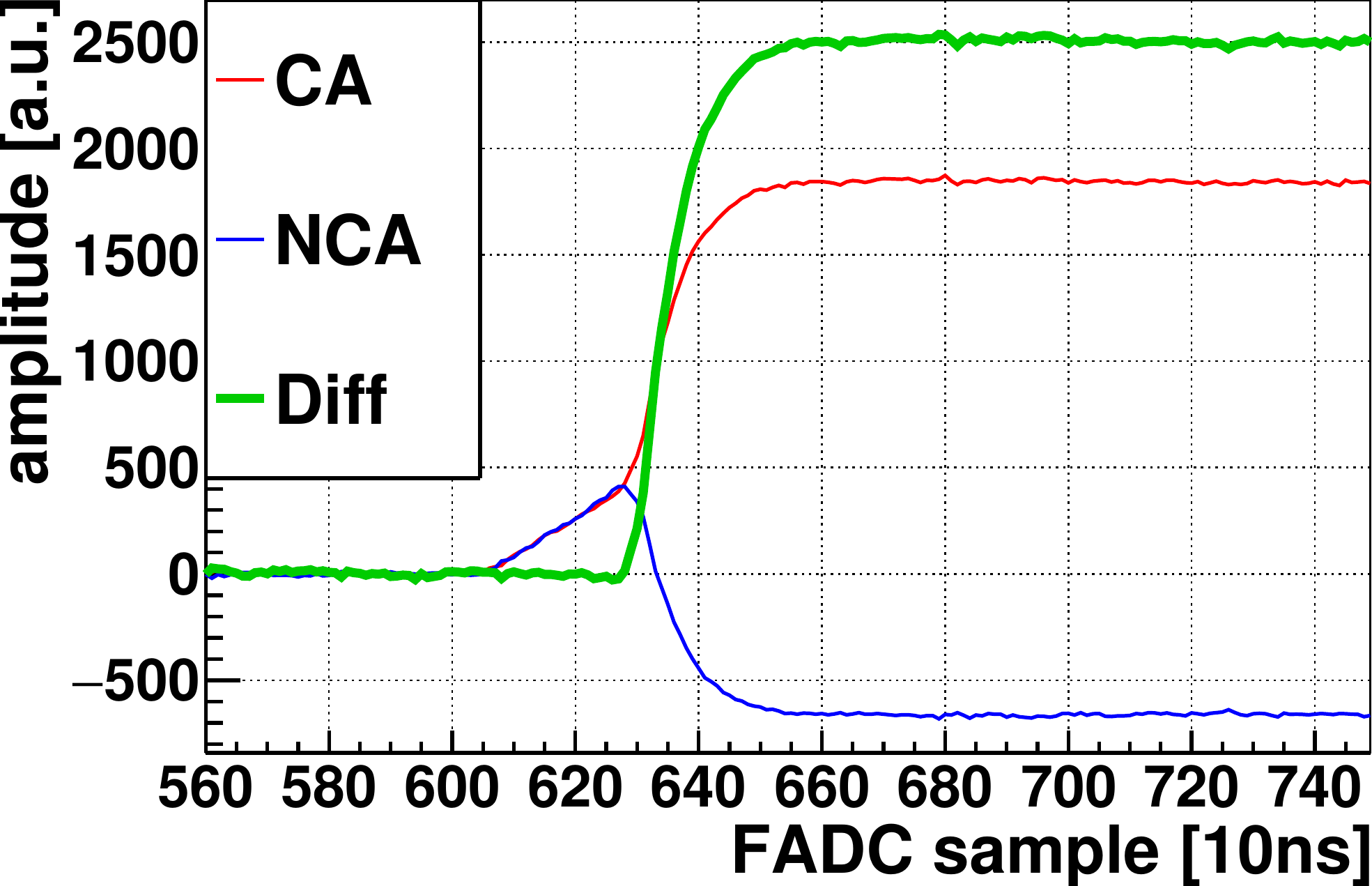}
  \includegraphics[width=0.325\textwidth]{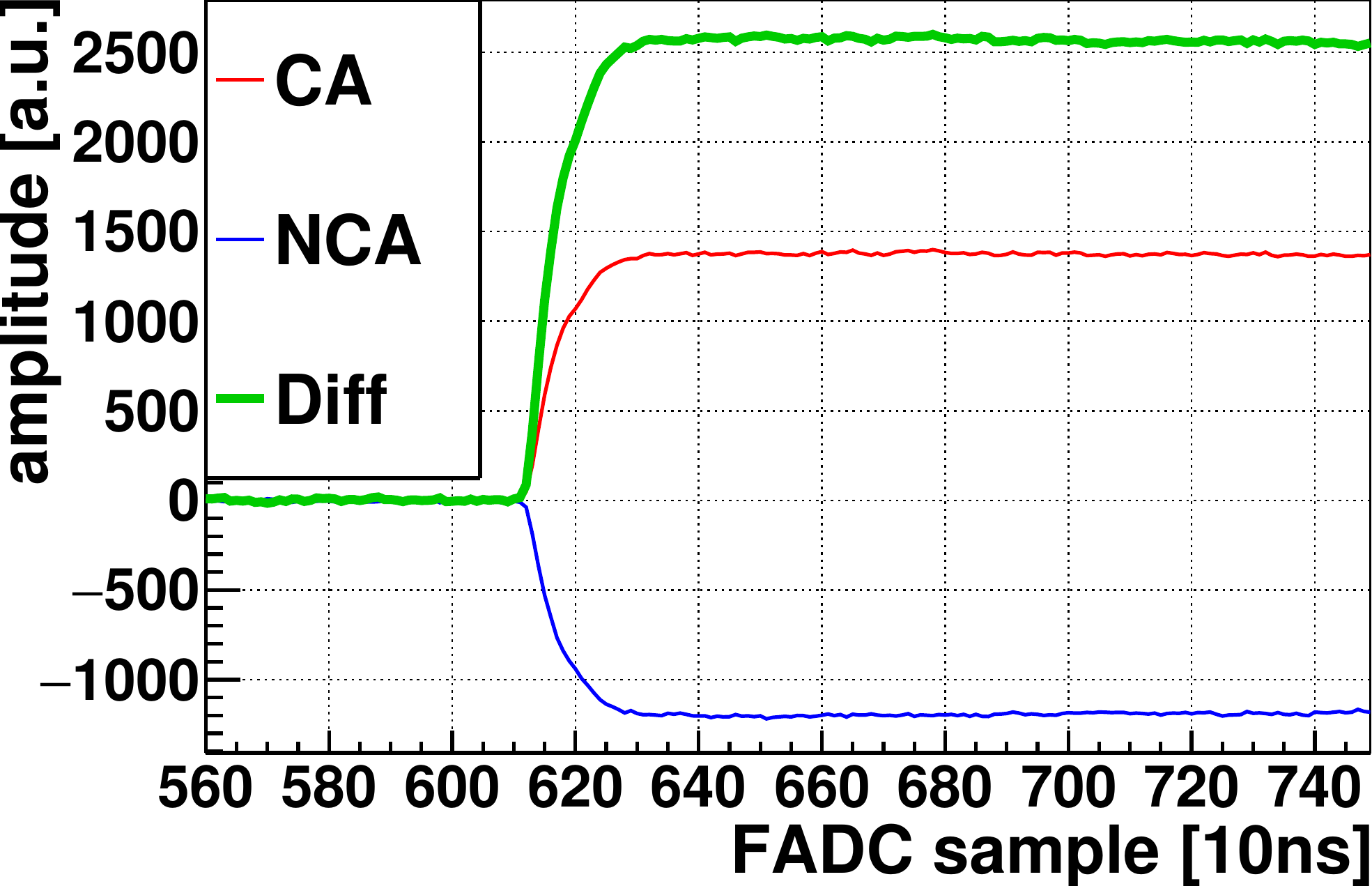}
  \captionsetup{width=0.9\textwidth} 
  \caption[Typical pulse shapes of a CPG detector]
	  {Typical pulse shapes of events with approximately \SI{2.5}{MeV} from a \isotope[232]Th source. From left to right: near-cathode, central and near-anode events. Note the decreasing length of the common rise of the CA and NCA signals, resulting from the decreasing drift length of the electrons towards the anodes. }
  \label{fig_cpg_normal_pulse_shapes}
\end{figure}

In \autoref{fig_cpg_normal_pulse_shapes}, typical pulse shapes of events in different interaction depths are shown:
The pulse shapes of a measured particle consist of a common rise of the NCA and CA signals, and after that of a decline and rise for the NCA and CA signal, respectively.
The common rise stems from the drift of the charge cloud through the detector volume due to the BV, the GB (\textbf{g}rid \textbf{b}ias) has no influence. 
Hence, both anodes measure the same signal.
Only in the near-anode region, the influence of the GB becomes important and directs the charge cloud to the CA and away from the NCA. 
Consequently, the CA signal shows a sharp rise, while the NCA signal falls.
The difference signal of CA - NCA shows a sharp rise to the full amplitude, which is proportional to the deposited energy.\\
The common rise due to the drift of the charge cloud towards the anodes is depending on the length the charge could has to drift.
Hence, the common rise depends on the interaction depth of the incident particle.
Detailed discussions and calculations to this are presented in \autoref{sec_drift_time}.

%%%%%%%%%%%%%%%%%%%%%%%%%%%%%%%%%%%%%%%%%%%%%%%%%%%%%%%%%%%%%%%%%%%%%%%%%%%%%%%%%%%%%%%%%%%%%%%%%%%%%%%%%%%%%%%%%%%%%%%%%%%%
%%%%%%%%%%%%%%%%%%%%%%%%%%%%%%%%%%%%%%%%%%%%%%%%%%%%%%%%%%%%%%%%%%%%%%%%%%%%%%%%%%%%%%%%%%%%%%%%%%%%%%%%%%%%%%%%%%%%%%%%%%%%
\FloatBarrier
\section{Data acquisition and analysis principles}
\label{sec_daq_analysis}

The data-taking is done using the DAqCORE (\textbf{D}ata \textbf{a}c\textbf{q}uisition for the \textbf{CO}B\textbf{R}A-\textbf{e}xperiment) software. One can define trigger thresholds for all channels independently. 
If the measured pulse exceeds the trigger threshold, it is saved to disk.
Furthermore, the readout of all channels is possible, if one channel triggers. 
The assignment which of the anodes is the CA and which the NCA, is defined by the applied voltage on the anodes.
In laboratory measurements, this can be done by swapping the cables at the preamplifier devices.\\ 

The digital off-line data-processing is done by ParamEst (\textbf{Param}eter \textbf{est}imation)
and MAnTiCORE (\textbf{M}ultiple-\textbf{An}alysis \textbf{T}oolk\textbf{i}t for the \textbf{CO}B\textbf{R}A \textbf{E}xperiment) \cite{Schulz, quante_bachelor, homann_master}.

ParamEst is used to determine several parameters: 
The electronics components used in each channel differ from each other, i.a. due to production tolerances. Consequently, the amplification of each channel is different as well. This is compensated in a gain-correction. 
Furthermore, the optimal weighting factor $w$ of \autoref{eq_energy_calculation} is calculated.

MAnTiCORE includes data-cleaning methods, these pulses are marked as such. 
These methods are optimized for the detectors of a good quality used in the COBRA demonstrator, and for the high energy ROI above \SI{2}{MeV}. 
As a consequence, these methods are not used at analyses where these conditions are not met.

Dedicated algorithms are used to calculate the amplitudes of the signals, from which quantities like deposited energy and interaction depth are calculated. 
Events whose amplitudes are below the trigger threshold are marked as such.
Furthermore, detailed pulse properties are computed to be used for the PSA (\textbf{p}ulse-\textbf{s}hape \textbf{a}nalysis),
like LSE (\textbf{l}ateral \textbf{s}urface \textbf{e}vents) and A/E (\textbf{A} divided by \textbf{E}). MSE (\textbf{m}ulti-\textbf{s}ite \textbf{e}vent) are tagged as such, as well \cite{zatschler_diplom}.

%%%%%%%%%%%%%%%%%%%%%%%%%%%%%%%%%%%%%%%%%%%%%%%%%%%%%%%%%%%%%%%%%%%%%%%%%%%%%%%%%%%%%%%%%%%%%%%%%%%%%%%%%%%%%%%%%%%%%%%%%%%%
\FloatBarrier
\subsection[Lateral surface events recognition]{Lateral surface events recognition}
\label{sec_LSE_recognition}
Events on the anode and cathode surfaces can be vetoed using the interaction depth calculation.
A PSA cut was developed in \cite{Fritts:2014qdf} to discard events on the other four lateral surfaces. 
The author of this thesis was one of the main authors of that paper.
This so called LSE cut consists of two quantities, one for the CA surfaces (ERT (\textbf{e}arly \textbf{r}ise \textbf{t}ime)), and one for the NCA surfaces (DIP (\textbf{dip})).
The following description of the LSE analysis is based on that publication.\\ 

\begin{SCfigure}
 \centering
  \includegraphics[width=.4\textwidth]{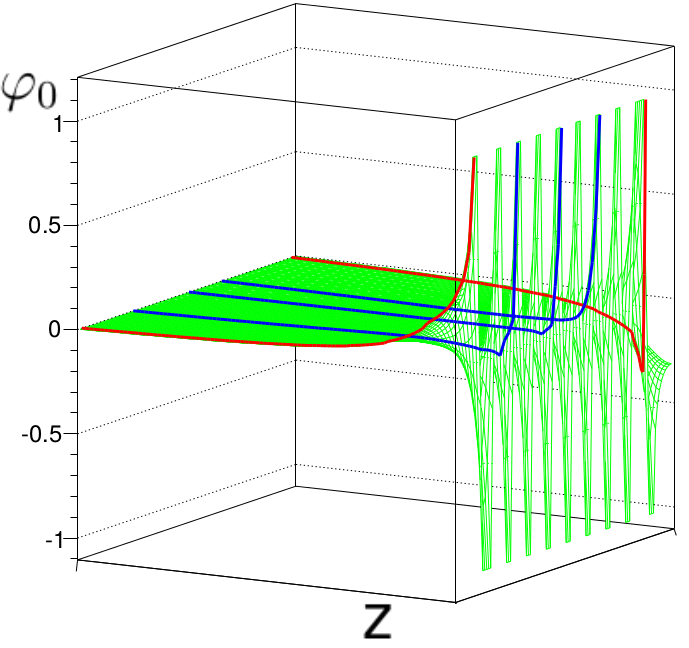}
  \caption[Weighting potential and LSE-areas]
          {Difference weighting potential, similar to \autoref{fig_weighing_potentials}.
          Indicated is the weighting potential for central events in blue, and for lateral surface events in red. Adapted from \cite{Fritts:2014qdf}}
  \label{fig_LSE_weighting_potential}
\end{SCfigure}

A detailed investigation of the difference weighting potential in a CPG detector shows distortions from the usual characteristics: 
The weighting potential close to the lateral surfaces has deviations compared to the typical weighting potential form in the center.
At one surface, it rises earlier, while it dips at the other.
This is shown in \autoref{fig_LSE_weighting_potential}.

\begin{figure}
 \centering
  \includegraphics[width=.32\textwidth]{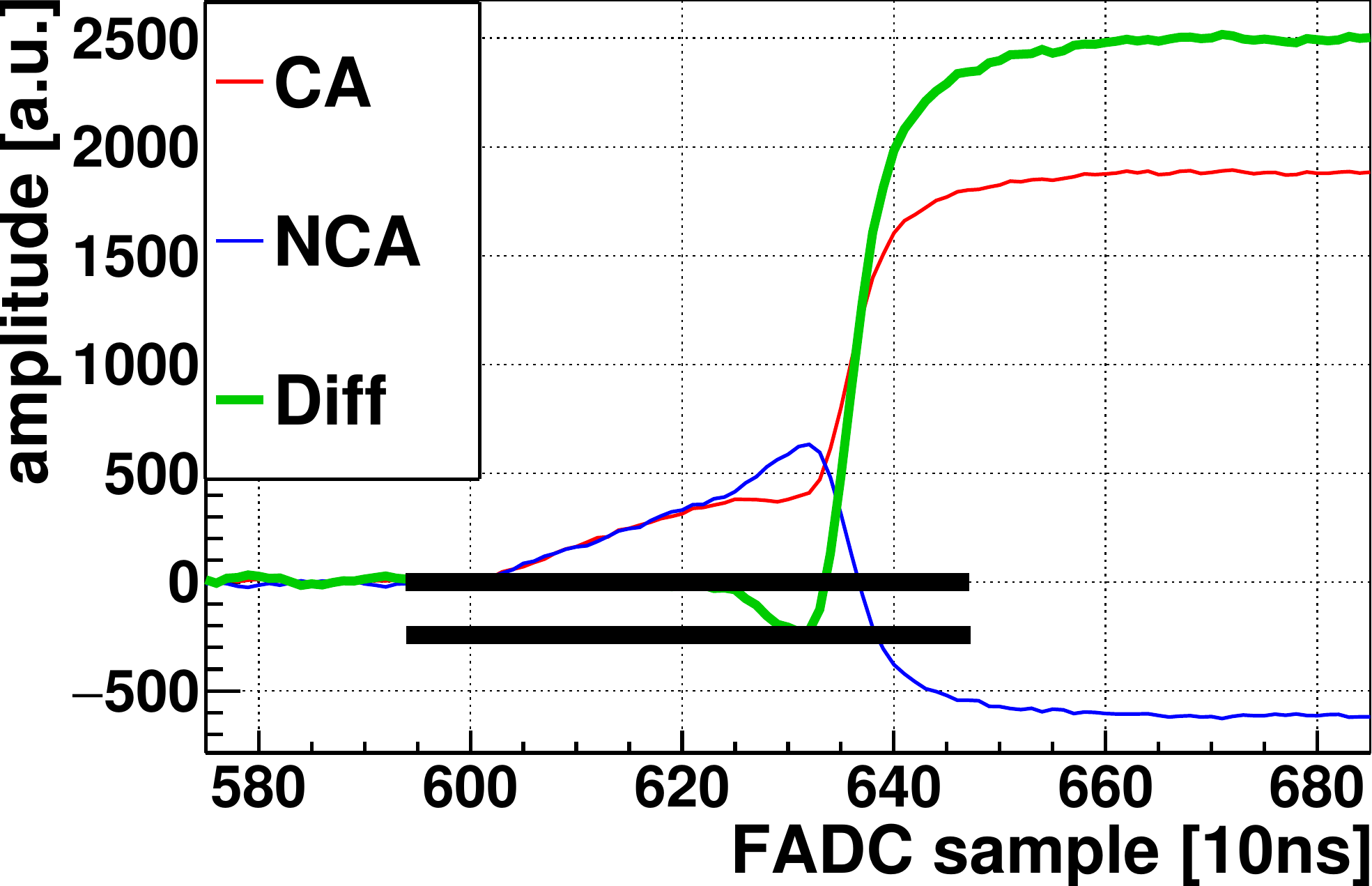}
  \includegraphics[width=.32\textwidth]{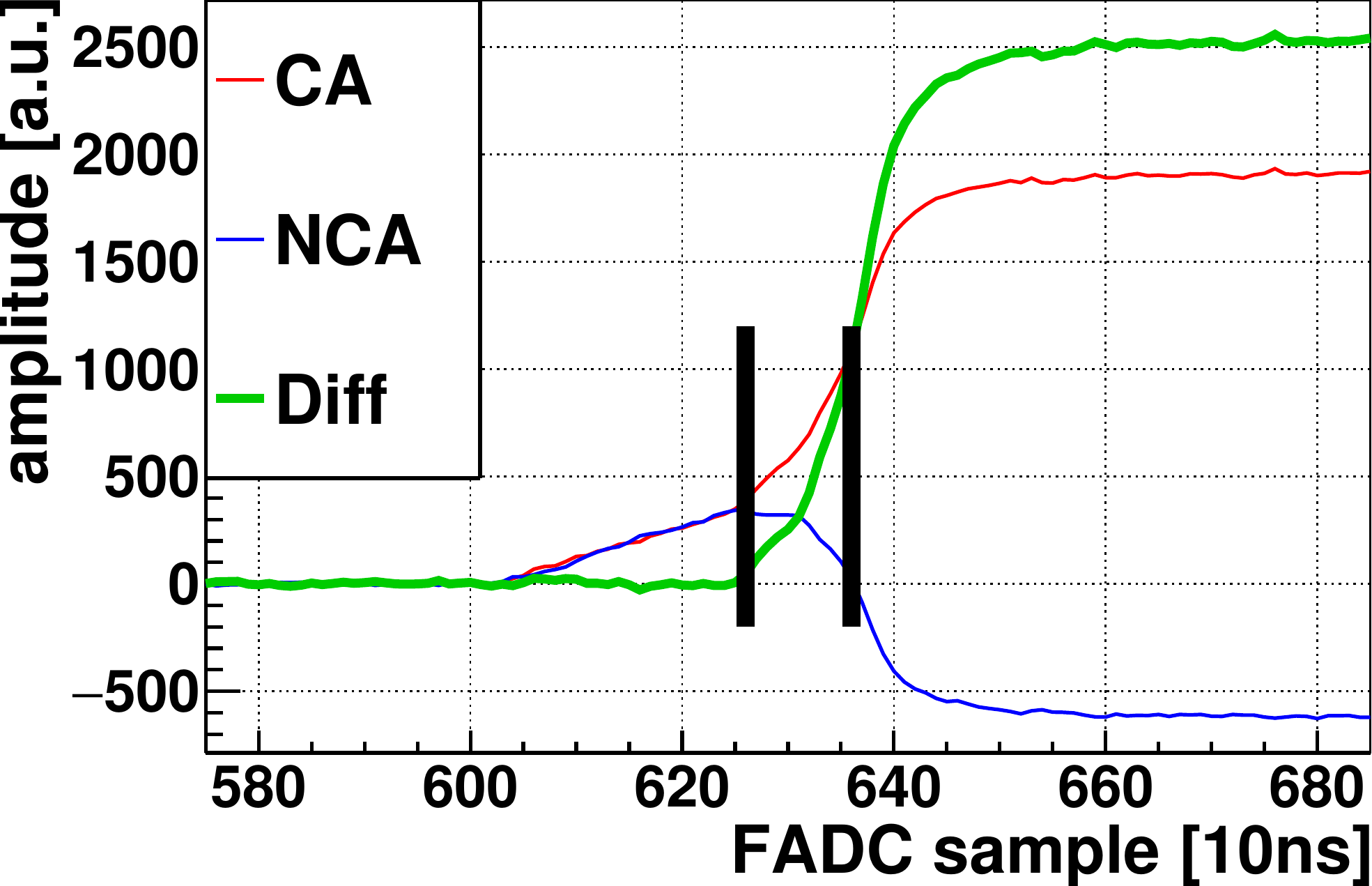}
  \includegraphics[width=.32\textwidth]{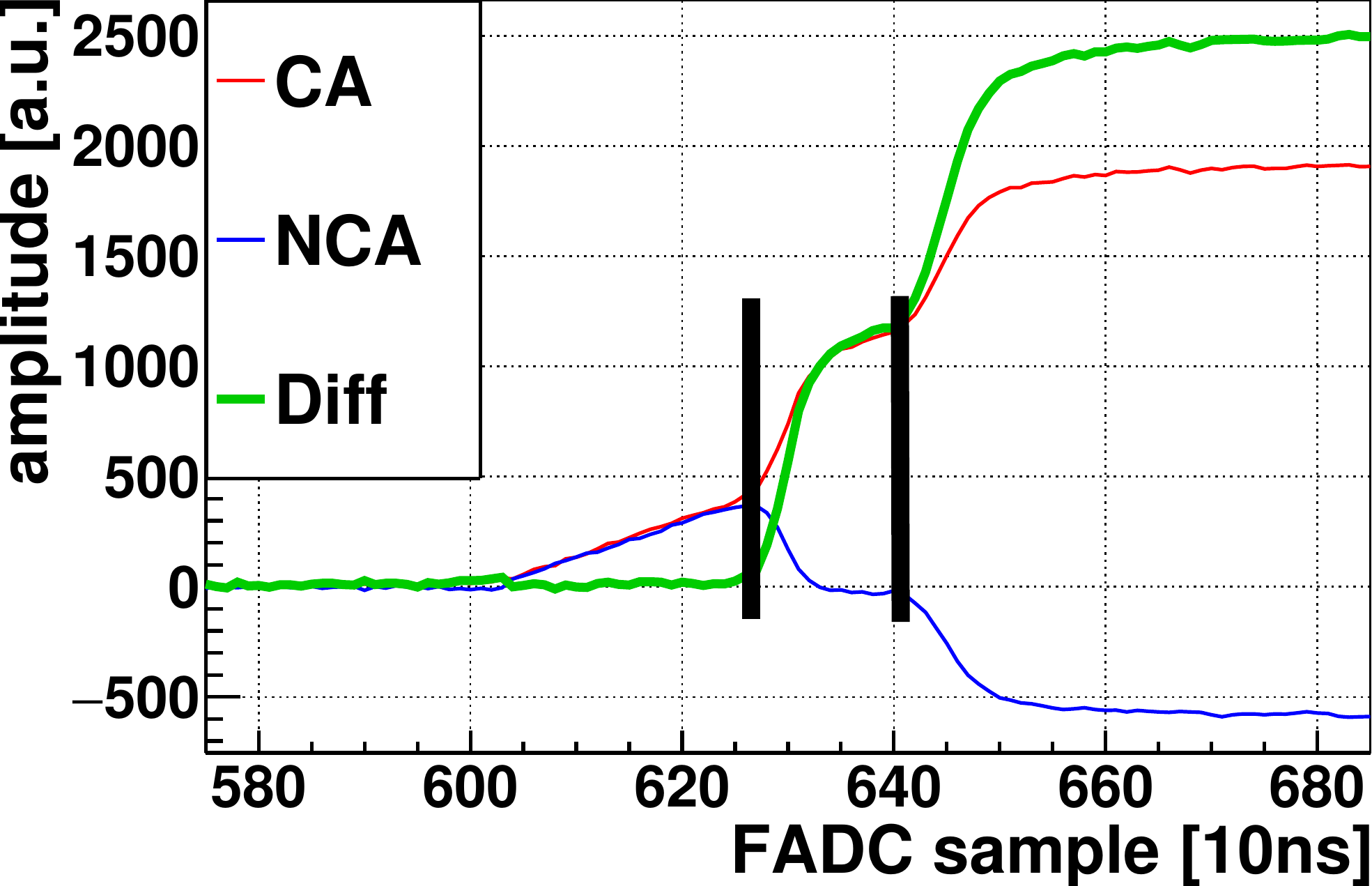}
  \captionsetup{width=0.9\textwidth} 
  \caption[Typical pulse shapes of DIP, ERT, and MSE]
          {Typical pulse shapes of DIP, ERT and MSE (from left to right) with \SI{2.5}{MeV} at an interaction dept of \num{0.5}. The black lines indicate the baseline and largest dip below it (left), and \SI{50}{\%} and \SI{3}{\%} of the amplitude of the difference pulse (center and right). 
          MSE-events have large ERT values too, but have a different origin and have to be excluded before applying the LSE cut.}
  \label{fig_LSE_examples}
\end{figure}

According to the Shockley-Ramo theorem (\autoref{eq_Shockly-Ramo}), the induced charge is proportional to the weighting potential. 
Hence, the pulse shapes of events at the detector surface differ from those of central events.\\
For the CA-adjacent surfaces, the quantity ERT is defined as the rise time of the difference pulse. 
It is found by starting at the \SI{50}{\%} level of the amplitude of the difference pulse. 
From that point, the \SI{3}{\%} level to the left is searched for. This horizontal distance is the ERT value. It is measured in FADC (\textbf{f}lash \textbf{a}nalog to \textbf{d}igital \textbf{c}onverter) samples, 1 sample corresponds to \SI{10}{ns} at \SI{100}{MHz} sampling frequency. 
By looking backwards from the \SI{50}{\%}-level, effects of pre-pulse fluctuations are ignored.
ERT is also sensitive to MSE. These are two interactions within the recorded \SI{10.24}{\mu s} timespan. 
Consequently, there is a rise of the difference pulse for each of them, which is tagged as an \textbf{e}arly \textbf{r}ise \textbf{t}ime then.
However, these events can be tagged by an own dedicated algorithm before the LSE-identification.\\
The corresponding quantity at the NCA-adjacent surfaces is called DIP. It is the maximum amount by which the difference pulse falls below its baseline before it rises. 
It is counted as a positive value, measured in FADC channels. 
To minimize pre-pulse fluctuations here, the DIP is searched for in a time interval of 30 samples (=\SI{300}{ns}) with the right edge at the \SI{50}{\%}-level of the difference pulse. 
This time-span is approximately one third of the maximal drift time of the electron charge cloud of cathode events, which is typically \SI{1}{\mu s}, see e.g. \autoref{fig_cpg_normal_pulse_shapes} or \autoref{sec_drift_time}.
\autoref{fig_LSE_examples} shows the definition of DIP and ERT at the difference pulse, as well as an example for a MSE.

All pulses have an ERT and DIP value, but for surface events, these are significantly larger.
A single threshold for each cut can discriminate LSE from central events.
The cuts are designed to be conservative, especially at lower energies, where PSA in general is likely to fail due to a poor SNR (\textbf{s}ignal-to-\textbf{n}oise \textbf{r}atio)\footnote{Examples for the non-functionality of PSA at low energies is shown in \autoref{fig_ps_60keV_Am241} on page \pageref{fig_ps_60keV_Am241}.}.
The cut thresholds can be chosen as desired. A standard way is to tune the thresholds such that \SI{90}{\%} of the central events are kept by DIP and ERT, resulting in approximately \SI{80}{\%} for the combined LSE cut.

The quantity DIP is connected to the amplitude of the pulse. 
As the amplitude is proportional to the deposited energy, DIP is roughly proportional to energy. 
An energy-independent way, for example by normalizing the DIP to the amplitude of the pulse, would reduce the conservative lower-energy behavior.\\
The quantity ERT is a matter of timing. By this it has no connection to the amplitude, and is constant with energy.

In the physics data-taking of the COBRA demonstrator, it cannot be determined at which surface the events happen. 
Consequently, the LSE cut comprises both the DIP and ERT cut. This means they have to be applied independently of the other at each analysis.\\

\begin{figure}
 \centering
  \includegraphics[width=.99\textwidth]{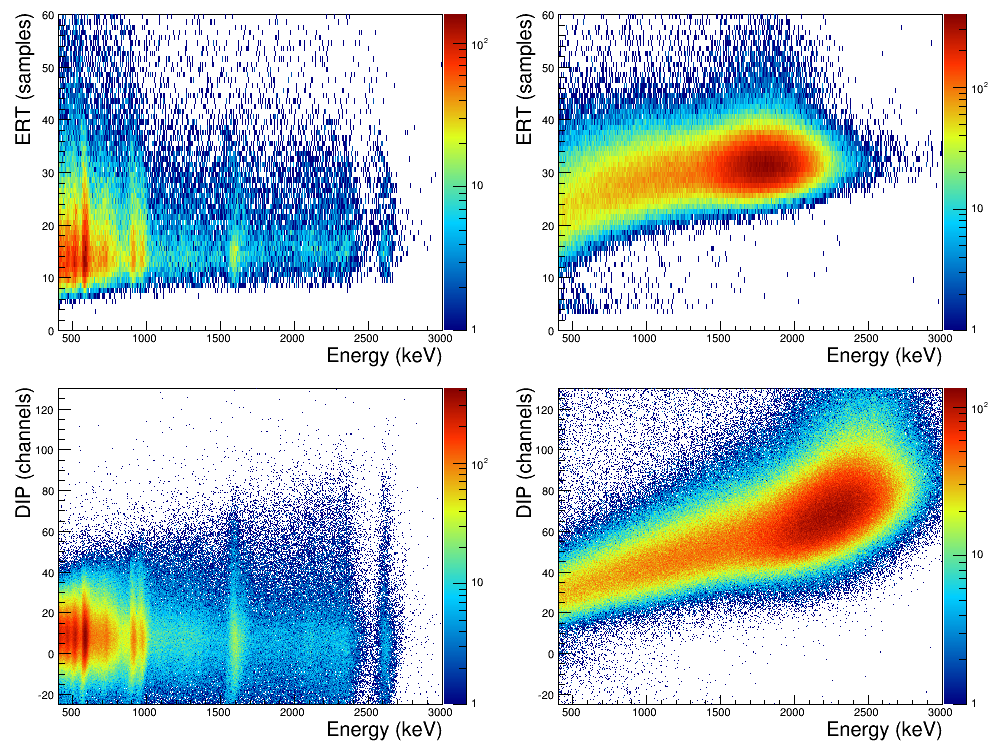}
  \includegraphics[width=0.49\textwidth]{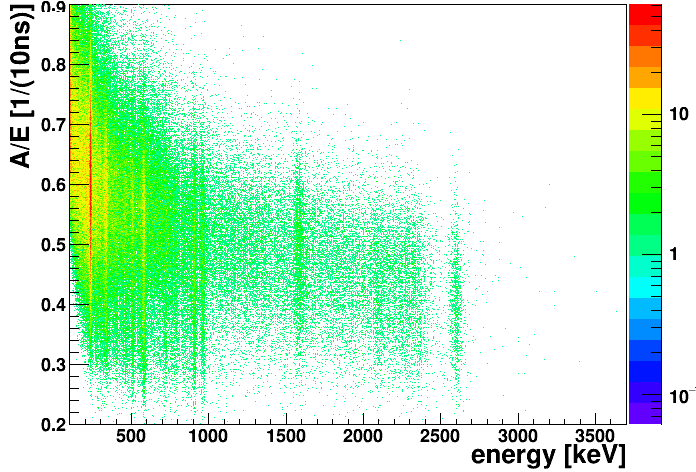}
  \includegraphics[width=0.49\textwidth]{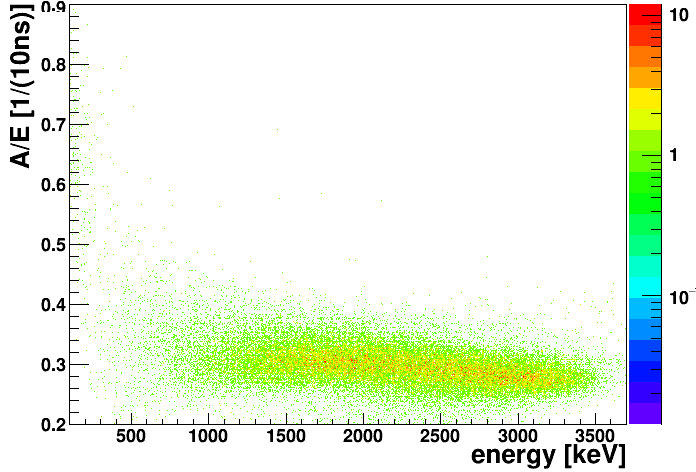}
  \captionsetup{width=0.9\textwidth} 
  \caption[Energy dependence of LSE and A/E]
          {Energy dependence of ERT (top line) and DIP (middle line) for gamma and alpha radiation (left and right, respectively), taken from \cite{Fritts:2014qdf}. 
           Each line has the same axes ranges. Bottom: Energy dependence for the A/E PSA criterion, discussed in the following \autoref{sec_A/E}. These plots have the same mapping as above. A/E is approximately constant with energy.}
  \label{fig_LSE_A/E_energy_dependence}
\end{figure}

\begin{figure}
 \centering
  \includegraphics[width=.49\textwidth]{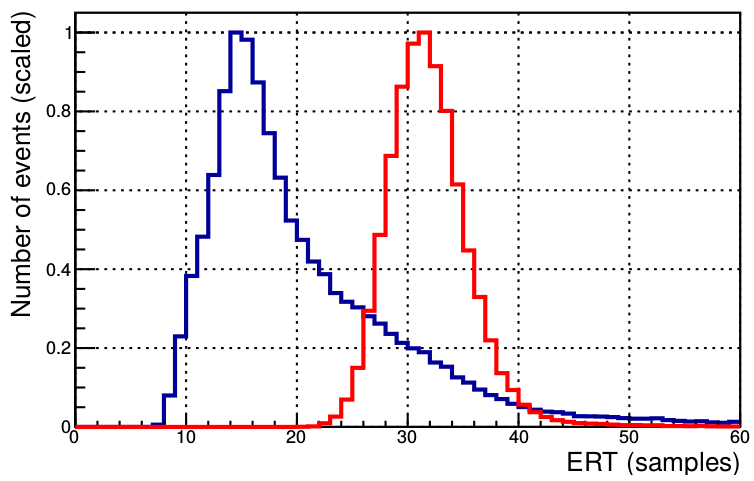}
  \includegraphics[width=.49\textwidth]{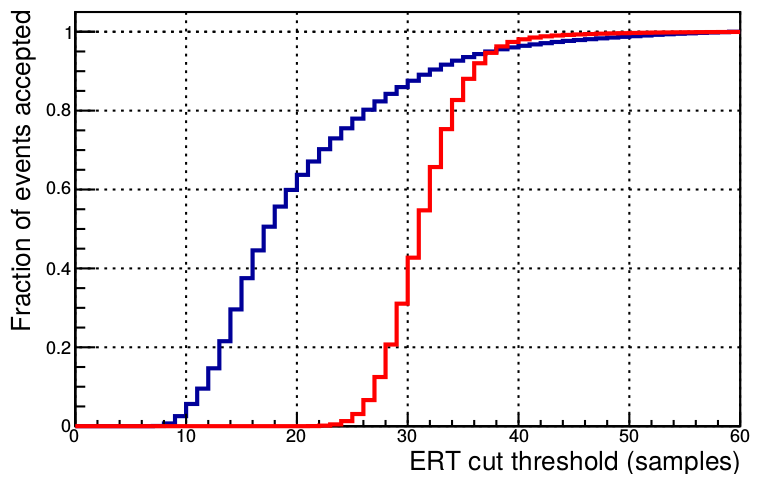}
  \includegraphics[width=.49\textwidth]{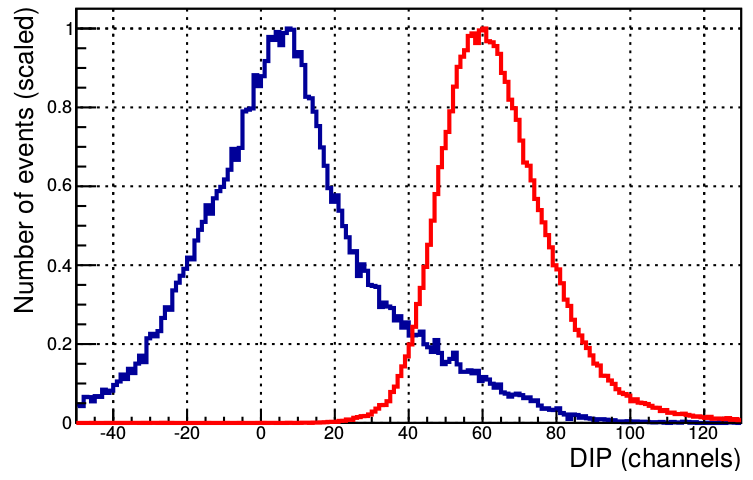}
  \includegraphics[width=.49\textwidth]{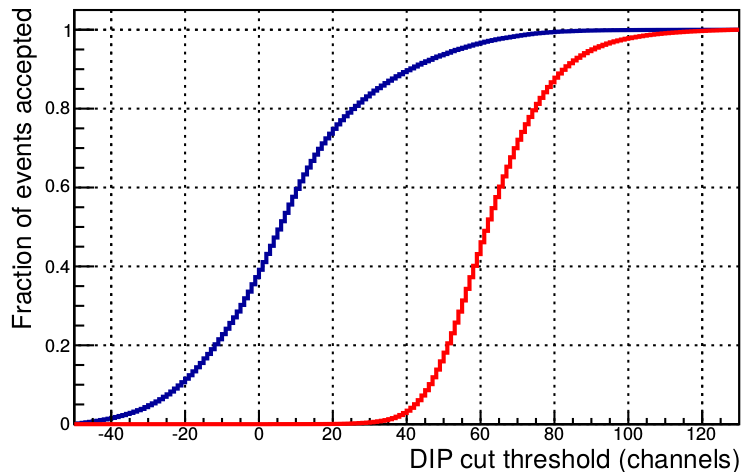}
  \includegraphics[width=.49\textwidth]{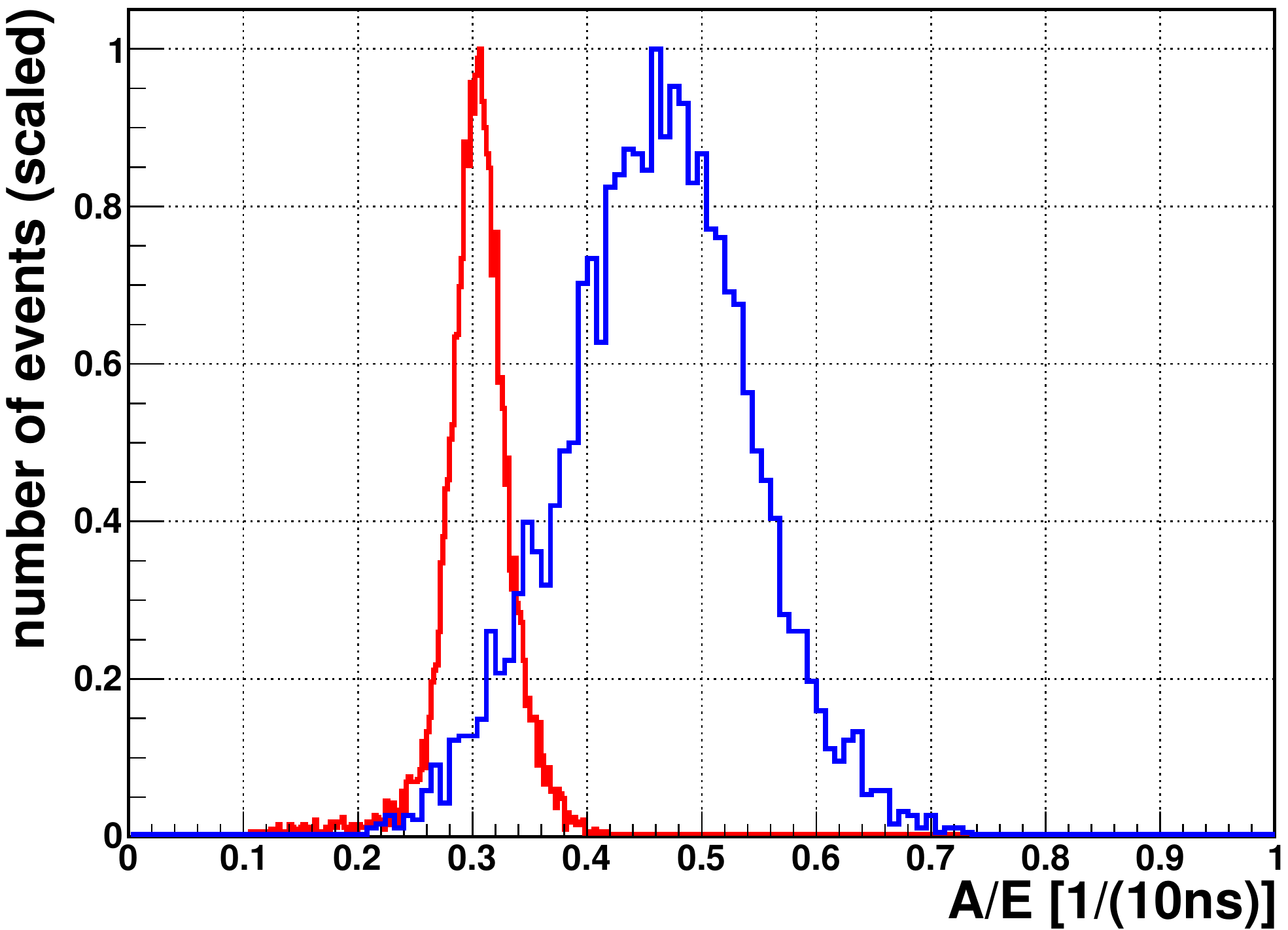}
  \includegraphics[width=.49\textwidth]{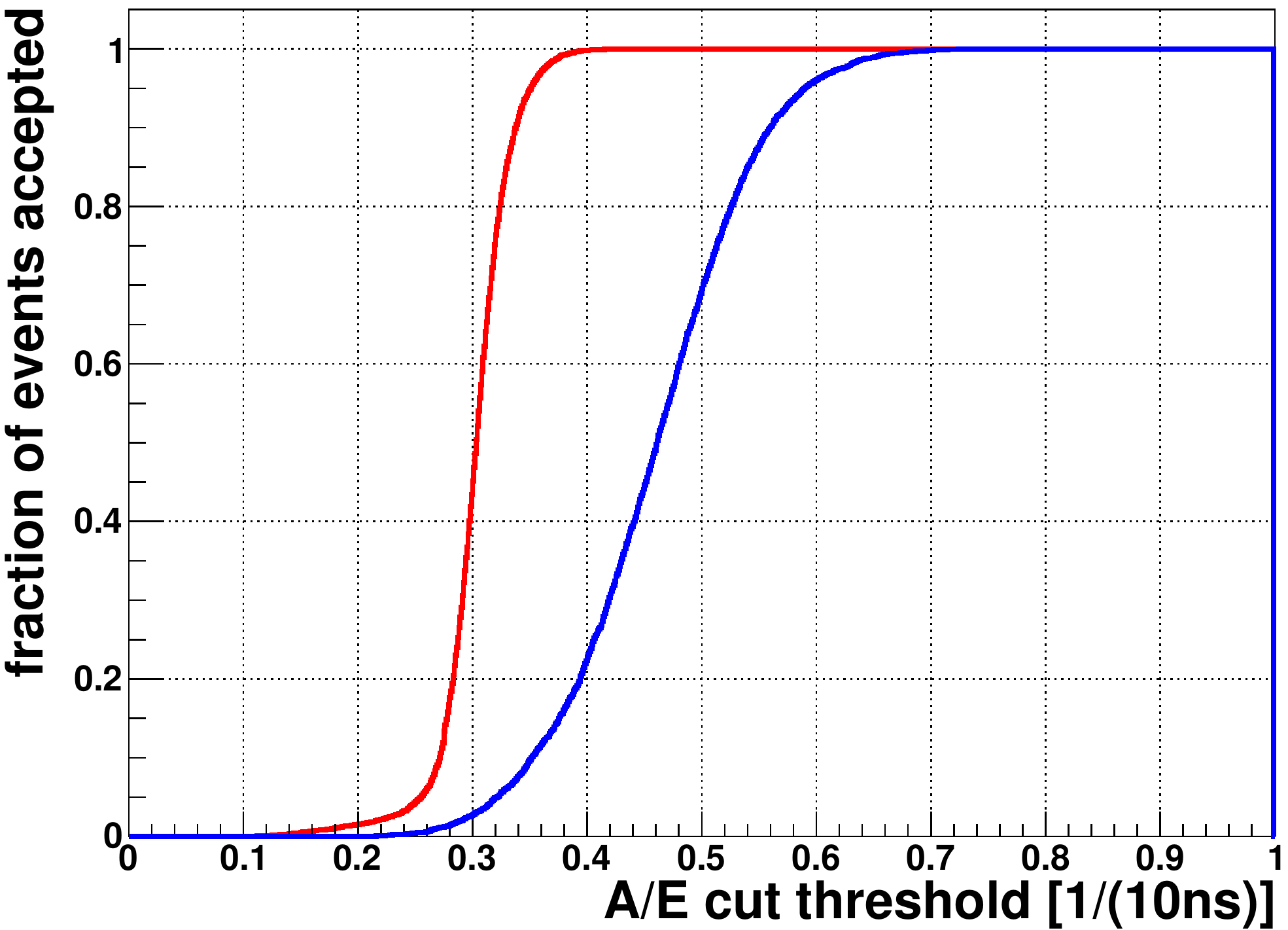}
  \captionsetup{width=0.9\textwidth} 
  \caption[Distribution and fraction of accepted events of LSE and A/E]
          {Distribution and fraction of accepted events of LSE and A/E identification quantities for gamma (blue) and alpha radiation
 	  (red) in an energy range of 1.7-2.0 MeV. 
	  Adapted from \cite{Fritts:2014qdf}.
	  Bottom: Analog representations for the A/E criterion.}
  \label{fig_LSE_efficiency}
\end{figure}

An experimental verification of the LSE quantities was done in laboratory measurements with a \isotope[232]Th and a \isotope[241]Am alpha source producing central and surface events, respectively.
The different characteristics and energy dependence of the LSE values for surface and central events is shown in \autoref{fig_LSE_A/E_energy_dependence}.
These plots indicate the use of a single threshold to discriminate surface events above a certain level. 
This can be seen in \autoref{fig_LSE_efficiency}, where the distribution of the LSE quantities and the fraction of accepted events as a function of the threshold is plotted. 
For comparison, these diagrams include the same representations for the A/E pulse shape criterion introduced in the next \autoref{sec_A/E}.

The NCA and CA surfaces meet at two edges of the detectors. 
As DIP and ERT have an opposed pulse shape, these two effects might cancel each other out. 
As a consequence, it is expected that there are transition zones at the edges, where the LSE tagging does not work.
Preliminary detector simulations indicate that this happens especially at the NCA surface.
It is unclear, if this happens at all, and to which extend. 
This is investigated in laboratory tests by alpha scanning measurements, discussed in \autoref{sec_alpha_scanning}.

%%%%%%%%%%%%%%%%%%%%%%%%%%%%%%%%%%%%%%%%%%%%%%%%%%%%%%%%%%%%%%%%%%%%%%%%%%%%%%%%%%%%%%%%%%%%%%%%%%%%%%%%%%%%%%%%%%%%%%%%%%%%
\FloatBarrier
\subsection[A divided by E pulse shape criterion]{A/E pulse shape criterion}
\label{sec_A/E}
The $0\nu\beta\beta$-decay experiments GERDA and MAJORANA \cite{1742-6596-606-1-012004} using Germanium detectors developed a PSA criterion called A/E. Extensive investigations of this method have been published in \cite{Agostini:2013jta} and \cite{Mertens:2015hwa}.
It is used to discriminate surface events from central events, and MSE from SSEs (\textbf{s}ingle-\textbf{s}ite \textbf{e}vents).\\ 
A/E is the ratio of the maximum current (A) and the amplitude (E) of the pulse. The current pulse is obtained by differentiating the measured charge pulse.
In other words, A/E is the largest slope of the pulse, normalized by its amplitude.
By this, it has the dimensions of an inverse time (in FADC bins), which is $\frac{1}{\SI{10}{ns}}$.

SSE and MSE discrimination is more important for Germanium experiments, as the $Q$-value of \isotope[76]Ge is only at \SI{2039}{keV}, which is well in the naturally occurring gamma background.
High-energetic photons are likely to do Compton-scattering producing MSE, especially in these detectors with a larger volume. 

\begin{SCfigure}
 \centering
  \includegraphics[width=0.5\textwidth]{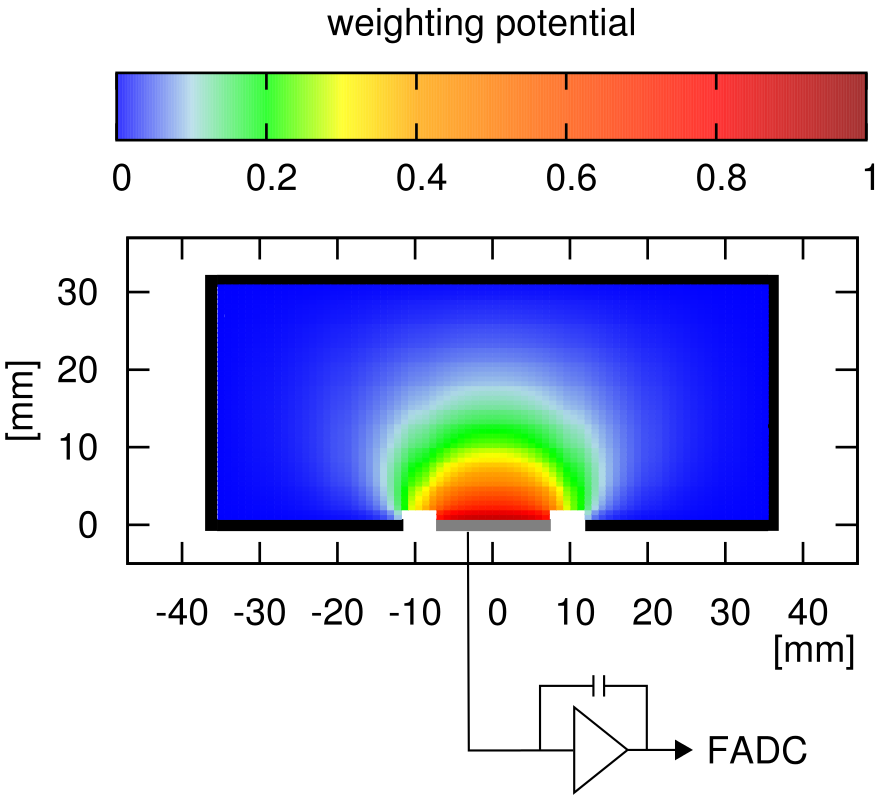}
  \caption[Sketch of GERDA BEGe (broad energy Germanium) detector]
          {Sketch of the cross-section of a GERDA BEGe (\textbf{b}road \textbf{e}nergy \textbf{Ge}rmanium) detector indicating the weighting potential and readout electronics. The readout of the $p^+$-electrode (grey) consists of a CSA (\textbf{c}harge-\textbf{s}ensitive pre\textbf{a}mplifier) and a FADC sampling with \SI{100}{MHz}, so it is comparable to COBRA. Adapted from \cite{Agostini:2013jta}.}
  \label{fig_A_over_E_gerda}
\end{SCfigure}

The principle of the A/E cut criterion is based on a large gradient of the weighting potential in the detector center, causing charge induction only in a very small volume. In \autoref{fig_A_over_E_gerda}, this is close to the $p^+$-electrode.
This principle is fulfilled at the COBRA CPG detectors (see the weighting potential in \autoref{fig_LSE_weighting_potential}), so the use of the A/E criterion can be possible for COBRA as well and is investigated in \autoref{sec_surface_events}.

According to the Shockley-Ramo theorem (\autoref{eq_Shockly-Ramo}), the measured pulse is proportional to weighting potential. 
Applied to the COBRA CPG detectors, the A/E criterion uses the fact that lateral surface events are influenced by a weaker weighting potential, and show hence a lower slope than central events.

The quantities A and E are calculated by MAnTiCORE by default, so the A/E criterion can be tested easily.

Analog to the LSE quantities, the characteristics of the A/E distribution is shown energy-dependent in \autoref{fig_LSE_A/E_energy_dependence} in the bottom line.
For the distribution and fraction of accepted events as a function of the threshold see \autoref{fig_LSE_efficiency}.

\chapter{The COBRA demonstrator at LNGS}
\label{sec_COBRA}

%%%%%%%%%%%%%%%%%%%%%%%%%%%%%%%%%%%%%%%%%%%%%%%%%%%%%%%%%%%%%%%%%%%%%%%%%%%%%%%%%%%%%%%%%%%%%%%%%%%%%%%%%%%%%%%%%%%%%%%%%%%%
%%%%%%%%%%%%%%%%%%%%%%%%%%%%%%%%%%%%%%%%%%%%%%%%%%%%%%%%%%%%%%%%%%%%%%%%%%%%%%%%%%%%%%%%%%%%%%%%%%%%%%%%%%%%%%%%%%%%%%%%%%%%
\label{sec_cobra_demonstrator}
The COBRA demonstrator is an experimental setup at the LNGS, made of 64 CdZnTe CPG detectors in a $4\times4\times4$ array, housed in a complex shielding architecture. Its goals are demonstrating the feasibility of operating the detectors on a low-background level, monitor the long-term stability, identify potential background sources and to obtain physics results, amongst others.\\

The status of the demonstrator at the beginning of this thesis in January 2012 was the following: The (re-)commissioning after moving to a new location within the LNGS has started three months before with the installation of the first 16 detectors (layer 1). The read-out electronics and DAQ (\textbf{d}ata \textbf{a}c\textbf{q}uisition)-system were functional, developed by the author (diploma thesis \cite{tebruegge}) and others (\cite{Schulz, koettig_2012_phd}). Nevertheless, several improvements and additions had to be done.\\
In the course of this work, the author spent six shifts with more than seven working-weeks on site at LNGS.

A complete description of the actual final COBRA demonstrator can be found in a recent publication by the author \cite{Ebert:2015pgx}. 
The following \autoref{sec_cobra_demonatrator_final_stage} describing the COBRA demonstrator in its final stage is taken from that publication. 
In \autoref{sec_addition_work_lngs}, additional work of the author at LNGS in completing and improving the demonstrator setup is described, which is not part of the publication.

%%%%%%%%%%%%%%%%%%%%%%%%%%%%%%%%%%%%%%%%%%%%%%%%%%%%%%%%%%%%%%%%%%%%%%%%%%%%%%%%%%%%%%%%%%%%%%%%%%%%%%%%%%%%%%%%%%%%%%%%%%%%
\section{The COBRA demonstrator in its final stage}
\label{sec_cobra_demonatrator_final_stage}
The COBRA demonstrator at the LNGS in Italy employs CdZnTe semiconductor detectors.
It is covered by about \SI{1400}{m} of rock which corresponds to a shielding of approximately \SI{3800}{m} of water equivalent. This reduces the flux of muons from cosmic rays by six orders of magnitude to \SI{3.41(1)e-4}{m^{-2}s^{-1}} \cite{Bellini:2012te}.
The COBRA demonstrator is located between halls A and B in a part of the building that was 
formerly used by the Heidelberg-Moscow experiment \cite{KlapdorKleingrothaus:2000sn}.  
Preliminary studies were done with similar setups prior to 2011 in other locations at LNGS, indicating that the operation of CdZnTe detectors should be further investigated \cite{Bloxham:2007aa, Dawson:2009ni, Dawson:2009cc, Goessling:2005jw, Kiel2003499}. 

\begin{SCfigure}
 \centering
  \includegraphics[width=0.7\textwidth]{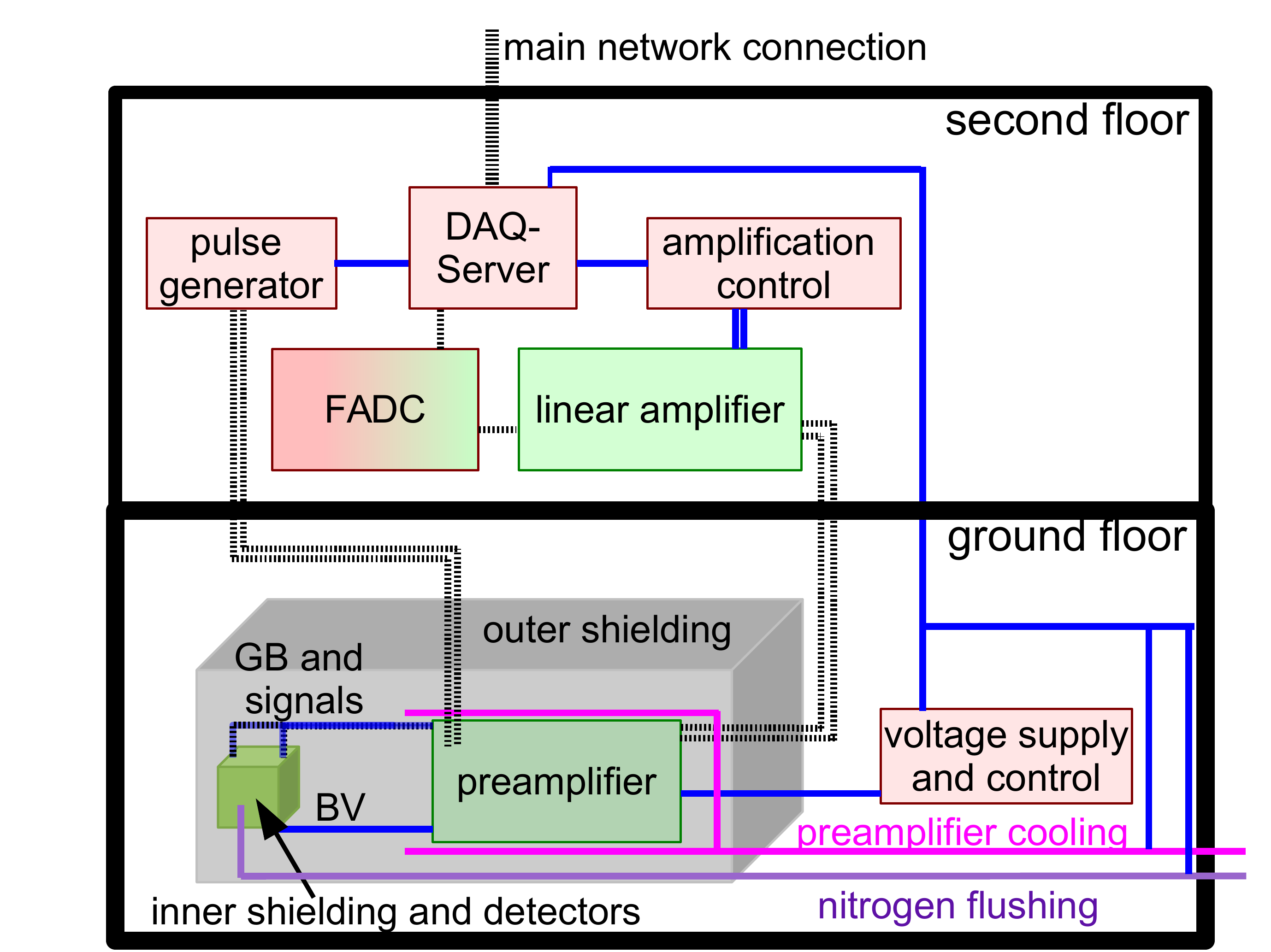}
  \caption[Principle sketch of the COBRA demonstrator in its final stage]
          {Principle sketch of the COBRA demonstrator in its final stage, taken from \cite{Ebert:2015pgx}. Dotted lines indicate the signal flow, solid ones the supply and control. Double lines symbolize differential signaling. Green and red boxes show the analog and digital part of the system.}
  \label{fig_lngs_demonstrator_sketch}
\end{SCfigure}

The building COBRA uses consists of two floors. The experimental setup and the analog part of the readout electronic system are located on COBRA's part of the ground floor. 
The rest of the electronics and the DAQ system are  
installed in an air-cooled room on the second floor, since it can dissipate waste heat more effectively.
\autoref{fig_lngs_demonstrator_sketch} shows the detailed design and location of the experiment's components on the two floors.

%%%%%%%%%%%%%%%%%%%%%%%%%%%%%%%%%%%%%%%%%%%%%%%%%%%%%%%%%%%%%%%%%%%%%%%%%%%%%%%%%%%%%%%%%%%%%%%%%%%%%%%%%%%%%%%%%%%%%%%%%%%%
\subsection{Detectors and shielding}

The demonstrator consists of 64 detectors, each with a size of (1$\times$1$\times$1)\,cm$^3$ and a weight of about \SI{5.9}{g}. \num{16} detectors are mounted on one layer, arranged in a symmetric $4\times4$ structure with a horizontal separation of \SI{4}{\milli\metre}. 
Two precisely cut Delrin frames provide the mechanical holding system for one layer. Four such layers are arranged on top of each other in the copper support structure with a vertical distance of \SI{1.2}{\centi\metre} in between.
The electrical contacting of the electrodes is achieved with a \SI{50}{\micro\metre} thin gold wire which is fixed to each of the electrode contact pads with conductive silver lacquer LS200. A glue spot next to it stabilizes it mechanically. 
Studies showed that there is no significant contribution of silver isotopes, especially \isotope[110m]Ag, to the background level \cite{neddermann_2014_phd, hillringhaus}.
Alternative contacting schemes, such as pressure-contacts or soldering, were studied in Ref. \cite{koettig_2008_dipl}, but failed. The gold wires are connected to an RG178 shielded coaxial cable on the cathode side providing the BV. 
For the most recently installed detectors, special low-background high voltage cables are used.\\
The two gold wires on the anode side are glued to a Kapton ribbon cable which provides the constant GB and ground contact, respectively. Furthermore, the anode signals are transmitted on the Kapton cable to the preamplifier devices. 
All cables run on a wavy path through the lead shielding, so that no direct opening through the shielding is pointing towards the detectors.\\ 

\begin{figure}
 \centering
  \includegraphics[width=0.9\textwidth]{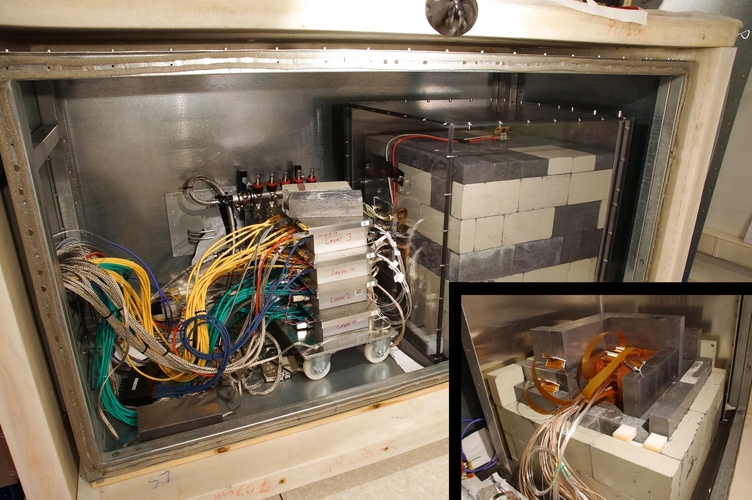}
  \captionsetup{width=0.9\textwidth} 
   \caption[Photograph of the COBRA demonstrator setup]
           {Photograph of the setup with opened outer shielding. All cables enter the setup from the chute on the left, the stacked preamplifier devices are in the center. On the right side the inner shielding can be seen; the inset shows it partly opened. Taken from \cite{Ebert:2015pgx}. }
  \label{fig_setup_lower_floor}
\end{figure}

The complex shielding structure used for the COBRA demonstrator consists of an outer and an inner shield. The outer one comprises a neutron shield and a shield against EMI (\textbf{e}lectro\textbf{m}agnetic \textbf{i}nterference) with dimensions of approximately $(2\times1\times1)\,\text{m}^3$. 
The neutron shield is made of \SI{7}{cm} borated polyethylene with \SI{2.7}{wt\%} of boron content. Hydrogen-rich plastics, such as polyethylene, moderate neutrons to thermal energies very effectively, while \isotope[10]B has a very high cross section to capture such neutrons \cite{muenstermann_2007_phd, neddermann_2014_phd}.
The EMI shield is a construction of iron sheets with a thickness of \SI{2}{mm} which are carefully welded or connected with mesh bands to ensure proper electrical contacts between the different sheets \cite{neddermann_2014_phd}. 
This material was chosen because of its ferromagnetic properties which are highly desirable to shield also against the magnetic component of the EMI.
A chute filled with copper granulate acts as a cable feed-through. 
All cables entering the setup pass through this chute in a way that either their outer insulation is stripped off, or an extra metal mesh shielding is added, so that their shielding is connected tightly to the electrical potential of the EMI shield. 
This ensures that none of these cables act as an antenna and transport EMI to the inside of the shielding. 
The functionality of the EMI shield was tested by the independent competence center EMC Test NRW \cite{bernau}.
The chute is also carefully surrounded by a polyethylene construction, so that there are no gaps in the neutron shield.
All preamplifier devices, their cooling plates and the inner shielding are surrounded by the outer shielding.\\
The inner shielding consists of an air-tight sealed housing of metal and polycarbonate plates, which is constantly flushed with evaporated nitrogen creating a slight overpressure. 
This prevents dust, and especially radon and its decay products, to settle on the detector surfaces. Within this housing, the shield against radioactive radiation is situated, which is a cube with a volume of $(60\times60\times60)\,\text{cm}^3$ with the detectors in the very center. It is built from \SI{15}{cm} of standard lead at the outside, followed by \SI{5}{cm} of ultra-low-activity lead with a \isotope[210]Pb activity of less than \SI{3}{Bq/kg}. 
The innermost shielding level (\SI{5}{cm}) and the support structure of the detectors are made of ultra-pure OFHC (\textbf{o}xygen-\textbf{f}ree \textbf{h}igh-\textbf{c}onductivity) electroformed copper. 
A total view of the setup including the shielding layers can be seen in \autoref{fig_setup_lower_floor}, the inset shows the partly opened lead shielding.      

%%%%%%%%%%%%%%%%%%%%%%%%%%%%%%%%%%%%%%%%%%%%%%%%%%%%%%%%%%%%%%%%%%%%%%%%%%%%%%%%%%%%%%%%%%%%%%%%%%%%%%%%%%%%%%%%%%%%%%%%%%%%
\subsection{Readout electronics}
 \label{sec:readout_electronics}
The electronic readout system consists of several components: directly outside of the lead shielding, the preamplifiers convert the sensitive detector charge signals into differential voltage signals. This ensures a robust and stable transmission. 
The signals are amplified with the linear amplifiers and converted to single-ended signals. 
The FADCs digitize the signals.\\
The electronic readout system is separated between the two floors of the building. The devices which dissipate a large amount of waste heat, especially the FADCs with approximately \SI{1}{kW}, are located in an air-cooled side room on the second floor. Consequently, the analog part of the readout electronics had to be separated between the two floors. 
As these devices have been custom-made, the issues arising from this, such as a potentially degraded signal quality due to noise and interference, were taken into account completely, i.e. by using differential signal transmission \cite{Schulz, muenstermann_2007_phd, tebruegge}.

\subsubsection{Preamplifiers}
\label{sec:preamplifiers}
Cremat CR110 charge sensitive preamplifier components convert the charge signals coming from the detectors into voltage signals. These signals 
are transformed to differential signals directly afterwards on the same printed circuit board.\\
One preamplifier device is used for each layer, where each device can handle up to 32 signal inputs (two signal channels for each of the 16 detectors). Each CR110 component is surrounded by a metal shield to suppress crosstalk. 
The differential signal output is fed to RJ-45 connectors, as
matching high-quality differential cables are easily available. Eight ethernet network cables per preamplifier device are needed to transmit all 16 differential detector signals. \\
Separate input lines for signals from a pulse generator are also present for each CR110 component.
The pulse generator signals are supplied via differential signaling to ensure a high signal quality. The conversion to single-ended signals is done directly on the preamplifier printed circuit board.\\ 
Furthermore, the bias voltage and grid bias are also applied via the preamplifier device: the grid bias on the same board, the bias voltage on a separate printed circuit directly beneath the other. The voltages are filtered by an RC low-pass filter.\\
To process all 64 detectors of the COBRA demonstrator, four of such preamplifier devices are needed. These are placed on top of each other next to the inner shielding.
\begin{verbatim}
 
 
\end{verbatim}

\subsubsection{Differential signaling}
 \label{sec:differential_signaling}
Differential signal transmission is widely used in computer and communication products. It allows for stable and robust transmission as it minimizes crosstalk, electromagnetic interference and noise.
For the COBRA demonstrator, it is mainly used for signal transmission between the preamplifier and linear amplifier, as well as between pulse generator and preamplifier. The length of each cable is \SI{25}{\metre} and covers the passage between the two floors.\\
Standard category 6 network cables are used for the transmission. These are commercially available and are appropriate to transmit sensitive differential signals due to the shielded twisted-pair wires surrounded by an additional shielding layer.
In laboratory tests, cable lengths of \SI{50}{\metre} were tested, even passing areas with enhanced electromagnetic interference, such as housings of running computers and network switches, without a significant deterioration of the signal quality or noise increase.

\subsubsection{Linear amplifiers}
\label{sec:linear_amplifierts}
The custom-made linear amplifiers are the last element of the analog readout chain. These devices amplify the
detector signals to match the input range of the FADCs, and convert the differential signals to single-ended signals. The amplification can be adjusted in 16 steps of \SI{3}{dB}, which is equivalent to a factor of $10^{3/20} = 1.41$
per step. The gain ranges from 0.5 to $0.5\cdot10^{15\cdot3/20} = 89$.
All operations are remotely controllable via an Arduino micro-controller board, acting as an SPI (\textbf{S}erial \textbf{P}eripheral \textbf{I}nterface Bus) gateway.
The linear amplifiers are operated in 19$''$ NIM crates. Each linear amplifier processes signals of four detectors. It has 16 differential input and eight signal output channels to match the eight input channels of the following FADCs. 

\subsubsection{FADC}
\label{sec:FADCs}
The FADCs are Struck SIS3300 FADC devices. These are VME bus modules with a maximum sampling 
frequency of \SI{100}{MHz} and a 12-bit resolution. Additional 3-bit resolution are gained by averaging over 128 samples. Hence, the dynamic resolution is 15 bit \cite{Schulz}.
Two memory banks with $2^{17}$ samples per channel are used. \num{1024} samples are acquired per event, 
resulting in a capacity of 128 events per memory bank. The two-memory-bank mode allows for a dead-time free operation. At \SI{100}{MHz} sampling frequency, this results in a timespan of \SI{10.24}{\mu s} per event that is recorded. As the longest physical signal spans are of the order of \SI{1}{\mu s}, there is enough pre- and post-trigger information for later off-line analysis methods.\\
Each module has eight input channels. Firmware and hardware were modified to match the specification of the COBRA demonstrator, such as choosing a peak-to-peak input range of \SI{2}{V} and adjusting timing precision.

%%%%%%%%%%%%%%%%%%%%%%%%%%%%%%%%%%%%%%%%%%%%%%%%%%%%%%%%%%%%%%%%%%%%%%%%%%%%%%%%%%%%%%%%%%%%%%%%%%%%%%%%%%%%%%%%%%%%%%%%%%%%
\subsection{Experimental infrastructure}
\label{sec:infrastructure}
The experimental infrastructure of the COBRA demonstrator is designed to be remotely controllable. Special care has been taken to include the infrastructure into the shielding concept and the low-background operation.

\subsubsection{Uninterruptible power supply}
Several UPS (\textbf{u}ninterruptible \textbf{p}ower \textbf{s}upply) units are used to ensure stable operation of the experiment, protecting it from rare external voltage breakdowns. Furthermore, these units stabilize and clean the voltage levels.
The UPS units are installed on both floors.
All aspects of them can be controlled remotely. Fully loaded, they can run the experiment for at least 20 minutes if necessary.
Extra UPS power lines without circuit breakers were installed to supply the UPS units because when supplied by the LNGS main line, a high residual current caused occasional power cut-offs. 

\subsubsection{Voltage supplies}
Different voltages are needed to run the experiment. Therefore several ISEG and WIENER low noise voltage supply devices working in an MPOD module frame are placed close to the experimental setup on the ground floor.\\
These cables are carefully shielded and fed through the chute included in the electromagnetic shielding. Other voltages, such as the supply of the thermal resistor in the nitrogen dewar are less critical. Nevertheless a shielded cable is used here as well to prevent a coupling of distortions to the other voltages. All voltage supply modules can be controlled remotely.

\subsubsection{Nitrogen-flushing} 
\label{sec:nitrogen_flushing}
\begin{figure}
 \includegraphics[width=0.99\columnwidth]{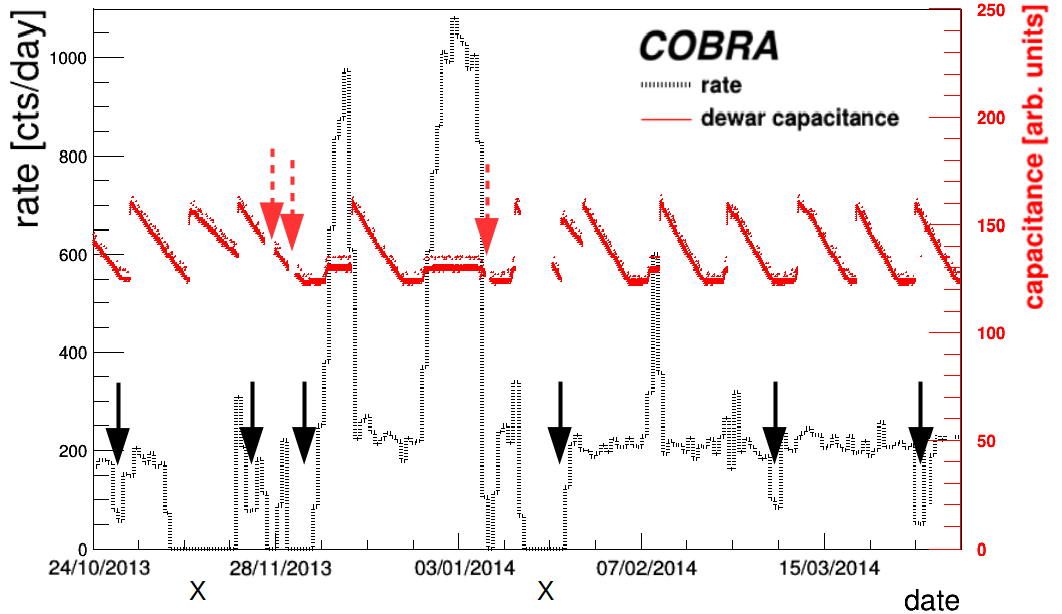}
  \captionsetup{width=0.9\textwidth} 
  \caption[Effect of nitrogen-flushing on the data rate of the demonstrator]
         {Effect of nitrogen-flushing on the measured data rate of the whole COBRA demonstrator. The solid curve shows the filling level of the nitrogen dewar, for details see text. The rate of events over \SI{500}{keV}surviving data cleaning cuts is shown in dashed. 
  The dashed arrows show power cut-offs due to residual currents, the solid ones calibration measurement periods. The X indicate major upgrade works on the setup. Taken from \cite{Ebert:2015pgx}}
 \label{fig_nitrogen_flushing}
\end{figure}

To prevent dust and especially radon settling on the detector surfaces, the inner part of the shielding is constantly flushed with evaporated purified dry nitrogen at a typical rate of \SI{5}{l/min}. The dewar of liquid nitrogen is located outside the building. It has a thermal resistor in the liquid phase of 
the nitrogen that can be heated remotely. The nitrogen used for flushing the setup is taken from the gas phase in the dewar, filtered, and fed through tubes into the setup. 
The filtration of the gaseous nitrogen is done by an activated carbon filter which is installed at the bottom of the dewar. 
The filling level is measured with a long pipe-shaped capacitor whose capacitance is directly proportional to the dielectric within (i.e. nitrogen filling level). This capacitor does not reach the bottom of the vessel, because of the activated carbon filter. Consequently, there is still some 
liquid nitrogen left in the vessel, even if the measured capacitance has reached its lowest level. When the liquid nitrogen is used up completely, the temperature of the dewar vessel increases from the temperature of liquid nitrogen to ambient temperature. This will cause the capacitance to increase a little due to thermal effects \cite{muenstermann_2007_phd, alex}.
The effect of the nitrogen-flushing and its rare failure on the measured data rate can be seen in \autoref{fig_nitrogen_flushing}.
The rate is fairly constant over time, but it rises abruptly when the nitrogen-flushing fails. If the flushing is working again, the data rate falls to its lower level quickly as the nitrogen removes radon, its decay products and other contamination from the detectors. 
Another effect of the nitrogen-flushing is a very dry atmosphere of typically \SI{2}{\%} relative humidity, which allows to cool the preamplifier devices without the risk of condensation.
During the upgrade shift in November 2013, the voltage supplies, and hence the nitrogen heating, had been switched off so the liquid nitrogen boiled off slower, which can be seen in a lower slope of the red curve. During the shift in January 2014, no filling level data was measured because the whole experiment was shut down completely for the installation of the UPS power lines. The dewar vessel is refilled bi-weekly by a service provider on site. 

\subsubsection{Preamplifier cooling}
The preamplifiers produce approximately \SI{30}{W} of waste heat. This heat cannot dissipate passively and results in a higher temperature of those devices which deteriorates their signal 
quality. To prevent this warming, and to even cool the devices below ambient temperature, a cooling system was installed. Metal plates that are being flushed with cooled water are placed above all preamplifier devices (i.e. in between them as they are placed on top of each other). The 
water cooling system Julabo FL 601 is placed outside the building so that the waste heat does not affect the experiment. The actual working condition of the cooling system can be controlled remotely.

\subsubsection{Pulse generator}
A Berkeley Nucleonics PB-5 Precision Pulse Generator is installed, which can inject defined electric pulses into each of the 128 preamplifier channels at a time. This device is used for two main reasons.
One is to monitor the functionality and long-term stability of the electronic readout system. Therefore, a sequence of defined pulses is used in each data-taking run.
The other reason is to inject pulses to synchronize the FADCs. This timing precision is needed for coincidence analysis, which is used to identify events happening at the same time in several detectors.
The pulse generator is located on the second floor. The signals are transmitted differentially directly to the preamplifier devices, where they are converted to single-ended signals. All pulse generator pulses are flagged as such in the meta data by the FADCs, so that they can be discriminated against pulses from physics events.

\subsubsection{Calibration}
\begin{SCfigure}
 \includegraphics[width=0.8\columnwidth]{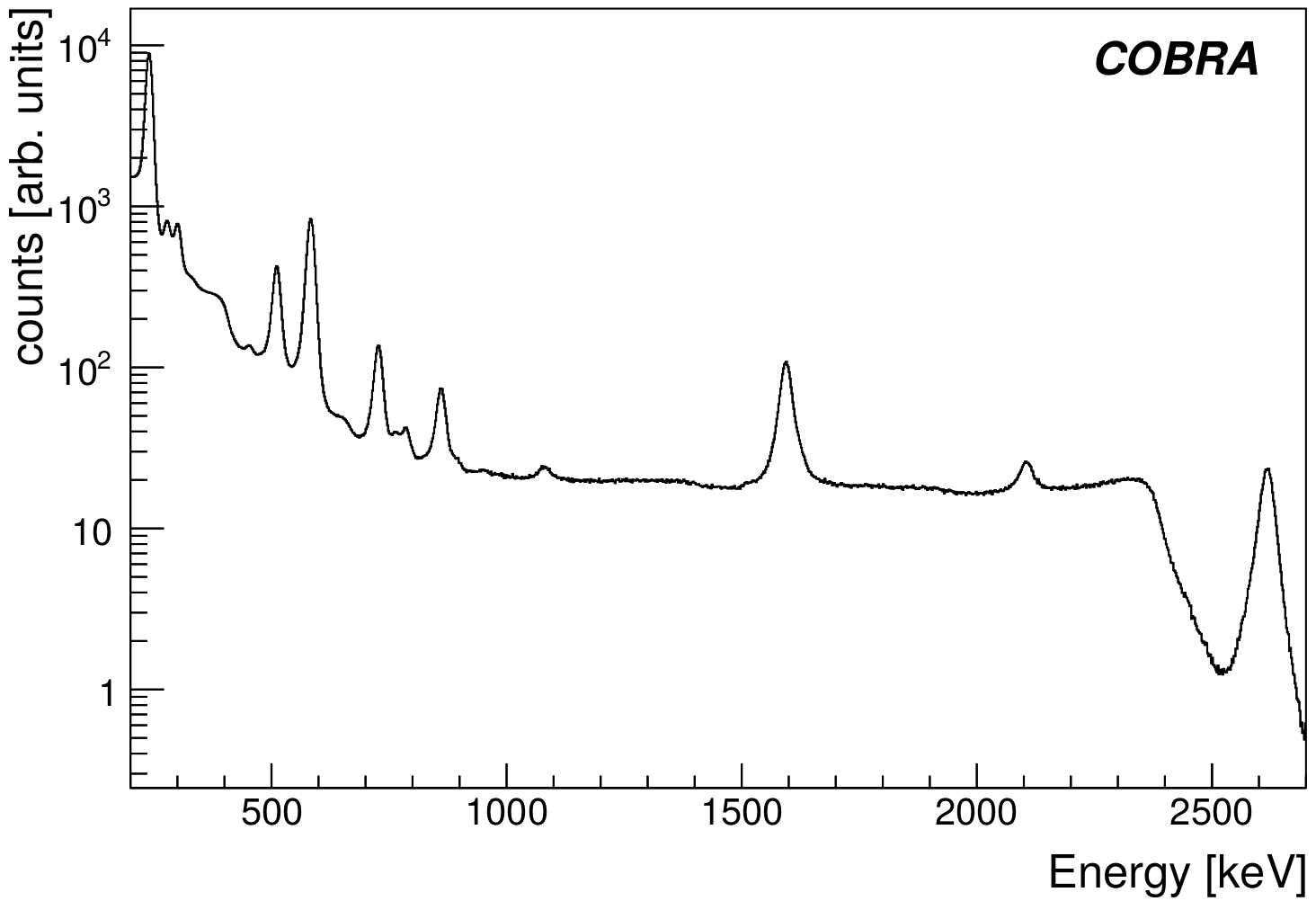}
 \caption[Calibration spectrum of the COBRA demonstrator]
         {Exposure-weighted \isotope[228]Th calibration spectrum collected with the COBRA demonstrator. Taken from \cite{Ebert:2015pgx}.}
 \label{fig_calibration}
\end{SCfigure}

Radioactive wire sources of \isotope[228]Th and \isotope[22]Na are used regularly to calibrate the detectors. Five Teflon tubes run through the shielding layers to the detectors, so that the calibration sources can be placed in the center above and below all four detector layers.
All tubes are arranged in a curved way, so that the shielding layers are not perforated in a straight line pointing directly to the detectors.
Each data-taking period has its own pre- and post calibration at the beginning and end, respectively. An example of a spectrum taken with a \isotope[228]Th source is shown in \autoref{fig_calibration}. The contributions from each detector are weighted according to their exposure collected during the physics runs.
 
\subsubsection{DAQ-server}
The main computer used for the experiment is located on the second floor. It can be accessed from outside via a network, and it provides all main services, like running the data acquisition and communication to all sub-systems.
The typical duration of each data-taking run is \SI{4}{h}. The data is stored in the ROOT (\textsc{root}: c++-based
data analysis framework by CERN (European Organization for Nuclear Research)) file format and is copied and backed up regularly to external servers. 

\subsubsection{Monitor system}
A set of sensors are installed to monitor and log the environmental conditions of the experiment.
Several sensors measure the temperature, humidity and pressure in the ground floor of the building, within the outer shielding on top of the preamplifiers, and in the inner shielding on top of the lead bricks. 
In particular, the preamplifier cooling and nitrogen-flushing are monitored.
Furthermore, the preamplifier's supply voltage is monitored, as well as the filling level and temperature of the nitrogen dewar vessel. Several cameras allow for real-time images of the setup.

%%%%%%%%%%%%%%%%%%%%%%%%%%%%%%%%%%%%%%%%%%%%%%%%%%%%%%%%%%%%%%%%%%%%%%%%%%%%%%%%%%%%%%%%%%%%%%%%%%%%%%%%%%%%%%%%%%%%%%%%%%%%
\subsection{Performance of the demonstrator}
\label{sec:performance}
\begin{figure}
 \includegraphics[width=0.99\columnwidth]{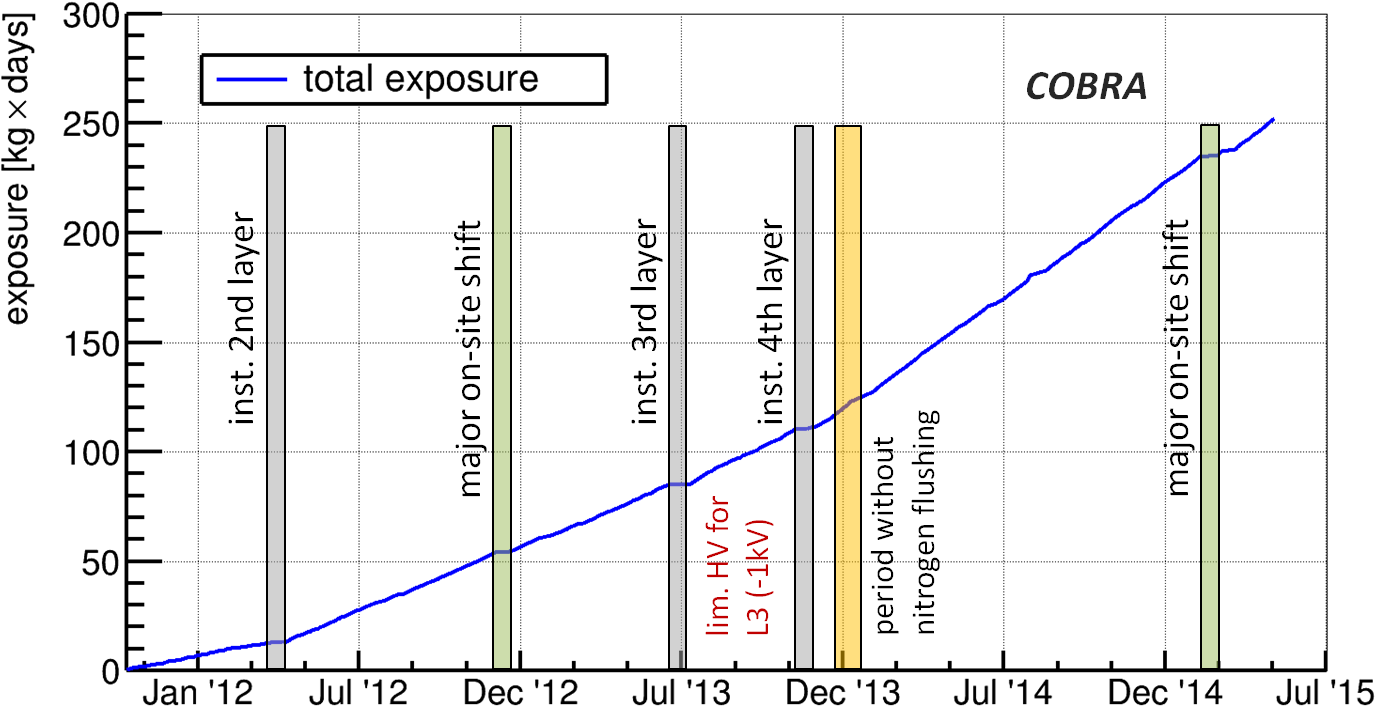}
  \captionsetup{width=0.9\textwidth} 
  \caption[Exposure of the COBRA experiment]
         {Exposure of the COBRA experiment, indicated are the upgrade works on site, notable periods where the nitrogen-flushing failed and detector layer 3 had to be operated with lower bias voltage. Taken from \cite{Ebert:2015pgx}}
 \label{fig_exposure}
\end{figure}

Currently 61 of the 64 detectors are functional, which corresponds to a yield of more than \SI{95}{\%}. Two of those three detectors suffer from faulty voltage contacts.\\
Under normal data-taking conditions, i.e. without periods of larger external power cuts, the demonstrator's life-time is only reduced by calibration measurements. These take typically one day per month, so the data taking efficiency is above \SI{95}{\%}.
In summer 2015, more than \SI{250}{kg\,d} of high quality low-background physics data have been collected, which are used for physics analysis.
\autoref{fig_exposure} shows the collected exposure as a function of time. Periods without data-taking are indicated.
The average count rate of the COBRA demonstrator above \SI{500}{keV} is \SI{225}{counts/day}, corresponding to a rate of \SI{3.6}{counts/(day \cdot detector}), see \autoref{fig_nitrogen_flushing}. In comparison, if one such detector is operated unshielded on the earth's surface under similar conditions, the rate is approximately \SI{2e4}{counts/day}.
This corresponds to a reduction of four orders of magnitude and is a result of operation in an underground laboratory, the shielding concept and selecting low-background materials.\\
Auxiliary measurements are performed to quantify the quality of the different components of the readout chain. A pulse generator injects pulses to the FADCs directly, then the pulses are converted to differential signals and vice versa before being measured by the FADCs. 
Finally, the whole DAQ chain of preamplifier, differential signaling, linear amplifier and FADC is tested. The amplitudes of the pulse heights are filled into a histogram. The FWHM of the resulting distribution is a measure for the resolution of the devices under test. The resolution of the FADC has a FWHM of \SI{0.05}{\%}. The conversion to differential signaling and vice versa results in a resolution of \SI{0.1}{\%}.
The resolution of the preamplifier devices cannot be measured directly, only in conjunction with the whole DAQ chain. The total resolution varies between \SI{0.4}{\%} FWHM and \SI{1.0}{\%} FWHM due to the variation of the different components used in the different channels. 
The resolution of the preamplifier channels can hence be estimated to be between \SI{0.3}{\%} and \SI{0.9}{\%}.\\
As can be seen in \autoref{fig_calibration}, the energy resolution of the \isotope[208]Tl peak at \SI{2615}{keV} is \SI{1.4}{\%} FWHM. The best detector has a resolution of \SI{0.8}{\%}.

%%%%%%%%%%%%%%%%%%%%%%%%%%%%%%%%%%%%%%%%%%%%%%%%%%%%%%%%%%%%%%%%%%%%%%%%%%%%%%%%%%%%%%%%%%%%%%%%%%%%%%%%%%%%%%%%%%%%%%%%%%%%
%%%%%%%%%%%%%%%%%%%%%%%%%%%%%%%%%%%%%%%%%%%%%%%%%%%%%%%%%%%%%%%%%%%%%%%%%%%%%%%%%%%%%%%%%%%%%%%%%%%%%%%%%%%%%%%%%%%%%%%%%%%%
\section{Additional work at LNGS}
\label{sec_addition_work_lngs}
The work at the demonstrator during thesis comprised completing and improving the experiment.\\
The main topics were the installation and commissioning of detector layers two to four and improving the performance, especially concerning the signal quality and noise reduction.

At the final stage of the COBRA demonstrator, the detector layer three suffers from distortions.
It is not clear why. By exchanging all parts of the readout electronics, the reason could be excluded there. Most probably, the problems arise at the detector layer itself.

\subsection{Signal quality: Ground connections and shielding\\ against electromagnetic interferences}
To improve the signal quality, special care had to be taken to ground-connections and potential ground-loops.
The main (master) ground is at the electronics rack next to the setup. All cables at the rack have a shielding layer set on this ground potential. 
An extra metal-ground connection and all cables going into the setup are installed in a big cable harness to the chute of the EMI shielding.
Within this shielding, the cables go to the housings of the preamplifier devices. All these are electrically tightly connected to each other, and via the other cable shields to the main ground at the chute. 
The PCB (\textbf{p}rinted \textbf{c}ircuit \textbf{b}oard) of the preamplifier devices consist of four layers, one is a dedicated ground layer. 
It is tightly connected to the housing, but only at one side to avoid uncontrolled currents on the PCB.\\ 

\begin{SCfigure}
 \centering
  \includegraphics[height=0.25\textheight]{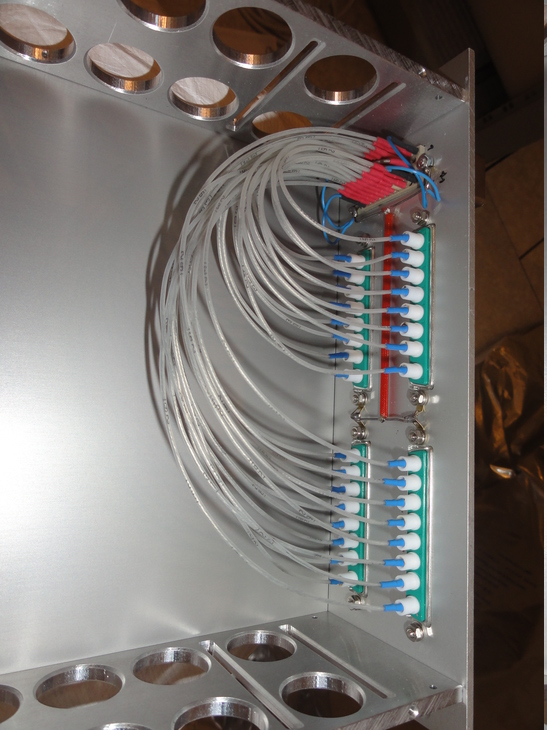} 
  \includegraphics[height=0.25\textheight]{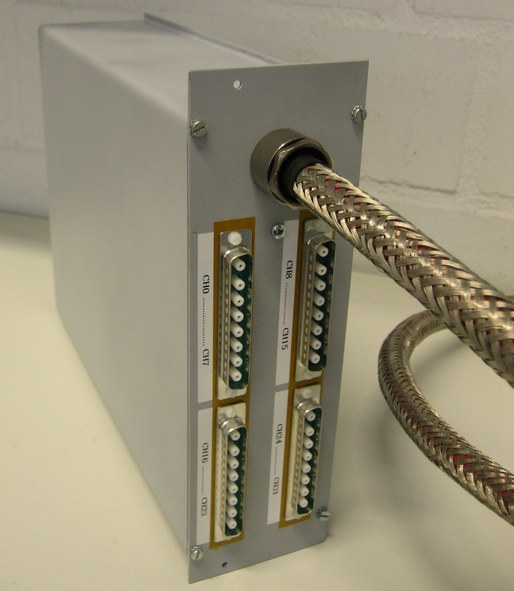} 
  \caption[Weakest part of the shielding]
	  {Left: weakest part of whole shielding-chain: unshielded adapter for bias supply in the electronics rack. It was replaced by a fully-shielded version (right). Note the inflexible Redel-cable with the extra metal-mesh shielding.}
  \label{fig_golden_box}
\end{SCfigure}

Nearly all cables that are used are coaxial or twisted pair cables. Some of them have only one metal-shield for signal (back-)flow. 
At all these cables an extra metal-mesh shielding was added. It is a dedicated shielding element to differentiate between the signal flow and shielding.
Extensive investigations were done to test where the shielding of the cables shall be connected to improve the shielding quality without allowing uncontrolled signal flows.
The only unshielded part of the whole EMI-shielding architecture was the adapter for the cables of the voltage supply device (``Redel-cable'') to match the COBRA-standard (8W8):
These adapters are put into the rack in a NIM-housing, because of mechanical stability, as the Redel-cables are very inflexible.
The adapters were totally unshielded: the housings were made of aluminum with big holes for ventilation. This is unnecessary, as this adapter contains only soldered cables between two connectors without any electronics. 
The complete shielding architecture is seriously threatened by this unshielded element directly in the rack. There are many active electronic devices which produce much EMI, and the whole EMI shielding functionality is dominated by the weakest element.
As a consequence, the unshielded adapters were replaced by fully shielded ones, see \autoref{fig_golden_box}, resulting in a much better noise behavior.

\subsection{Signal quality: Heat-dissipation}
Another issue is the heat-dissipation of the setup.\\
For two detector layers, the temperature in the ground floor building was around \SI{30}{}{\celsius}, in the surrounding of the preamplifier devices even more. 
An increased temperature affects the signal quality.
The waste heat of the completed demonstrator would be too much. Hence, improvements had to be made before installing layers three and four to solve this problem. 
A cooling of the ground floor building was not possible. Consequently, it was decided to use the air-cooled room in the upper floor level for a part of the DAQ system. 
Additionally, a water cooling-system for the preamplifier devices was installed to keep them at ambient temperature or slightly below (water cooling-system built by \cite{Oldorf:222041}). 
The cooling device is standing outside of the ground floor building, so that its waste heat does not affect the experiment.
There is the risk of condensation at the cooling plates (especially in the preamplifier devices), but it is mitigated by the nitrogen-flushing which reduces the humidity drastically to some percent.

The largest power-consumer is the VMEbus (\textbf{V}ersa \textbf{M}odule \textbf{E}urocard-\textbf{bus})-crate with its FADCs and VMEbus-computer. It was tested if the signal quality is unharmed if the linear amplifiers and the following VMEbus-crates are moved to the upper floor.
The signals are transmitted using differential signal transmission between the preamplifier devices and linear amplifier, which allows long distances.
To test the transmission, the DAQ electronics needed for layers three and four were installed in the upper floor, without installing the layers.
Furthermore, extra signal cables were installed to the upper floor.
By this, one could test the signal quality in the lower floor with signals going to the upper floor and back. 
Furthermore, a direct comparison between the signal quality of the DAQ completely in the ground floor building, and separated between both floors was possible.

A source of signal quality deterioration was the use of the normal mains at the upper floor level for supplying the electronics rack there. A potential difference of the ground and neutral connections between the two floors resulted in uncontrolled currents causing distortions. 
The installation of dedicated power lines from the lower floor to supply the electronics rack in the upper floor solved this problem.\\
Eventually, it was possible to run the experiment with the DAQ electronics separated between the two floors without any deterioration of the signal quality.

To install the cables running to the upper floor, a feedthrough in the walls was needed. It is used for the tubes for the preamplifier cooling system as well.
A hole in wall of the ground floor building was found. Only minor enlargements were necessary to use it. The wall is ca. \SI{40}{cm} thick. It consists of several layers of clay bricks, fabric and even lead.\\

Later, the UPS devices were installed to supply the electronics racks on both floors. 
The UPS units need dedicated power lines without residual current circuit breakers. These lines, capable of \SI{32}{A}, were built from the fuse box next to the upper floor building to the UPS units at both racks. 
By this, the experiment is protected from external voltage supply problems. Furthermore, all devices in the rack have the same ground and neutral potentials. 
The cables supplying the upper floor rack with the ground and neutral potential from the lower floor rack are obsolete.

\subsection{Signal quality: Raw signals}
\begin{SCfigure}
 \centering
  \includegraphics[width=0.69\textwidth]{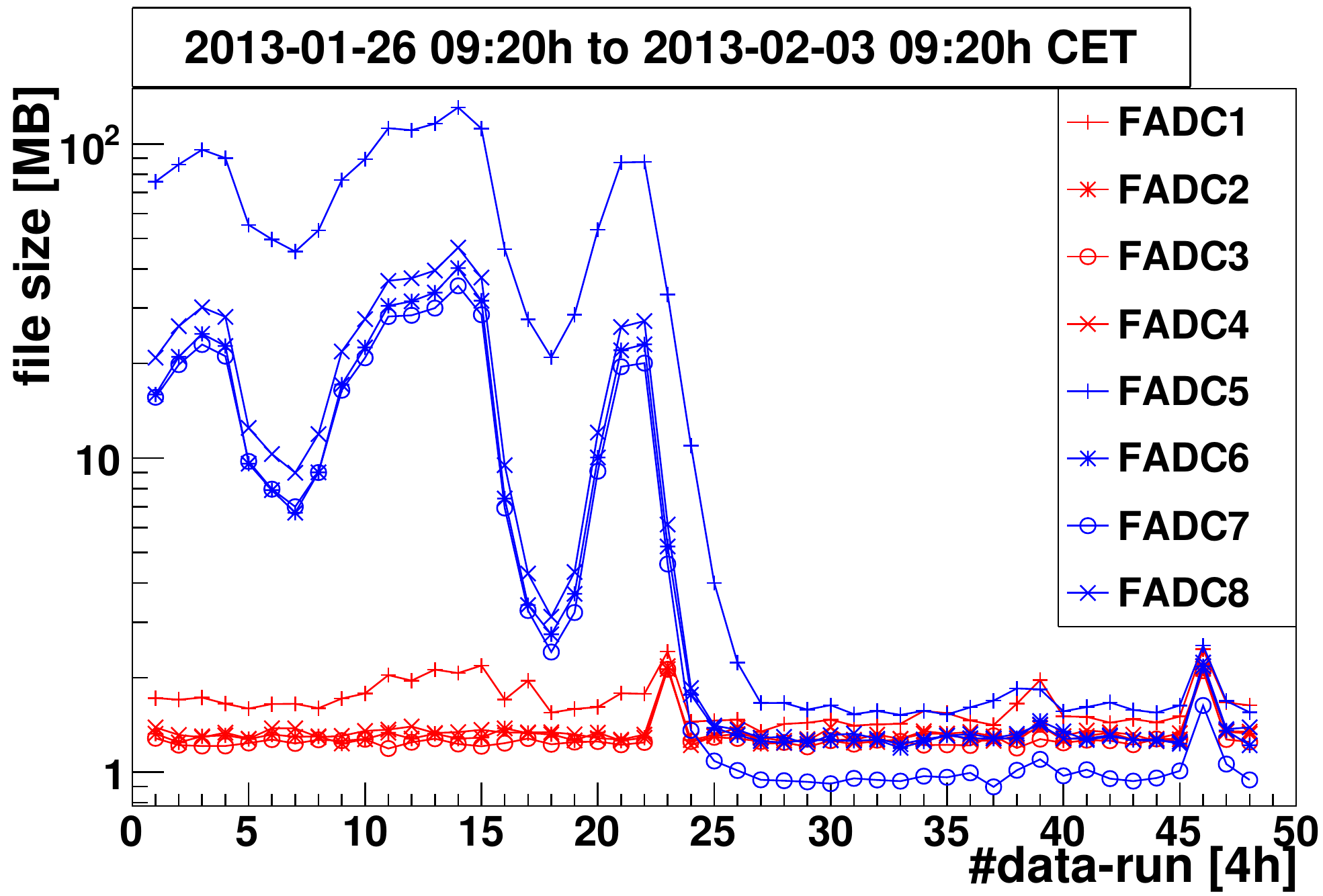}
  \caption[Variation of data rate after LNGS-shift in January 2013]
	  {Variation of the measured data rate after an LNGS-shift in January 2013. The red curves are the FADCs of layer one, the blue ones of layer two. The lines are just for better orientation.}
  \label{fig_lngs_data_rate_after_shift}
\end{SCfigure}

The demonstrator suffered from noise and distortions at certain times that could not be explained, as the complete experiment was operated without any changes. 
The data rate is a measure for the signal quality, as no significant variations are expected in the low-background physics-mode.
The variation of the file size is shown exemplarily in \autoref{fig_lngs_data_rate_after_shift} for the first data-taking runs (\SI{4}{h} each) after a working shift at the demonstrator.
At that time, only the detector layer one (FADCs one to four) and layer two (FADCs five to eight) were installed. 
An oscillation with a periodic time of 9-11\,bins = 36-44\,h occurred. The file size changed up to a factor of 20.
Layer two is affected much more than layer one, while the first FADCs of each layer (number one and five) are affected significantly more than the others. 
The trigger thresholds were set to \SI{28}{ch} and \SI{38}{ch} for detector layers 1 and 2, respectively. 
Without any apparent reason, the dramatic variation settled down.
This behavior complicates the works at site, because possible hardware changes or extra installations are validated often using the measured data rate. This procedure is not valid if the data rate oscillates that strong.

\begin{figure}[b!]
 \centering
  \includegraphics[width=0.49\textwidth]{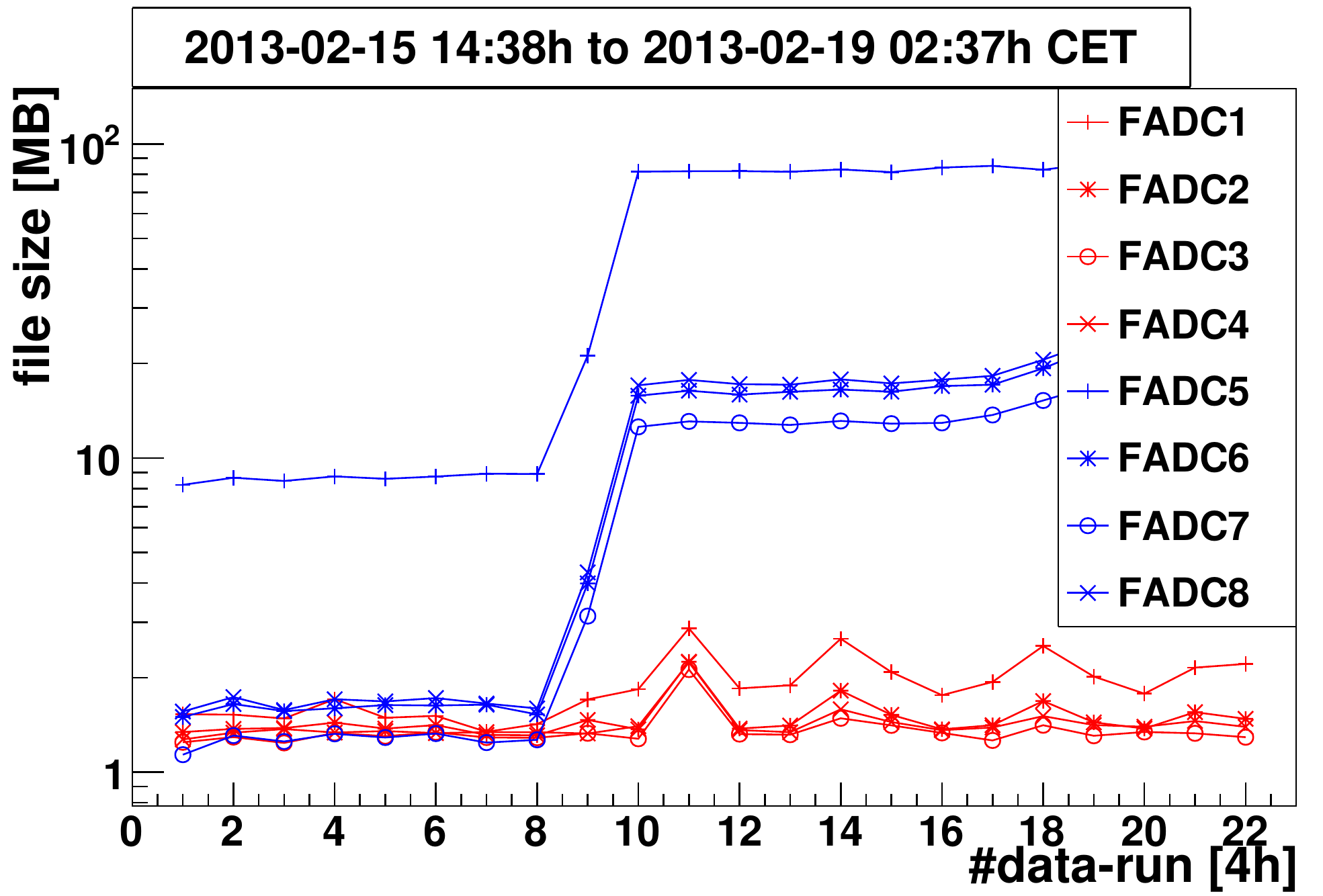}
  \includegraphics[width=0.49\textwidth]{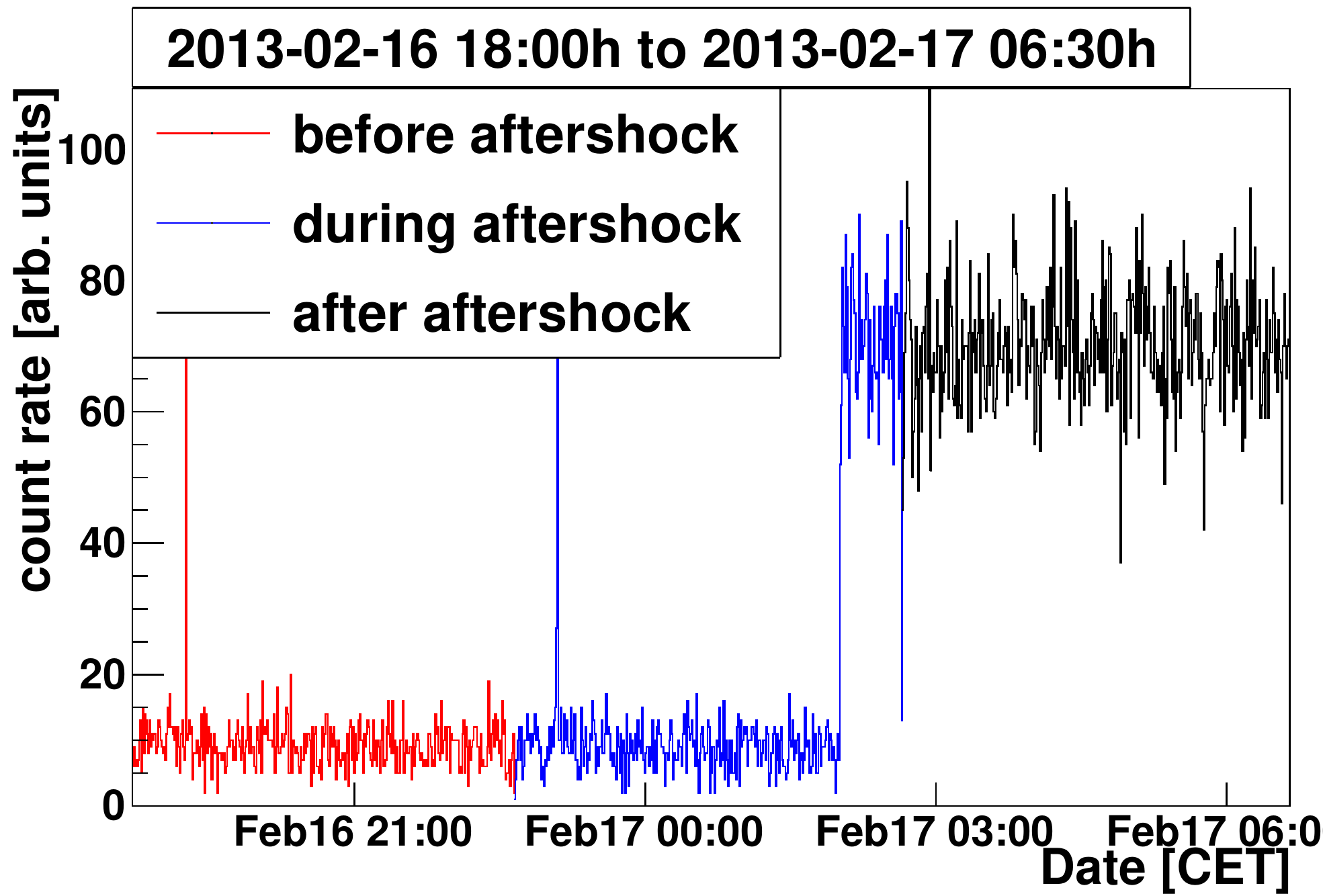}
  \captionsetup{width=0.9\textwidth} 
  \caption[Effect of earthquakes on data rate]
	  {Effect of earthquakes on the measured data rate. Left: file size of each data-taking run, analog to \autoref{fig_lngs_data_rate_after_shift}. Right: Count-rate of the data-taking runs of FADC 5 before, during and after the aftershock. The spike in each run stems from the pulse generator. The decline between the runs is due to the beginning of a new run. The earthquake at 22:16h cannot be seen, the aftershock at 02:00h results in the sharp rise.}
  \label{fig_lngs_data_rate_earthquake}  
\end{figure}

About two weeks later, a series of earthquakes happened in the Abruzzo region. These changed the data rate drastically again, although no obvious effects of the earthquake could be seen at the setup.
The effect was once again seen most clearly at detector layer two:
An earthquake occurred at 22:16h in the village of Sora, some \SI{80}{km} away from the demonstrator, with a moment magnitude scale of 4.9 in a depth of \SI{10}{km} \cite{earthquake_2013-02-16_21-16-09}, which cannot be seen in the data rate. 
A strong aftershock happened at 02:00h, \SI{10}{km} away from the LNGS with a Richter scale magnitude of 3.8 in a depth of \SI{16}{km} \cite{earthquake_2013-02-17_01-00-07}. 
A sudden rise and a periodic file size variation occurred due to this aftershock, especially at layer 2, most prominently at FADC 5.
The periodic time is now only about \SI{12}{h}, and file-size variation is much smaller, which can be seen clearly in \autoref{fig_lngs_data_rate_earthquake}.

\begin{figure}
 \centering
  \includegraphics[height=0.415\textwidth]{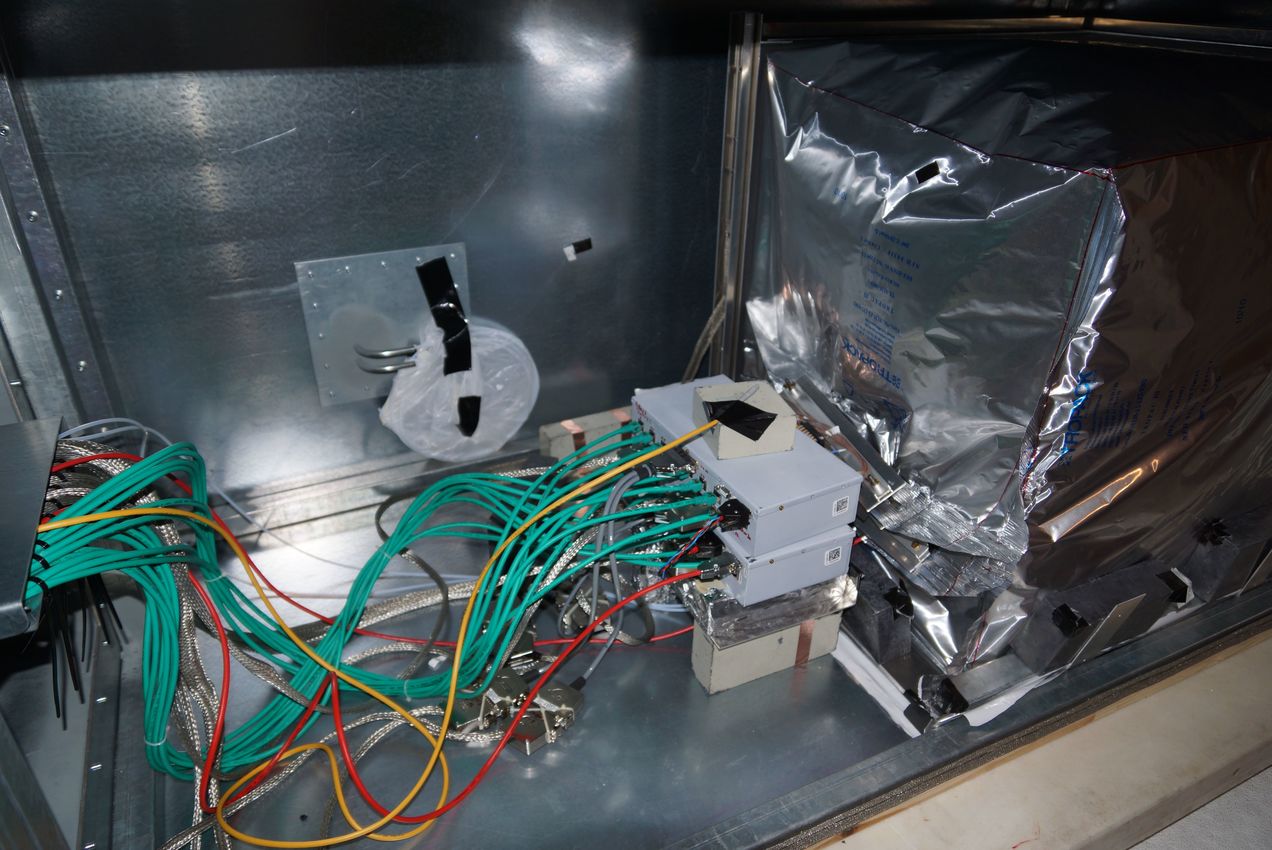}
  \includegraphics[height=0.415\textwidth]{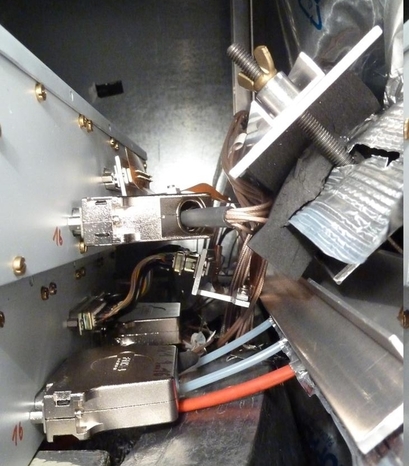}
  \captionsetup{width=0.9\textwidth} 
  \caption[View of the COBRA demonstrator at the stage of two installed layers]
	  {Left: Total view of the COBRA demonstrator at the stage of two installed layers. 
           Right: Details of the noise-critical areas are at the feedthroughs of the air-tight sealed compartment, especially if the bloating changes.}
  \label{fig_demonstrator_L2_radon_shield}
\end{figure}

Furthermore, one was often faced with non-repeatable results when working on the setup (in addition to the oscillating data rate). 
This happened especially when changing the cabling and positioning of the preamplifier devices and their cables to the detectors.\\
Concerning this issue, the main weak points were identified as the raw detector signals until reaching the preamplifier devices.
Furthermore, the arrangement of those devices and the feedthrough of the cables through the air-tight sealed compartment were critical, as well. 
This consisted of a metalized radon-tight foil, the feedthrough was made of gas-tight rubber foam with pressure contacts, see \autoref{fig_demonstrator_L2_radon_shield}. 
The complete compartment was bloating because of the nitrogen flushing. If the flushing changes, the bloating changes as well. As a consequence, the positions of the cables for the raw-signals changes.

\begin{SCfigure}
 \centering
  \includegraphics[width=0.6\textwidth, angle =-90]{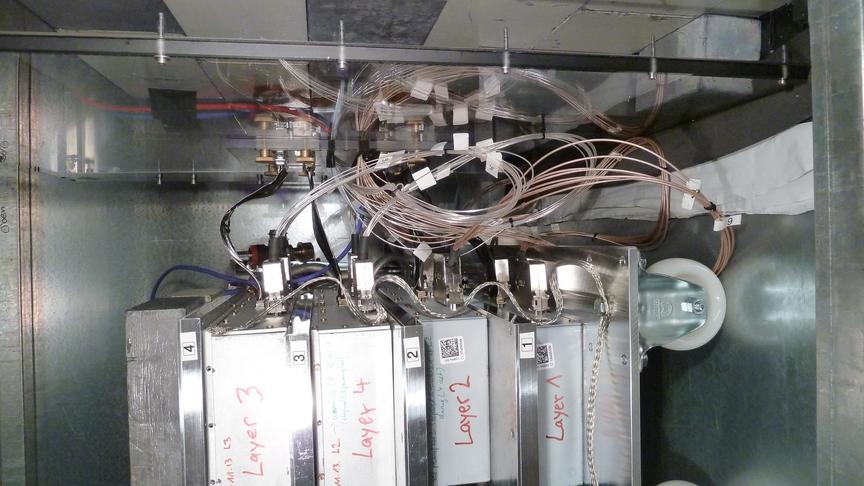}
  \includegraphics[width=0.6\textwidth, angle =-90]{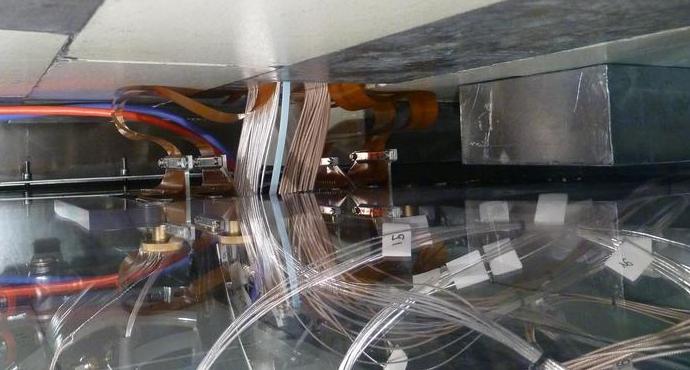}
  \caption[Detailed view of the noise-critical areas at final stage]
	  {Detailed view of the noise-critical areas at the final stage. Left: From the preamplifier devices to the polycarbonate box: Long BV cables are curled up. 
          Right: From the polycarbonate box (producing mirror images) to the lead castle: The BV cables are running straight here, the Kapton cables winding.}
  \label{fig_demonstrator_L4_radon_shield}
\vspace{2cm}
\end{SCfigure}

A major improvement was the replacement of the radon-tight foil with a polycarbonate glass housing with connectors as feedthroughs (constructed in \cite{gehre_diss}), see \autoref{fig_demonstrator_L4_radon_shield}. By this, the position of the cables is in a fixed defined order.

The raw detector signals are transmitted on Kapton cables. Due to restrictions in the production process, these have a maximal length of approximately \SI{57}{cm}, which is just long enough to leave lead-shielding. The BV cables (RG178 shielded coaxial cables) are much longer to reach the preamplifier devices safely, approximately \SI{1}{m}.
As only layer 1 was installed, the Kapton cables could be connected directly to the preamplifier devices, going through the metalized foil. 
As later more layers were installed, the Kapton cables did not reach all preamplifier devices, so that extension ribbon-cables had to be used.
After the installation of the polycarbonate glass shielding, the Kapton cables had to go only to the inner connector of the glass wall. Ribbon-cables in different lengths were used to connect the connectors from the glass-shielding to the preamplifier devices. 
The extensions were improved to be in a reliable cable-ordering.\\ 
The BV cables are too long, so they had to be coiled up. This may also change the position and distortions when working at the setup and cause non-repeatable results.
Consequently, the arrangement of those cables had to be done with care.

%%%%%%%%%%%%%%%%%%%%%%%%%%%%%%%%%%%%%%%%%%%%%%%%%%%%%%%%%%%%%%%%%%%%%%%%%%%%%%%%%%%%%%%%%%%%%%%%%%%%%%%%%%%%%%%%%%%%%%%%%%%%
\FloatBarrier
\section{Results of the COBRA demonstrator}
\label{sec_results_COBRA}
Two papers analyzing the data measured with the COBRA demonstrator were published lately:
One paper investigates the stability of the performance of the detectors using \isotope[113]Cd \cite{Ebert2016}.
The other uses a Bayesian analysis to estimate the signal strength of the $0\nu\beta\beta$-decays \cite{Ebert:2015rda}.\\
The two papers are summarized here very briefly.\\

Natural Cd contains \SI{12.2}{\%} of \isotope[113]Cd, which is a non-unique, fourfold forbidden beta-decay:
\begin{equation}
   \ce{^{113}Cd^{1/2+} -> ^{113}In^{9/2+}}
\end{equation}
Consequently, the half-life of the decay is very long: \SI{8.00(35)E15}{yr} \cite{Dawson:2009ni}.
The Q-value of the decay is \SI{322.2(12)}{keV} \cite{Dawson:2009ni}. This decay is the strongest signal, causing about \SI{98}{\%} of all measured events. 
Due to the low Q-value, this decay can only be measured, if the trigger thresholds are as low as possible. Therefore, the electric noise level has to be under control.\\
\isotope[113]Cd is distributed homogeneously within the detector volume. 
Furthermore, its half-life is long compared to the experiment life-time. Consequently, the decay-rate of \isotope[113]Cd can be considered as constant.
Concluding, one can state that it is a good measure for the stability of detector performance.\\
For this analysis, an experimental life-time of \SI{3.5}{yr} with a collected exposure of \SI{218}{kg\,d} data was used.
A variation of the count rate was found in the lower percent level per year ($\Delta_{\text{mean}} = \SI{0.5 \pm 0.4}{\%})$ for 45 out of 48 detectors.
This fulfills the desired stability requirements of less than $\SI{\pm1}{\%}$ per year.\\

The results of a peak-search for the five $0\nu\beta\beta$-decay ground state to ground state transitions (\isotope[114]Cd, \isotope[128]Te, \isotope[70]Zn, \isotope[130]Te and \isotope[116]Cd) were published:\\
A dataset of \SI{234.7}{kg\,d}, collected between September 2011 and February 2015, was used for a Bayesian analysis to estimate the signal strength of the $0\nu\beta\beta$-decay.
No signal was found. Consequently, half-life limits at \SI{90}{\%} credibility were calculated, shown in \autoref{tab_results_cobra_demonstrator}:
\begin{table}[h!]
 \centering
  \begin{tabular}{|c|c|c|c|c|}
      \hline
      \makecell{isotope}  & \makecell{N\\{[}\num{E23} atoms/kg]} & \makecell{$B$\\ {[}\cpkky]} & \makecell{$T_{1/2}$ \SI{90}{\%} C.L. \\ {[}\SI{E21}{yr}]} & \makecell{K} \\\hline
      \isotope[114][]{Cd} & 6.59                                 & $ 213.9^{+1}_{-1.7}$        & 1.8                                                        & 0.07 \\
      \isotope[128][]{Te} & 8.08                                 & $65.5^{+0.5}_{-1.6}$        & 2.0                                                        & 0.17 \\
      \isotope[70][]{Zn}  & 0.015                                & $45.1^{+0.6}_{-1}$          & \num{7.4E-3}                                               & 0.06 \\
      \isotope[130][]{Te} & 8.62                                 & $3.6^{+0.1}_{-0.3} $        & 6.7                                                        & 0.14 \\
      \isotope[116][]{Cd} & 1.73                                 & $ 2.7^{+0.1}_{-0.2}$        & 1.2                                                        & 0.27 \\\hline %\hline
  \end{tabular}
  \captionsetup{width=0.9\textwidth} 
  \caption[Results of the COBRA demonstrator for the $0\nu\beta\beta$-decay]
          {Results of the Bayesian signal estimation including all systematic uncertainties. The second column shows the number of atoms per kg (natural abundance). In the third column the background index $B$ for the different ROIs is presented. The fourth column reports the limit at \SI{90}{\%} credibility, the last column shows the calculated Bayes factor. A value below one indicates that the signal hypothesis is disfavored against the hypothesis for background only. Taken from \cite{Ebert:2015rda}.}
 \label{tab_results_cobra_demonstrator}
\end{table}

\chapter[Interaction and range of particles as input for surface sensitivity investigations]{Interaction and range of particles in matter as input for surface sensitivity investigations}
\chaptermark{Interaction and range of particles in matter} 
\label{sec_particles_matter_ranges}

In this chapter the interaction of particles in matter is discussed \cite{Knoll_radiation_detection_2000}.
For the following \autoref{sec_surface_events}, details to the interaction principles and the determination of the range of particles in matter is of importance.
Of special interest are alpha and gamma radiation, as well as electrons with the energies of \SIlist{480; 976; 1680}{keV}, which are the energies of the IC (\textbf{i}nternal \textbf{c}onversion) electrons of \isotope[207]Bi.\\ 
That is why several methods to obtain these values are presented here. Two different principles are used:
First, calculations using the NIST (\textbf{N}ational \textbf{I}nstitute of \textbf{S}tandards and \textbf{T}echnology)-database are shown, and secondly the following MC (\textbf{M}onte \textbf{C}arlo) simulation programs:
\begin{enumerate}
   \item VENOM (\textbf{v}icious \textbf{e}vil \textbf{n}etwork of \textbf{m}ayhem) for electrons and protons, standard COBRA interface based on GEANT4 (\textbf{ge}ometry \textbf{an}d \textbf{t}racking 4), suitable for all kind of particles 
   \item SRIM (\textbf{S}topping and \textbf{r}ange of \textbf{i}ons in \textbf{m}atter) for ions (alpha  and proton radiation) \cite{SRIM}
   \item PENELOPE (\textbf{p}enetration and \textbf{ene}rgy \textbf{lo}ss of \textbf{p}ositrons and \textbf{e}lectrons) for electron simulation \cite{PENELOPE}
   \item CASINO (monte \textbf{ca}rlo \textbf{si}mulation of electro\textbf{n} trajectory in s\textbf{o}lids) for electron simulation \cite{CASINO}
\end{enumerate}
At the end a brief comparison of the obtained values is given with respect to the particles and energies used in \autoref{sec_surface_events}.

%%%%%%%%%%%%%%%%%%%%%%%%%%%%%%%%%%%%%%%%%%%%%%%%%%%%%%%%%%%%%%%%%%%%%%%%%%%%%%%%%%%%%%%%%%%%%%%%%%%%%%%%%%%%%%%%%%%%%%%%%%%%
%%%%%%%%%%%%%%%%%%%%%%%%%%%%%%%%%%%%%%%%%%%%%%%%%%%%%%%%%%%%%%%%%%%%%%%%%%%%%%%%%%%%%%%%%%%%%%%%%%%%%%%%%%%%%%%%%%%%%%%%%%%%
\section[Interaction of particles in matter]{Physical interaction principles of particles in matter}
\label{sec_interaction_of_particles_in_matter}

%%%%%%%%%%%%%%%%%%%%%%%%%%%%%%%%%%%%%%%%%%%%%%%%%%%%%%%%%%%%%%%%%%%%%%%%%%%%%%%%%%%%%%%%%%%%%%%%%%%%%%%%%%%%%%%%%%%%%%%%%%%%
\FloatBarrier
\subsection{Alpha radiation}
\label{sec_alpha_radiation}
Alpha particles, being considered as fast heavy charged particles, interact with matter mainly through Coulomb force of their positive charge and the negative charge of atomic shell electrons. Interactions with nuclei can be ignored.
When entering matter, heavy charged particles interact instantly with many surrounding shell electrons by energy transfer to electrons.
This results either in an excitation of electrons (raising their energy level to another shell) or ionization (removing electrons from an atom). 
The maximum energy-transfer $E_{\text{max,\ transf}}$ in a single collision is 
\begin{equation}
  E_{\text{max,\ transf}}= 4\ E_{kin}\frac{m_e}{m_{\alpha}} \approx 3\,\text{keV}\,, 
\end{equation}
where $E_{kin}$ and $m_{\alpha}$ are the kinetic energy and mass of the alpha particles, and $m_e$ the electron rest mass \cite{turner_1995}.
The transferred energy is only a very small fraction of the charged particle's initial energy of typically some \si{MeV}.
By this, the heavy charged particle gradually loses its energy and comes to rest. 
As these interactions happen isotropically and due to the high mass of the incident particle, it is constantly slowed down without large deflections from its original direction. Hence, heavy charged particles have a defined range in matter.

The stopping power $S$ is the differential energy loss $dE$ of the particle in matter, divided by the differential path $dx$. It can be calculated using the Bethe formula 
\begin{equation}
S = -\frac{dE}{dx} =  \frac{4 \pi e^2 z^2}{m_e v^2} n Z \left[ \ln \frac{2 m_e v^2}{I} - \ln \left( 1- \frac{v^2}{c^2} \right) - \frac{v^2}{c^2} \right]\,,
\label{eq_bethe}   
\end{equation}
where $v$ and $ze$ being the velocity and charge of the primary particle, $n$ and $Z$ the number density and atomic number of the absorber, $I$ its average excitation and ionization potential, and $m_e$ and $e$ the electrons' mass and charge \cite{Knoll_radiation_detection_2000}.\\

The range of alpha radiation in solids is very short, some \SI{20}{\mu m} in CdZnTe, see discussions later in this chapter.\\
One other thing one has to keep in mind is the expansion of the charge cloud as it drifts through the detector. This is especially important if the expansion is larger than the original range of particles, which is the case for alpha radiation.

A complete analytic expression for the expansion of a drifting charge cloud is difficult, an extensive discussion was done in \cite{rebber_master}. 
It is summarized here, calculations are done for the maximal values for the whole drift length of \SI{1}{cm} for near-cathode events in the COBRA standard \SI{1x1x1}{cm} CdZnTe detectors.
The results are used to obtain the approximate maximal dimension of the charge cloud, as these calculations are simplified estimations only.

The electron cloud expands as it drifts through the detector volume towards the anodes due to two main reasons:
 
One is the random thermal motion of the electrons. The spread $\sigma_{diff}$ due to this diffusion can be calculated as
\begin{equation}
 \sigma_{diff} = \sqrt{\frac{2 k_B T \mu t}{e}} \approx \SI{70}{\mu m},
\end{equation}
where $k_B$ is the Boltzmann constant, $T$ the absolute temperature, $e$ the electric unit charge, $\mu = \SI{1000}{\frac{cm^2}{Vm}}$ the charge mobility (\autoref{tab_properties_CdZnTe}), and $t=\SI{1}{\mu s}$ the drift time. The last two quantities are also calculated in \autoref{sec_drift_time}. 

The other main effect is the electrostatic repulsion due to the mutual electron charge. The maximum repulsion $R_{rep}$ for the whole drift time for near-cathode events can be approximated to
\begin{equation}
 R_{rep} = \sqrt[3]{ \frac{3 \mu N e\, t} {4 \pi \epsilon_0 \epsilon_r} } \approx 340\mu m\,, 
\end{equation}
with the number of charge carriers $N \approx \num{E6}$ for a \SI{5}{MeV} alpha particle, and the relative and vacuum permittivity $\epsilon_r=10.9$ \cite{eV_detector_material} and $\epsilon_0$.

The initial size of the charge cloud cannot be more than the range of the alpha radiation of \SI{20}{\mu m}, which can be neglected.

The total effect $R_{tot}$ is not the sum of the two individual effects. If the effects of repulsion and diffusion are independent from each other, one can use the quadratic sum \cite{1596583}:
\begin{equation}
 R_{tot} = \sqrt{(\SI{70}{\mu m})^2 + (\SI{340}{\mu m})^2} \approx \SI{350}{\mu m}.
\end{equation}

\cite{2009NIMPA.606..508B} found using GEANT4 MC simulations, that the quadratic sum overestimates the real charge cloud expansion. This is because the mutual repulsion gets smaller as the charge cloud enlarges due to diffusion.\\

Concluding, one can state that alpha radiation travels on a straight line for some ten $\mu m$ in solids. 
In CdZnTe, the expansion of the drifting charge cloud is smaller than \SI{350}{\mu m} even for the maximal drift length of \SI{1}{cm} for near-cathode events in the COBRA standard detectors.

%%%%%%%%%%%%%%%%%%%%%%%%%%%%%%%%%%%%%%%%%%%%%%%%%%%%%%%%%%%%%%%%%%%%%%%%%%%%%%%%%%%%%%%%%%%%%%%%%%%%%%%%%%%%%%%%%%%%%%%%%%%%
\FloatBarrier
\subsection{Beta radiation}
\label{sec_beta_radiation} 
Electrons must not be taken as heavy charged particles. Because they interact with other shell electrons as collision partners which have the same mass, large angular deflections and energy losses are possible. 
Furthermore, they may interact with the nuclei, which leads to drastic changes in directions. 
The rate of energy loss is much smaller than for alpha radiation.
In addition to these collision losses $\left(\frac{dE}{dx}\right) _{\text{coll}}$ due to excitation and ionization, radiative losses via bremsstrahlung $\left(\frac{dE}{dx}\right) _{\text{rad}}$ may also occur:
\begin{equation}
  \left( \frac{dE}{dx} \right) _{\text{tot}} = \left( \frac{dE}{dx} \right) _{\text{coll}} + \left( \frac{dE}{dx} \right) _{\text{rad}}.
\end{equation}

The ratio of radiative  and collisional stopping powers for an electron with the energy $E$ (in MeV) can be approximated to \cite{turner_1995}
\begin{equation}
  \frac{ \left(\frac{dE}{dx}\right) _{\text{rad}} } { \left( \frac{dE}{dx}\right) _{\text{coll}} } = \frac{Z \cdot E}{800}\,. 
\end{equation}
The ratios are \SIlist{3;6;10}{\%} for the energies of the three \isotope[207]Bi IC electrons using the mean proton number $Z=49.1$ for CdZnTe (see \autoref{tab_properties_CdZnTe}).
Furthermore, it can be calculated that radiative losses are the dominant energy loss above \SI{16}{MeV}.
This is approximately one order of magnitude above the typical electron energy in these studies.\\
As a consequence, radiative losses via bremsstrahlung can be ignored here.\\

Another issue for beta radiation in contrast to mono-energetic electrons is the continuously distributed beta spectrum. 
Although the nominal energy of the endpoint of the spectrum may be high, many beta particles have much lower energy. These soft beta particles are being absorbed much faster.

As a result, the path of electrons in matter is much more tortuous and can even result in electrons being scattered out of the material towards their source, called backscattering. 
Hence, electrons do not have a defined range in matter. 
Graphical representations of this are shown e.g. in \autoref{sec_penelope_simulation}.

%%%%%%%%%%%%%%%%%%%%%%%%%%%%%%%%%%%%%%%%%%%%%%%%%%%%%%%%%%%%%%%%%%%%%%%%%%%%%%%%%%%%%%%%%%%%%%%%%%%%%%%%%%%%%%%%%%%%%%%%%%%%
\FloatBarrier
\subsection{Gamma radiation}
\label{sec_gamma_radiation}
Interactions of photons differ fundamentally from interactions of charged particles described before. If interactions occur at all, then only via single separate events.
This explains why photons can travel through matter without any interactions at all.
Two main effects, whose cross-sections are energy-dependent, are important for radiation measurements here: Photoelectric absorption and Compton scattering. Pair production and Rayleigh scattering are shortly discussed for completeness.
A graphical representation of the different effects is shown in \autoref{fig_nist_gammas_electrons} on page \pageref{fig_nist_gammas_electrons}.

\subsubsection{Photoelectric effect}
In the process of photoelectric absorption, a photon is absorbed completely by an atom, which emits a photo-electron (mostly from the K-shell) with an energy of 
$$ E_{\text{electr}} = E_{\gamma} - E_{\text{bind}},$$
where $E_{\gamma}$ is the energy of the incident photon, and $E_{\text{bind}}$ is the binding energy of the electron in its shell. The ionized atom quickly captures a free electron and often rearranges its shell electrons, so that characteristic X-rays are emitted as well. These are reabsorbed in close proximity of their origin in most cases due to their energy of some ten keV.

Photoelectric absorption is the dominant process for low-energy gamma radiation below \SI{300}{keV} in CdZnTe, see \autoref{fig_nist_gammas_electrons}.
There is no general analytic expression for the interaction probability for all gamma energies $E_{\gamma}$ and proton numbers $Z$, but it can roughly be estimated to be proportional to $Z^{4\dots5}/E_{\gamma}^{3.5}$ \cite{Knoll_radiation_detection_2000}.

\subsubsection{Compton scattering}
Compton scattering is an incoherent scattering process of the incoming gamma radiation with a shell electron with the rest mass $m_ec^2$, which is considered initially to be at rest. All scattering angles $\theta$ between \ang{0} and \ang{180} are possible. The energy loss of the photon, which is equal to the gain of energy of the so-called recoil electron, is depending on the direction of the recoil electron:
\begin{equation}
 E_{\gamma'}=\frac{E_{\gamma}}{1 + \frac{E_{\gamma}}{m_e c^2} \left( 1-\cos(\theta) \right) }
 \label{eq_backscatter}
\end{equation}
As Compton scattering is mainly dominated by the density of shell electrons as scattering partners, this effect is roughly proportional to $Z/E_{\gamma}$.
Compton scattering is the dominant effect from \SI{300}{keV} to \SI{7}{MeV} in CdZnTe (see \autoref{sec_NIST_data}), which are most of the energies used in this thesis.

\subsubsection{Pair production}
In the process of pair production in the nuclear field, the incident gamma ray is converted into an electron-positron-pair. This is only possible, if the gamma energy exceeds twice the electron mass $2 \cdot m_e = \SI{1022}{keV}$. If the gamma energy is above that energy, the difference is transferred into kinetic energy of the electron and positron. As the positron normally annihilates with an electron after slowing down, two \SI{511}{keV} photons are emitted as secondary particles.
The probability of pair production varies roughly with $Z^2$.
As pair production is dominant only from several \si{MeV} on (ca. \SI{7}{MeV} for CdZnTe, see \autoref{fig_nist_gammas_electrons}), it does not have a major effect for typical radiation detectors.

\subsubsection{Rayleigh-scattering}
Another scattering process, called Rayleigh-scattering, is a low energy coherent process significant only below some hundred \si{keV}. Here, the incident gamma ray does not excite or ionize atoms, so that it retains all its energy, but angular deflections occur. As no energy is transferred, this effect can be ignored in most cases for radiation detection considerations, unless a complete gamma ray transport model is needed.

%%%%%%%%%%%%%%%%%%%%%%%%%%%%%%%%%%%%%%%%%%%%%%%%%%%%%%%%%%%%%%%%%%%%%%%%%%%%%%%%%%%%%%%%%%%%%%%%%%%%%%%%%%%%%%%%%%%%%%%%%%%%
\FloatBarrier
\subsection{Proton radiation}
\label{sec_proton_radiation}
Protons are heavy charged particles, so their interaction mechanisms are similar to those of alpha radiation.
They travel on a straight path without angular deflections.
Their main energy deposition at the end of the track is known as Bragg-peak. 
Due to their typical energy loss (``Bragg-curve''), they have a defined, energy-dependent range.
These characteristics make them a good method to do scanning measurements of a probe, even three-dimensional. 
With these defined interaction positions in the detector volume, one could do exhaustive tests of the interaction depth calculation or investigate the areas of the surface events recognition. 
Difficulties could arise for the following reasons:\\
One needs an in-situ measurement while the accelerator is irradiating the detector. Typical rates of accelerators are several orders of magnitude too high and cannot be reduced. 
The CdZnTe detectors and COBRA DAQ are capable of only several hundred events per second.
Therefore a setup with a target (gold foil on a substrate of silicon-nitrite) to use only the backscattered protons could be used. The irradiation could be done at the Helmholtz Zentrum Dresden Rossendorf, for example.
Furthermore, at least the detectors and preamplifiers need a good shielding against EMI. This might be difficult to include in an accelerator measurement.

%%%%%%%%%%%%%%%%%%%%%%%%%%%%%%%%%%%%%%%%%%%%%%%%%%%%%%%%%%%%%%%%%%%%%%%%%%%%%%%%%%%%%%%%%%%%%%%%%%%%%%%%%%%%%%%%%%%%%%%%%%%%
%%%%%%%%%%%%%%%%%%%%%%%%%%%%%%%%%%%%%%%%%%%%%%%%%%%%%%%%%%%%%%%%%%%%%%%%%%%%%%%%%%%%%%%%%%%%%%%%%%%%%%%%%%%%%%%%%%%%%%%%%%%%
\FloatBarrier
\section{Penetration depth of particles in matter}
\label{sec_penetration_depth_of_particles_in_matter}

%%%%%%%%%%%%%%%%%%%%%%%%%%%%%%%%%%%%%%%%%%%%%%%%%%%%%%%%%%%%%%%%%%%%%%%%%%%%%%%%%%%%%%%%%%%%%%%%%%%%%%%%%%%%%%%%%%%%%%%%%%%%
\FloatBarrier
\subsection{NIST-database}
\label{sec_NIST_data}
The NIST-databases offer comprehensive data of interactions of particles in matter, like cross-sections or ranges.
For heavy charged particles (alpha and proton radiation), there is only a limited choice of target materials, CdZnTe could not be chosen. 
Tin (Sn)\footnote{in the metallic form (beta-allotropy) with a tetragonal crystal structure} is used here, as \isotope[][50]Sn is right between \isotope[][48]Cd and \isotope[][52]Te in the periodic table. 
The density of Tin of \SI{7.3}{g/cm^3} is \SI{25}{\%} larger than the density of CdZnTe with \SI{5.8}{g/cm^3}, other parameters are similar.

Two often used ranges are the CSDA (\textbf{c}ontinuously \textbf{s}lowing \textbf{d}own \textbf{a}pproximation)- and projected range.
The CSDA range is an approximation of the average path length traveled by a charged particle until it is at rest \cite{CSDA_wiki}. 
The basic assumption is that the particle slows down continuously. The rate of energy loss at every point is equal to the total stopping power. Energy-loss fluctuations are ignored. 
The CSDA range is calculated by integrating the inverse total stopping power with respect to energy:
\begin{equation}
   R_{CSDA}(E_0) = \int_0^{E_0} \left( -\frac{dE}{dx} \right)^{-1}dE\, .
   \label{eq_CSDA_range}
\end{equation}
In contrast to this, the projected range is the average penetration depth of a charged particle until it is at rest, measured in its initial direction of motion.
Consequently, the CSDA-range is always larger than the projected range, especially for multiple-deflected particles like electrons. 
For particles moving on a straight line like protons or alphas, there are no big differences between CSDA and projected range.\\

\begin{figure}
 \centering            
  \includegraphics[width=0.49\textwidth]{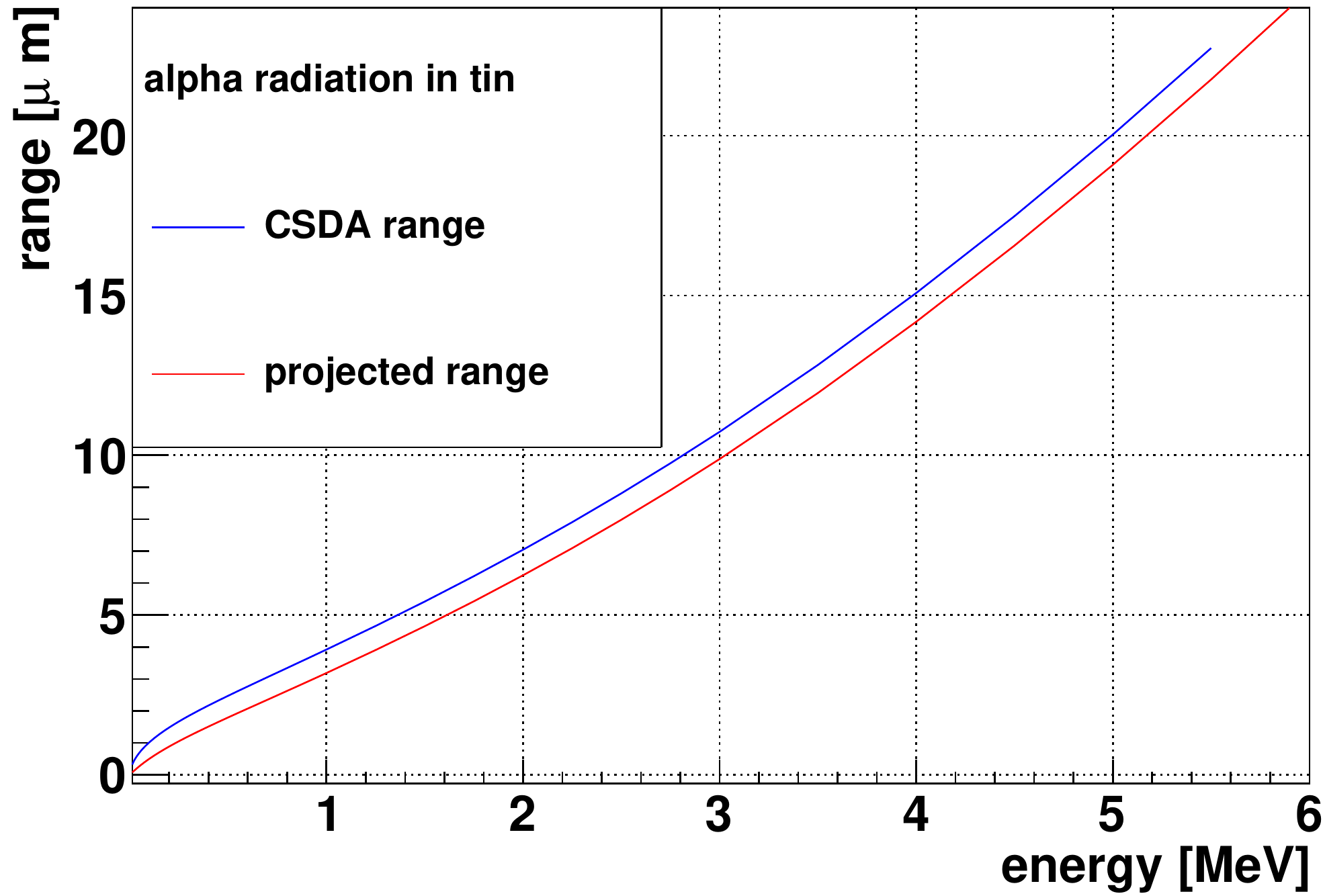}
  \includegraphics[width=0.49\textwidth]{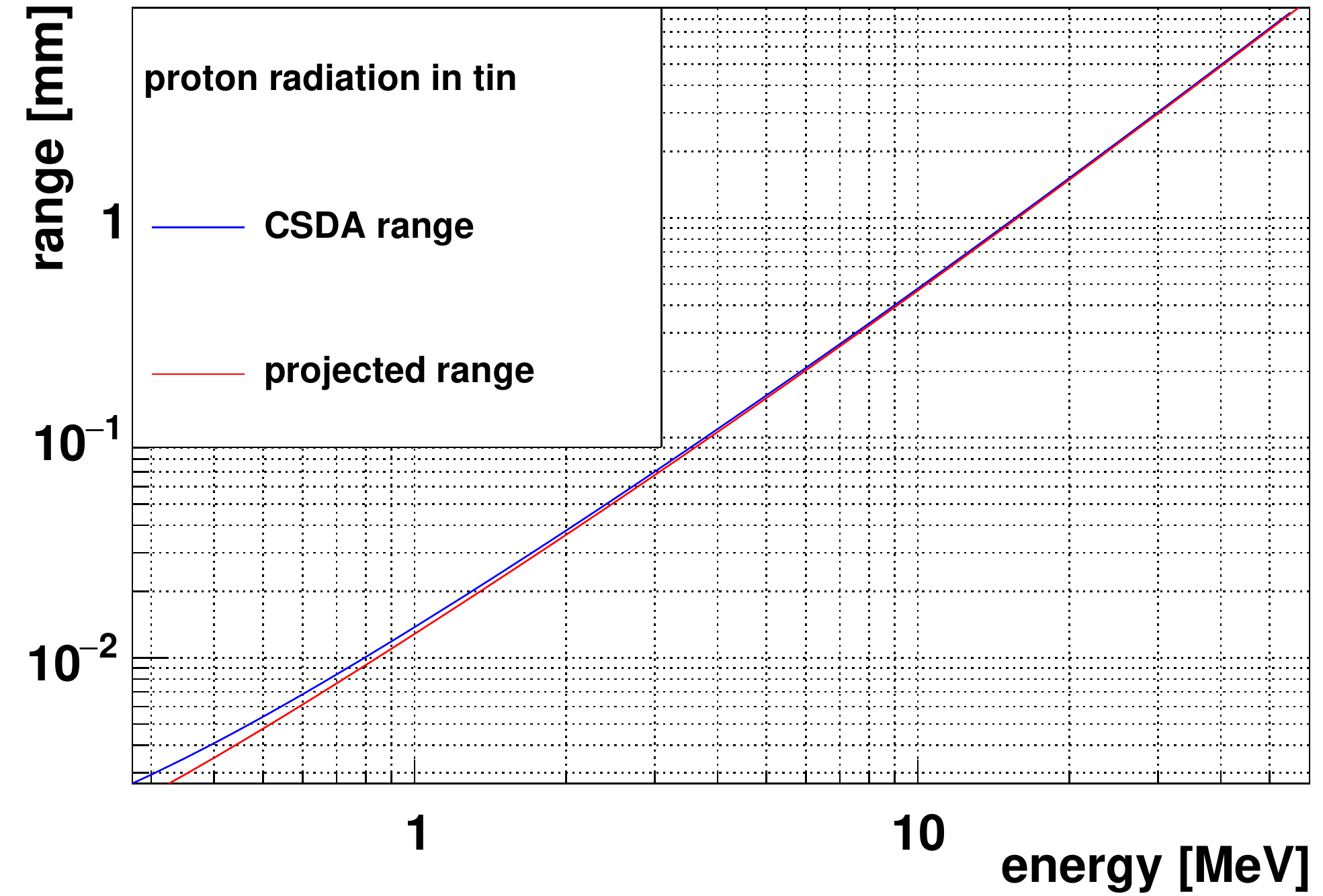}
  \captionsetup{width=0.9\textwidth} 
  \caption[Range of alpha and proton radiation in Sn]
	  {CSDA- and projected range for alpha particles (left) and protons (right) in \isotope[]Sn, values taken from \cite{NIST_astar, NIST_pstar}. CdZnTe could not be chosen, that is why Sn is used as being similar concerning atomic properties.}
  \label{fig_nist_alphas_protons}
\end{figure}

\paragraph{Heavy charged particles}
The ranges of proton and alpha radiation in Sn are shown in \autoref{fig_nist_alphas_protons}.
\isotope[241]Am with an energy of \SI{5.5}{MeV} is used as alpha radiation source in \autoref{sec_surface_events}. 
The projected range, which is the quantity of interest rather than the CSDA range, is about \SI{22}{\mu m} for that energy. 
Protons travel on a straight line with a Bragg-peak-like energy deposition at the end, see simulated trajectories in \autoref{sec_srim_simulation}. By this, they are an ideal method to make a three-dimensional detector scanning, as discussed in  \autoref{sec_proton_radiation}. 
Proton energies between \SI{4}{MeV} and \SI{40}{MeV} are needed to access the detector volume from the surface regions up to the detector center.
For both particle types, there is no big difference between the projected and the CSDA range.

\paragraph{Gamma radiation}
There is no defined penetration depth for gamma radiation, because the intensity of electromagnetic particles in matter is decreasing exponentially. 
Using the NIST XCOM database for gamma radiation \cite{NIST_xcom}, one can calculate the total absorption coefficient $\mu$ of gamma radiation in matter, and according to \autoref{eq_relative_intensity} the relative intensities $I/I_0$ after passing the distance $x$ in matter:
\begin{equation}
 \frac{I}{I_0}=\exp(-\mu \cdot x).
 \label{eq_relative_intensity}
\end{equation}

The mean free path $\lambda$ is defined as the average distance $x$ which a particle can travel through matter before interacting. It can be calculated as 
\begin{equation}
  \lambda = \frac{\int\limits_0^{\infty} x \exp^{-\mu x}dx}{\int\limits_0^{\infty} \exp^{-\mu x}dx} = \frac{1}{\mu}\,.
\end{equation}
Both formulas are taken from \cite{Knoll_radiation_detection_2000}.
When calculating these values, one can state that for the typical gamma sources used in \autoref{sec_surface_events}, only the \isotope[241]Am source (\SI{60}{keV}) can produce surface events.
All other sources produce central events, see the calculated mean free paths and relative intensities in \autoref{tab_gamma_intensity}.
\begin{SCtable}[][b!] 
 \centering
  \begin{tabular}{|c|c|c|c|}
      \hline
      radiation                                            & $\lambda [mm]$        & distance x      & \makecell{relative intensity\\after distance x} \\ \hline              
      \multirow{4}{*}{\SI{60}{keV} of \isotope[241]{Am}}   & \multirow{4}{*}{0.27} & \SI{10}{\mu m}  & \num{0.96}    \\ 
                                                           &                       & \SI{100}{\mu m} & \num{0.69}    \\
                                                           &                       & \SI{1}{mm}      & \num{0.03}    \\ 
                                                           &                       & \SI{2}{mm}      & \num{6.7E-4}  \\ \hline
      \multirow{3}{*}{\SI{570}{keV} of \isotope[207]{Bi}}  & \multirow{3}{*}{21}   & \SI{1}{mm}      & \num{0.95}    \\ 
                                                           &                       & \SI{3}{mm}      & \num{0.86}    \\ 
                                                           &                       & \SI{1}{cm}      & \num{0.62}    \\ \hline
      \multirow{3}{*}{\SI{662}{keV} of \isotope[137]{Cs}}  & \multirow{3}{*}{23}   & \SI{1}{mm}      & \num{0.96}    \\ 
                                                           &                       & \SI{3}{mm}      & \num{0.88}    \\ 
                                                           &                       & \SI{1}{cm}      & \num{0.65}    \\ \hline
      \multirow{3}{*}{\makecell{\SI{2615}{keV} of \isotope[208]Tl\\(\isotope[232]Th source) }}& \multirow{3}{*}{46}   & \SI{1}{mm}      & \num{0.98}    \\ 
                                                           &                       & \SI{3}{mm}      & \num{0.94}    \\ 
                                                           &                       & \SI{1}{cm}      & \num{0.80}    \\ \hline
    \end{tabular}
    \caption[Calculation of the mean free path and relative intensities of gamma radiation in CdZnTe]
          {Calculation of the mean free path $\lambda$ and relative intensities after passing the distance $x$ in CdZnTe. The length of the detector edge is \SI{1}{cm}.
           Calculations only for the main energies of the sources used in \autoref{sec_surface_events}.}
  \label{tab_gamma_intensity}
\end{SCtable}

\begin{figure}
 \centering            
  \includegraphics[width=0.49\textwidth]{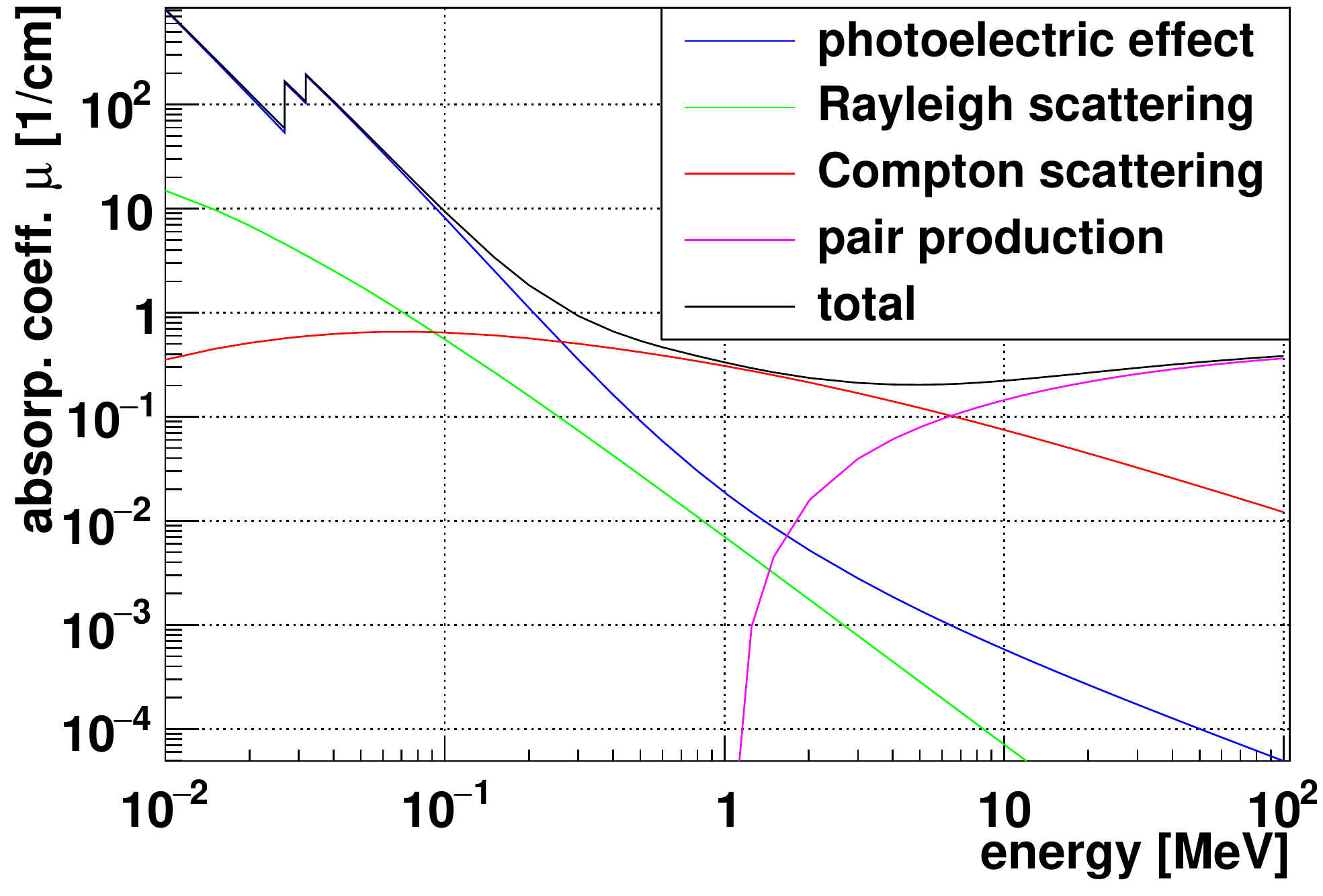}
  \includegraphics[width=0.49\textwidth]{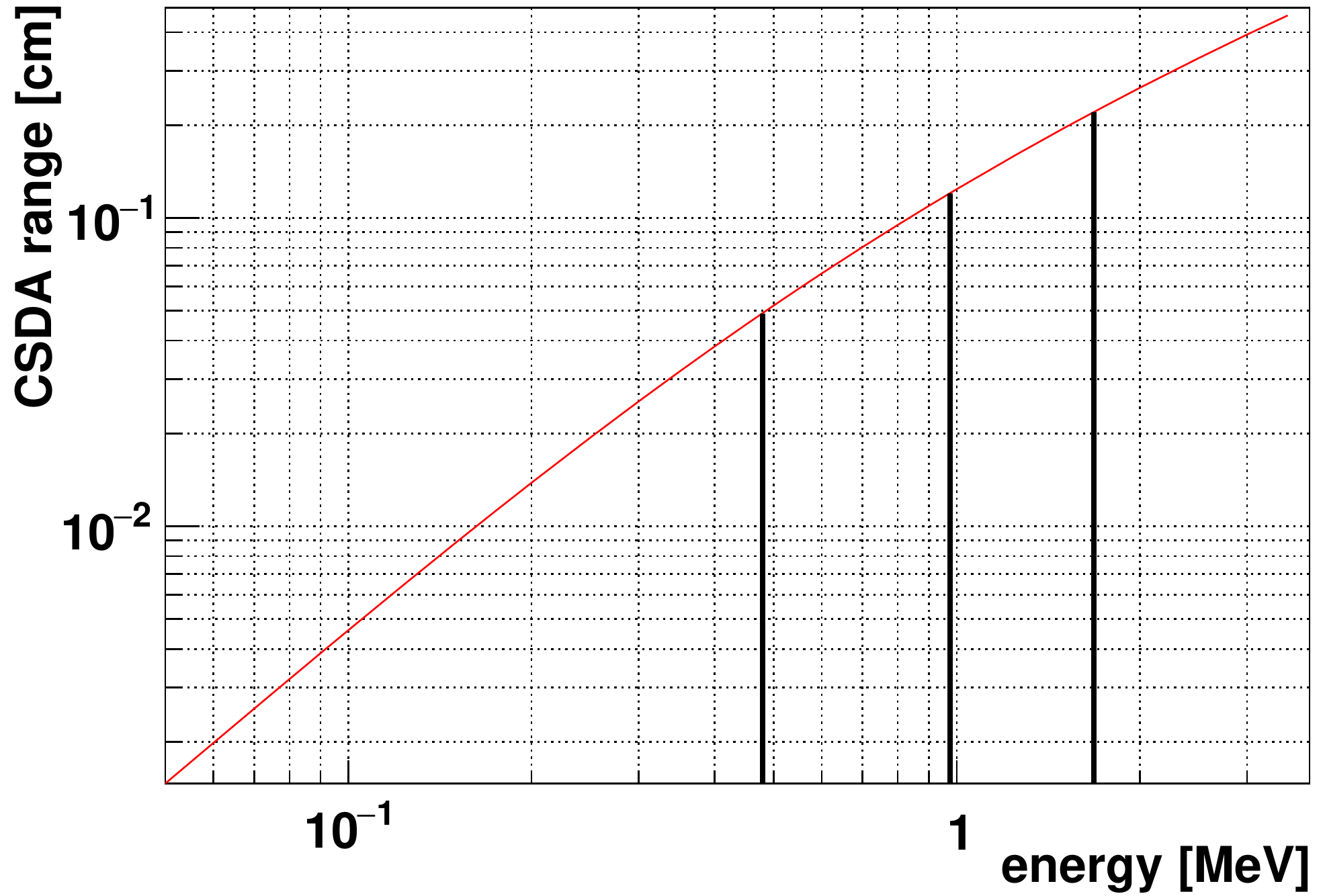}
  \captionsetup{width=0.9\textwidth} 
  \caption[Absorption coefficients of gamma radiation and range of beta radiation]
	  {Left: Absorption coefficients of gamma radiation in CdZnTe \cite{NIST_xcom}.
	   Right: CSDA range of electrons in CdZnTe \cite{NIST_estar}. The energies of the IC electrons of \isotope[207]Bi are marked with the black lines.}
  \label{fig_nist_gammas_electrons}
\end{figure}

The total absorption coefficient of gamma radiation in CdZnTe and its components are shown in \autoref{fig_nist_gammas_electrons} left.
The photoelectric effect is the dominant interaction at low energies up to about \SI{300}{keV}, while Compton scattering is dominant between \SI{300}{keV} and \SI{7}{MeV}. 
All gamma sources used in this thesis are in the energy range of the latter, except for \SI{60}{keV} gamma radiation of \isotope[241]Am.
At \SI{60}{keV}, the interaction probability for photoelectric absorption is a factor of 53 larger than Compton scattering, while for \SI{662}{keV}, Compton scattering is a factor of eight more likely.
The characteristic x-rays ($K_{\alpha}$ lines) can be seen at \SI{27}{keV} and \SI{32}{keV}.

\paragraph{Electron radiation}
The CSDA range of electrons in CdZnTe is plotted in \autoref{fig_nist_gammas_electrons} right. It is \SIlist{0.5; 1.1; 2.2}{mm} for the mono-energetic IC electrons of \isotope[207]Bi with \SIlist{480; 976; 1682}{keV}, which are used in \autoref{sec_surface_events}. 
It should be kept in mind, that the CSDA range is not the penetration depth of electrons in matter, as large angular deflections in the paths occur, as discussed before and will be seen in the following simulations.
The penetration depth is obtained using the simulation methods in the next section.
The CSDA range can be understood as the maximal value for the penetration depth.

%%%%%%%%%%%%%%%%%%%%%%%%%%%%%%%%%%%%%%%%%%%%%%%%%%%%%%%%%%%%%%%%%%%%%%%%%%%%%%%%%%%%%%%%%%%%%%%%%%%%%%%%%%%%%%%%%%%%%%%%%%%%
\FloatBarrier
\subsection{VENOM simulation}
\label{sec_venom_simulation}
VENOM is the COBRA standard simulation framework. It is based on GEANT4, which is the most common simulation tool in particle und nuclear physics. 
It is a very versatile tool, all kind of particles and geometries can be simulated.
The interaction parameters for this low-energy application are set in the physics list ``shielding''.
Many theses of the COBRA collaboration are based on VENOM, the most recent ones are \cite{rebber_master, heidrich_ph_d, koettig_2012_phd}.

%%%%%%%%%%%%%%%%%%%%%%%%%%%%%%%%%%%%%%%%%%%%%%%%%%%%%%%%%%%%%%%%%%%%%%%%%%%%%%%%%%%%%%%%%%%%%%%%%%%%%%%%%%%%%%%%%%%%%%%%%%%%
\subsubsection{Electron range}
\label{sec_venom_electron_simulation}
\begin{SCfigure}[][h!]
 \centering
  \includegraphics[width=0.7\textwidth]{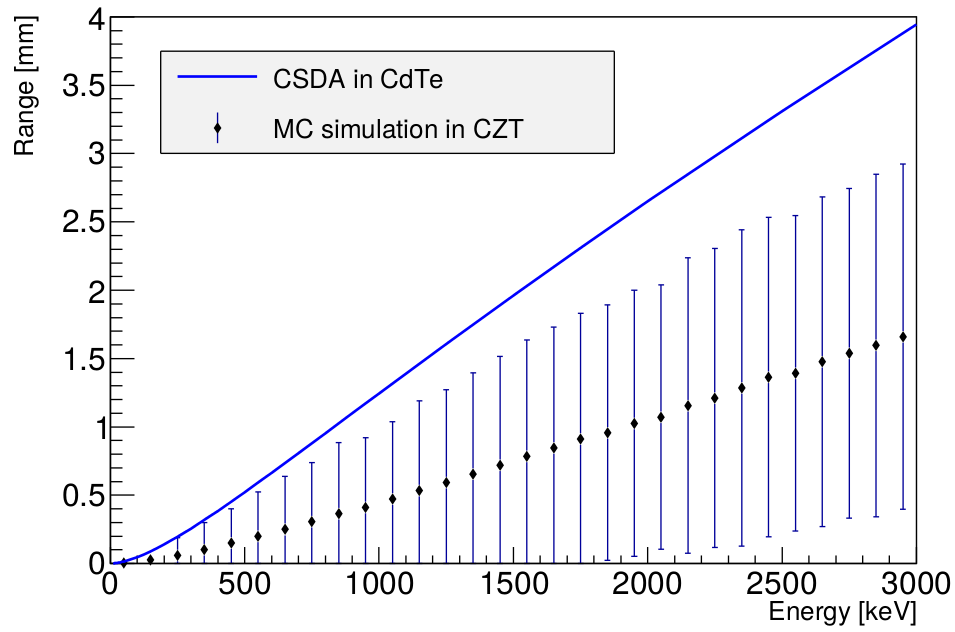}
  \caption[VENOM simulation of electron range]
	  {VENOM simulation of electron range. The CSDA range obtained by \cite{NIST_estar} is shown for comparison. Taken from \cite{rebber_master}.}
  \label{fig_venom_electron_range_rebber}
\end{SCfigure}

The penetration depth of electrons in CdZnTe was simulated lately in a master thesis \cite{rebber_master}:
For each energy in steps of \SI{100}{keV}, \num{10000} particle trajectories were simulated. 
For each of them, the distance from the detector surface to the outermost created charge was taken as range.
The mean value and standard deviation $\sigma$ of the range of all electrons trajectories are obtained.
The ranges have large standard deviations. This resembles the fact, that due to large angular deflections at the interactions, electrons perform a random walk.
Furthermore, the simulated ranges are well below the CSDA range and point out, that this value has to be understood as an upper limit for the range, as discussed in \autoref{sec_penetration_depth_of_particles_in_matter}.
The results are shown graphically in \autoref{fig_venom_electron_range_rebber}, the numbers are compiled in \autoref{tab_venom_electrons}.

\begin{SCtable}[][h!]
  \begin{tabular}{|c|c|c|}
    \hline
    energy [keV] & mean range [mm] & upper $\sigma$ value [mm] \\  \hline
    480 	 & \num{0.2}  & \num{0.4} \\
    976		 & \num{0.4}  & \num{1.0} \\
    1680	 & \num{0.9}  & \num{1.8} \\     \hline
  \end{tabular}
  \caption[Simulated range of electron irradiation using VENOM]
	  {Results of the electron range of the VENOM simulation, values from \autoref{fig_venom_electron_range_rebber}.}
  \label{tab_venom_electrons}
\end{SCtable}

%%%%%%%%%%%%%%%%%%%%%%%%%%%%%%%%%%%%%%%%%%%%%%%%%%%%%%%%%%%%%%%%%%%%%%%%%%%%%%%%%%%%%%%%%%%%%%%%%%%%%%%%%%%%%%%%%%%%%%%%%%%%
 \FloatBarrier
\subsubsection{Proton irradiation}
\label{sec_proton_irradiation}
\begin{figure}[h!]
 \centering
  \includegraphics[width=0.49\textwidth]{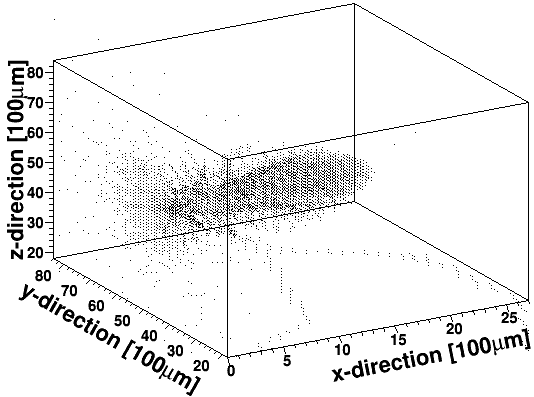}
  \includegraphics[width=0.49\textwidth]{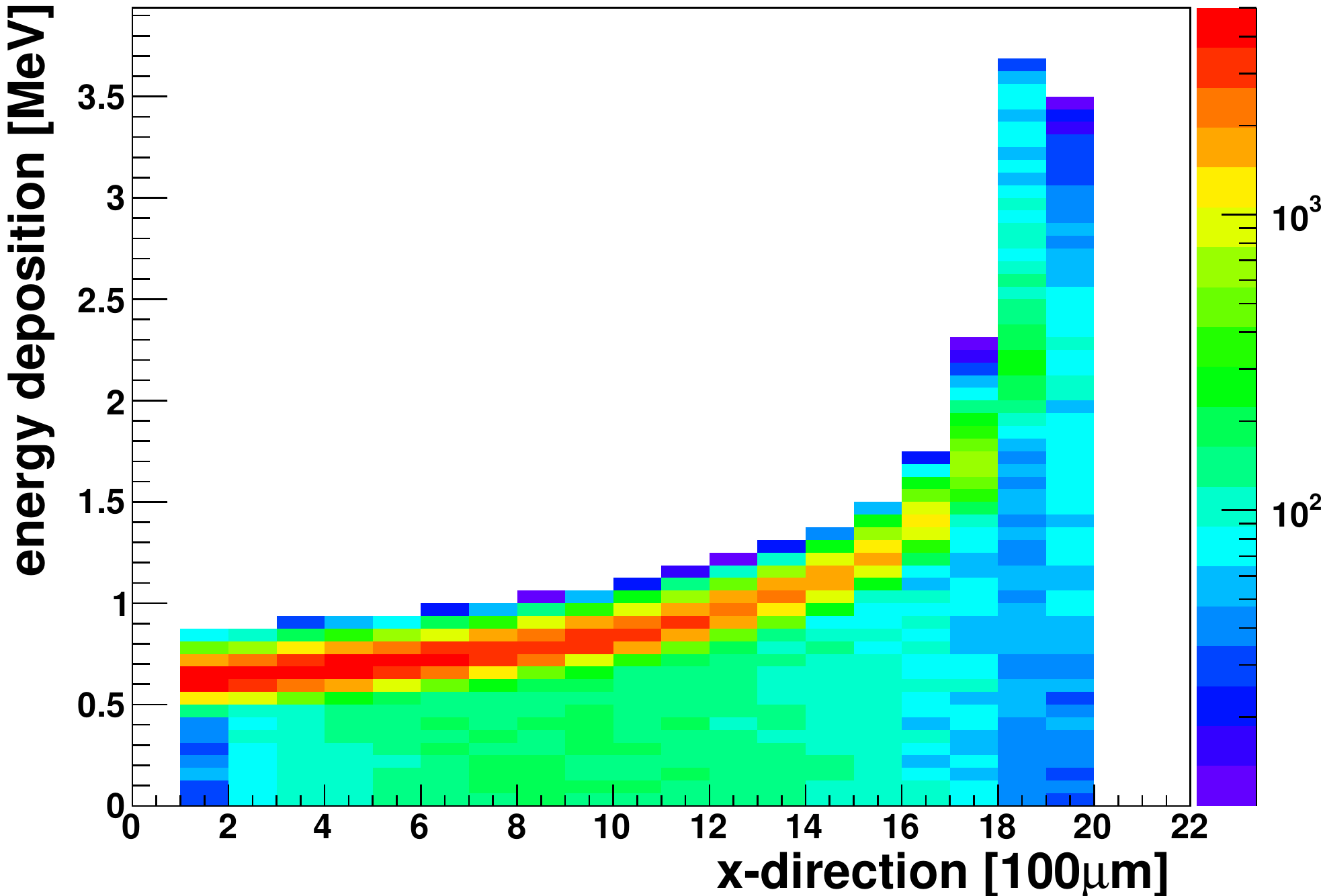}
  \captionsetup{width=0.9\textwidth} 
  \caption[Range of \SI{20}{MeV} protons in CdZnTe]
          {Range of \SI{20}{MeV} protons in CdZnTe simulated using VENOM. Left: Three-dimensional information of interactions, protons enter from the left. Right: Energy deposition projected onto the initial direction of motion.}
  \label{fig_20MeV_protons}
\end{figure}

As discussed in \autoref{sec_proton_radiation}, protons move on a straight path in matter without much scattering, and deposit their energy mostly at the end of their track (Bragg-peak). 
So they have a defined range depending on their initial energy.\\
For the simulations here, the detector is implemented as a pixel detector to get three-dimensional position information.
Each sub-detector has a volume of \SI{100}{\mu m} side length each. 
The beam is set perpendicular to the surface.\\
A graphical output for a proton beam with an energy of \SI{20}{MeV} is shown exemplarily in \autoref{fig_20MeV_protons}.
The number of interactions per pixel detector versus the pixel coordinates is plotted in a three-dimensional way. 
By this, one can see that only very few protons suffer from large angular deflections, most of them move on a straight line until being stopped at the Bragg peak.
This peak can be seen in an energy-resolved projection of the distribution onto the initial direction of motion.
The energy deposition per pixel detector (\SI{100}{\mu m}) is well below \SI{1}{MeV} at the beginning of the trajectory, while the depositions at the last two pixel detectors are up to \SI{3.5}{MeV}. 
The total energy deposition sums up correctly to \SI{20}{MeV} for each proton.
The resulting ranges for \SIlist{10; 20; 50}{MeV} protons are \SIlist{0.6; 1.9; 9}{mm}. Using these energies, one could investigate the detector completely.

%%%%%%%%%%%%%%%%%%%%%%%%%%%%%%%%%%%%%%%%%%%%%%%%%%%%%%%%%%%%%%%%%%%%%%%%%%%%%%%%%%%%%%%%%%%%%%%%%%%%%%%%%%%%%%%%%%%%%%%%%%%%
\FloatBarrier
\subsection{SRIM simulation}
\label{sec_srim_simulation}
\begin{figure}
 \centering
  \includegraphics[width=0.99\textwidth]{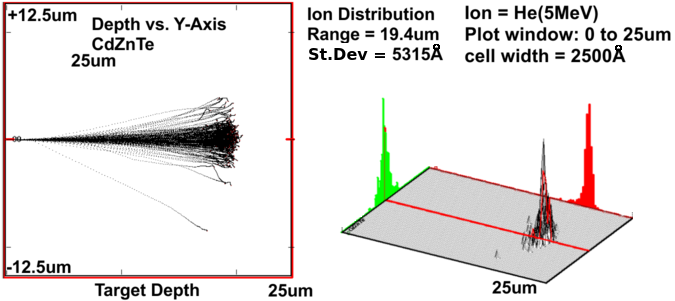}
  \captionsetup{width=0.9\textwidth} 
  \caption[Graphical output of an SRIM simulation of \SI{5}{MeV} alpha radiation]
	  {Graphical output of an SRIM simulation of \SI{5}{MeV} alpha radiation in CdZnTe. 
          Left: Trajectories of alpha particles entering from the left side. Right: Three-dimensional representation of the same simulation. The alpha particles enter from the left side, the beam axis is marked with the red line. The projections in the direction of the beam axis and perpendicular to it are shown in green and red, respectively. }
  \label{fig_srim_simulation_alpha_5MeV}
\end{figure}

The SRIM (\textbf{S}topping and \textbf{r}ange of \textbf{i}ons in \textbf{m}atter) simulation tool for ions is widely used in radiation material science \cite{SRIM}. Its core TRIM (\textbf{tr}ansport of \textbf{i}ons in \textbf{m}atter) uses a binary collision approximation to calculate interactions with matter.
Ions are approximated to travel through solids on straight paths, until interacting with nuclei in independent binary collisions. 
No loss of energy by collisions with nuclei is assumed, only by the stopping power of shell electrons. Needed input to run SRIM simulations are the target material, ion type and ion energy from \SI{10}{eV} to \SI{2}{GeV}.
The range of alpha particles and protons was simulated here. 
As target materials CdZnTe and \isotope[]Sn are used, because \isotope[]Sn is used for the calculations in the NIST-databases in case CdZnTe could not be chosen as material, compare \autoref{sec_NIST_data}. 
The simulation tool generates several plots showing the trajectories of the simulated particles. One is a three-dimensional representation, the other is a projection onto the direction of motion. 
Values like ranges are calculated automatically for each simulation. 
The lateral range shows a strict upper bound with a standard deviation at the percent level. 
In contrast, the radial ranges show relative large standard deviations, which stem from (few) particles that encounter larger angular deflections, which can be seen in the trajectories in \autoref{fig_srim_simulation_alpha_5MeV}.
The difference between the two materials Sn and CdZnTe is about \SI{20}{\%}, which can be explained probably mostly by the difference of their densities of about \SI{25}{\%}.
An example of the graphical output is shown in \autoref{fig_srim_simulation_alpha_5MeV} for \SI{5}{MeV} alpha radiation in CdZnTe, a compilation of all SRIM simulation results is given in \autoref{tab_srim_simulation_results}.

\begin{table}[htbp]
\centering
\begin{tabular}{|c|c|c|c|c|c|c|}
\hline
\multirow{2}{*}{material} & \multirow{2}{*}{radiation} & energy & longitudinal    & radial           & CSDA & projected\\%ected\\ 
                          &                            & [MeV]  & range           & range            & range      & range\\ \hline
\multirow{8}{*}{Sn}       & \multirow{2}{*}{alpha}     & 1      & \num{2.53(37)}  & \num{0.51(33)}   & 3.9        & 3.2 \\ 
                          &                            & 5      & \num{15.3(4)}   & \num{1.15(68)}   & 20.0       & 19.1\\ \cline{2-7}
                          & \multirow{6}{*}{proton}    & 1      & \num{10.4(6)}   & \num{1.53(82)}  & 14         & 13\\ 
                          &                            & 5      & \num{121(3)}    & \num{12.3(60)} & 150        & 150\\ 
                          &                            & 10     & \num{371(11)}   & \num{32.1(265)}  & 490        & 490\\ 
                          &                            & 20     & \num{1180(20)}  & \num{115(52)}  & 1550       & 1550\\ 
                          &                            & 50     & \num{5680(112)} & \num{405(273)}   & 7250       & 7250 \\ 
                          &                            & 100    & \num{1880(298)} & \num{1290(961)}  &            &       \\ \hline
\multirow{9}{*}{CdZnTe}   & \multirow{3}{*}{alpha}     & 1      & \num{3.18(35)}  & \num{0.66(38)}   &            & \\ 
                          &                            & 3      & \num{9.97(57)}  & \num{1.21(82)}  &            & \\ 
                          &                            & 5      & \num{19.4(5)}  & \num{1.55(106)}  &            & \\ \cline{2-6}
                          & \multirow{6}{*}{proton}    & 1      & \num{13.1(6)}   & \num{1.80(106)}  &            & \\       
                          &                            & 5      & \num{151(4)}    & \num{14.3(92)} &            & \\ 
                          &                            & 10     & \num{464(18)}   & \num{40.8(321)}  &            & \\ 
                          &                            & 20     & \num{1480(26)}  & \num{123(69)}     &            & \\ 
                          &                            & 50     & \num{7130(337)} & \num{537(383)}   &            & \\ 
                          &                            & 100    & \num{2370(370)} & \num{1650(1060)} &            & \\ \hline
\end{tabular}
\captionsetup{width=0.9\textwidth} 
\caption[Results of SRIM simulations]
	{Results of SRIM simulations, ranges in $\mu m$. For comparison, the CSDA and projected ranges for Sn are added (calculated in \autoref{sec_NIST_data}).}
\label{tab_srim_simulation_results}
\end{table}

%%%%%%%%%%%%%%%%%%%%%%%%%%%%%%%%%%%%%%%%%%%%%%%%%%%%%%%%%%%%%%%%%%%%%%%%%%%%%%%%%%%%%%%%%%%%%%%%%%%%%%%%%%%%%%%%%%%%%%%%%%%%
\FloatBarrier
\subsection{PENELOPE simulation}
\label{sec_penelope_simulation}
\begin{SCfigure}[][b!]
 \centering
  \includegraphics[width=0.75\textwidth]{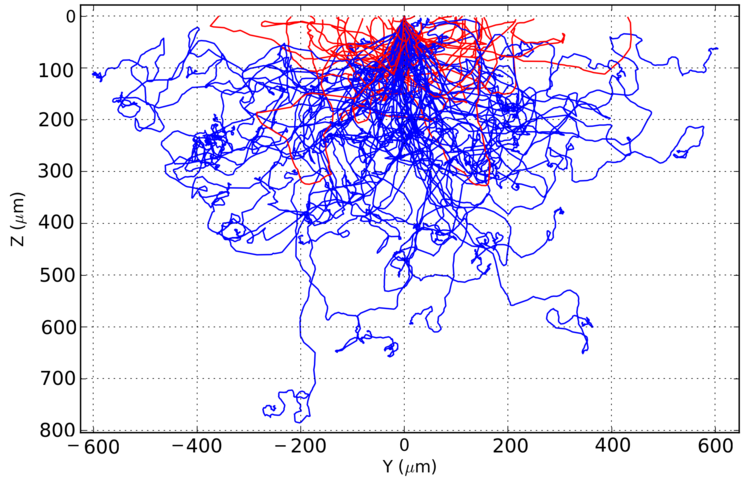}  
  \caption[PENELOPE simulation of an electron beam with \SI{976}{keV} in CdZnTe]
          {PENELOPE simulation of an electron beam with \SI{976}{keV} and diameter of \SI{10}{nm} in CdZnTe. The beam is pointed to (0,0) going downwards. 
           Red: Electrons being backscattered out of the material.}
  \label{fig_penelope_simulations}
\end{SCfigure}

PENELOPE (Python pyPENELOPE version 0.2.10) is a MC program to simulate coupled photon-electron transport in matter, from \SI{50}{keV} to \SI{1}{GeV} \cite{PENELOPE}. 
Interaction models are based on ``the most reliable information'' \cite{PENELOPE}, using first-principal calculations, semi-empirical models and databases.
As the penetration depths of \SIlist{480; 976; 1680}{keV} electrons from \isotope[207]Bi used in \autoref{sec_surface_events} are of interest here, PENELOPE seems to be an ideal simulation tool.
For simulating the penetration depth of electrons, the ``shower simulation'' task is used.
It simulates the trajectories of electrons penetrating a sample, losing energy through multiple scattering processes. The developers recommend to use standard settings, as inputs only material and beam parameters are needed.\\
The program generates plots of projections of the three-dimensional trajectories, but displays only a limited number of tracks, as otherwise no details can be seen.
\autoref{fig_penelope_simulations} demonstrates the random walk of a \SI{976}{keV} electron beam in matter, which results in a huge lateral spread of the trajectories to some hundred $\mu m$, while the initial beam has a diameter of only \SI{10}{nm}.
Furthermore, the difficulties in stating penetration depths can be seen by the longitudinal spread. 
The random walk even leads to electrons being backscattered out of the material.

\begin{figure}
 \centering
 \includegraphics[width=0.49\textwidth]{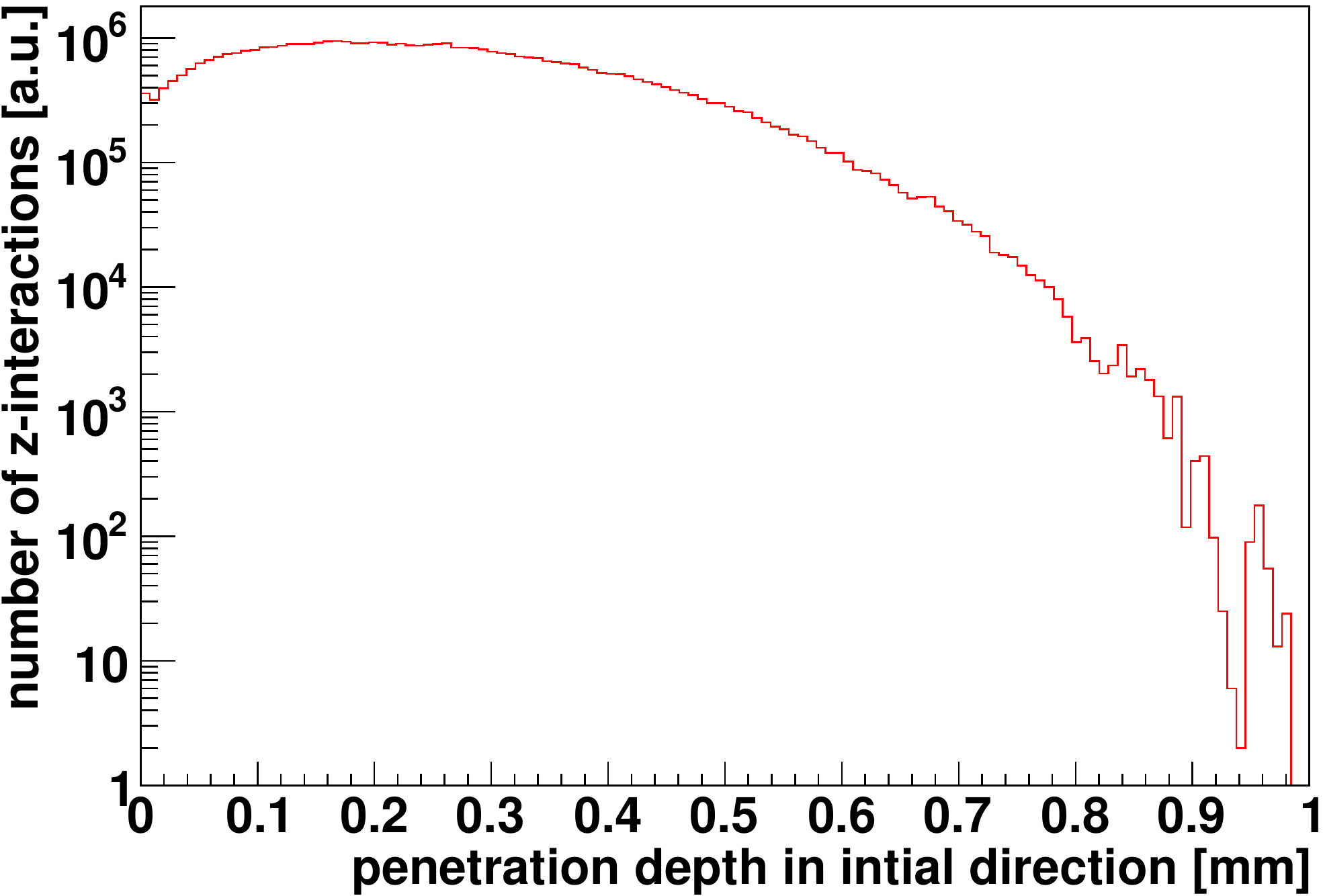}
 \includegraphics[width=0.49\textwidth]{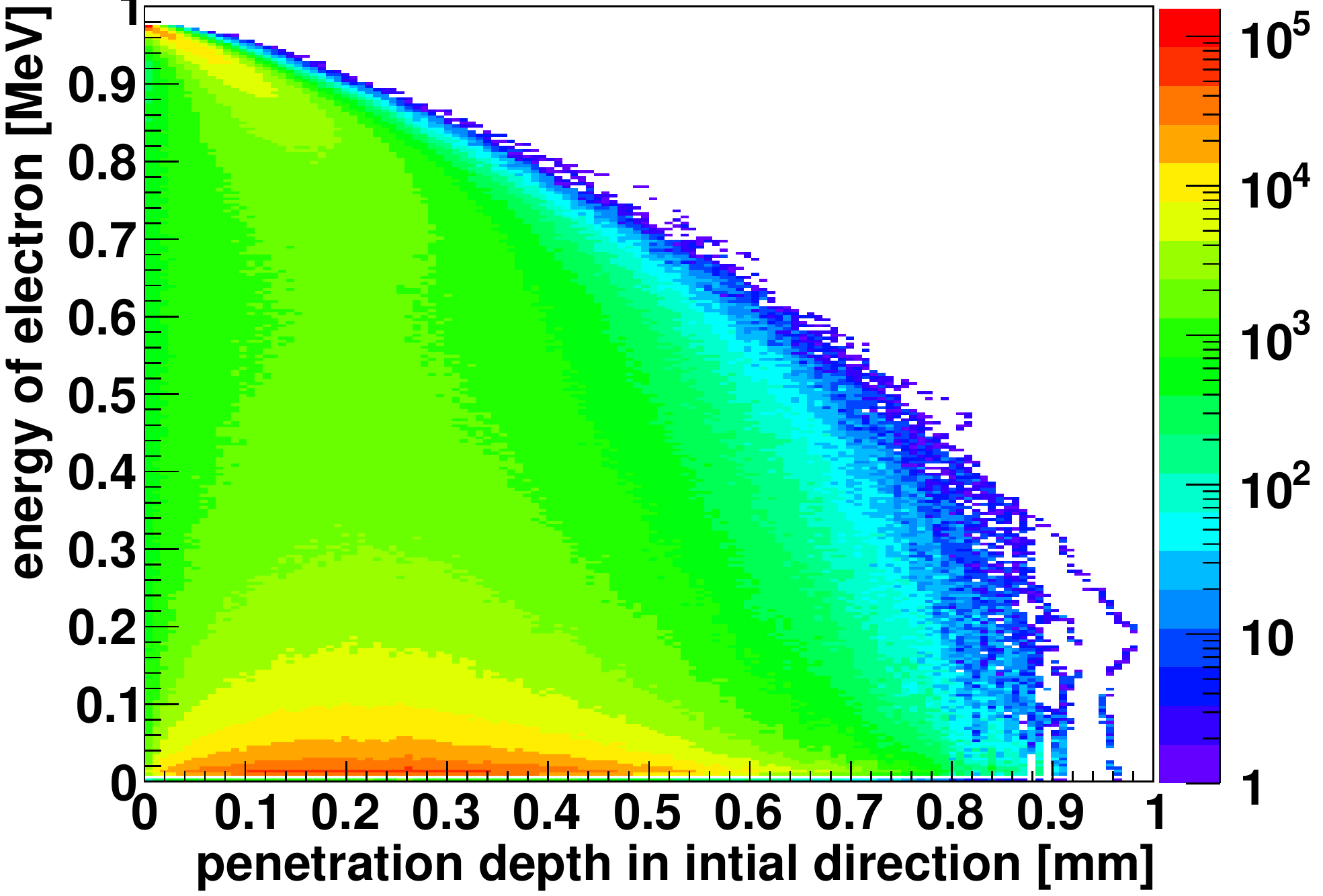}
 \captionsetup{width=0.9\textwidth} 
 \caption[Penetration depth of \SI{976}{keV} electrons in CdZnTe; PENELOPE]
         {Penetration depth of \SI{976}{keV} electrons in CdZnTe, simulated by PENELOPE. Left: Projection onto the initial direction of motion. Right: Same simulation, but energy-resolved representation.}
 \label{fig_penelope_simulations_max_penetration_depth}
\end{figure}

Apart from the graphical output, detailed information of the interactions like coordinates, energy depositions and particle type is saved to a text file. 
From these files, the simulation results can be obtained and transformed to the desired form using ROOT.
According to the chosen geometry in the simulation, the values of the penetration depth in the initial beam direction (z-direction) are of special interest to answer the question of the penetration depth of electrons.
Hence, the projection of the trajectories onto the initial beam direction (z-axis) is plotted.
Furthermore, the two-dimensional projection of depth and energy of the same distribution of the electrons is plotted. 
All electrons start with their initial energy of \SI{976}{keV} at z-position zero. 
The next following z-position and energy is filled to the histogram and so on. 
This distribution clearly shows, that the entries with the largest penetration depth stem from a single electron. 
It reaches nearly \SI{1}{mm} into the detector, probably without much angular deflection, before it travels slightly backwards and then perpendicular to the initial beam direction. 
After additional interactions it is slowed down to rest. If these interactions would lead to forward-scatterings like the ones before, the electron would reach a penetration depth of about \SI{1.1}{mm}, which is nearly the exact CSDA range value of that energy.\\
These two graphical representations are shown in \autoref{fig_penelope_simulations_max_penetration_depth} for \SI{976}{keV} electrons.

The mean value of the total interaction depth distribution is calculated, as well as the maximum penetration depth, and the maximum interaction density.
Additionally, the range, where the interaction density falls to 50\% and 10\% is calculated.
As a comparison, the CSDA range obtained in \autoref{sec_NIST_data} is also shown.
\num{100000} electrons were simulated for each energy. 
For a comparison to the CASINO simulations in the following (\autoref{tab_casino_simulation}), \num{10000} electrons were also simulated for an energy of \SI{976}{keV}.
The main difference between \num{10000} and \num{100000} simulated electrons is in the maximal penetration depth. 
This is plausible, as some electrons will be deflected only by small angles and travel nearly on a straight line to the maximal penetration depth. 
If more  electrons are simulated, the maximal penetration depth is enlarged by just a few or even a single electron.
All numbers are given in \autoref{tab_penelope_simulation}.

\begin{table}[b!]
\begin{tabular}{|c|c|c|c|c|}
\hline
						      &  \multicolumn{4}{c|}{electron energy}   \\ 
		    \multicolumn{1}{|c|}{} & \SI{480}{keV} & \makecell{\SI{976}{keV}\\ \SI{10}{k} $e^-$} & \SI{976}{keV} & \SI{1680}{keV} \\ \hline
average number of interactions per track & 536  & 471  & 472 & 476 \\ 
mean penetration depth [mm]		 & 0.10 & 0.26 & 0.26 & 0.51 \\ 
maximal interaction density [mm]	 & 0.07 & 0.17 & 0.17 & 0.43 \\ 
50\% maximal interaction density [mm]	 & 0.16 & 0.39 & 0.43 & 0.84 \\ 
10\% maximal interaction density [mm] 	 & 0.24 & 0.61 & 0.61 & 1.20 \\ 
maximal penetration depth [mm]		 & 0.43	& 0.84 & 0.98 & 1.85 \\ \hline
CSDA range [mm] (\autoref{sec_NIST_data}) & 0.5 & \multicolumn{2}{c|}{1.1} & 2.2 \\ \hline
\end{tabular}
\captionsetup{width=0.9\textwidth} 
\caption[Results of the PENELOPE simulations for \SIlist{480; 976; 1680}{keV} electrons]
	{Results of the PENELOPE simulations for \SIlist{480; 976; 1680}{keV} electrons, which are the energies of the \isotope[207]Bi conversion electrons used in \autoref{sec_bi_207_measurements}. \num{100000} electrons were simulated for each energy, additionally \num{10000} for \SI{976}{keV} to compare these values to \autoref{tab_casino_simulation}.}
\label{tab_penelope_simulation}
\end{table}

%%%%%%%%%%%%%%%%%%%%%%%%%%%%%%%%%%%%%%%%%%%%%%%%%%%%%%%%%%%%%%%%%%%%%%%%%%%%%%%%%%%%%%%%%%%%%%%%%%%%%%%%%%%%%%%%%%%%%%%%%%%%
\FloatBarrier
\subsection{CASINO simulation}
\label{sec_casino_simulation}
CASINO (Windows winCASINO v2.48(2.4.8.1)) simulates trajectories of electrons in solids, especially designed for low energy beam interactions in scanning electron microscopes, typically from \SIrange{0.1}{30}{keV} \cite{CASINO}. Nevertheless, CASINO is also used to cross-check other results.\\
As input, the material CdZnTe and a mono-energetic electron beam with a radius of \SI{10}{nm} is chosen. 
The simulations are performed until the electrons slowed down to rest (which was first defined at \SI{0.05}{keV}, later \SI{1}{keV} to save computation time).
Three energies were simulated: \SIlist{480; 976; 1680}{keV}, which are the energies of the conversion electrons of \isotope[207]Bi, which are needed in \autoref{sec_bi_207_measurements} for surface sensitivity considerations.

\begin{SCfigure}
 \centering
  \includegraphics[width=0.7\textwidth]{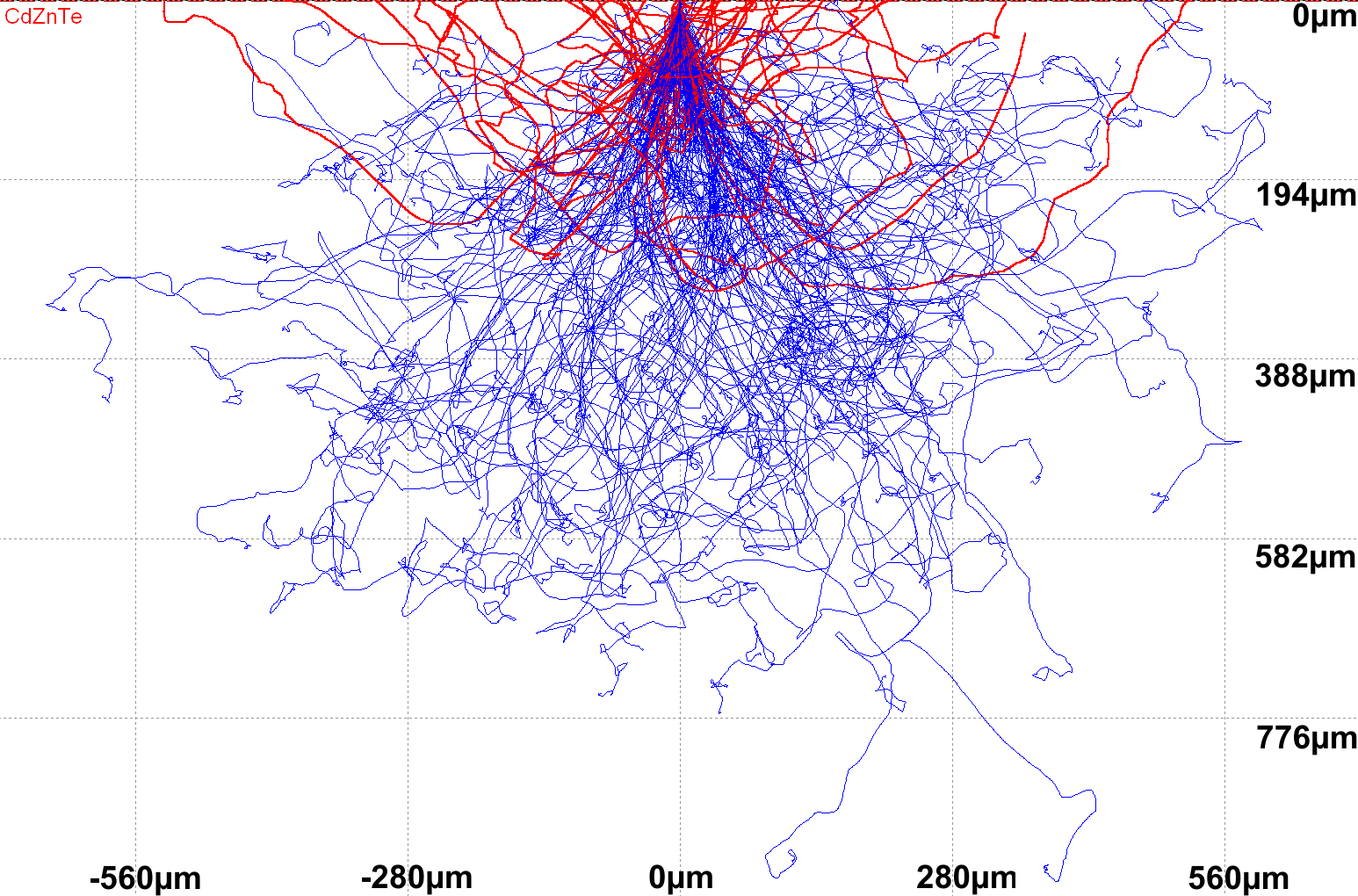}
  \caption[CASINO simulation of a \SI{976}{keV} electron beam in CdZnTe]
          {CASINO simulation of a \SI{976}{keV} electron beam in CdZnTe. The beam enters at (0,0) pointing downwards perpendicular to the surface. Red: Electrons being scattered out of the material.}
  \label{fig_casino_simulations}
\end{SCfigure}

The outputs of CASINO are analog to those of PENELOPE. 
A graphical illustration of the trajectories of electrons is shown exemplarily in \autoref{fig_casino_simulations} for a \SI{976}{keV} beam.
It shows qualitatively the same behavior as the PENELOPE result.

\begin{figure}
 \centering
 \includegraphics[width=0.49\textwidth]{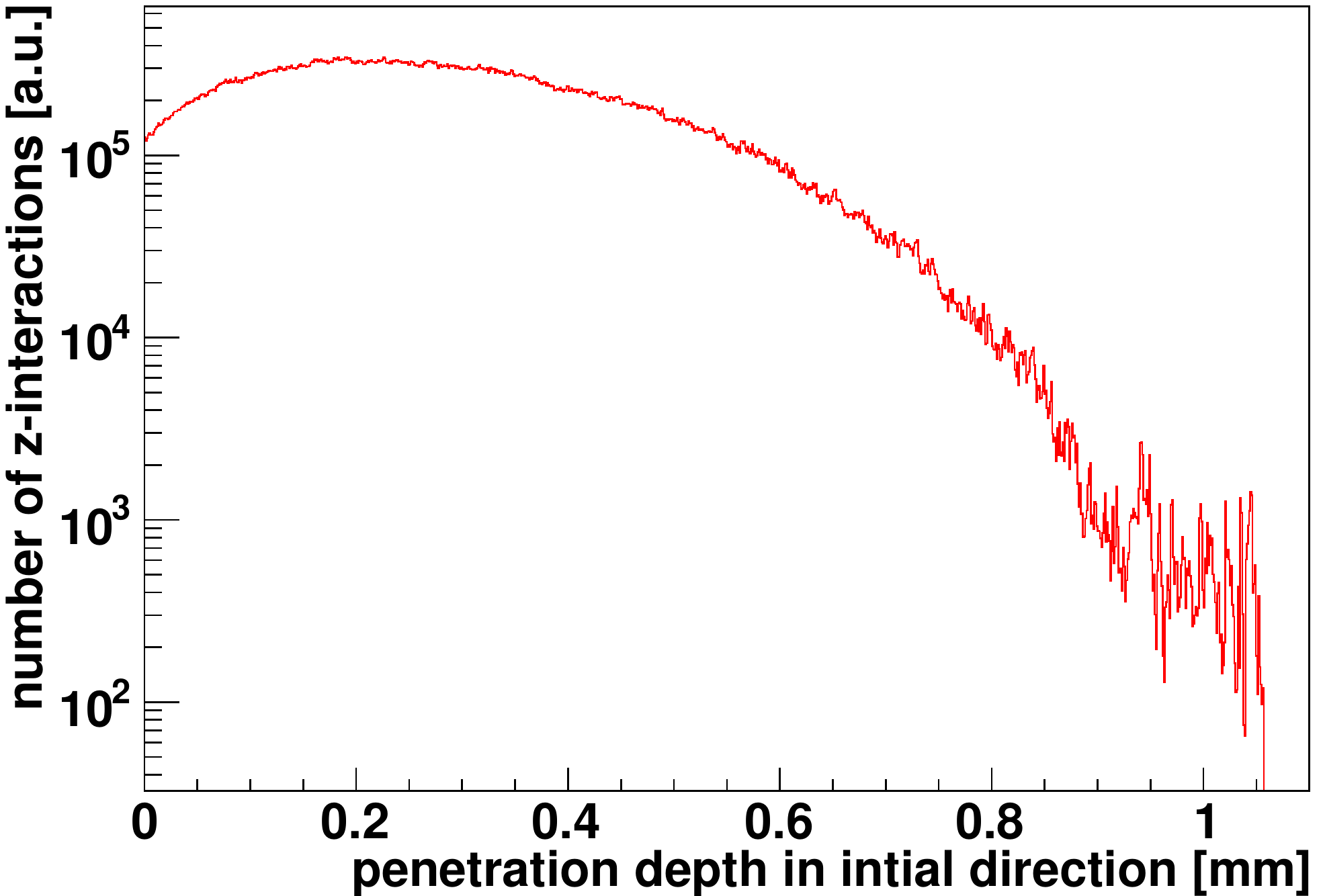}
 \includegraphics[width=0.49\textwidth]{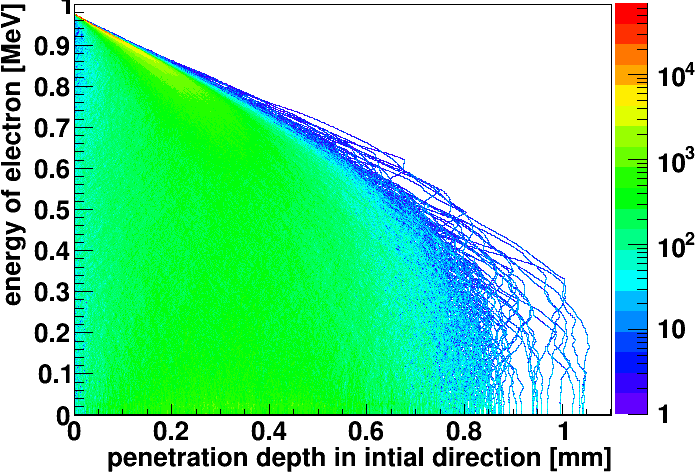}
 \captionsetup{width=0.9\textwidth} 
 \caption[Penetration depth of a\SI{976}{keV} electron beam in CdZnTe; CASINO]
	 {Penetration depth of a\SI{976}{keV} electron beam in CdZnTe, simulated by CASINO. Left: Projection onto the direction of motion. Right: Same simulation, energy resolved representation.}
 \label{fig_casino_simulations_max_penetration_depth}
\end{figure}

Like before, a detailed output of all interactions to a text-file is given. 
As CASINO is designed for lower energies down to \SI{0.1}{keV}, the trajectories are simulated in smaller energy steps and more details, resulting in much more output than PENELOPE. The average number of interactions per electron trajectory is larger by up to a factor of 40. 
Consequently, about \num{10000} electrons were simulated for \SI{480}{keV} and \SI{976}{keV}, but only \num{9000} for \SI{1680}{keV}, because this output file exceeded \SI{10}{GB}.\\
This output is imported and analyzed, analog to the PENELOPE-outputs.
\autoref{fig_casino_simulations_max_penetration_depth} shows the projection onto the initial direction of motion for a \SI{976}{keV} electron beam analog to the representation for PENELOPE.
The obtained maximal penetration depth is close to the CSDA value.
As can be seen in the right plot, the value for the maximal penetration depth is dominated by just a few electrons, like before.
These are only forward-scattered without much energy loss most of the time. Only at their maximal penetration depth values, they are moving perpendicular to the initial direction of motion.
If these last scattering processes would also lead to more forward scattering, these electrons could reach up to about \SI{1.2}{mm} into the material. 
This is more than the CSDA range, which was discussed in \autoref{sec_NIST_data} as the maximal possible range.

The complete results of the calculated quantities as before are given in \autoref{tab_casino_simulation}, the same arguments on the influence of the random walk on the penetration depth are valid.
\begin{table}[h!]
 \centering
 \begin{tabular}{|c|c|c|c|}
  \hline
  &  \multicolumn{3}{c|}{electron energy}   \\ 
  \multicolumn{1}{|c|}{}				 & \SI{480}{keV} & \SI{976}{keV} & \SI{1680}{keV} \\ \hline
  average number of interactions per track	 & 7.68E3 	 & 1.29E4 	 & 1.81E4 \\ 
  mean penetration depth [mm]			 & 0.11 	 & 0.30		 & 0.61 \\ 
  maximal interaction density [mm] 		 & 0.06 	 & 0.18 	 & 0.51 \\ 
  50\% maximal interaction density [mm]		 & 0.16 	 & 0.49 	 & 1.00 \\ 
  10\% maximal interaction density [mm]  	 & 0.26 	 & 0.69 	 & 1.37 \\ 
  maximal penetration depth [mm]		 & 0.47 	 & 1.06   	 & 2.13 \\ \hline
  CSDA range [mm] (\autoref{sec_NIST_data}) & 0.5 	 & 1.1	 	 & 2.2 \\ \hline
 \end{tabular}
 \captionsetup{width=0.9\textwidth} 
 \caption[Results of the CASINO simulations for \SIlist{480; 976; 1680}{keV} electrons]
	 {Results of the CASINO simulations for \SIlist{480; 976; 1680}{keV} electrons, which are the energies of the \isotope[207]Bi conversion electrons used in \autoref{sec_bi_207_measurements}.}
 \label{tab_casino_simulation}
\end{table}

%%%%%%%%%%%%%%%%%%%%%%%%%%%%%%%%%%%%%%%%%%%%%%%%%%%%%%%%%%%%%%%%%%%%%%%%%%%%%%%%%%%%%%%%%%%%%%%%%%%%%%%%%%%%%%%%%%%%%%%%%%%%
\newpage
\FloatBarrier
\subsection{Conclusion on penetration depths}
\label{sec_conclusion_penetration_depth}
The results of the databases and different simulation programs differ. 
A brief comparison and conclusion is shown here concerning the sources and energies used in \autoref{sec_surface_events}.

\subsubsection{Alpha radiation}
The values of the ranges obtained by NIST are larger than those of SRIM by about \SI{20}{\%}.
This comparison could only be done for Sn, not for CdZnTe, although the same behavior is assumed.
However, the absolute value of the range is so small, that possible differences can be ignored here.
The ranges can be stated as: \SI{10}{\mu m} for \SI{3}{MeV}, \SI{20}{\mu m} for \SI{5}{MeV} alpha radiation of \isotope[241]Am.

\subsubsection{Electron radiation}
The results of the penetration depth of CASINO are about \SI{15}{\%} larger than the ones of PENELOPE, although qualitatively they are very similar.\\
PENELOPE meets the requirements of the energy range better. 
Furthermore, its maximal penetration depth is compatible with the CSDA range from NIST, and also with upper values of the standard deviation of VENOM.\\
The maximal penetration depth can stem from very few electrons. 
Consequently, as penetration depth the range is taken, where the interaction density of the PENELOPE simulation falls to a low percentage level, i.e. between the \SI{10}{\%} density and maximal range.\\
This is \SIlist{0.3;0.8;1.5}{mm} for the \SIlist{480;976;1680}{keV} IC electrons of \isotope[207]Bi.

\subsubsection{Gamma radiation}
For gamma radiation, only the NIST databases were used. Due to the exponentially decreasing intensity in matter, no fixed penetration depth value can be given.
However, the results emphasize that all sources used here produce mainly central events, only \isotope[241]Am can produce a larger amount of surface events.

\subsubsection{Proton radiation}
The values of VENOM are approximately \SI{25}{\%} larger than those of SRIM.
The SRIM results for Sn are smaller by \SI{25}{\%} than the ones for CdZnTe. This arises probably mostly from the difference in density (also \SI{25}{\%}).\\
The CSDA ranges (for Sn) are even larger, especially if they are scaled by the factor 0.25 due to the density.
This cannot be resolved in this work.

\chapter{Exploring surface events}
\label{sec_surface_events}
One conclusion of the analysis of the COBRA demonstrator is that the dominant background sources are alpha particles on the detector surfaces.
This is because of several reasons:\\
First, no gamma lines can be seen in the spectra. 
Furthermore, the background does not show a significant decline above the highest prominently occurring gamma line of \SI{2.6}{MeV}, which should be the case if many gamma events were present.
Additionally, the rejection of MSE does not result in a large background reduction. 
First tests of a coincidence analysis discarding simultaneous multiple detector hits also showed no big improvement.

Alpha radiation has not been investigated in detail by COBRA so far.
Hence, a comprehensive investigation of surface events is done here, mostly using alpha radiation.

The detectors that have been investigated and their measurement methods are listed in \autoref{tab_detector_overview}.
The detectors F01 to F16 are reworked, former red-painted LNGS detectors. The mapping of the detector numbers of that layer is lost, so it is not possible to compare their LNGS and laboratory data.\\
Detector 667615-02 was used at the COBRA demonstrator, but was removed in  November 2013. Other detectors were removed in that shift, which were not working at all. 
The assumed problems of detector 667615-02 turned out to stem from the readout electronics in that channel, the detector is fully functional.\\
Detector 631972-06 is a detector for laboratory measurements, which is coated with Parylene, a light-material polymer that can be applied to very thin coatings through vapor deposition.\\ 
Detectors Det\_04 to Det\_107 were foreseen for being used in the COBRA demonstrator. 
All detectors have been measured with gamma radiation (flush and scanning) prior to installation at LNGS. The gamma scanning was performed in Dresden \cite{soerensen}. Some of the results of the gamma scanning are shown here as comparative measurements: These show typical behavior of central events, in contrast to the lateral surface events measurements done in this thesis.\\
The \textit{reference detector} performs reliably and is often used for laboratory test measurements.\\
Detector 6000-2 was investigated in the author's diploma thesis \cite{tebruegge} when first investigations to surface events were started.
\begin{table}
\begin{tabular}{|c|c|c|c|c|c|c|c|c|}
\hline
						 &            &                               \multicolumn{7}{c|}{measuring method} \\ \cline{3-9}
                                                 &            & \multicolumn{2}{c|}{scan} &                    \multicolumn{5}{c|}{flush} \\ \cline{3-9}
detector name                                    & comment    & $\alpha$ S   & $\gamma$ S & $\alpha$ F & $\beta$ F  & \makecell{mono-ener-\\getic $e^-$ F} & \makecell{\SI{60}{keV}\\$\gamma$ F} & $\gamma$ F \\ \hline
631972-06                                        & Parylene   & \checkmark   &            & \checkmark & \checkmark & \checkmark                           & \checkmark                          & \checkmark \\ \hline
667615-02                                        & LNGS & \checkmark   &            & \checkmark &            & \checkmark                           &                                     & \checkmark \\ \hline
\makecell{667467-03\\"Det\_04``}                 & LNGS &              & \checkmark &            &            &                                      &                                     & \checkmark \\ \hline
\makecell{681165-10\\"Det\_65``}                 & LNGS &              & \checkmark &            &            &                                      &                                     & \checkmark \\ \hline
\makecell{681165-10\\"Det\_102``}                & LNGS &              & \checkmark &            &            &                                      &                                     & \checkmark \\ \hline
\makecell{682872-04\\"Det\_107``}                & LNGS &              & \checkmark &            &            &                                      &                                     & \checkmark \\ \hline
F01                                              & LNGS &              &            & \checkmark & \checkmark &                                      &                                     & \checkmark \\ \hline
F05                                              & LNGS &              &            & \checkmark &            &                                      &                                     & \checkmark \\ \hline
F07                                              & LNGS & (\checkmark) &            & \checkmark &            &                                      &                                     & \checkmark \\ \hline
F12                                              & LNGS &              &            & \checkmark &            &                                      &                                     & \checkmark \\ \hline
F15                                              & LNGS & \checkmark   &            & \checkmark & \checkmark & \checkmark                           & \checkmark                          & \checkmark \\ \hline
F16                                              & LNGS &              &            & \checkmark &            &                                      &                                     & \checkmark \\ \hline
\makecell{\textit{reference}\\\textit{detector}} &            &              &            & \checkmark &            &                                      &                                     & \checkmark \\ \hline
6000-2                                           & Parylene   &              &            & \checkmark & \checkmark &                                      &                                     & \checkmark \\ \hline
\end{tabular}
 \captionsetup{width=0.9\textwidth} 
 \caption[List of tested detectors]
         {List of tested detectors. ``F'' indicates a flush measurement, ``S'' a scanning measurement.}
 \label{tab_detector_overview}
\end{table}

\begin{SCfigure}
  \centering
  \includegraphics[width=.35\textwidth]{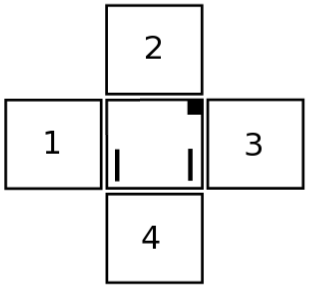}
  \includegraphics[width=.40\textwidth]{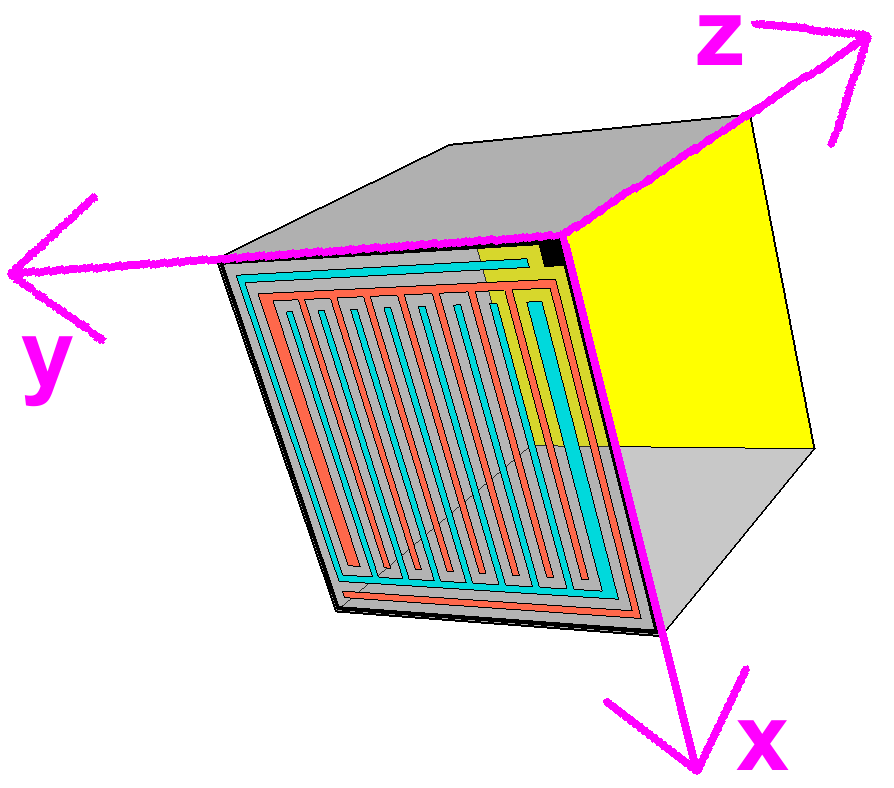}
  \caption[Definition of detector sides and axes]
          {Definition of detector sides and axes. The black rectangles (left picture only) and square on the anode surface symbolize the contacts for NCA/ CA an GR.}
  \label{fig_detector_sides_axis_conventions}
\end{SCfigure}

The conventions for detector sides and axes are shown in \autoref{fig_detector_sides_axis_conventions}.

%%%%%%%%%%%%%%%%%%%%%%%%%%%%%%%%%%%%%%%%%%%%%%%%%%%%%%%%%%%%%%%%%%%%%%%%%%%%%%%%%%%%%%%%%%%%%%%%%%%%%%%%%%%%%%%%%%%%%%%%%%%%
%%%%%%%%%%%%%%%%%%%%%%%%%%%%%%%%%%%%%%%%%%%%%%%%%%%%%%%%%%%%%%%%%%%%%%%%%%%%%%%%%%%%%%%%%%%%%%%%%%%%%%%%%%%%%%%%%%%%%%%%%%%%
\FloatBarrier
\section{Comparison: gamma scanning measurements}
\label{sec_gamma_scanning}
All detectors that were foreseen for installation at the COBRA demonstrator were scanned with gamma radiation at TU Dresden. 
The scanning was performed with a collimated \SI{100}{MBq} \isotope[137]Cs source. 
The collimator was made of lead, with a drill of \SI{0.5}{mm} in diameter and \SI{6}{cm} length, standing in a distance of \SI{2.7}{mm} to the detector.
The collimation was not as good as expected: No focused pure gamma beam was achieved.
The rate of SSE full energy peaks within the whole detector volume, arising from gamma rays not been blocked by the collimator, is approximately one third of the height at the target position.

$10\times10$ scanning points with a distance of \SI{1}{mm} were irradiated at the anode and at least one other lateral side.
The relative precision was very high, controlled by a step motor. 
The absolute calibration could not be done precisely:
The first scanning point was set manually to the position, where the gamma beam hits the detector fully. 
All other points were then accessed by the step motor with a very high relative precision.

\begin{SCfigure}[][b!]
 \centering
  \includegraphics[width=.7\textwidth]{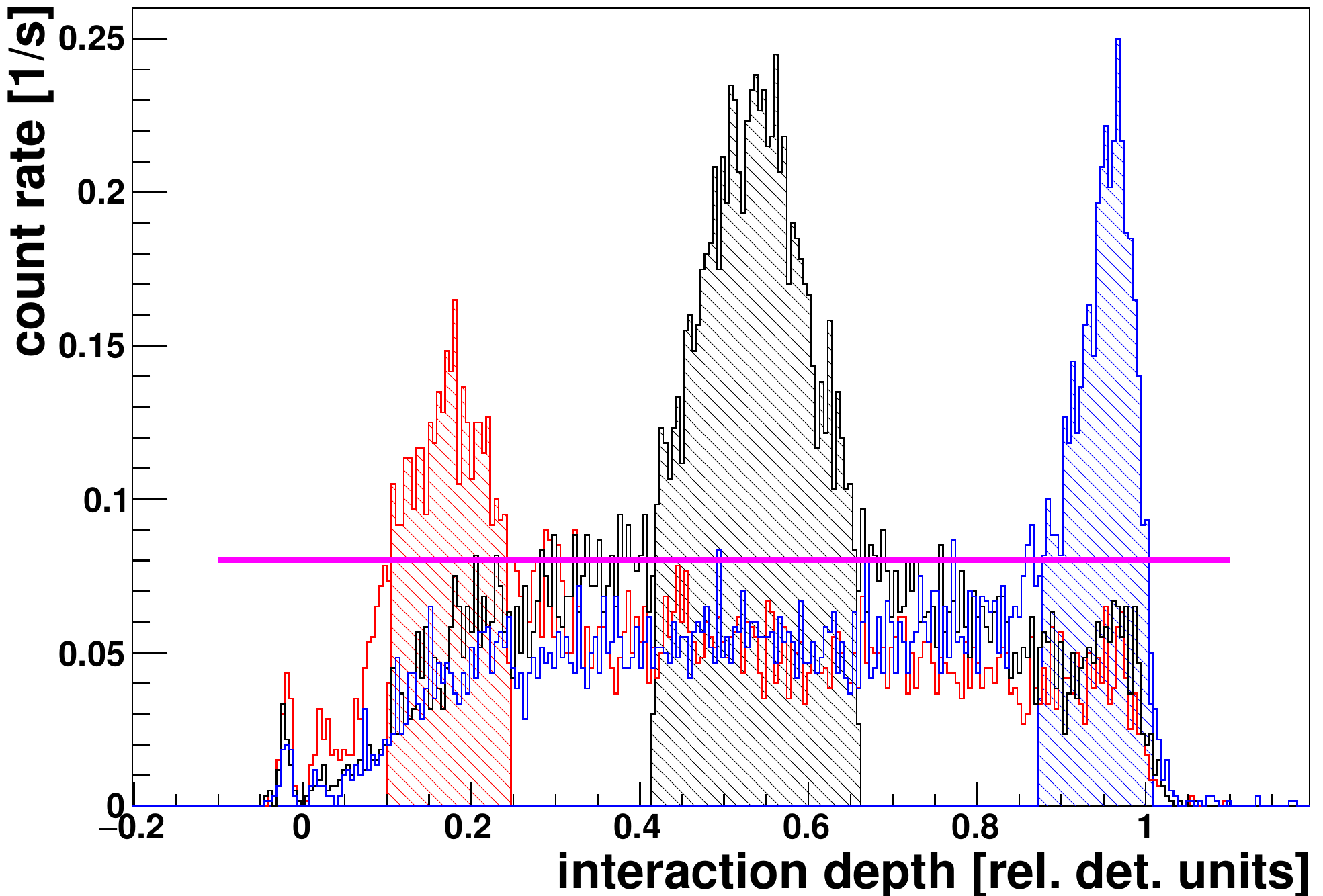}
  \caption[Interaction depth of gamma scanning]
          {Interaction depth for three scanning positions (only SSE): At the center (black), and the diagonal opposed corners (red and blue) of a detector side. The magenta line indicates the threshold to separate the peak from the background. The filled areas show the peak content used for the following analyses.}
  \label{fig_find_gamma_peak}
\end{SCfigure}

For the analyses, the interactions of the gamma radiation only at the targeted scanning point is needed.
These are found as follows:
The maximal value in the interaction depth plot is searched for. From this point, the interval to both sides is taken, until the background level is reached.
This stems from the gamma rays that should have been blocked off by the collimator.
A common threshold above the background level is chosen for all measurements of one detector side.
Only the events of this interval are used for the analyses.
\autoref{fig_find_gamma_peak} shows examples of this procedure.

The mean values of the interaction depth, LSE and A/E distributions are calculated using this peak-content.
In the following, these are used for detailed analyses to study the behavior of those quantities.\\
The gamma radiation of the last scanning point at the cathode, shown in blue in \autoref{fig_find_gamma_peak}, does not hit the detector fully. 
Consequently, this distribution is cut off, and hence the calculated mean values cannot be compared to the values of the other positions where the beam hits the detector fully.
These values are ignored for the following analyses.

\begin{figure}
 \centering
  \includegraphics[width=.49\textwidth]{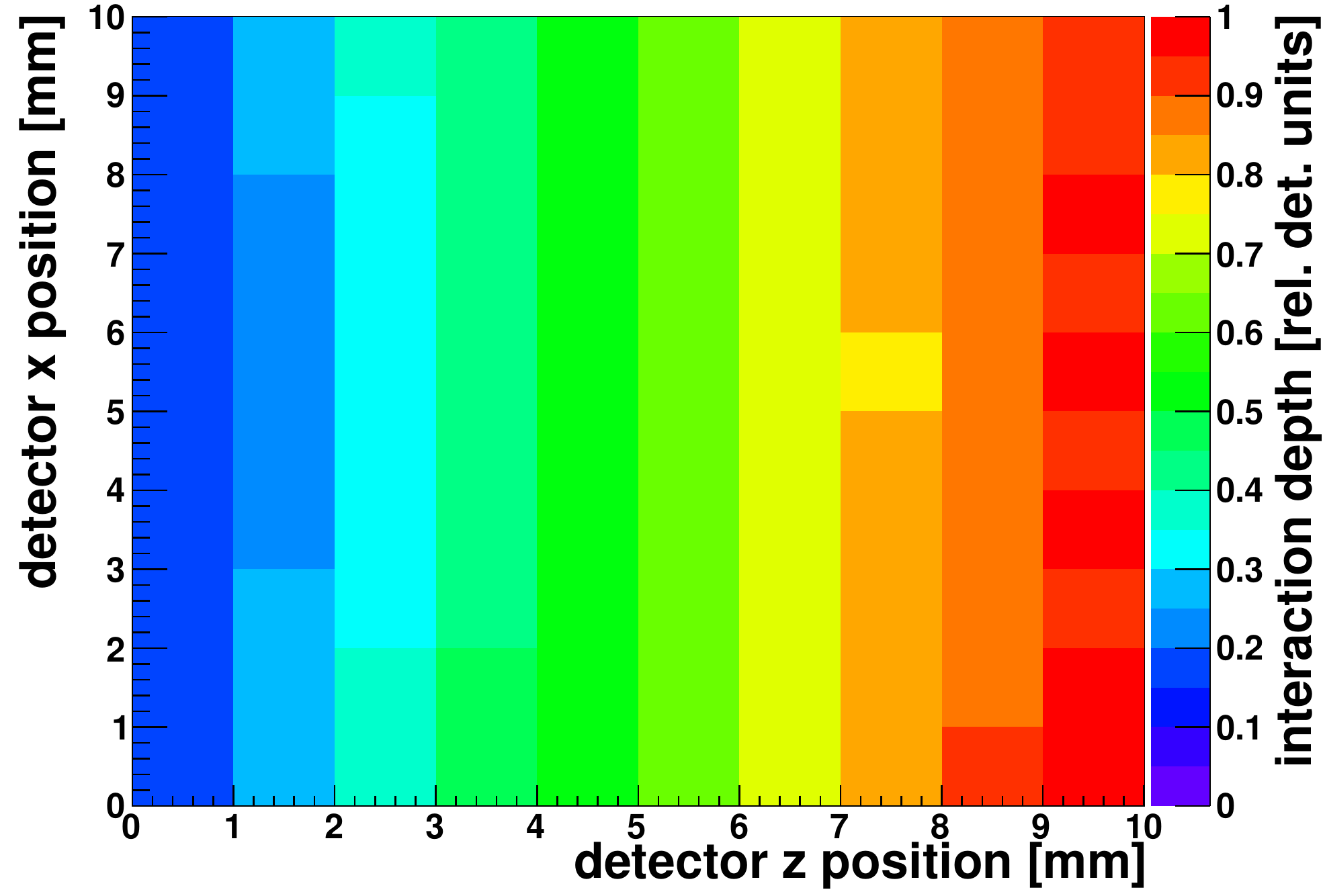}
  \includegraphics[width=.49\textwidth]{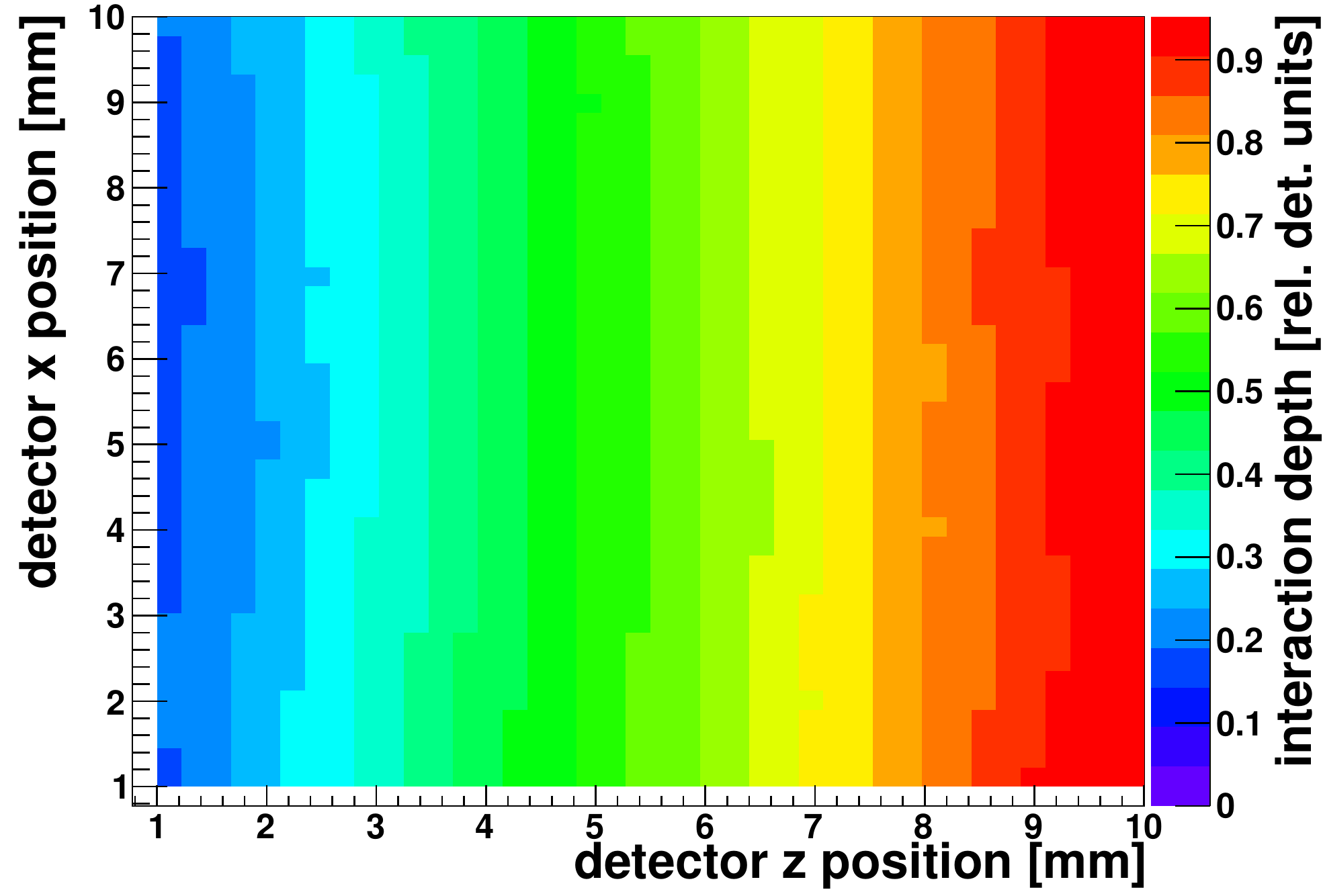}
    \captionsetup{width=0.9\textwidth} 
    \caption[Interaction depth of detector Det\_107 side 1.]
            {Interaction depth of detector Det\_107 side 1. Left: Calculated interaction depth of each scanning point shown at its position. Right: Interpolation.}
   \label{fig_gamma_scanning_interaction_depth_3D}
\end{figure}

Some of the results are shown here exemplarily, especially as a comparison for the alpha scanning measurements in \autoref{sec_alpha_scanning}.\\
Using this set of gamma scanning data, one can verify the calculation of the interaction depth.
\autoref{fig_gamma_scanning_interaction_depth_3D} shows the calculated interaction depth for all scanning points of the detector side.
The two-dimensional diagrams comprise a representation, where each calculated value is plotted versus its scanning position, and an interpolation of this point-density.
The interpolation is done by ROOT using the Delaunay triangulation method. This is a method to generate a mesh of triangles from a set of points in Euclidean space.
For details see the original paper (in French) \cite{delaunay_original_paper}, or a discussion in \cite{delaunay_secundary_lit}\label{text_delaunay}. 
The Delaunay triangulation is often used in computer graphics for modeling terrain or densities, as it minimizes rounding errors.
The calculated values show a linear behavior in z-direction, and constant values in x-direction, as expected.
However, at some positions, deviations from this pattern can be seen, e.g. at (7.5,5.5). 
This probably stems from local crystal defects.

Furthermore, it is possible to compare the two different formulas \autoref{eq_cal_ipos_z} based on the simple model ($z$), and \autoref{eq_cal_ipos_ztc} including trapping corrections ($z_{tc}$).
For each scanning point and both models separately, the mean value of the calculated interaction depth (as shown in \autoref{fig_find_gamma_peak}), the error of the mean value, and the FWHM of the distribution are obtained.
The error of the mean value of each scanning measurements is smaller than the mean value by approximately three orders of magnitude.
As the mean values are measured values $x_i$ with known statistical distributed uncertainties $\sigma_i$ (the mean error in this case), the weighted mean $x{_\text{mean}}$ and its standard deviation $\sigma_{\text{mean}}$ can be calculated as follows \cite{Blobel:3519032430}:
\begin{equation}
 \begin{split}
  x{_\text{mean}} = \frac{\sum\limits_i x_i \cdot \sigma_i^{-2}} {\sum\limits_i \sigma_i^{-2}}\\
  \sigma_{\text{mean}} = \frac{1}{ \sqrt{ \sum\limits_i\sigma_i^{-2}}}
 \end{split}
 \label{eq_weighted_mean}
\end{equation}

For each of the ten z-positions, the weighted mean values of the ten different x-positions are plotted versus their z-positions. 
The uncertainties based on the mean errors are so small, that they could not be seen in the graphical representation in \autoref{fig_gamma_scanning_interaction_depth}
Here, the width (standard deviation) of each calculated interaction depth distribution is plotted as error bar, as not only the mean position, but also the width of the distribution is of interest. 

A linear fit $f(x)=p0+p1\cdot x$ is performed for each model (discarding the measurements at z=\SI{10}{mm} as discussed before).
For the fit, the mean errors are used as uncertainties.
The resulting fit parameters $\chi^2$, NDF (\textbf{n}umber of \textbf{d}egrees of \textbf{f}reedom), and $\chi^2_{red}=\chi^2/NDF$ are given in \autoref{tab_gamma_scanning_interaction_depth_results}. 
The parameters are similar for both models. 
As the absolute calibration of the collimator was not done precisely, the offset ($p_0$) is not zero.
More important is the linear behavior of the calculated interaction depth.
The reduced $\chi^2$ are much larger than one.
This arises probably from the fact, that the uncertainties are very small, and that local systematic variations occur due to crystal defects.\\
Overall, no large systematic differences between the two models can be seen in the bulk of the detector volume.
A comparison for the interaction depth calculation of anode- and cathode surface events is done \autoref{sec_alpha_scanning}.\\

 \vspace{0.5cm}
\begin{minipage}[]{0.59\textwidth}
  \includegraphics[width=.99\textwidth]{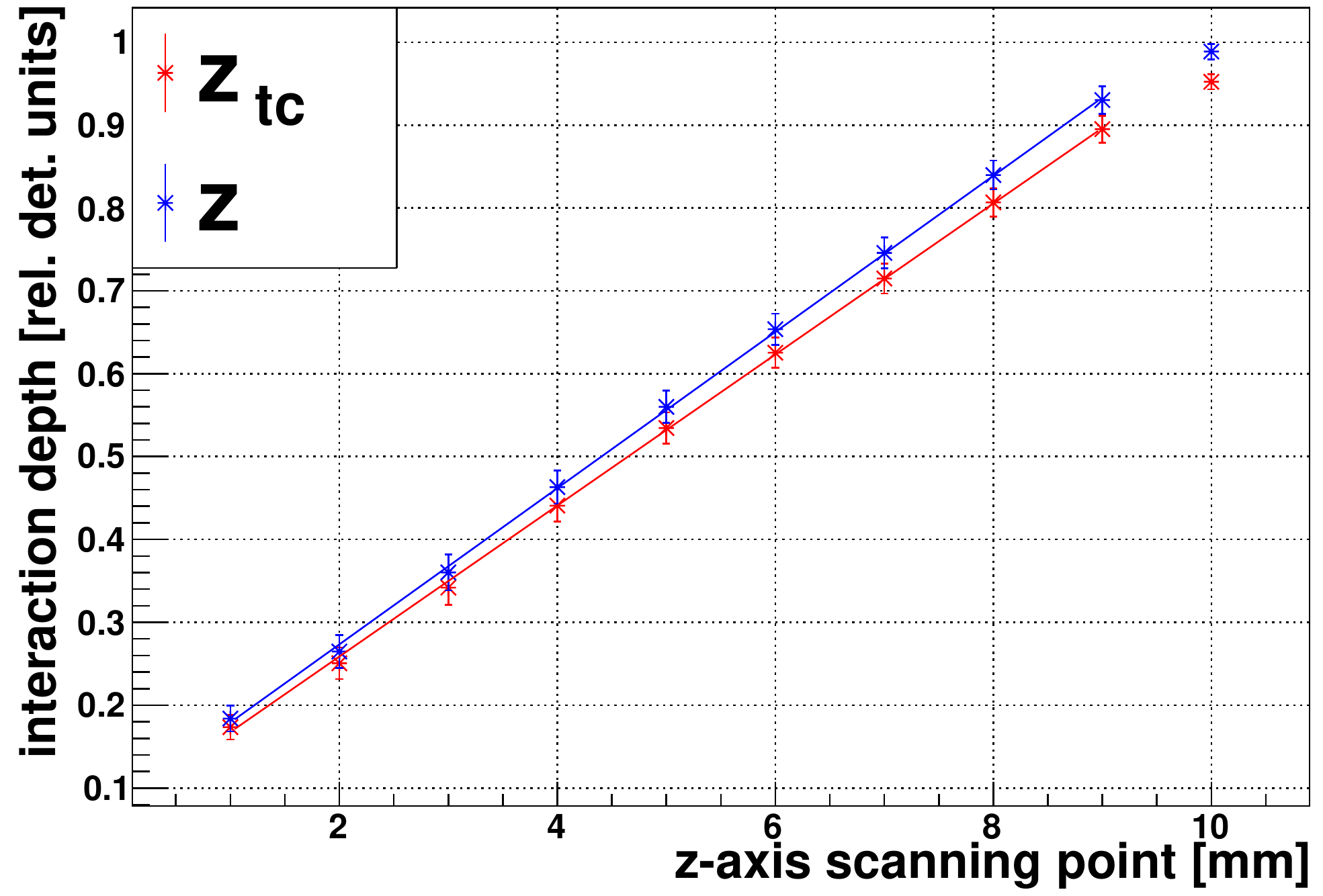}
  \captionof{figure}[Comparison of interaction depth calculations]
                    {Comparison of the two interaction depth calculations $z$ and $z_{tc}$. The FWHM of each distribution in \autoref{fig_find_gamma_peak} is used for the error bars here.
                    A linear fit is performed using the errors of the mean value as uncertainty.  The results are compiled in \autoref{tab_gamma_scanning_interaction_depth_results}.}
   \label{fig_gamma_scanning_interaction_depth}
\end{minipage}
\hfill
\begin{minipage}[]{0.35\textwidth}
  \begin{tabular}{|c|cc|}
    \hline
        fit param.      & $z$                                                    & $z_{tc}$                                            \\ \hline
    p0                  & \num[round-mode=places,round-precision=3]{0.073827}  & \num[round-mode=places,round-precision=3]{0.0824444} \\ \hline 
    p1                  & \num[round-mode=places,round-precision=3]{0.0913926} & \num[round-mode=places,round-precision=3]{0.0945102} \\ \hline
    $\chi^2$            & \num[round-mode=places,round-precision=3]{2263}      & \num[round-mode=places,round-precision=3]{2214}      \\ \hline
    NDF & 7                                                    & 7                                                    \\ \hline
    $\chi^2_{red}$      & \num[round-mode=places,round-precision=3]{323}       & \num[round-mode=places,round-precision=3]{316}       \\ \hline   
  \end{tabular}
  \captionof{table}[Results of the comparison of the interaction depth calculations]
                   {Results of the comparison of the interaction depth calculations $z$ and $z_{tc}$.}             
  \label{tab_gamma_scanning_interaction_depth_results}
\end{minipage}
\vspace{0.5cm}

\begin{figure}%[b!]
 \centering
  \includegraphics[width=.49\textwidth]{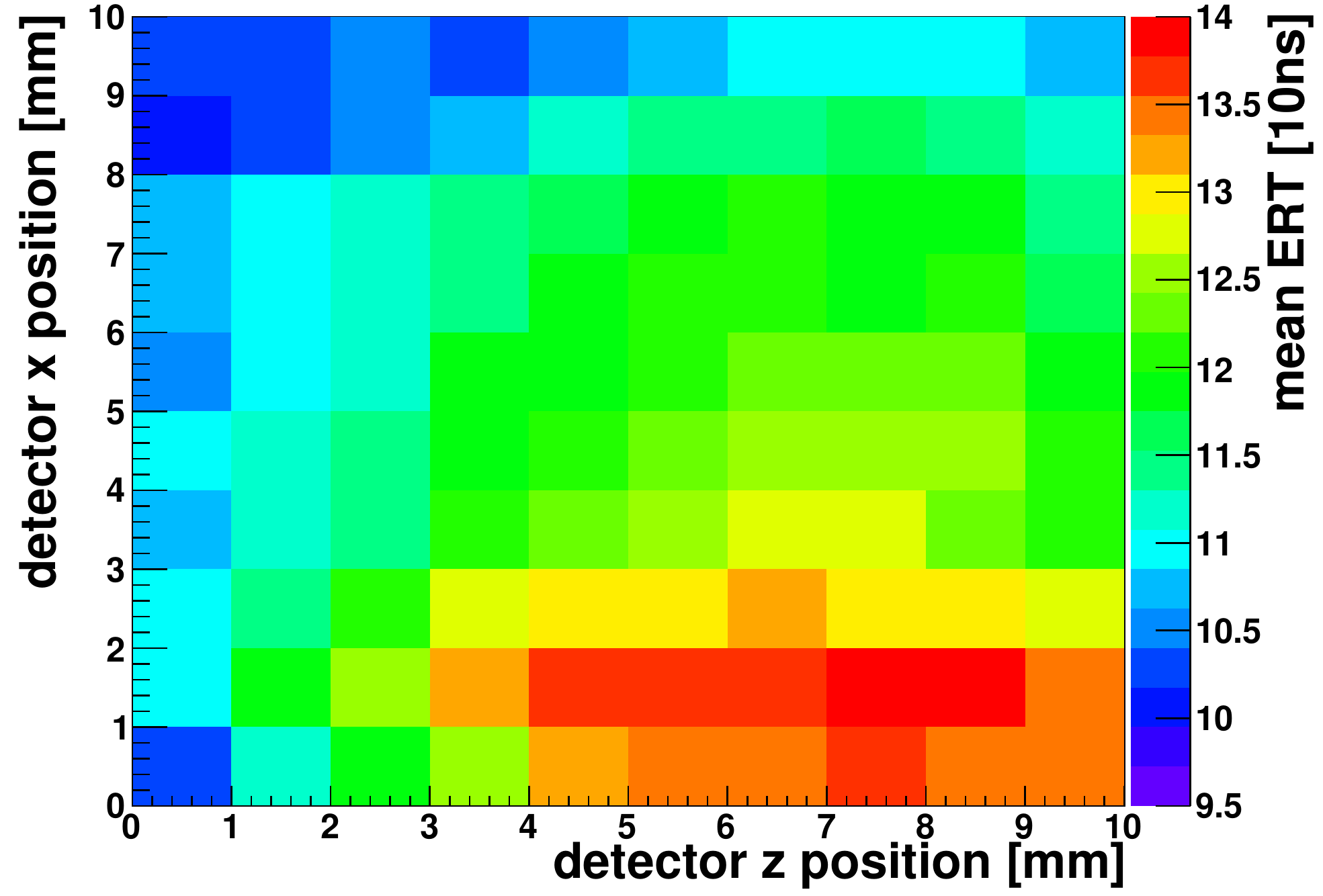}
  \includegraphics[width=.49\textwidth]{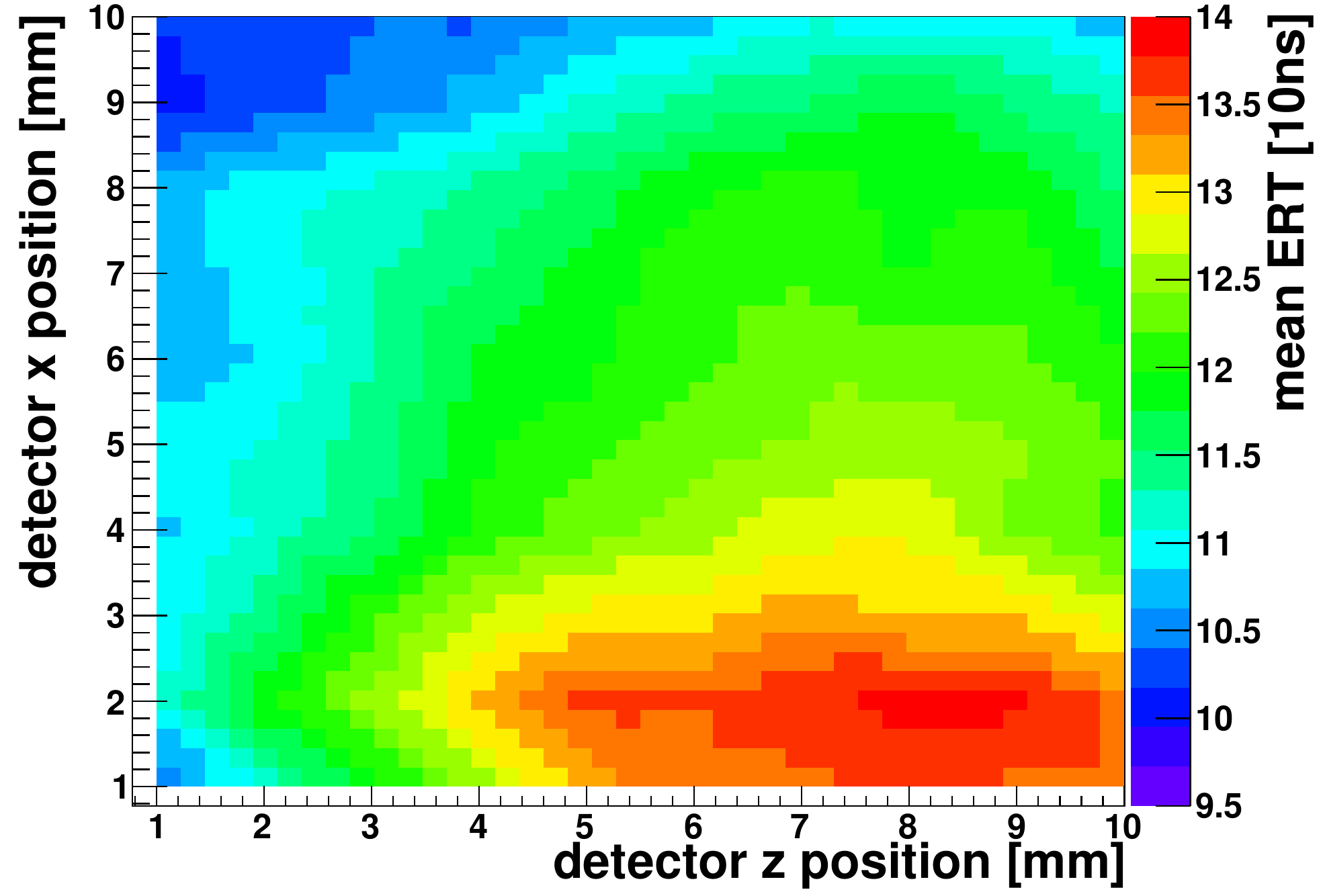}
    \captionsetup{width=0.9\textwidth} 
    \caption[Mean ERT values of detector Det\_107 side 1]
            {Mean ERT values of detector Det\_107 side 1. Left: Mean values at their scanning positions. Right: Interpolation. An area of significant deviations from the otherwise constant values for central events can be seen.}
   \label{fig_gamma_scanning_interaction_depth_ERT}
\end{figure}

In \autoref{fig_gamma_scanning_interaction_depth_ERT}, the distribution of the ERT values of the same measurement is shown exemplarily. 
The mean ERT value of each scanning point is calculated as described above. It is plotted as a function of its position directly, and interpolated.
A relative homogeneous distribution can be seen, especially if compared to the alpha scanning measurements in \autoref{sec_alpha_scanning}. 
Nevertheless, an area is found where higher values occur, which seems to stem from a localized crystal defect.
Other quantities like DIP or A/E show qualitatively a similar behavior of relative homogeneous distributions with local variations.

%%%%%%%%%%%%%%%%%%%%%%%%%%%%%%%%%%%%%%%%%%%%%%%%%%%%%%%%%%%%%%%%%%%%%%%%%%%%%%%%%%%%%%%%%%%%%%%%%%%%%%%%%%%%%%%%%%%%%%%%%%%%
%%%%%%%%%%%%%%%%%%%%%%%%%%%%%%%%%%%%%%%%%%%%%%%%%%%%%%%%%%%%%%%%%%%%%%%%%%%%%%%%%%%%%%%%%%%%%%%%%%%%%%%%%%%%%%%%%%%%%%%%%%%%
\FloatBarrier
\section{General surface sensitivity}
\label{sec_general_surface_sensitivity}

First hints to a lower surface sensitivity occurred at the author's diploma thesis \cite{tebruegge} and in the publication to LSE \cite{Fritts:2014qdf}.

Sensitivity problems were also published in several papers:
Bolotnikov et al. stated in \cite{botonikov_laser_czt}:
\begin{quote}
``[\dots] [it is] important to know the width of the dead area near the edges of the CZT [CdZnTe] detector (the
area from which the collection efficiency is significantly degraded) [\dots]  For this geometry [\dots] [the dead area] [\dots] can exceed \SI{100}{\mu m} or more.''
\end{quote}

In a later publication by the same author et al. \cite{5075926}, the electric field lines of a CdZnTe pixel detector were investigated using a high-spatial resolution X-ray mapping technique with a beam spot of less than $(\SI{25}{nm}\times\SI{25}{nm})$.
A large scale variation of the field lines was found, bending away or towards the sides, resulting in focusing and de-focusing effects. Additional local defects (Te-inclusions and twin boundaries) and strains were found within the crystal.\\
A lateral variation of the detector properties was measured, a lower drift-velocity at the edges and a higher concentration of traps were found. This results in a poor charge-collection efficiency, called 'edge effect'.\\
In the following, some relevant statements of the publication are cited:
\begin{quote}
``[\dots] clearly reveals that the electric field inside the thick CZT [CdZnTe] crystals is far from uniform. 
[\dots] the field lines are so strongly bent that some of the edge pixels do not collect any charges. 
[\dots] the charges generated by incident particles might be driven towards wrong electrodes, or become trapped in 'dead' regions, e.g., near the side surfaces. 
[\dots] Our goal is to emphasize the surprisingly strong discrepancies between the expected and actual field-line distributions in high-quality commercial CZT detectors.''
\end{quote}
As a possible solution, the authors propose to extend the cathode metalization \SI{2}{mm} to the sides to have a focusing effect. 
In these studies pixel detectors were used. These only have one BV, but no GB. Consequently, no dependencies of this quantity could be investigated (which is done in this thesis).\\

\begin{SCfigure}
 \centering
  \includegraphics[width=0.7\textwidth]{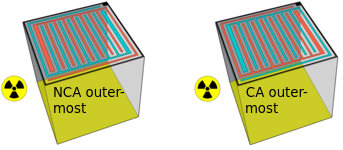}
  \caption[Sketch indicating the swapping of NCA and CA]
          {Sketch indicating the swapping of the wiring of NCA and CA without changing source and detector position (not to scale). Radioactive symbol taken from \cite{radioactive_symbol}. }
  \label{fig_NCA_CA_swapping}
\end{SCfigure}

In this work, the investigations of the surface sensitivity of NCA and CA as outermost anode rails were conducted with the following types of radiation:\\
\isotope[241]Am as source of alpha  and gamma radiation. The main alpha energies of \SI{5486}{keV} (\SI{84.5}{\%}) and \SI{5443}{keV} (\SI{13}{\%}) cannot be separated. The main gamma line is at \SI{60}{keV}. 
Furthermore, \isotope[204]Tl beta radiation (\SI{763}{keV}), and \isotope[207]Bi IC electrons (\SIlist{482; 976; 1680}{keV}) were used.\\
The procedure was as follows: A detector was irradiated with a source at a certain setup geometry. 
Then the wiring of CA and NCA was swapped by exchanging the relevant cables, without altering the position of source or detector to have comparable measurement settings. 
Differences in the sensitivity between them originate only from effects of the detector biases, not from detector lacquer or similar.\\
The outermost anode rail at each (lateral) detector side can be supplied with NCA or CA bias.
In the following, a description like ``side 1 CA'' means that the outermost anode rail on side 1 is on CA potential, for example.\\
A sketch of this measurement procedure is shown in \autoref{fig_NCA_CA_swapping}.\\

Typical results of the different measurements are shown in the following, a complete results compilation is shown in \autoref{sec_results_of_general_surface_sensitivity}.

%%%%%%%%%%%%%%%%%%%%%%%%%%%%%%%%%%%%%%%%%%%%%%%%%%%%%%%%%%%%%%%%%%%%%%%%%%%%%%%%%%%%%%%%%%%%%%%%%%%%%%%%%%%%%%%%%%%%%%%%%%%%
\FloatBarrier
\subsection{Alpha flush measurements with \isotope[241]Am}
\label{sec_5MeV_alpha_radiation}

In contrast to the scanning measurements described in \autoref{sec_alpha_scanning}, a flush measurement irradiates the whole detector uncollimated.
Alpha radiation of \isotope[241]Am is used to see if the surfaces are fully sensitive.
As possible effects of alpha radiation at the surfaces are unclear, events that are removed by the data-cleaning algorithm (discussed in \autoref{sec_analysis_principle}) are shown as well. 

\begin{figure}[t!]
 \centering
  \includegraphics[width=0.49\textwidth]{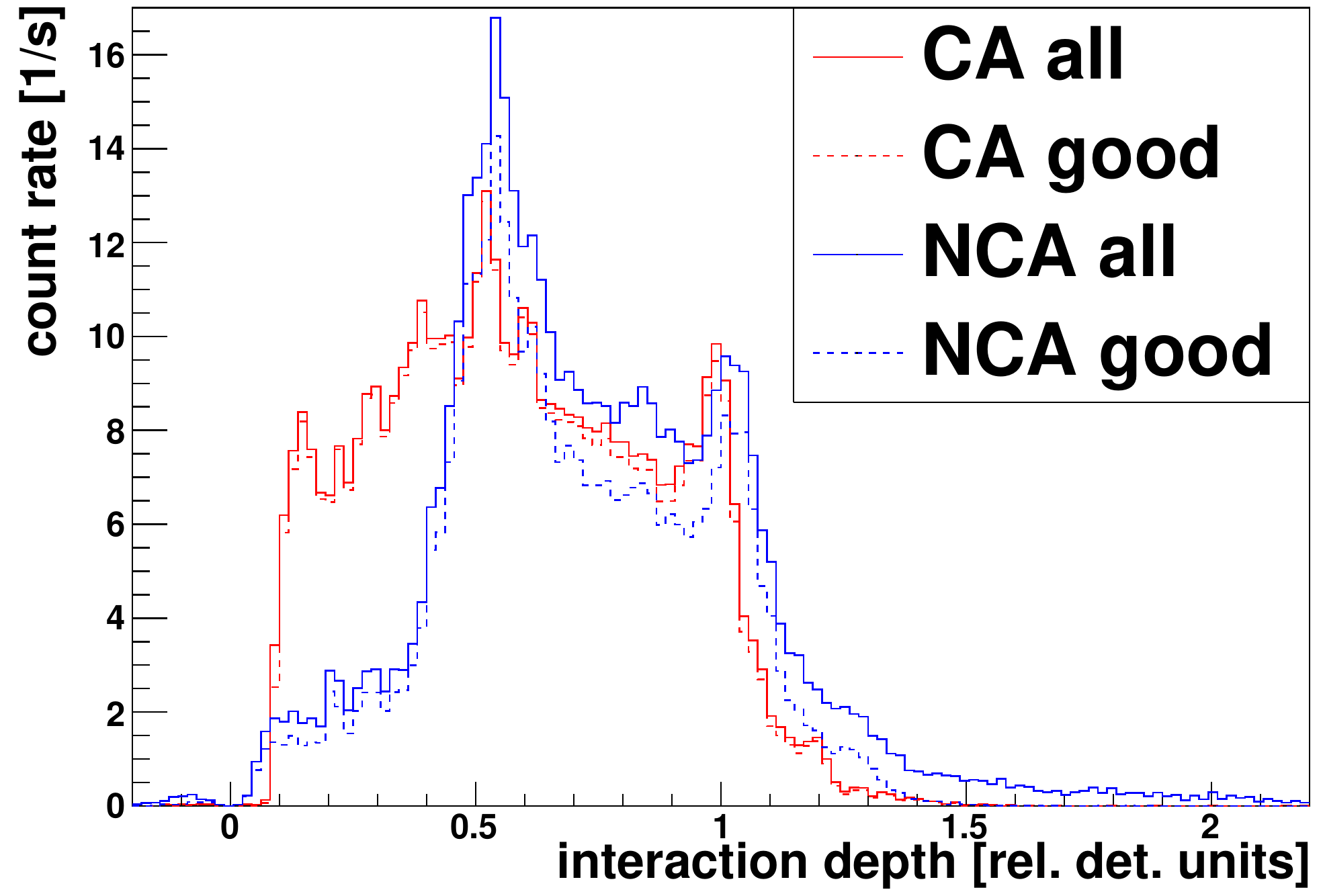}
  \includegraphics[width=0.49\textwidth]{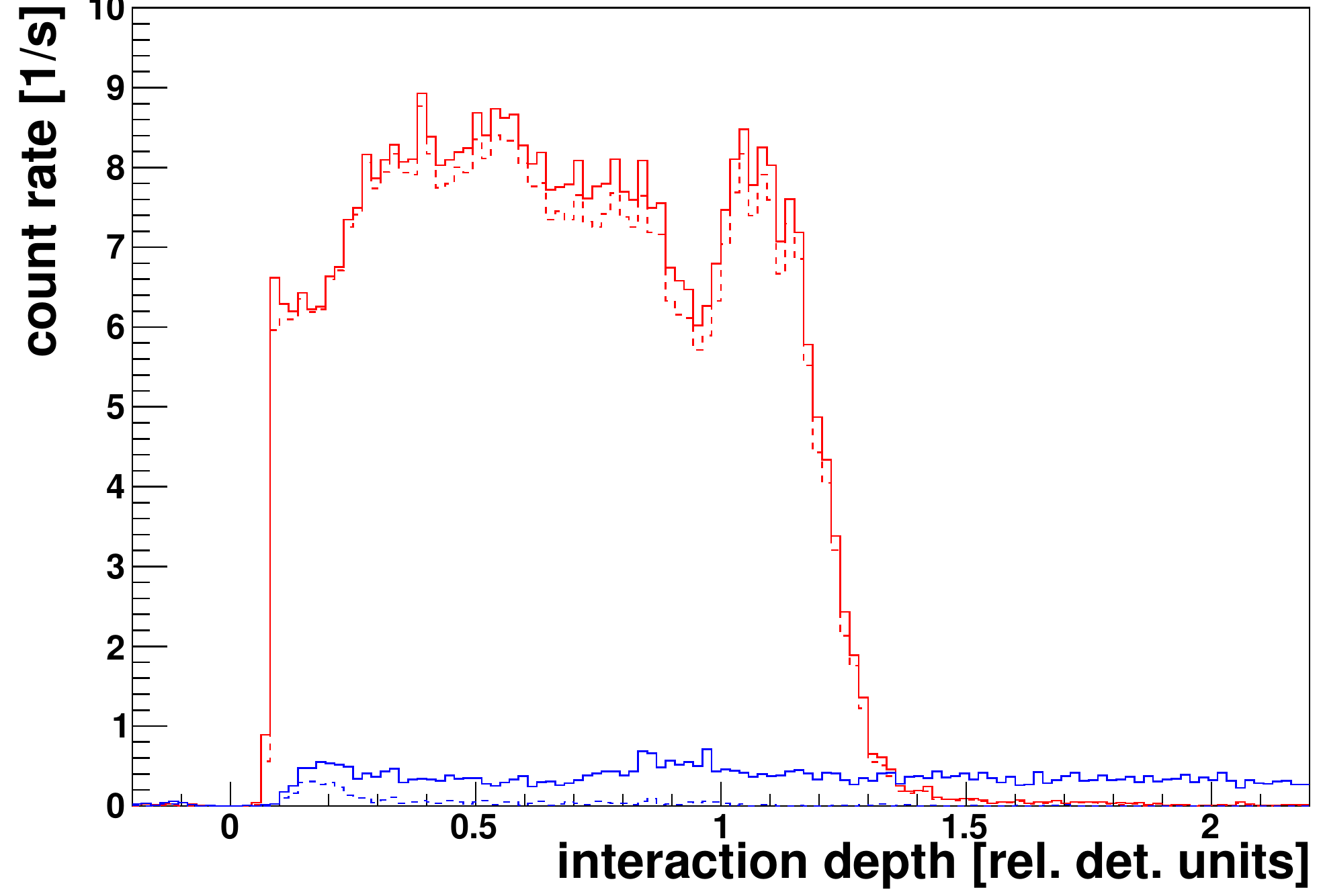}
  \captionsetup{width=0.9\textwidth} 
      \caption[Surface sensitivity to alpha radiation]
              {Surface sensitivity of detectors F16 side 1 (left) and F05 side 4 (right) to alpha radiation.
              Left: The detector is nearly fully sensitive, but near-anode events at NCA are reconstructed wrongly.
              Right (same color code): The NCA sensitivity is strongly suppressed versus the CA sensitivity.}
      \label{fig_alpha_surface_sensitivity}
\end{figure}

\begin{SCfigure}[][b!]
 \centering
  \includegraphics[width=0.75\textwidth]{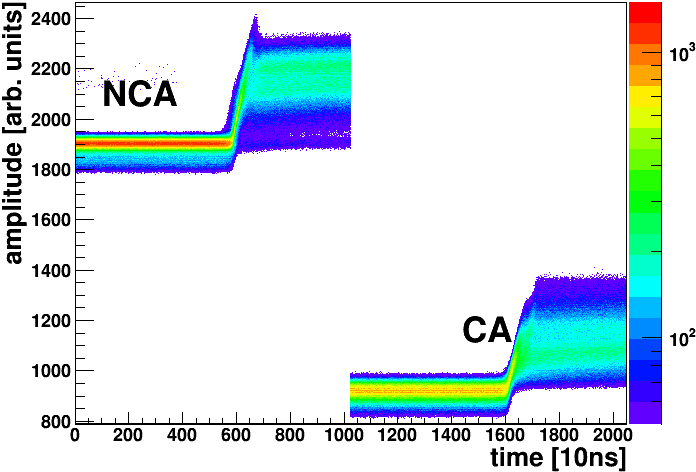}
  \caption[Pulse shapes of alpha measurements at NCA surface]
          {Superposition of all pulse shapes of the measurement at the NCA surface. The NCA and CA signals are plotted successively, but happen at the same timespan of \SI{1024}{bins} which corresponds to \SI{10.24}{\mu s} at \SI{100}{MHz}.}
      \label{fig_alpha_surface_sensitivity_pulse_shapes}
\end{SCfigure}

The measurements show that the surface sensitivity can differ strongly for each detector.
The detectors can be fully sensitive for both biases, or show only slight insensitivities, as
\autoref{fig_alpha_surface_sensitivity} left shows: Minor problems with the interaction depth calculation can occur, especially at a low interaction depth for NCA as outermost anode rail. 
The detector is not fully sensitive there, and the interaction depth calculation reconstructs values to higher z-positions. 
Hence, the interaction depth calculation is not completely valid for surface alpha radiation here.
The NCA sensitivity can also be suppressed more, or nearly totally, as can be seen in \autoref{fig_alpha_surface_sensitivity} right.
There, the CA shows an almost rectangular interaction depth shape, which is expected (however, a reconstruction of interaction depth values above one occurs).
In contrast to that, the distribution for the NCA as outermost anode rail shows very few events (above threshold), which are distributed to high values above one. 
Additionally, most of them are removed by the data-cleaning, so that only less than \SI{1}{\%} of the events remain.
In all other cases, the data-cleaning does not remove many events, the level of the surviving events is mostly constant below the non-cleaned distribution.\\
An explanation, why the NCA sensitivity is heavily suppressed, can be obtained by investigating the measured pulse shapes shown in \autoref{fig_alpha_surface_sensitivity_pulse_shapes}.
These do not show the typical CPG behavior. The signals on both anodes look very similar. Furthermore, the characteristic drift and (de-)charging part of the pulses cannot be seen. 
This indicates, that the moving charge cloud is not affected by the CPG anode structure, but only by the outermost anode rail, in this case only by the NCA. 
As a consequence, no useful energy deposition can be calculated.
Hence, nearly all of these events have reconstructed energies around zero and are removed by the data-cleaning and threshold mechanisms.
These non-CPG like pulse shapes have been identified by the author when working on the COBRA demonstrator. 
Not knowing this as a typical distortion of surface events, the reason for this was sought at the read-out electronics at that time.\\
Obviously, alpha (surface) events are not measured correctly at the NCA surface at this detector surface, so the detector has a dead-layer there.
This dead-layer is not an effect of the lacquer, but of the applied bias type.

%%%%%%%%%%%%%%%%%%%%%%%%%%%%%%%%%%%%%%%%%%%%%%%%%%%%%%%%%%%%%%%%%%%%%%%%%%%%%%%%%%%%%%%%%%%%%%%%%%%%%%%%%%%%%%%%%%%%%%%%%%%%
\FloatBarrier
\subsection{\SI{60}{keV} gamma radiation of \isotope[241]Am}
\label{sec_60keV_gamma_radiation}

\begin{figure}[]
 \centering
  \includegraphics[width=0.505\textwidth]{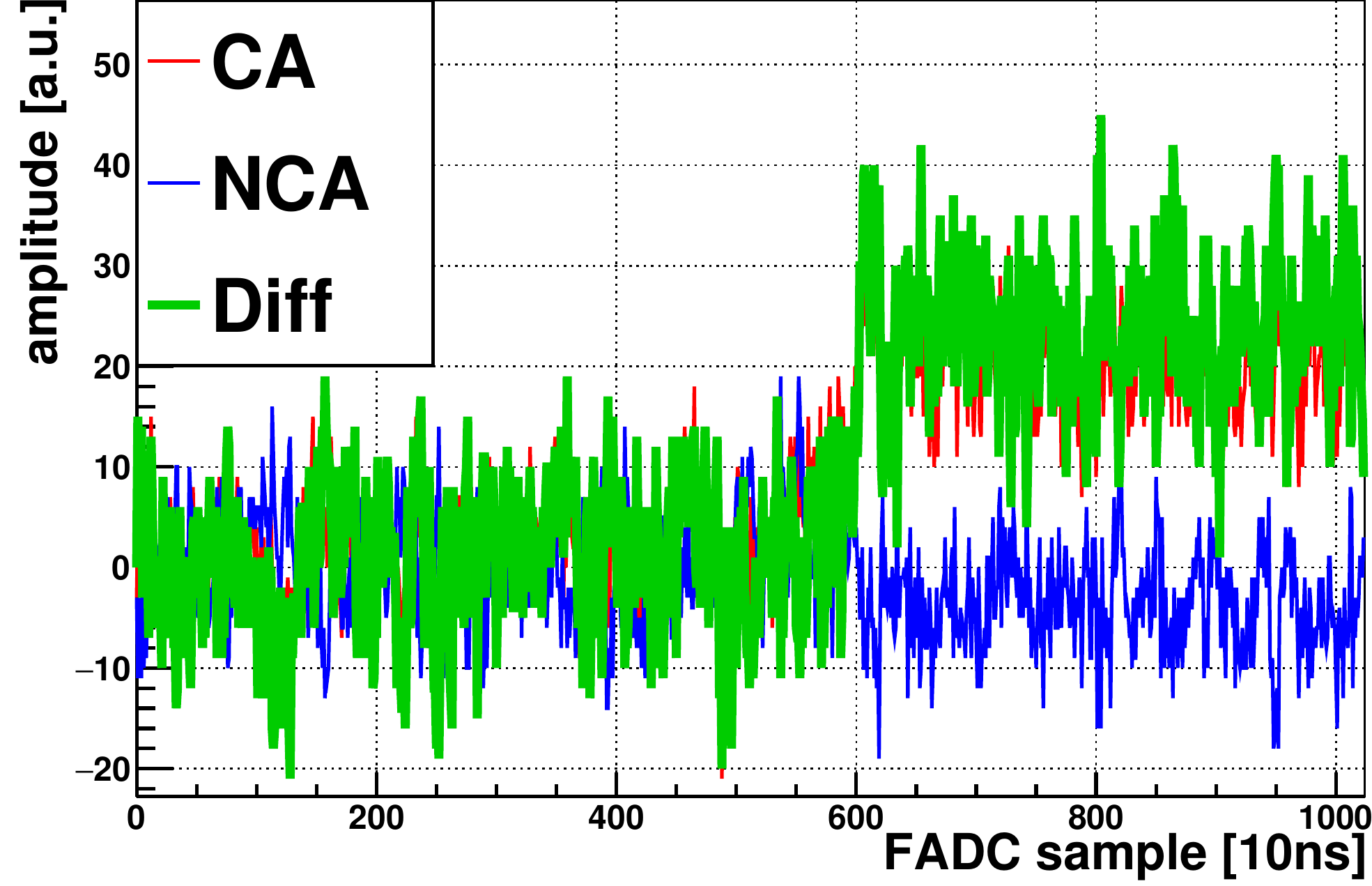}
  \includegraphics[width=0.485\textwidth]{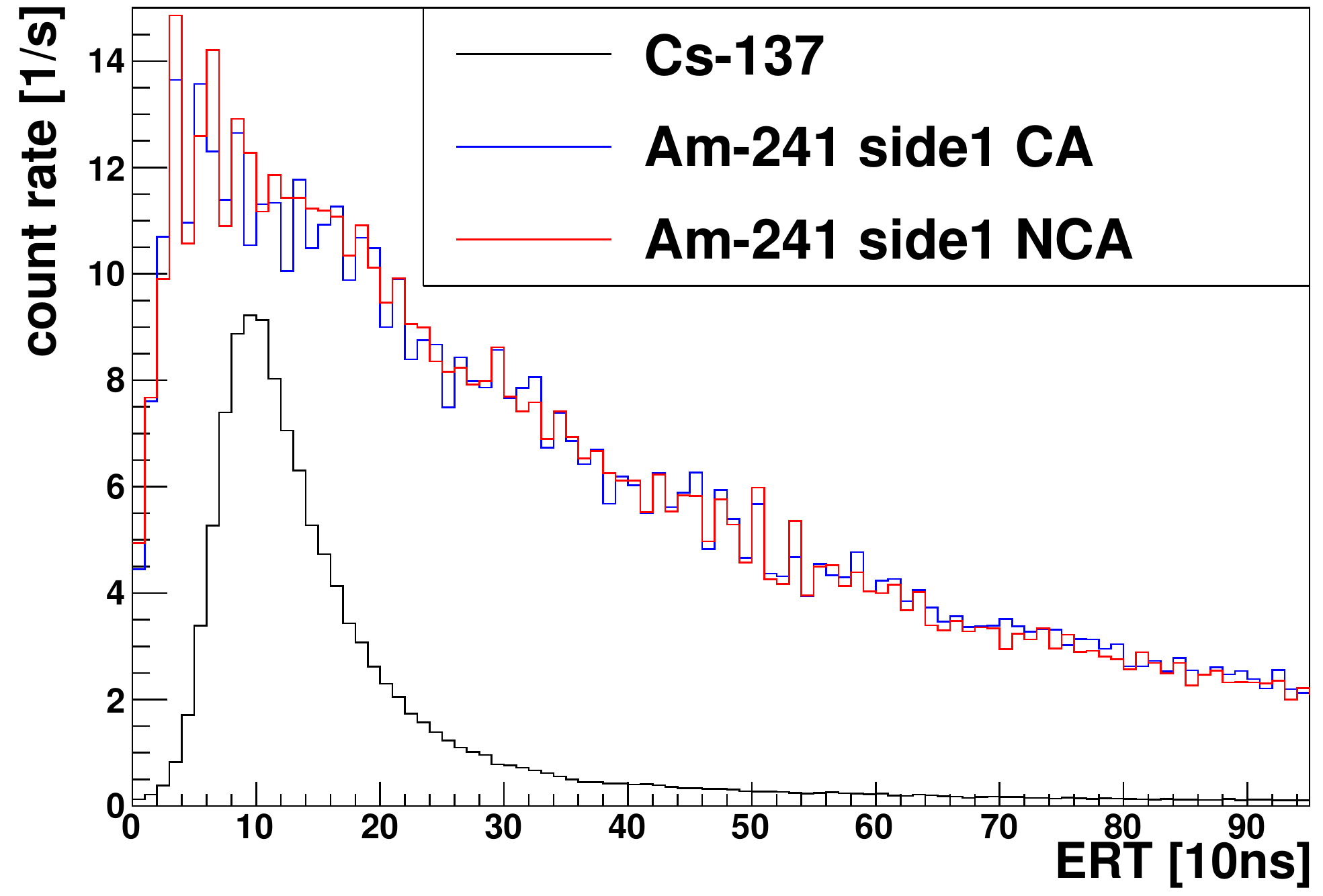}
  \captionsetup{width=0.9\textwidth} 
  \caption[Pulse shapes and ERT distribution of \SI{60}{keV} gamma event.]
          {Left: Pulse shape of a \SI{60}{keV} gamma event. Right: ERT distribution. An example for a meaningful distribution (of central events) is shown in the black curve for comparison.}
  \label{fig_ps_60keV_Am241}
\end{figure}

An advantage of the \SI{60}{keV} gamma radiation of \isotope[241]Am for surface events test measurements is the low penetration depth, as calculated in \autoref{tab_gamma_intensity}. 
Furthermore, no angular deviation is expected, as these low-energy gamma rays interact only via  the photoelectric-effect, see \autoref{fig_nist_gammas_electrons} left. 
But the SNR is poor, and \SI{60}{keV} is often hardly above the noise level, depending on the detector quality and experimental setup.\\
An example of a pulse shape and a measured ERT distribution is shown in \autoref{fig_ps_60keV_Am241}.
The amplitude of the difference pulse can be seen, but PSA is not meaningful due to a poor SNR.
The spectral form of the ERT distribution is not like of typical central (or surface) events.
Additionally, the fact that the ERT distribution is the same for both NCA and CA as outermost anode rail, clearly demonstrates that PSA is not useful here.
Concluding, non-PSA methods like energy and interaction depth reconstruction work, while PSA is not meaningful.

\begin{figure}[h!]
 \centering
  \includegraphics[width=0.49\textwidth]{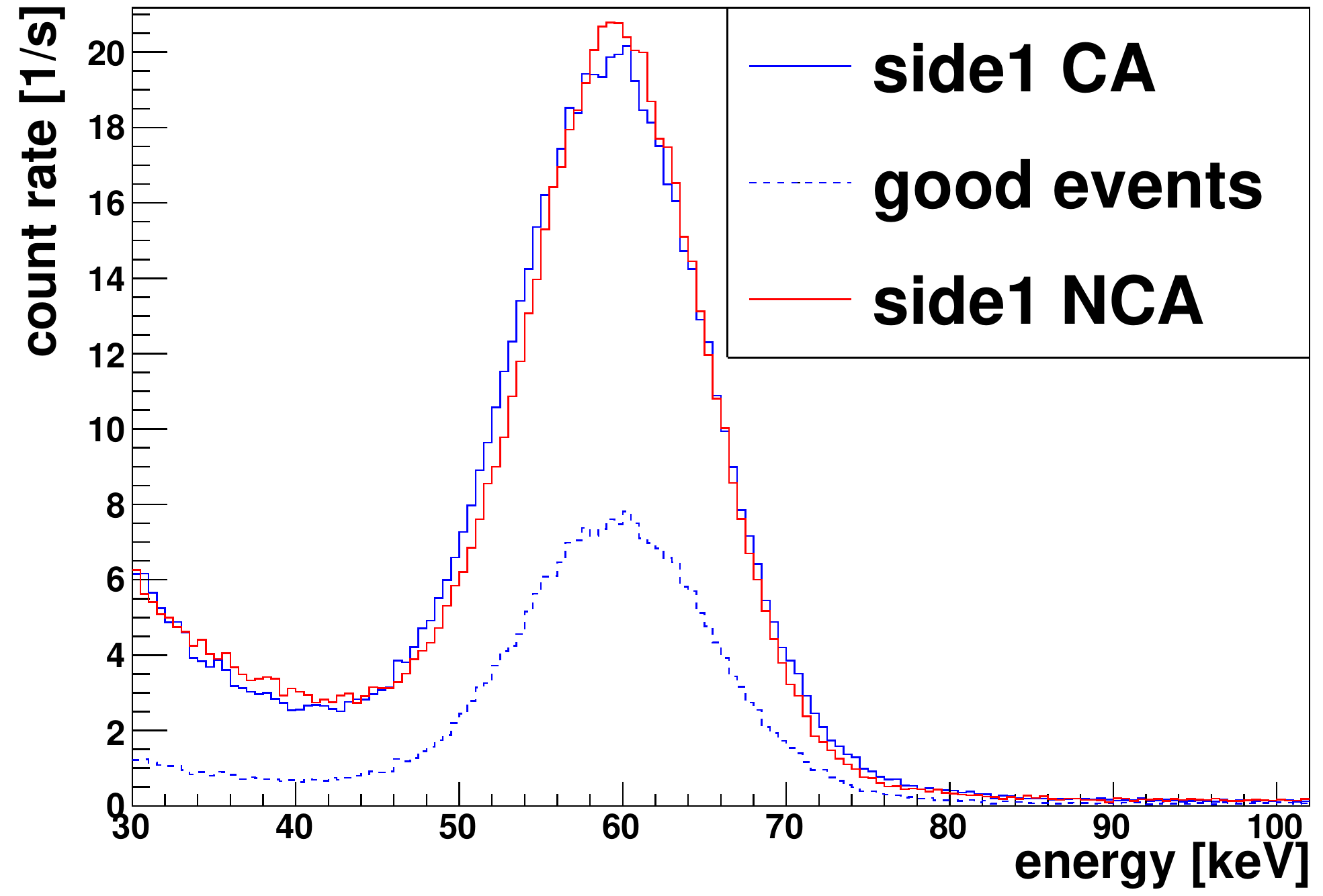}
  \includegraphics[width=0.49\textwidth]{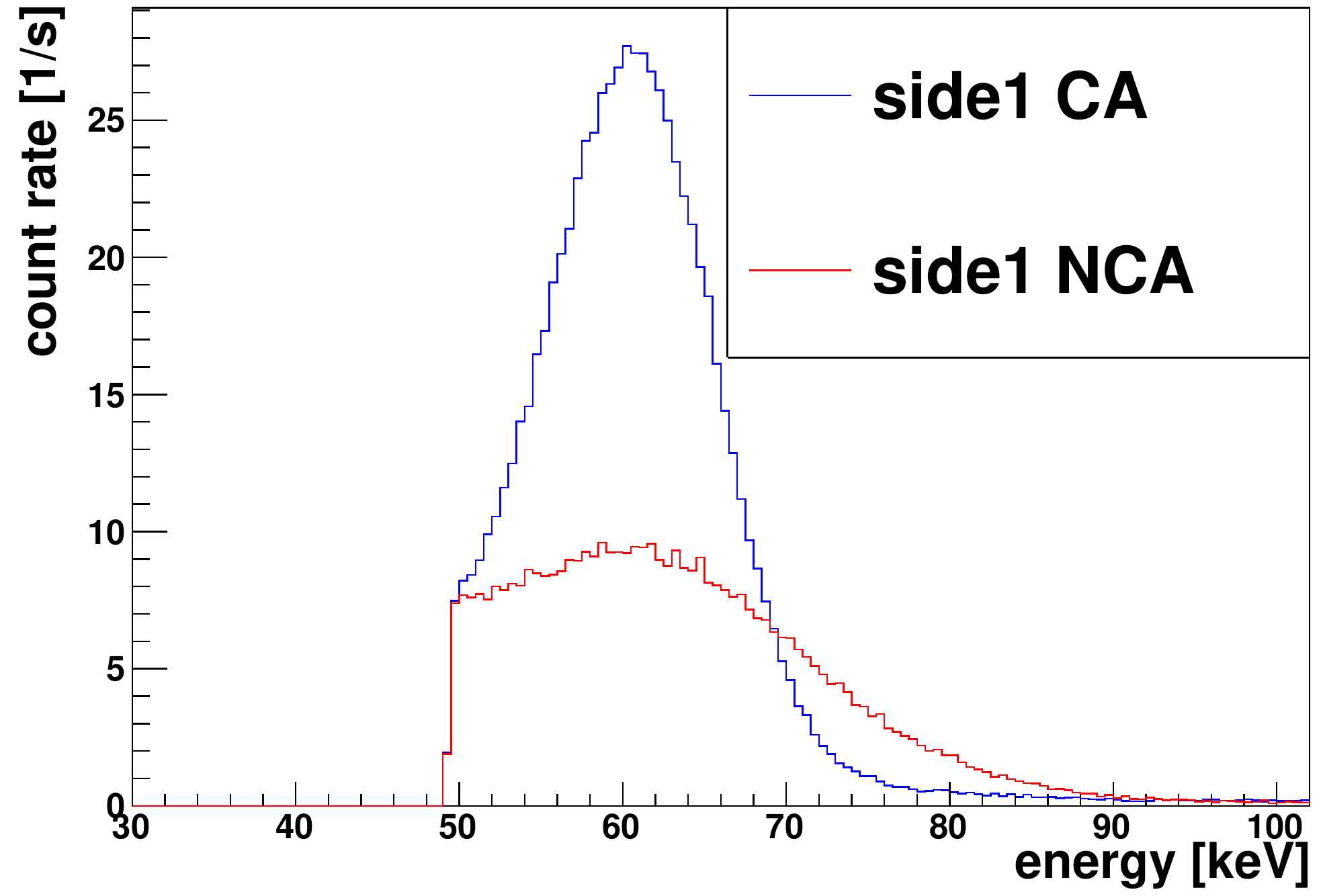}
  \captionsetup{width=0.9\textwidth} 
  \caption[Energy spectra of \SI{60}{keV} gamma radiation of 631972-06 and F15]
          {Energy spectra of \SI{60}{keV} gamma radiation. Detector 631972-06 (left) has the same sensitivity for both biases. The dotted curve shows the events that survived the data-cleaning cut. 
          Detector F15 (right) shows a significant difference between the two biases. The threshold for this detector is higher due to more distortions. In the results compilation \autoref{tab_detector_surface_sensitivity_results}, detector 631972-06 is marked as \checkmark and \checkmark for NCA and CA, detector F15 as $\sim$ and \checkmark, respectively.}
  \label{fig_60keV_surface_sensitivity}
\end{figure}

\SI{60}{keV} gamma rays can be used to test the surface sensitivity at both biases. Detector F15 shows a bias-dependent difference, 631972-06 does not.
Furthermore, it can be seen, that at this low energy and bad SNR, the data-cleaning algorithm obviously does not work. 
Only \SI{28}{\%} of all events survive the data-cleaning. The events that are discarded have the correct shape, like the good events, see \autoref{fig_60keV_surface_sensitivity}.\\

\begin{SCfigure}
 \centering
  \includegraphics[width=0.6\textwidth]{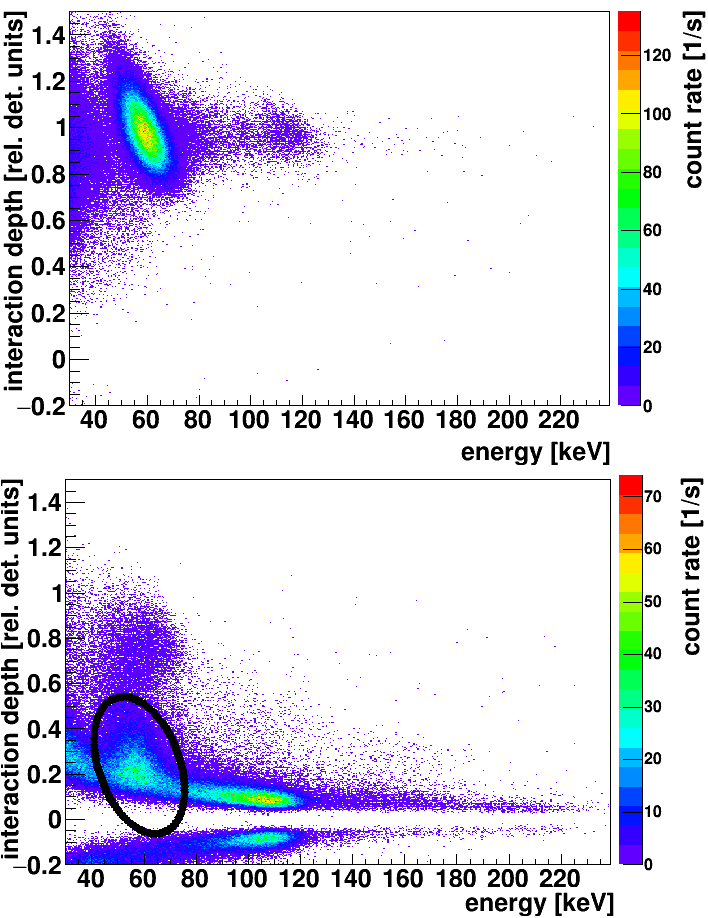}
 \captionsetup{width=0.9\textwidth} 
 \caption[Near-anode distortion and energy-doubling effect]
         {Near-anode distortion and energy-doubling effect measured with \SI{60}{keV} gamma radiation. 
         Top: Irradiation to the cathode. Bottom: Irradiation to the anodes. 
         The measurement at the cathode shows a correctly reconstructed energy, while this is true only for a small fraction for anode events, indicated by the black ellipse. 
         Most of the anode events suffer from the near-anode energy-doubling effect, as can be seen by the main peak around \SI{105}{keV} at an interaction depth of \num{0.1}).}
  \label{fig_60keV_Am241_energy_doubling}
\end{SCfigure}

\newpage
An effect already described in the author's diploma thesis \cite{tebruegge} (for \SI{5.5}{MeV} alpha and \SI{763}{keV} beta radiation) is the near-anode energy-doubling effect:\\ 
In the vicinity of the anodes, the holes are collected by the NCA, and do not drift towards the cathode.
This results in symmetric pulse shapes of CA and NCA. 
As the energy calculation is the weighted difference between the two of them (\autoref{eq_energy_calculation}), the calculated energy is spuriously (almost) doubled\footnote{Due to the weighting factor $w \approx 0.7$, the energy is increased by $(1+w) \approx 1.7$.}.\\
This effect is seen here as well. Interestingly, some events are reconstructed correctly, but most suffer from the energy-doubling.
A possible conclusion to this is, that the range of the \SI{60}{keV} gamma rays can be long enough to leave the near-anode region, where these distortions happen.\\
Furthermore, the symmetric distortion around the interaction depth of zero is visible, see \autoref{fig_60keV_Am241_energy_doubling}. 
This is a known CPG effect of the near-anode region due to calculating the difference of two numbers of approximately the same value in \autoref{eq_cal_ipos_z}.
As a consequence of this, all events below an interaction depth of \num{0.2} are discarded in the standard COBRA physics analyses.
Especially at the irradiation to the cathode can be seen, that some random coincidences happen, where two gamma rays are measured within the \SI{10.24}{\mu s} sampling time.
These sum up to \SI{120}{keV}. This is not the energy-doubling effect discussed here.\\ 

The energy-doubling behavior can be used to estimate the size of the near-anode region:\\
At a flush measurement to the anodes, \SI{20(3)}{\%} of the events are reconstructed correctly, indicated by the black ellipse in \autoref{fig_60keV_Am241_energy_doubling}. The assumption is, that these travel far enough to leave the near-anode region.
According to \autoref{eq_relative_intensity}, the intensity of the \SI{60}{keV} gamma rays is reduced to \SI{20(3)}{\%} after \SI{440(40)}{\mu m}, if assumed that the gamma rays vanish at their interaction via the photoelectric effect. 
Hence, the near-anode region is estimated to have a length of \SI{440(40)}{\mu m}. 
This is larger, but in the same range, as the rule of thumb that the region is as large as the CPG pitch of \SI{300}{\mu m} (see a dimensioned sketch of the CPG design in \autoref{fig_anode_grid_with_dimensions} on page \pageref{fig_anode_grid_with_dimensions}).
The transition zone between the region of energy-doubling and normal reconstruction is considered to be negligible in size. 
Either the holes move to the NCA, or to the cathode, but not to both of them.

%%%%%%%%%%%%%%%%%%%%%%%%%%%%%%%%%%%%%%%%%%%%%%%%%%%%%%%%%%%%%%%%%%%%%%%%%%%%%%%%%%%%%%%%%%%%%%%%%%%%%%%%%%%%%%%%%%%%%%%%%%%%
\FloatBarrier
\subsection{Beta measurements with \isotope[204]Tl}
\label{sec_beta_tl-204_measurements}
\begin{figure}
 \centering
  \includegraphics[width=0.49\textwidth]{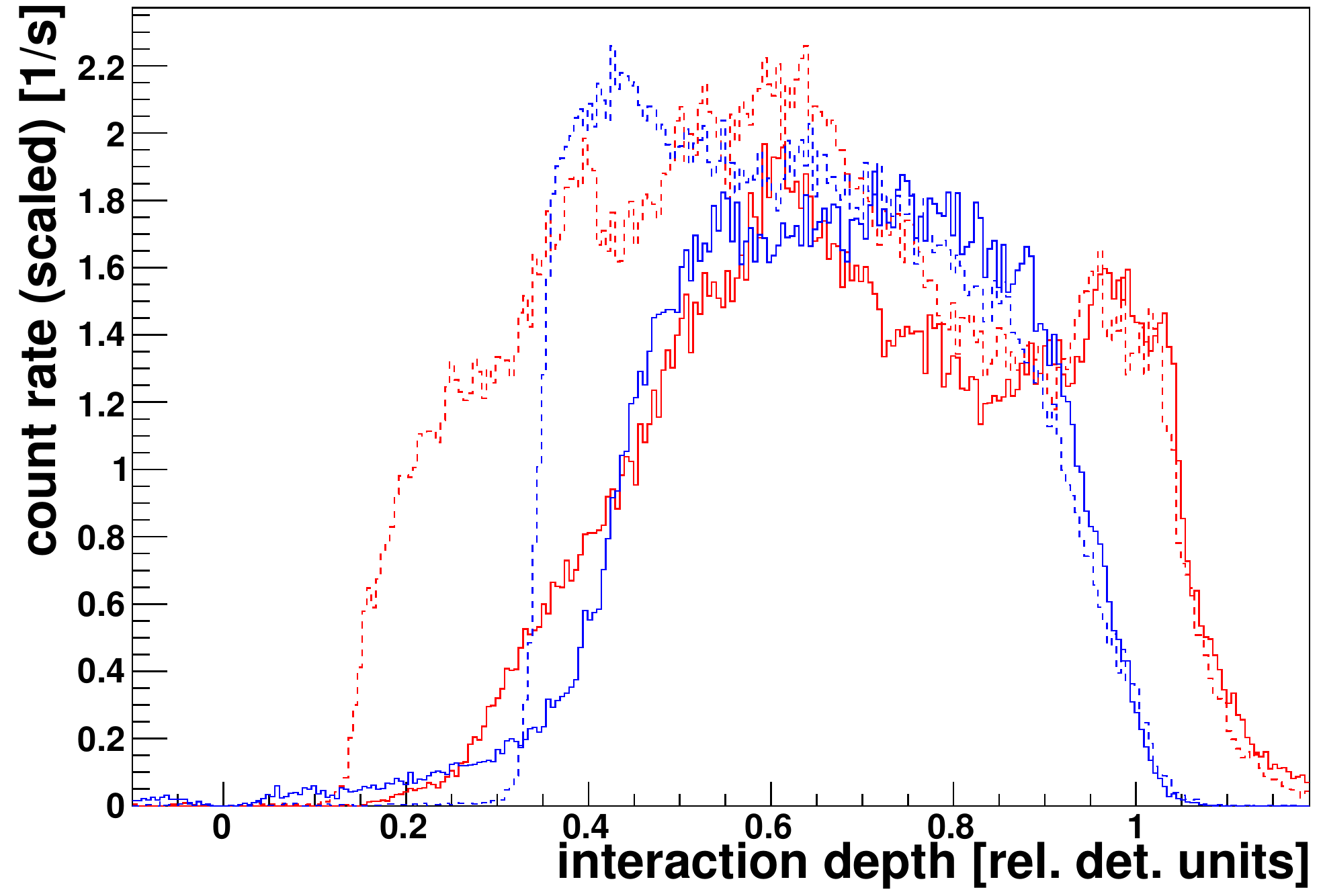}
  \includegraphics[width=0.49\textwidth]{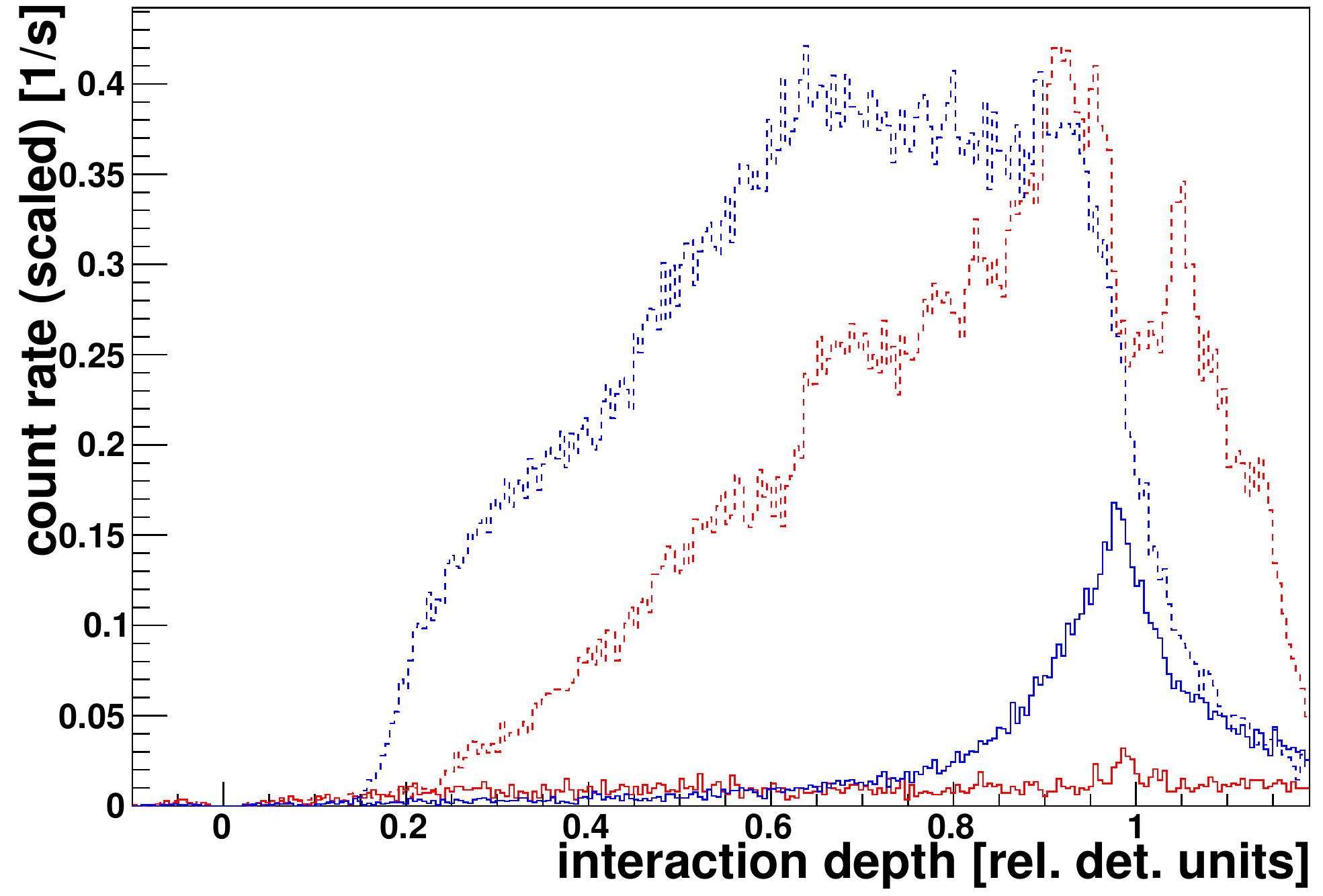}
  \captionsetup{width=0.9\textwidth} 
  \caption[Interaction depth of beta radiation of detectors 631972-06 and F15]
          {Interaction depth of beta radiation (blue) at side 2 of detectors 631972-06 (left) and F15 (right). For comparison, alpha radiation is shown in red. Solid lines: NCA outermost anode rail, dotted lines: CA outermost.
          Left: The sensitivity of the NCA is lower, especially towards the anodes. Right: The spectral shape is distorted, in particularly the events at the NCA are strongly suppressed.}
  \label{fig_beta_alpha_interaction_depth}
\end{figure}

\begin{figure}
 \centering
  \includegraphics[width=0.49\textwidth]{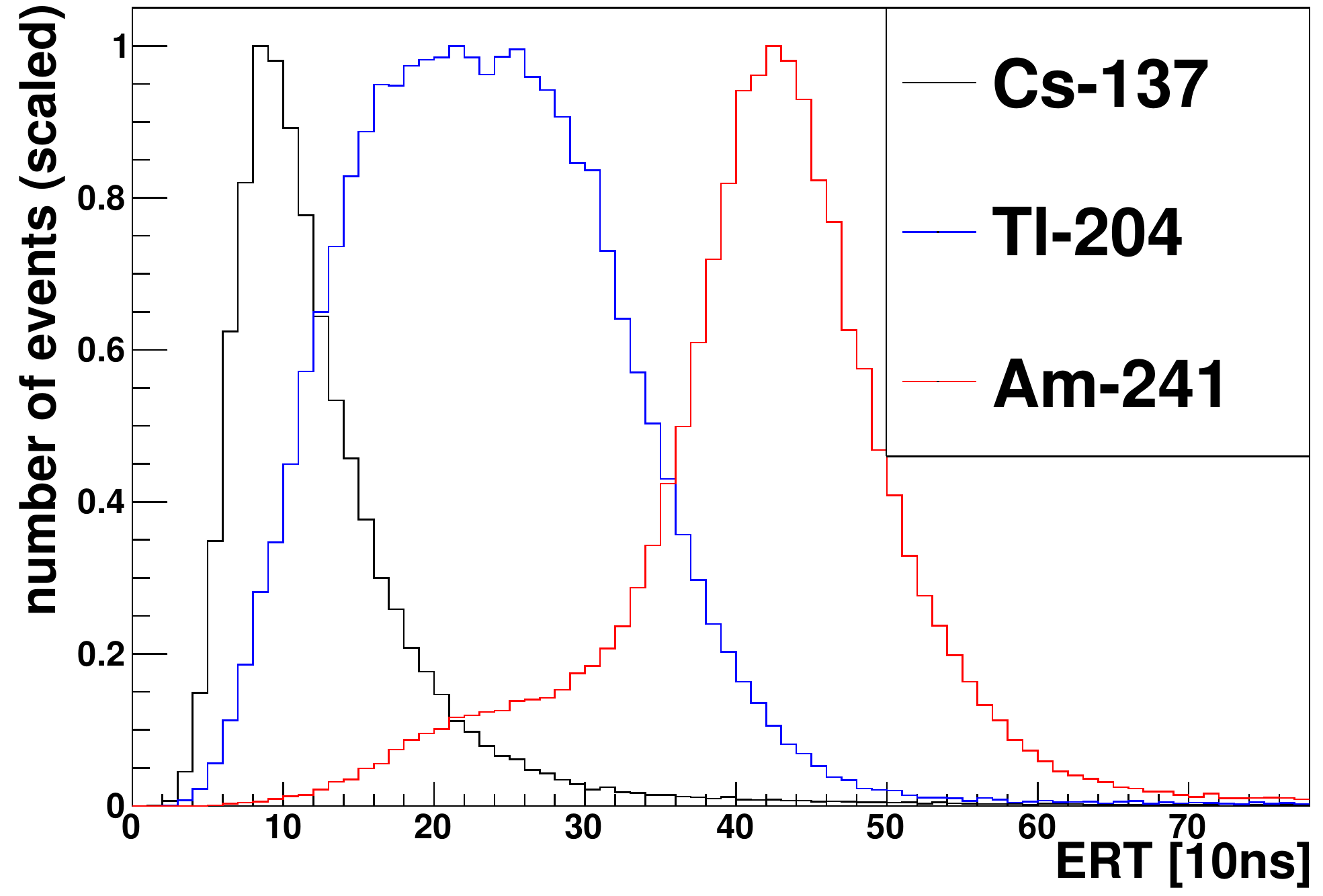}
  \includegraphics[width=0.49\textwidth]{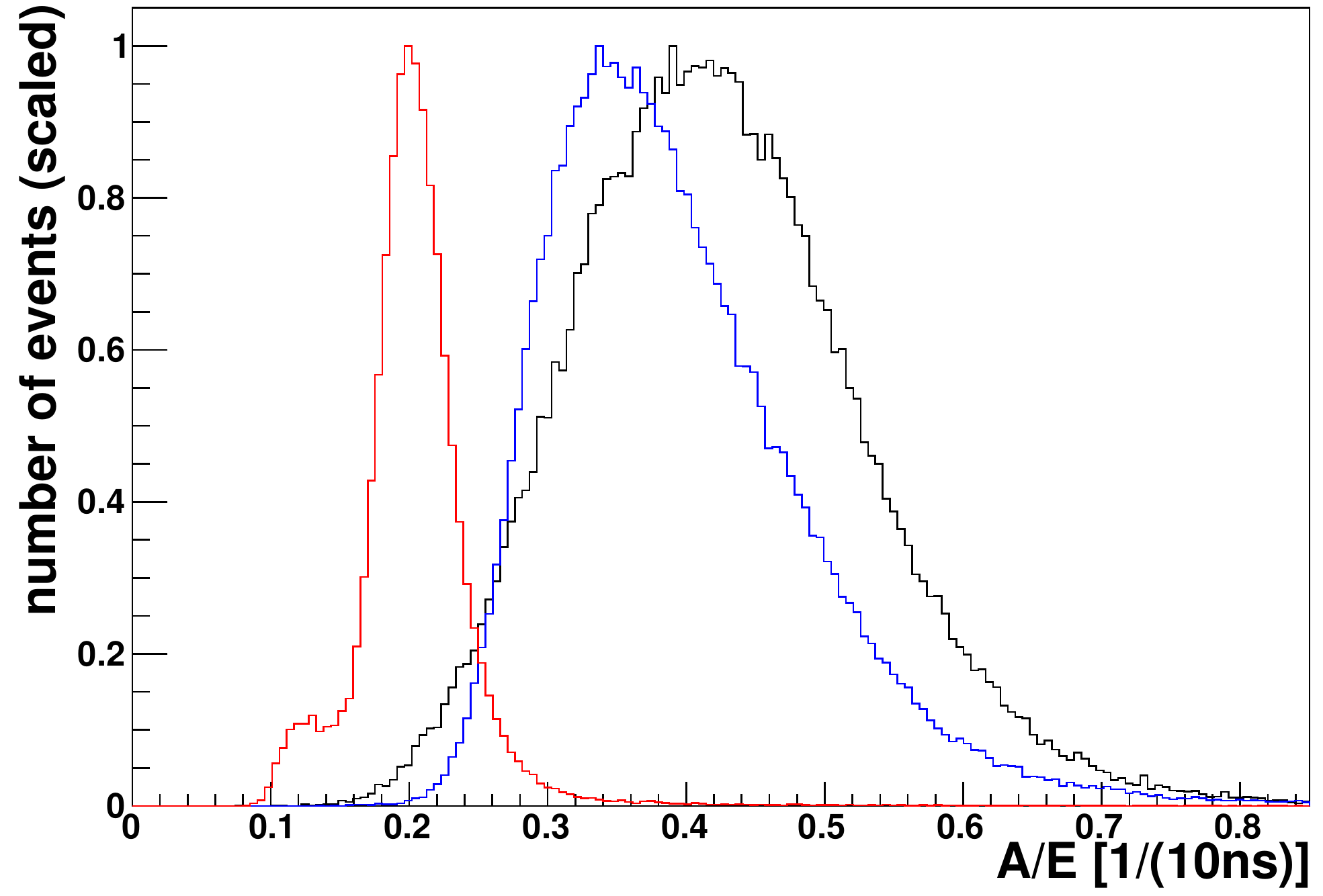}
  \captionsetup{width=0.9\textwidth} 
  \caption[ERT and A/E distribution of $^{204}$Tl beta radiation] % \isotope[204]Tl produced error here: \protect did not solve problem
	  {The ERT and A/E distribution of the same measurement of detector 631972-06 side 1 CA with \isotope[204]Tl beta radiation (blue) are between those of alpha (red) and central gamma events from \isotope [137]Cs (black), as expected.}
  \label{fig_tl204_beta_lse}
\end{figure}

The investigations with beta radiation were done with a \isotope[204]Tl source:
\begin{equation}
  \isotope[204]Tl \rightarrow \isotope[204]Pb, E_{\text{max}}=\SI{763}{keV}, E_{\text{aver}}=\SI{244}{keV}\,.
\end{equation}
The CSDA range (discussed in \autoref{eq_CSDA_range}) of these electrons is \SI{0.1}{mm} and \SI{0.9}{mm} for the average and maximal energy. 
In general, the range is in between the range of alpha and gamma radiation.
But the continuously distributed beta-spectrum and the random-walk of electrons in matter make it difficult to give realistic penetration depths.
Nevertheless, the surface-sensitivity and bias-dependency of the detectors can be tested.
In \autoref{fig_beta_alpha_interaction_depth}, examples are shown for a detector that is nearly fully sensitive (left), and a detector where the NCA sensitivity is strongly suppressed (right).
For comparison, the results from an irradiation with \isotope[241]Am alpha radiation is shown there as well. 
To be able to plot the results for both types of radiation into one histogram, the count rates of the measurements using beta radiation are scaled. 
The relative count rates of NCA and CA each are not scaled.\\
Furthermore, it is possible to veto beta radiation with PSA-cuts. As expected, they have values between those of surface and central events.
For the results of detector 631972-06 side 1 CA shown in \autoref{fig_tl204_beta_lse}, the quantity ERT shows a much better separation than A/E.
At other detectors, this behavior is vice versa.
A comparison of efficiency of the PSA methods LSE and A/E is done in \autoref{sec_comparison_A_over_E_LSE_threshold}.

%%%%%%%%%%%%%%%%%%%%%%%%%%%%%%%%%%%%%%%%%%%%%%%%%%%%%%%%%%%%%%%%%%%%%%%%%%%%%%%%%%%%%%%%%%%%%%%%%%%%%%%%%%%%%%%%%%%%%%%%%%%%
\FloatBarrier
\subsection{Measurements with \isotope[207]Bi}
\label{sec_bi_207_measurements}

The test methods used so far for the investigation of the surface insensitive areas had some drawbacks. 
The penetration depth of alpha particles is small ($\approx \SI{20}{\mu m}$), \SI{60}{keV} gamma radiation is not completely usable, 
and beta radiation is difficult because of the continuously distributed spectrum.
A good choice is to use a mono-energetic electron beam, obtained by IC.
This method was invented by the author and applied for the first time here.

IC is a radioactive decay process, where the decay energy of the nucleus $E_{nucl}^i - E_{nucl}^f$ is transferred completely to a shell electron minus the binding energy $E_{bind}$ of that electron in its shell:
\begin{equation}
  E_{electr} = (E_{nucl}^i - E_{nucl}^f) - E_{bind} .
  \label{eq_conversion_electron}
\end{equation}
This electromagnetic process can occur, if the probability density of a shell electron has a sizable value inside the nucleus itself (mostly for s-electrons of the K-shell). 
The electron takes the de-excitation energy directly without an intermediate gamma (contradictory to the name ``internal conversion''). 
The resulting vacancy in the shell will most likely be filled by another electron from higher shells, so that characteristic X-rays or Auger electrons are emitted as well.
IC is possible for all gamma emission processes, its probability depends on the decay energy and spin, amongst others. 
For $0^+ \rightarrow 0^+$ transitions (zero-spin and positive parity), it is the favored decay process, as a single gamma emission ($J^{P} = 1^-$) is forbidden by angular momentum conservation. 

Here \isotope[207]Bi is used, which undergoes EC with a maximal decay energy of \SI{806}{keV} to \isotope[207]Pb. 
Relevant decays are listed in \autoref{tab_decay_energies_bi-207}, the Q-values are rounded to full keV, the intensities to significant values. 
The decay scheme is shown in \autoref{fig_bi_207_decay_scheme}.

\vspace{0.5cm}
\begin{minipage}[]{0.35\textwidth}
  \begin{tabular}{|c|c|c|}
    \hline
          & \makecell{Q-value\\{[}keV]} & \makecell{intensity\\{[}\%]} \\ \hline
    CE K  & 482  & 1.5 \\ \hline
    CE L  & 554  & 0.44 \\ \hline
    CE M  & 566  & 0.11 \\ \hline
    gamma & 570  & 97.8 \\ \hline
    CE K  & 976  & 7.1 \\ \hline
    CE L  & 1048 & 1.8 \\ \hline
    gamma & 1064 & 75 \\ \hline
    CE K  & 1682 & 0.02 \\ \hline
    gamma & 1770 & 6.9 \\ \hline
  \end{tabular}
  \captionof{table}[Decay energies of $^{207}$Bi]
                   {Decay energies of \isotope[207]Bi (rounded), taken from \cite{NNDC_decay_data_bi-207}. CE stands for conversion electrons.}
  \label{tab_decay_energies_bi-207}
\end{minipage}
\hfill
\begin{minipage}[]{0.59\textwidth}
  \vspace{1.1cm}
  \includegraphics[width=0.99\textwidth]{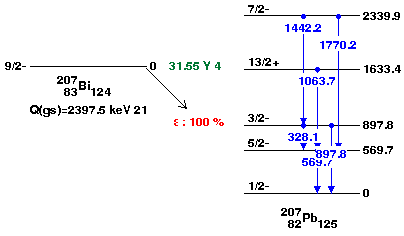}
  \captionof{figure}[Decay scheme of $^{207}$Bi]
                   {Decay scheme of \isotope[207]Bi, taken from \cite{NNDC_decay_data_bi-207}}
  \label{fig_bi_207_decay_scheme}
\end{minipage}
\vspace{0.5cm}

\begin{figure}[t!]
 \centering
 \includegraphics[width=0.49\textwidth]{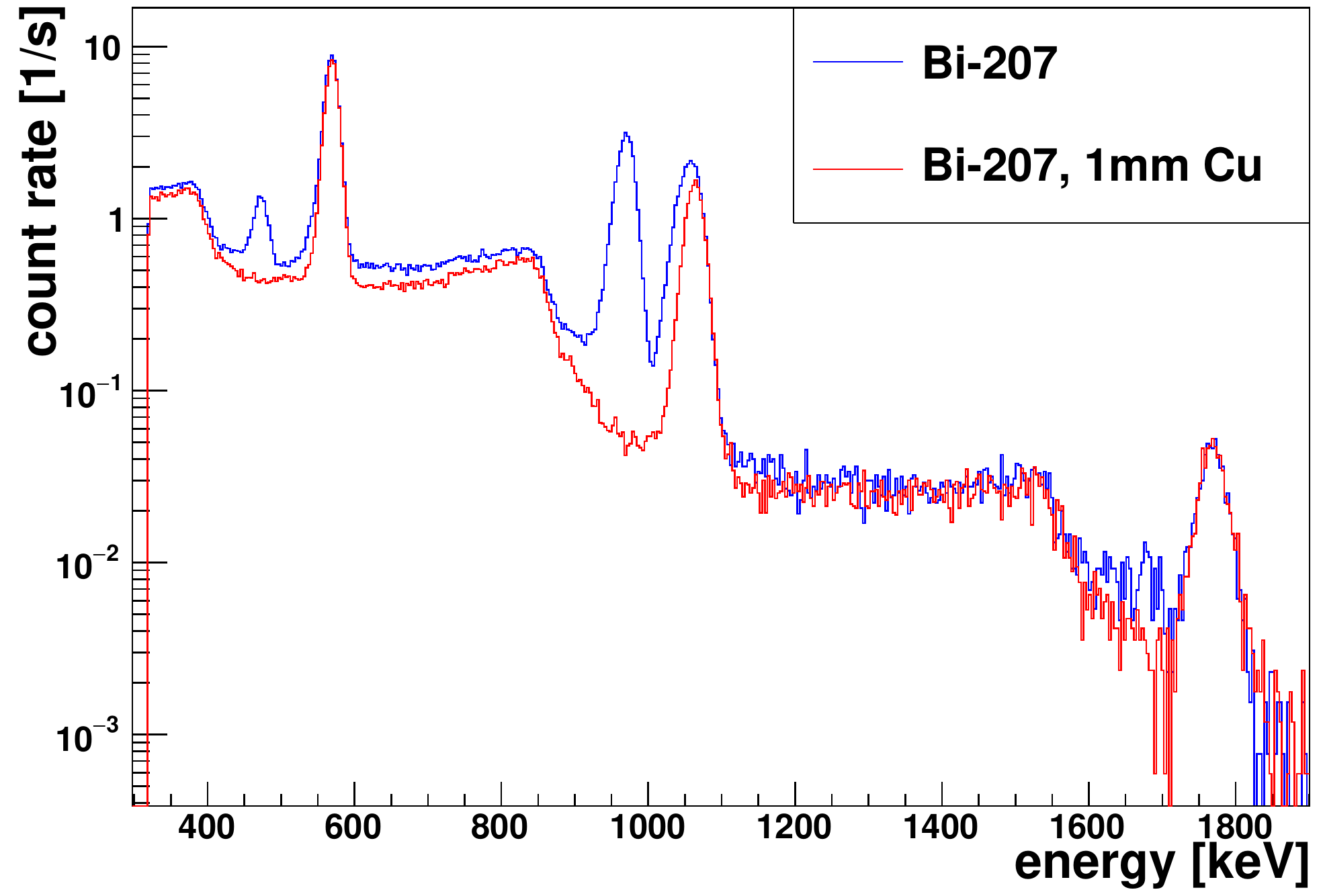}
 \includegraphics[width=0.49\textwidth]{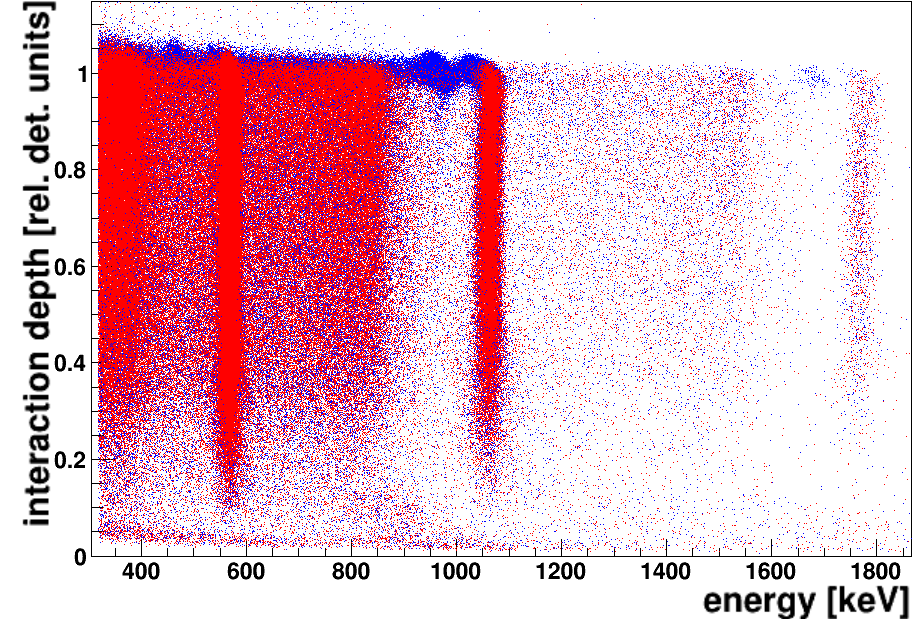}
 \captionsetup{width=0.9\textwidth} 
 \caption[Detector 631972-06 irradiated with $^207$Bi to cathode]
         {Blue: Detector 631972-06 irradiated with \isotope[207]Bi to the cathode. Red: Additional \SI{1}{mm} of copper between source and detector to prevent electrons reaching the detector. The additional blue populations in comparison to the red ones stem only from IC-electrons. This can especially be seen at the interaction depth plot (right), where the electrons populate only areas close to one due to their small penetration depth. } 
 \label{fig_spectra_bi_207}
\end{figure}

The electron lines have the energy of the corresponding gamma ray, minus the binding energy, see \autoref{eq_conversion_electron}.
Consequently, they can be clearly distinguished from the electrons, as these are in the Compton-valley of the gamma rays.
Because electrons lose energy continuously in contrast to gamma rays (as explained in \autoref{sec_particles_matter_ranges}), one can reduce the measured energy of the electron in the detector by enlarging the distance of source and detector, or by adding additional material in between.
This can be used to block off the electrons completely to distinguish between electron and gamma interactions.
The possible effects of bremsstrahlung can be ignored for two arguments:
First, as calculated in \autoref{sec_particles_matter_ranges}, those radiative energy losses occur only at a percent level compared to the collision losses.
Second, the bremsstrahlung is stopped mostly within the copper itself.\\
The electron lines that can be seen most prominently are \SIlist{976; 482; 1048; 1682; 544; 566}{keV}.
However, the one at \SI{1048}{keV} is partly within the gamma line at \SI{1064}{keV}, and the two lines around \SI{560}{keV} are within the line at \SI{570}{keV}, as shown in \autoref{fig_spectra_bi_207}.

The penetration depth of the different conversion electrons was investigated in \autoref{sec_penetration_depth_of_particles_in_matter} to 
\SI{0.3}{mm} for \SI{482}{keV}, \SI{0.8}{mm} for \SI{976}{keV}, and \SI{1.5}{mm} for \SI{1680}{keV}.
This can be used now to test the surface sensitivity of the detector in three different energy ranges.

\begin{figure}[b!]
 \centering
  \includegraphics[width=0.49\textwidth]{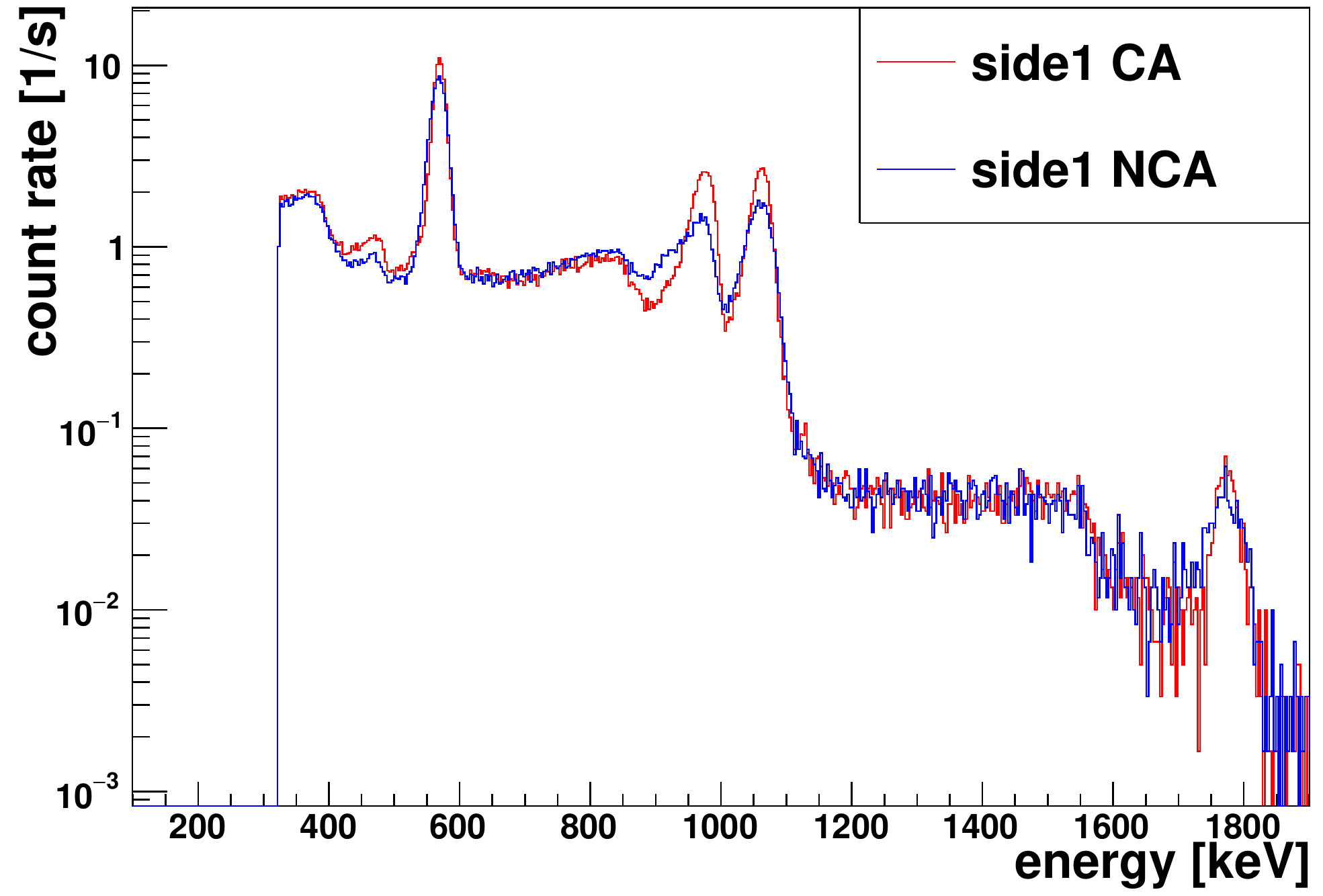} 
  \includegraphics[width=0.49\textwidth]{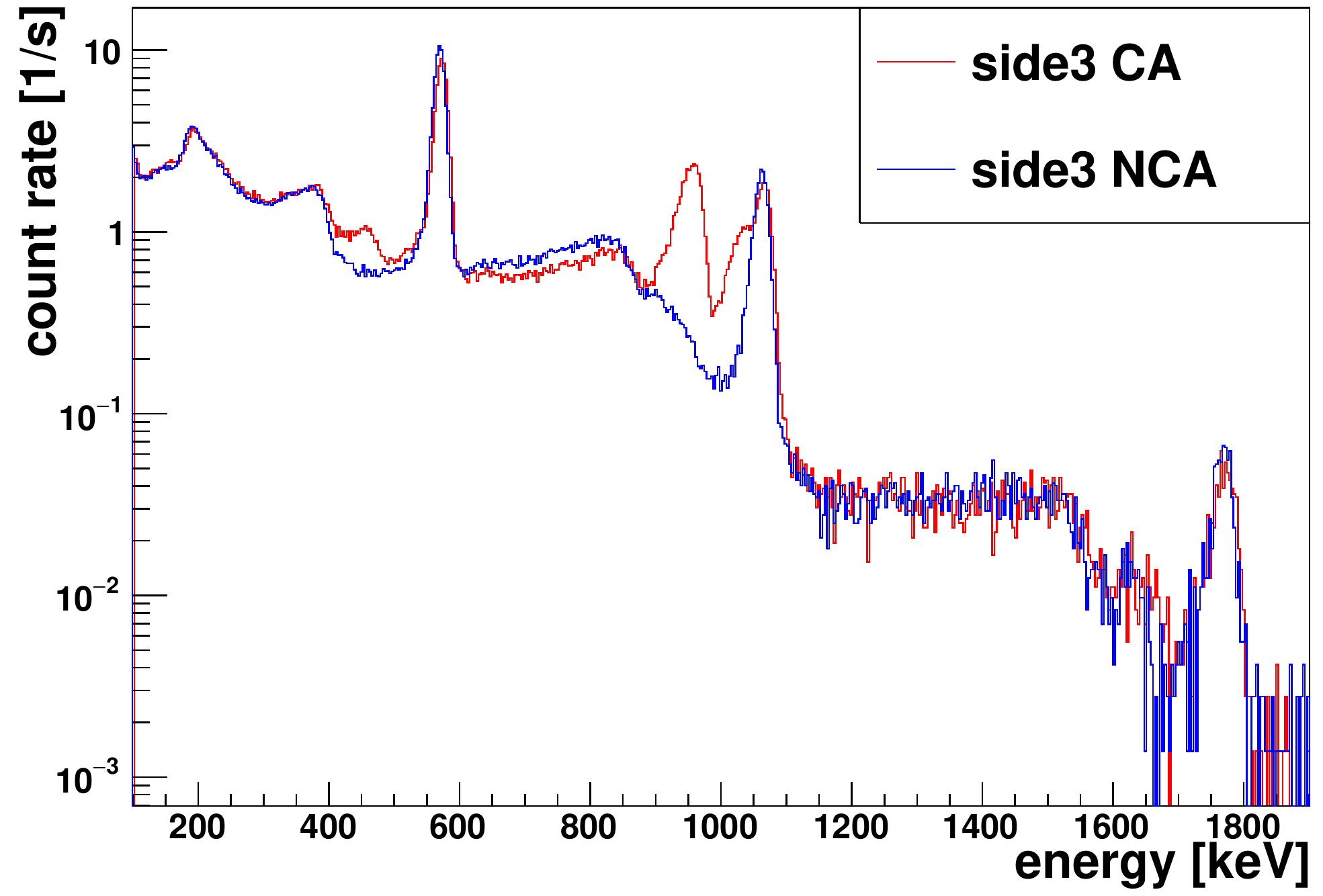} 
  \captionsetup{width=0.9\textwidth} 
  \caption[Surface sensitivity to IC electrons]
          {Surface sensitivity of detectors 631972-06 and 667615-02. Left: Detector 631972-06 is fully sensitive for both biases on side 1, a little less for NCA.
           Right: Detector 667615-02 is sensitive on side 3 CA, but for NCA only for the \SI{1680}{keV} electrons. }
  \label{fig_bi207_631972-06_side1_3_cal_edep}
\end{figure}

Detector 631972-06 is (nearly) fully sensitive for both biases at side 1. 
Consequently, a possible dead-layer is smaller than \SI{0.3}{mm}. 
Contrary, detector 667615-02 is sensitive for CA outermost, but if the NCA is outermost, only the highest energy of \SI{1680}{keV} is measured, see \autoref{fig_bi207_631972-06_side1_3_cal_edep}.
By this one can conclude that the dead-layer here has a size between \SI{0.8}{mm} and \SI{1.5}{mm}.
 
\begin{SCfigure}
 \centering
  \includegraphics[width=0.7\textwidth]{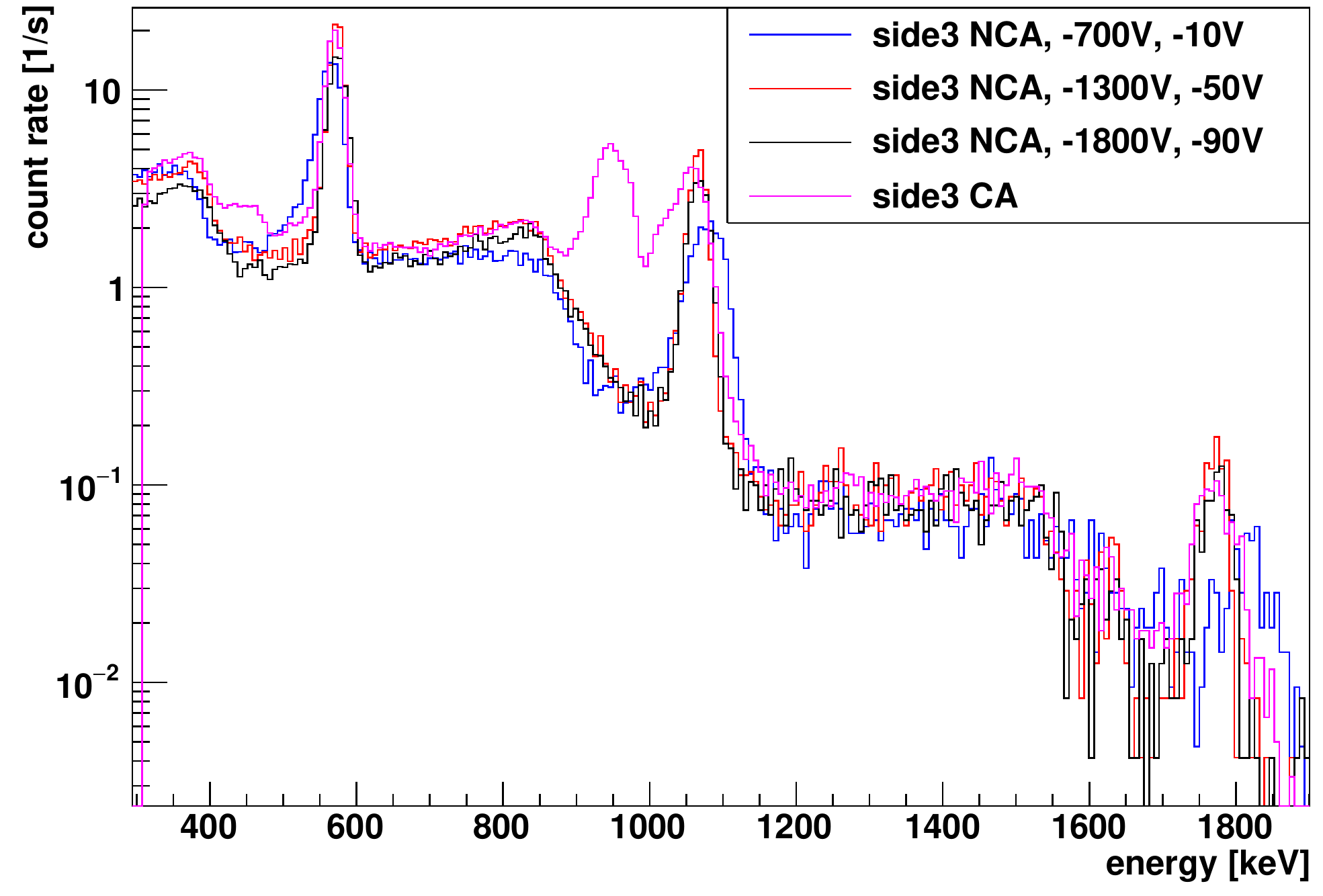}
  \caption[Voltage-dependency of insensitivity of detector 667615-02 IC of $^207$Bi.]
          {The insensitivity of detector 667615-02 on side 3 NCA is not an effect of too low voltages. The values for BV and GB are shown in the legend.
          For comparison, the measurement CA is shown in magenta, where the electron lines appear clearly.}
  \label{fig_bi207_667615-02_side3_cal_edep_different_voltages}
\end{SCfigure}

The surface insensitivity is not an effect of too low voltages. It was tested by varying BV and GB. 
Though the general performance and efficiency of the detector depends on the applied voltages, variations in the biases had no effect on the surface sensitivity, as shown in \autoref{fig_bi207_667615-02_side3_cal_edep_different_voltages}.\\

\begin{SCfigure}[][b!]
 \centering
  \includegraphics[width=0.7\textwidth]{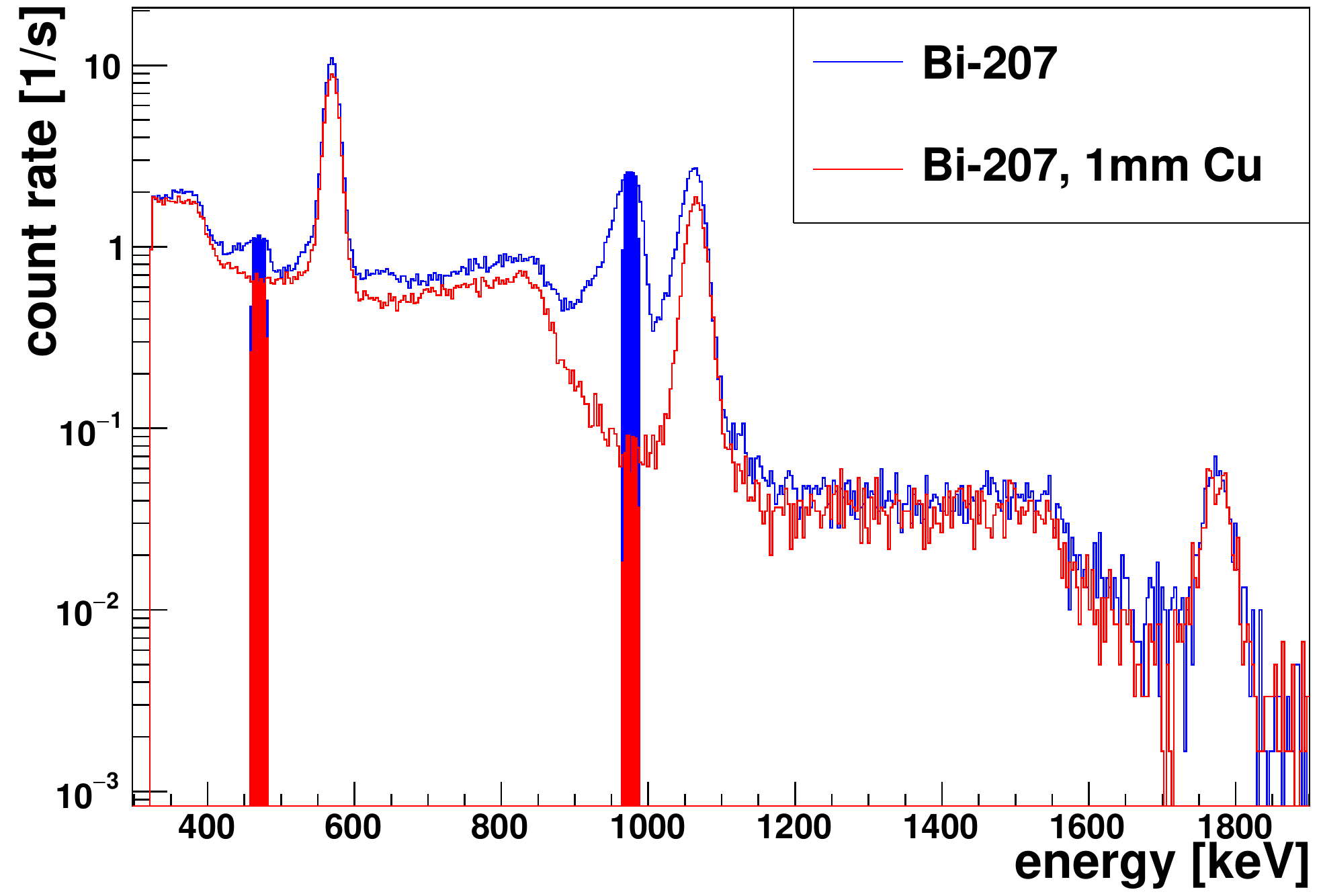} 
  \caption[Detailed measurements of electrons at detector 631972-06 side 1 CA]
	  {Detailed measurements to the surface sensitivity of electrons at detector 631972-06 side 1 CA. Blue: Measurement with electrons. Red: Measurement without electrons. The filled areas comprise the IC electrons and the gamma rays at the same energy range.}
  \label{fig_bi207_631972_06_side1_CA_w_wo_copper_cut_areas}
\end{SCfigure}

% \newpage
It is possible to study the surface sensitivity of electrons in detail. For that purpose, a set of dedicated measurements was conducted. 
These include two measurements with the same conditions, except that the first was done with a copper block in the beam to block off electrons, while the second one had no copper. 
The difference between these two measurements arises only from electrons. 
Analysis cuts to an energy interval of $\pm\SI{10}{keV}$ of the IC electrons are applied in both measurements, to have enough statistics and a good separation of the two measurements.
At \SI{976}{keV}, the electrons exceed the gamma events by a factor of nearly 30, consequently, this is nearly a pure electron beam.
At \SI{482}{keV}, the ratio is only a factor of two. 
Due to low statistics, the electron distribution at \SI{1682}{keV} is not taken into account here.
The spectra indicating the chosen areas are shown in \autoref{fig_bi207_631972_06_side1_CA_w_wo_copper_cut_areas}.

\begin{SCfigure}[][t!]
 \centering
  \includegraphics[width=0.49\textwidth]{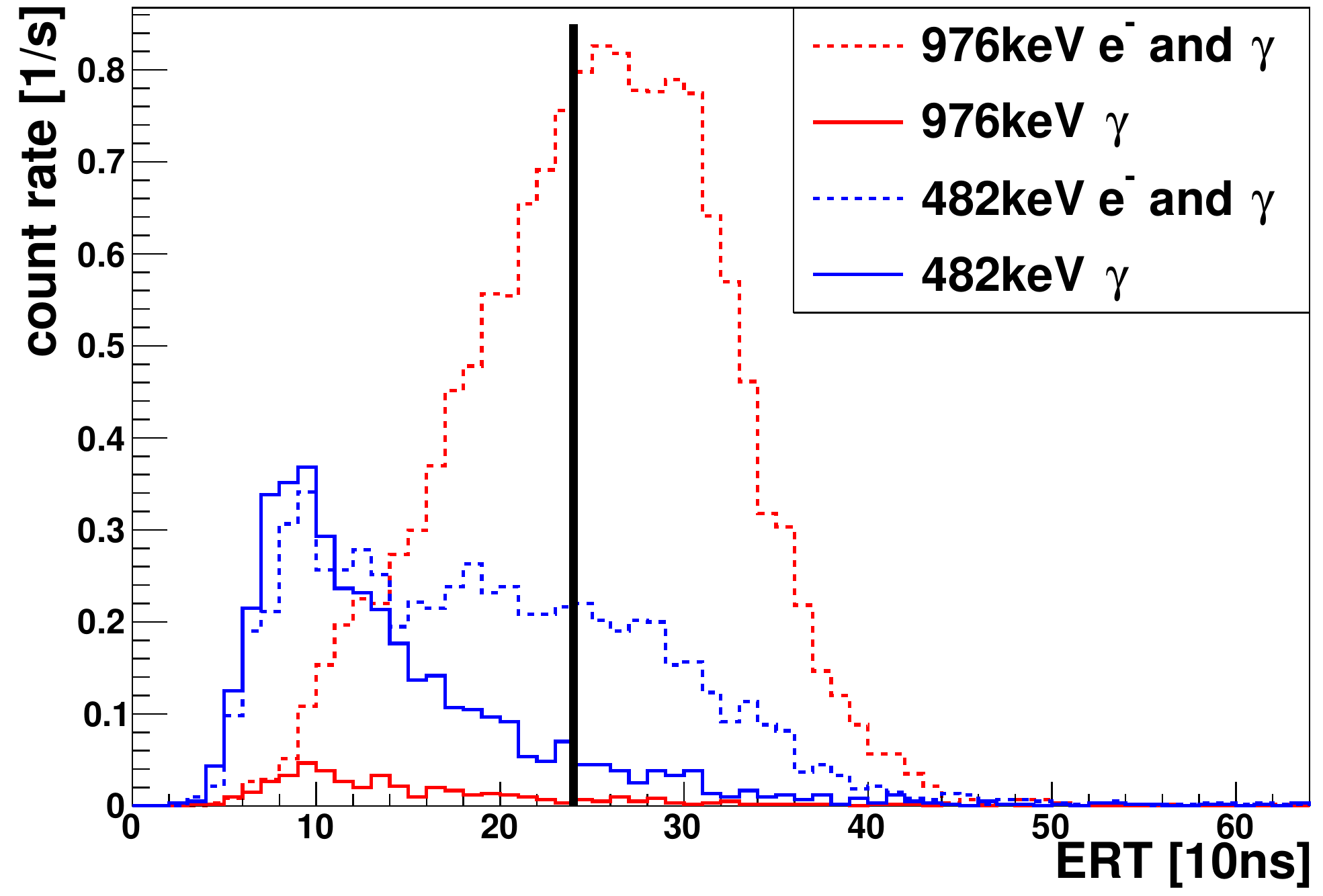} 
  \captionsetup{width=0.9\textwidth} 
  \caption[Surface cut quantities of marked areas of \autoref{fig_bi207_631972_06_side1_CA_w_wo_copper_cut_areas}]
	  {Surface cut quantities of marked areas of \autoref{fig_bi207_631972_06_side1_CA_w_wo_copper_cut_areas}. Solid lines: only gamma radiation, dotted lines: gamma and electron radiation.
	  The distribution of the \SI{482}{keV} electrons starts to differ from its gamma distribution around bin 15, the \SI{976}{keV} already below bin 10. The lower energetic electrons have surface character, the higher energetic ones much more central character.}
  \label{fig_bi207_631972_06_side1_CA_w_wo_copper_cut_areas_ert_values}
\end{SCfigure}

One can estimate the size of the LSE and A/E surface-sensitivity using these events.
The interactions of the gamma rays (of the measurement without electrons) are distributed homogeneously within the detector volume. Consequently, these have values typical for central events.
The electron distribution can also have central-character -- if the penetration depth of electrons is larger than the surface-sensitivity. 
On the other hand, the electron distribution has surface character, if the penetration depth is smaller. 
This is discussed here exemplarily for detector 631972-06 side 1 CA for the quantity ERT, see \autoref{fig_bi207_631972_06_side1_CA_w_wo_copper_cut_areas_ert_values}: 
The \SI{482}{keV} electrons show typical surface character, while the \SI{976}{keV} electrons also have central character. 
Hence one can conclude that the area where ERT tags events as ``surface'' is smaller than the maximum penetration depth of the \SI{976}{keV} electrons, but larger than the one from the \SI{482}{keV} electrons. 
According to the values obtained for the ranges in \autoref{sec_particles_matter_ranges}, this is between \SI{0.3}{mm} and \SI{0.8}{mm}.

\autoref{fig_bi207_667615_06_side1_ipos_zt__edep_wrong_reconstruction} shows two different types of reconstruction artifacts of detector 667615-02, that occurred as well at detector F15, but not at 631972-06:
\begin{figure}[t!]
 \centering
  \includegraphics[width=0.99\textwidth]{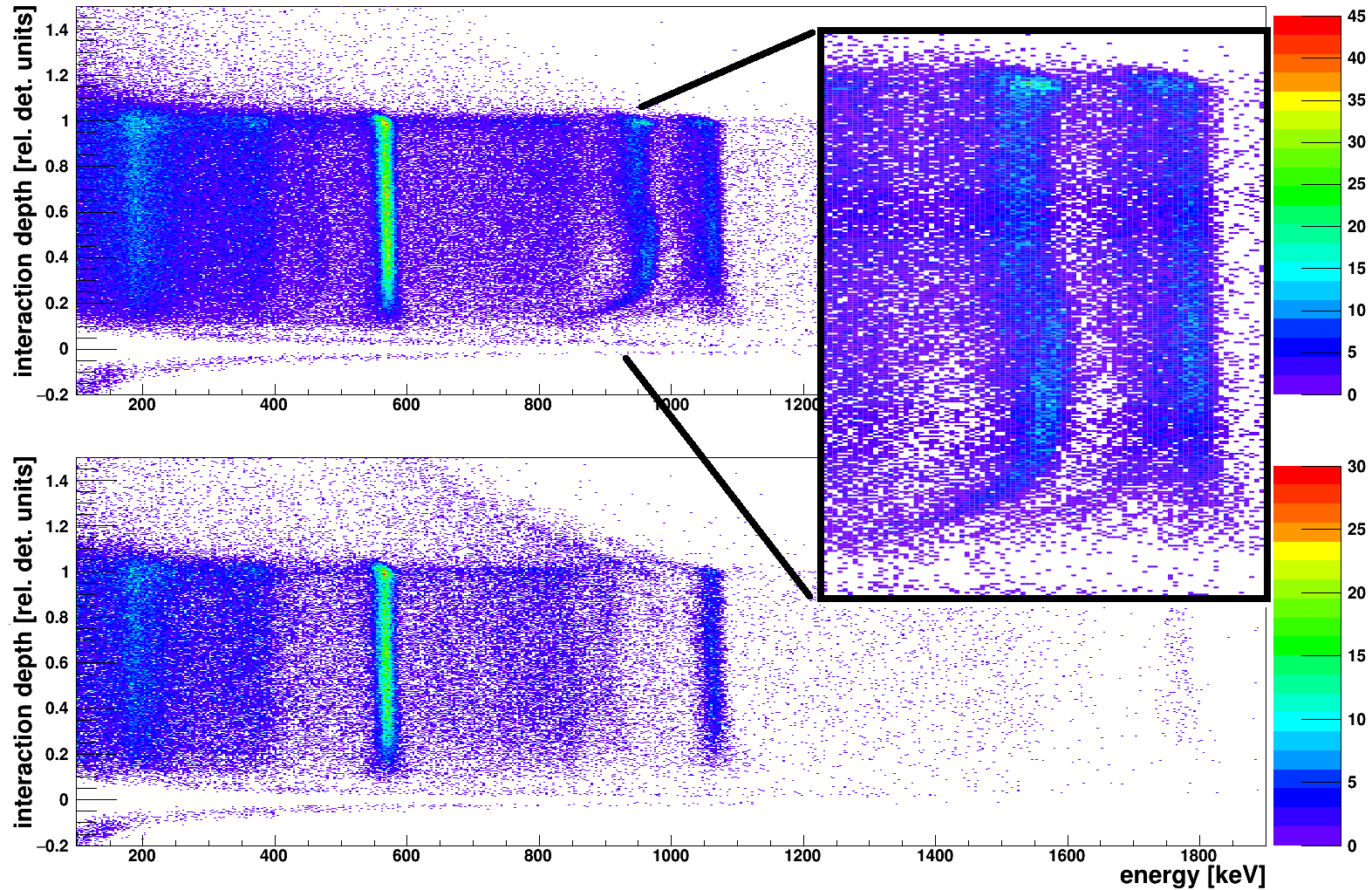} 
  \captionsetup{width=0.9\textwidth} 
  \caption[Reconstruction artifacts of detector 667615-02 side 1]
          {Reconstruction artifacts of detector 667615-02 side 1. Top (CA): mono-energetic electron events are not on a vertical line, as neighboring gamma lines are. Bottom (NCA): Hyperbolic population to interaction depth values above one, especially from higher-energetic electron events of \SI{976}{keV} and \SI{1048}{keV}.}
  \label{fig_bi207_667615_06_side1_ipos_zt__edep_wrong_reconstruction}
\end{figure}
The ``mono-energetic'' events should lie on a vertical line at constant energy, like the lines of the full energy deposition of gamma radiation. 
If the CA is the outermost anode rail, the electron lines do not: the reconstructed energy depends on the interaction depth, the shape looks more like an ``S''. 
This can be seen most prominently at the main electron energy of \SI{976}{keV} in the inset. 
The electron line at \SI{1048}{keV} which is very close to the gamma line at \SI{1064}{keV} shows this behavior as well. The same seems to be true for the \SI{482}{keV} electron line.
Due to low statistics, this distortion cannot be checked at the \SI{1682}{keV} line.
This effect cannot arise from the beam divergence, as that kind of distortion should be somewhat symmetric, which is not the case here.
Artifacts of the source can also be ruled out, as this behavior does not appear at all detectors, which were measured with the same source position.
Mono-energetic electrons have not been investigated so far. 
As the $0\nu\beta\beta$-decay COBRA is searching for relies on measuring the electrons of the beta-decay, this reconstruction artifacts should be investigated further.\\
At the NCA as outermost anode, the electron lines are not visible (discussed before). 
However, some electrons of \SI{976}{keV} and \SI{1048}{keV} can be seen, but these are reconstructed in a hyperbolic shape to higher interaction depth values. This behavior is discussed in more detail in \autoref{sec_hyperbolic}.

%%%%%%%%%%%%%%%%%%%%%%%%%%%%%%%%%%%%%%%%%%%%%%%%%%%%%%%%%%%%%%%%%%%%%%%%%%%%%%%%%%%%%%%%%%%%%%%%%%%%%%%%%%%%%%%%%%%%%%%%%%%%
\FloatBarrier
\subsection{Results of general surface sensitivity}
\label{sec_results_of_general_surface_sensitivity}

The results of the surface sensitivity of the laboratory measurements of this chapter are compiled in \autoref{tab_detector_surface_sensitivity_results}. 
Each measured detector side is rated qualitatively as fully, partial, and non-sensitive for both NCA and CA as outermost anode-rail with the symbols \checkmark, $\sim$ and \ding{53}.
For alpha surface events, the rating is obtained by comparing the ratios of the measured events for the NCA as outermost anode rail, compared to the CA as outermost anode rail.
For ratios between \SI{75}{\%} and \SI{100}{\%}, \SI{25}{\%} and \SI{74.9}{\%}, and zero to \SI{24.9}{\%} the symbols \checkmark, $\sim$ and \ding{53} are given.

Treating these rates as \numlist{1; 0.5; 0}, one can calculate the ratios of the relative sensitivity of CA and NCA as outermost anode rail for alpha radiation to \num{3.3}.

The ratio of the number of all entries of the measurement at CA bias compared to NCA bias is calculated to \num{3.7}.

This ratio can also be calculated for the low-background physics data from the COBRA demonstrator:\\
In the \SI{82.3}{kg\,d} dataset from the COBRA demonstrator (2013) used in the LSE-publication \cite{Fritts:2014qdf}, \num{1978} events remain in the energy range between \SI{2}{MeV} and \SI{4}{MeV} (after cleaning-cuts). 
\SI{53.1}{\%} of them are tagged as events with a high ERT value (CA detector surface), and \SI{23.0}{\%} as high DIP events (NCA surface). The ratio between events on CA side to NCA is hence \num{2.7}.\\
In a \SI{150}{kg\,d} dataset from summer 2015, \num{16241} events remain after the same cuts, with \SI{57.5}{\%} ERT and \SI{16.3}{\%} DIP tags. So the ratio between CA and NCA events is \num{3.5}.
There are more events (relative), because the detector layers 3 and 4 were installed in the meantime, which have a much higher background rate, see \autoref{sec_addition_work_lngs}.\\
If this dataset is reduced to detector layers 1 and 2 only, as in the \SI{82.3}{kg\,d} dataset, \num{5575} events remain. Of these, \SI{57.7}{\%} have ERT tags, and \SI{19.3}{\%} DIP, so the ratio is \num{3.3}.\\

As a conclusion, one can estimate that for the detectors under study, the CA surfaces are typically about a factor of three more sensitive than the NCA surfaces.
From this one can conclude that the detectors often have a dead-layer if the NCA is the outermost anode rail. 
This dead-layer cannot be removed by increasing the applied biases, as shown in \autoref{fig_bi207_667615-02_side3_cal_edep_different_voltages}.

The size of the dead-layer can be determined using IC electrons of \isotope[207]Bi. 
At some detector surfaces and biases, there was no dead layer at all, while at others, the largest dead-layer was smaller than the penetration depth of \SI{1.7}{MeV} electrons which is \SI{1.5}{mm}.
If the size of the dead-layer is small $(\sim \SI{10}{\mu m})$, so that it affects only alpha radiation, this would not be a disadvantage concerning background reduction.
However, a larger dead-layer in the range of mm seriously reduces the active detector volume, which is incorporated in the detection efficiency $\epsilon$ in \autoref{eq_half_life_sensitivity}.\\

Is the (in)sensitivity dependent on the position on the detector surface? There are indications for less sensitivity on the sides towards the anode (see e.g. \autoref{fig_beta_alpha_interaction_depth}). 
A lower sensitivity could also be expected at the lateral edges, especially where the CA and NCA sides meet, as described in the beginning of this chapter on page \pageref{sec_general_surface_sensitivity}.\\
To investigate this, alpha scanning measurements were done, discussed in \autoref{sec_alpha_scanning}.

\begin{table}
\begin{tabular}{|c|c|c|c|c|c|c|c|c|c|}
\hline
				&    & \multicolumn{2}{c|}{alpha} & \multicolumn{2}{c|}{\SI{763}{keV} beta}    & \multicolumn{2}{c|}{\isotope[207]Bi IC $e^-$}    & \multicolumn{2}{c|}{\SI{60}{keV} gamma }\\ 
detector name                   &    & NCA  & CA   & NCA  & CA   & NCA                        & CA                         & NCA  & CA   \\ \hline
\multirow{3}{*}{631972-06}      & S1 & $\sim$     & \checkmark & \checkmark & \checkmark & \checkmark \checkmark \checkmark & \checkmark \checkmark \checkmark & \checkmark & \checkmark \\ 
				& S2 & \checkmark & \checkmark & \checkmark & \checkmark & 				    &                                  & \checkmark & \checkmark \\ 
				& S3 & $\sim$     & \checkmark & \checkmark & \checkmark & $\sim$ $\sim$ \checkmark         & \checkmark \checkmark \checkmark & \checkmark & \checkmark \\ \hline
\multirow{3}{*}{667615-02}      & S1 & \ding{53}  & \checkmark &            &            & \ding{53} \ding{53}  $\sim$      & $\sim$ \checkmark \checkmark     &            &            \\ 
				& S2 &            &            &            &            &                                  &                                  &            &            \\ 
				& S3 & \ding{53}  & \checkmark &            &            & \ding{53} \ding{53} \ding{53}    & $\sim$ \checkmark \checkmark     &            &            \\ \hline
\multirow{3}{*}{F01}            & S1 & \ding{53}  & \checkmark & $\sim$     & \checkmark &                                  &                                  &            &            \\ 
				& S2 & \ding{53}  & \checkmark &            &            &                                  &                                  &            &            \\ 
				& S3 & \ding{53}  & \checkmark & $\sim$     & \checkmark &                                  &                                  &            &            \\ 
				& S4 & \ding{53}  & \checkmark &            &            &                                  &                                  &            &            \\ \hline
\multirow{4}{*}{F05}            & S1 & \ding{53}  & \checkmark &            &            &                                  &                                  &            &            \\ 
				& S2 & \ding{53}  & \checkmark &            &            &                                  &                                  &            &            \\ 
				& S3 & \ding{53}  & \checkmark &            &            &                                  &                                  &            &            \\ 
				& S4 & \ding{53}  & \checkmark &            &            &                                  &                                  &            &            \\ \hline
\multirow{4}{*}{F07}            & S1 & $\sim$     & \checkmark &            &            &                                  &                                  &            &            \\
				& S2 & \checkmark & \checkmark &            &            &                                  &                                  &            &            \\ 
				& S3 & $\sim$     & \checkmark &            &            &                                  &                                  &            &            \\ 
				& S4 & $\sim$     & \checkmark &            &            &                                  &                                  &            &            \\ \hline
\multirow{4}{*}{F12}            & S1 & \ding{53}  & \checkmark &            &            &                                  &                                  &            &            \\
				& S2 & \ding{53}  & \checkmark &            &            &                                  &                                  &            &            \\ 
				& S3 & \ding{53}  & \checkmark &            &            &                                  &                                  &            &            \\ 
				& S4 & \ding{53}  & \checkmark &            &            &                                  &                                  &            &            \\ \hline
\multirow{3}{*}{F15}            & S1 & \ding{53}  & \checkmark & $\sim$     & \checkmark & \ding{53} \ding{53}  $\sim$      & \ding{53} $\sim$ \checkmark      & $\sim$     & \checkmark \\ 
				& S2 & \ding{53}  & \checkmark & $\sim$     & \checkmark &                                  &                                  & \checkmark & \checkmark \\ 
				& S3 & \ding{53}  & \checkmark & $\sim$     & \checkmark & \ding{53} \ding{53}  $\sim$      & \ding{53} \checkmark \checkmark  & \checkmark & \checkmark \\ \hline
\multirow{4}{*}{F16}            & S1 & \checkmark & \checkmark &            &            &                                  &                                  &            &            \\ 
				& S2 & $\sim$     & \checkmark &            &            &                                  &                                  &            &            \\ 
				& S3 & $\sim$     & \checkmark &            &            &                                  &                                  &            &            \\ 
				& S4 & $\sim$     & \checkmark &            &            &                                  &                                  &            &            \\ \hline
\multirow{2}{*}{6000-02}        & S1 & \checkmark & \checkmark & \checkmark & \checkmark &                                  &                                  &            &            \\ 
				& S2 & \checkmark & \checkmark &            &            &                                  &                                  &            &            \\ \hline

\end{tabular}
 \captionsetup{width=0.9\textwidth} 
 \caption[Results of the surface sensitivity of the tested detectors]
         {Results of the surface sensitivity of the tested detectors. ``S'' is abbreviated for ``side''. \checkmark, $\sim$ and \ding{53} mean fully, half and non-sensitive. The three columns for the \isotope[207]Bi IC electrons are for the energies of \SIlist{482;976;1682}{keV}. Only relative differences in the sensitivities between the NCA and CA as outermost anode rail could be tested, not absolute. Only detectors with a thin coating (Parylen) or no lacquer at the sides were used, as a thick lacquer blocks the radiation too much.}
 \label{tab_detector_surface_sensitivity_results}
\end{table}

%%%%%%%%%%%%%%%%%%%%%%%%%%%%%%%%%%%%%%%%%%%%%%%%%%%%%%%%%%%%%%%%%%%%%%%%%%%%%%%%%%%%%%%%%%%%%%%%%%%%%%%%%%%%%%%%%%%%%%%%%%%%
%%%%%%%%%%%%%%%%%%%%%%%%%%%%%%%%%%%%%%%%%%%%%%%%%%%%%%%%%%%%%%%%%%%%%%%%%%%%%%%%%%%%%%%%%%%%%%%%%%%%%%%%%%%%%%%%%%%%%%%%%%%%
\FloatBarrier
\section{Comparison of LSE and A/E}
\label{sec_section_comparison_A_over_E_LSE}
The different PSA methods LSE and A/E are compared here concerning their stability of thresholds and their efficiency.
Both comparisons use the laboratory data discussed in the last section.

%%%%%%%%%%%%%%%%%%%%%%%%%%%%%%%%%%%%%%%%%%%%%%%%%%%%%%%%%%%%%%%%%%%%%%%%%%%%%%%%%%%%%%%%%%%%%%%%%%%%%%%%%%%%%%%%%%%%%%%%%%%%
\FloatBarrier
\subsection{Variation of cut thresholds}
\label{sec_variation_cut_thresholds}
Constant global LSE cut thresholds are used for the detectors of the COBRA demonstrator at LNGS: \num{8} for ERT and \num{53} for DIP. 
The detectors there are of a high quality, and are operated under constant conditions. Furthermore, they are supplied with bias voltages nearly all the time.
Nevertheless, it is known that some detectors have a different timing-behavior and have other optimal ERT thresholds.

\begin{figure}[b!]
 \centering
  \includegraphics[width=0.49\textwidth]{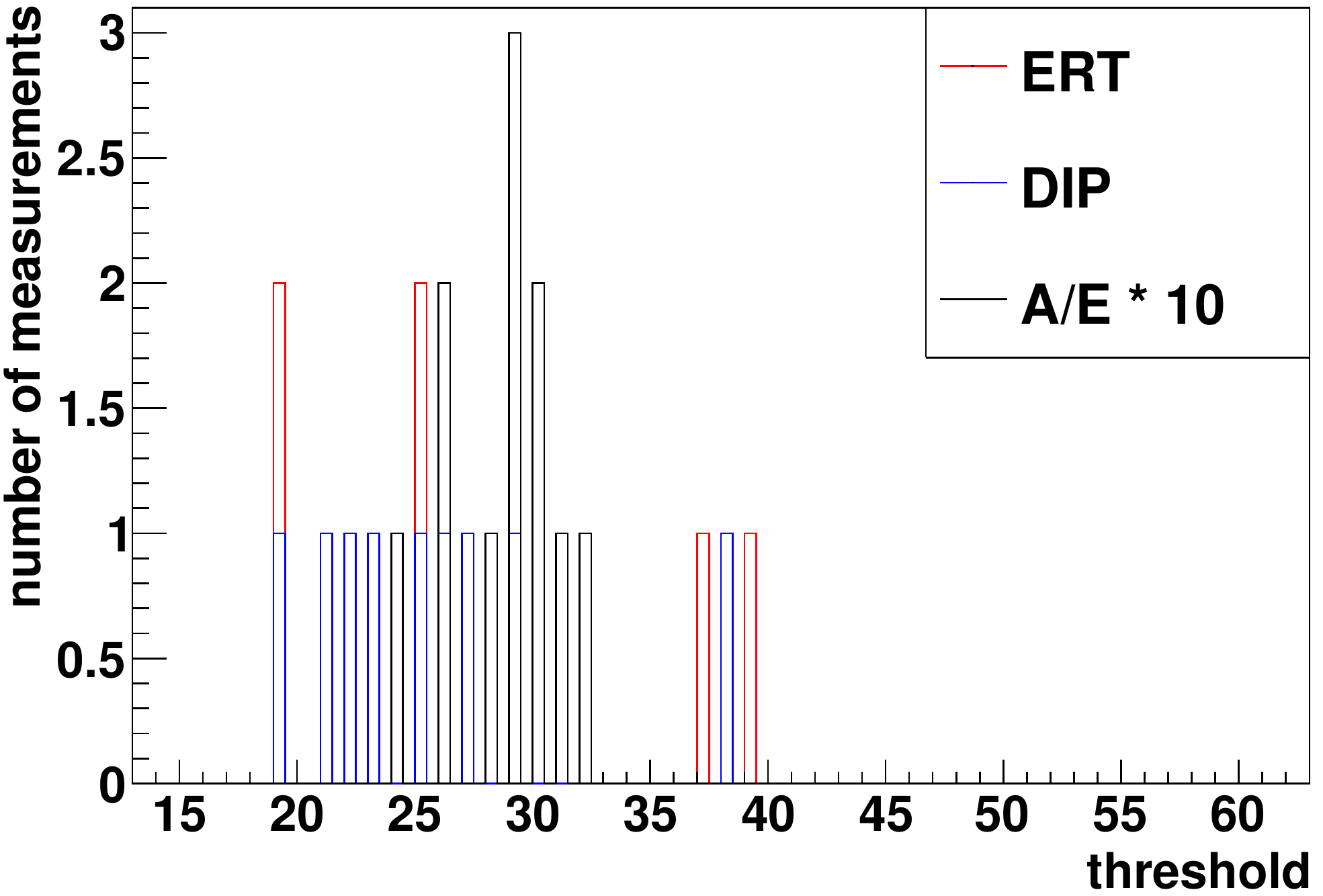}
  \includegraphics[width=0.49\textwidth]{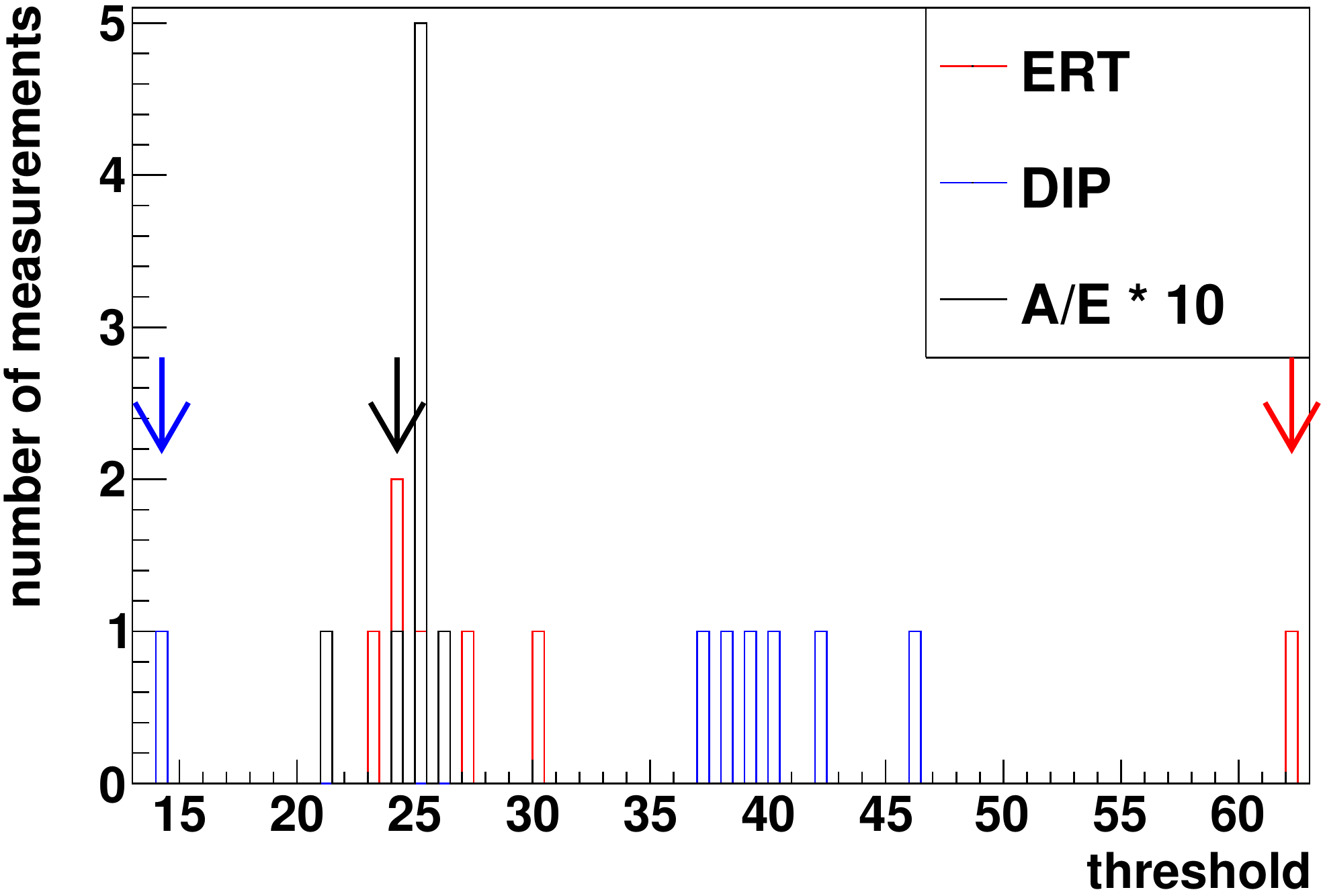}
  \captionsetup{width=0.9\textwidth} 
  \caption[Variation of thresholds of A/E and LSE]
          {Variation of thresholds of A/E and LSE of several measurements. Left: Full energy peak of \isotope[137]Cs. Right: Gamma radiation with a lower energy cut of \SI{1500}{keV} of \isotope[232]Th and \isotope[207]Bi. One measurement produces drifted thresholds. These are marked with the arrows. Note that the A/E value is not affected by this.}
  \label{fig_surface_cut_variation}
\end{figure}

The quality of the detectors used in laboratory test measurements is not that good. Additionally, the voltages had to be switched off and on for each new measurement configuration like changed source or source position.
It is known that the energy calibration may change by a few percent by this. But the surface cut thresholds can change also. 
This is probably an effect of the electric field within the detector, which may not always form up exactly the same way when switching on the voltages.\\ 
Furthermore, the surface cuts behave differently with energy, so the thresholds can be energy-dependent as well \cite{theinert_master}. 
This has to be kept in mind, when comparing thresholds determined with different radioactive sources.\\
The variation of the cut thresholds of several measurements is shown in \autoref{fig_surface_cut_variation} for the full energy events of \isotope[137]Cs (left), and higher energetic gamma events above \SI{1500}{keV} of different measurements of \isotope[232]Th and \isotope[207]Bi (right). 

\begin{SCfigure}[][t!]
 \centering
  \includegraphics[width=0.6\textwidth]{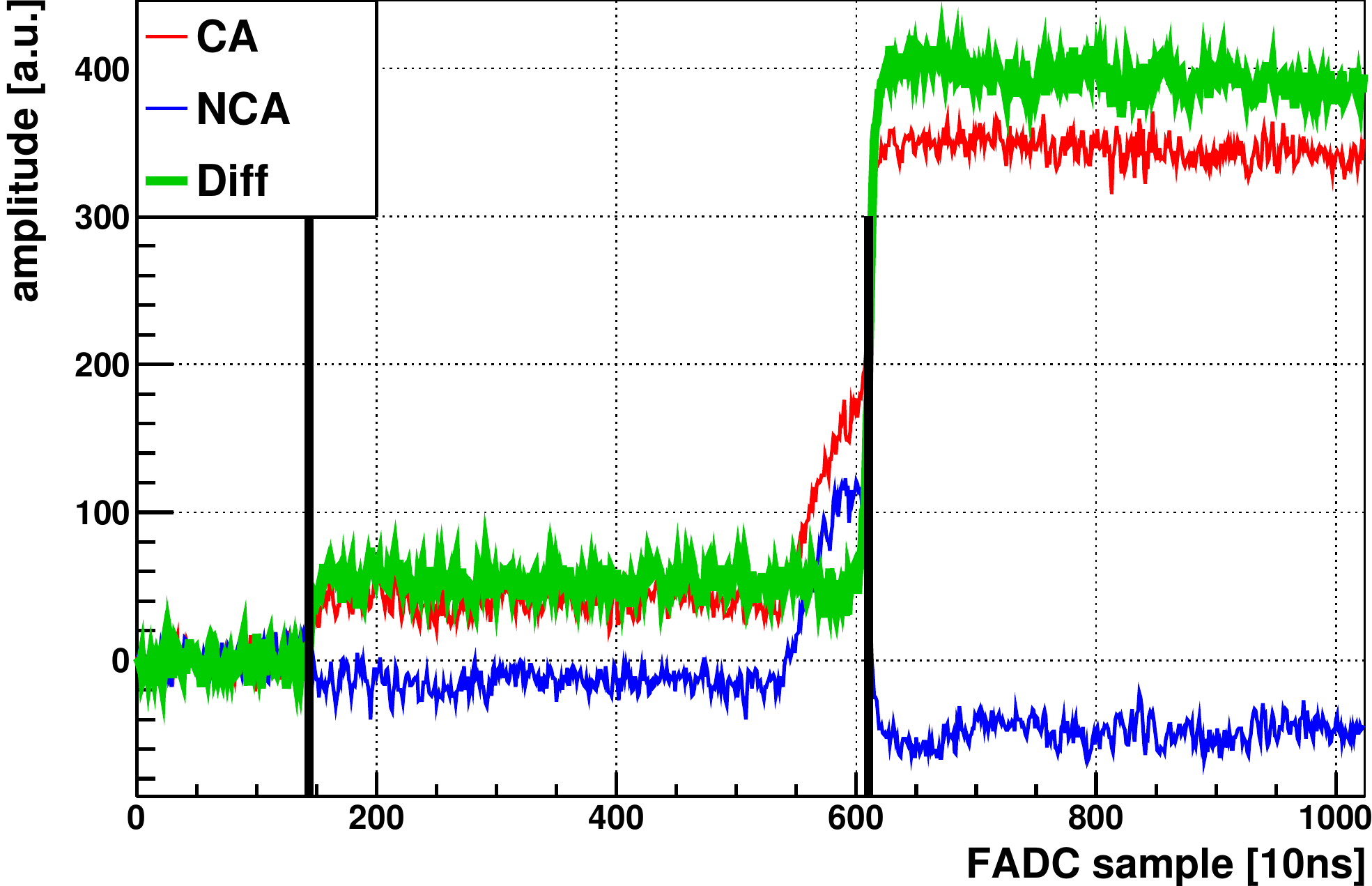}
  \captionsetup{width=0.9\textwidth} 
  \caption[Explanation for too high ERT values]
          {Explanation for too high ERT-values. The \SI{50}{\%} and \SI{3}{\%} levels of the difference pulse, which define the ERT-value, are marked with the black lines. The A/E value (steepest part of the difference pulse) is at bin 612, close to the black line.}
  \label{fig_too_high_ert_values}
\end{SCfigure}

The thresholds fluctuate less for higher energies. Due to a better SNR the mechanisms of PSA work much better there. 
Furthermore, the A/E thresholds are more stable than the ones for the LSE cut. 
For one measurement, the calculated LSE thresholds vary dramatically from the expected ones. The A/E threshold is not affected.
This can be explained, as the MSE-recognition did not tag MSE-events (or two independent events) correctly at this particular measurement. 
By this, the spectral shapes of the LSE quantities are distorted. As the thresholds are calculated as relative fractions of the shape, these can drift a lot.
The first interaction raises the pulse above the \SI{3}{\%}-level, which is the starting point for the ERT calculation. The second interaction provides the \SI{50}{\%} level and stops the ERT calculation.\\ 
In \autoref{fig_too_high_ert_values}, this is shown exemplarily for a single event.
It has an ERT-value of 466, corresponding to time of \SI{4.66}{\mu s}, which is much more than the maximal drift time of about a \SI{1}{\mu s}, so it have to be two independent interactions (a detailed calculation of the drift times is done in \autoref{sec_drift_time}).
As DIP is somehow complementary to ERT, this event has a DIP value of \num{-58}, which is too low.\\ 
The A/E values do not suffer from not-tagged MSEs, as these are based on the largest slope of the pulse, which is less affected by noise or disturbances.

Concluding, the optimal cut thresholds should be calculated at each calibration measurement, instead of using constant global values.
This is especially true for the detectors of the COBRA demonstrator, where each physics data-taking run has dedicated pre- and post-calibration measurements.

%%%%%%%%%%%%%%%%%%%%%%%%%%%%%%%%%%%%%%%%%%%%%%%%%%%%%%%%%%%%%%%%%%%%%%%%%%%%%%%%%%%%%%%%%%%%%%%%%%%%%%%%%%%%%%%%%%%%%%%%%%%%
\FloatBarrier
\subsection{Comparison of LSE and A/E efficiency}
\label{sec_comparison_A_over_E_LSE_threshold}
The laboratory measurements discussed in the last section can also be used to compare the efficiency of the LSE and A/E cuts. 
The detectors 631972-06, 667615-02 and F15 were irradiated with a \isotope[241]Am alpha source and a \isotope[232]Th gamma source to have defined alpha and gamma populations. 
This is done for nine detector configurations (detector sides). 
The number of comparisons is limited, because there are not more appropriate measurements with the two mentioned sources at the same measurement setup available.
As shown in \autoref{sec_daq_analysis}, a threshold is used for each of the PSA methods to discriminate surface from central events.
The cut thresholds are tuned as follows:\\
The number of events higher than \SI{1500}{keV} without any surface cuts is counted for the gamma population.
For ERT and DIP separately, the threshold is increased in steps of one bin (integer value), and the events are counted, that are below that threshold. 
The threshold is increased until the ratio of events below the threshold reaches \SI{90}{\%} of all events. 
The DIP and ERT events may be correlated positively:
In the LSE publication \cite{Fritts:2014qdf}, \SI{1.1}{\%} of the events were reported to have values above threshold for both DIP and ERT.
Hence, the final LSE ratio is obtained by applying the two DIP and ERT cuts independently, and calculating the ratio of events below threshold and all events.
For the ratio at the alpha measurements, the same thresholds are applied as for the gamma measurement.\\
The ratios for the A/E criterion are obtained in an analog way.
An exception is that the threshold is decreased in steps of \num{0.001} starting at one, as central events have smaller values than surface events.
Furthermore, the tuned threshold is found, if the ratio of tagged events has reached the same value as for the combined LSE criterion for this particular measurement.

\begin{figure}
 \centering
  \includegraphics[width=1.0\textwidth]{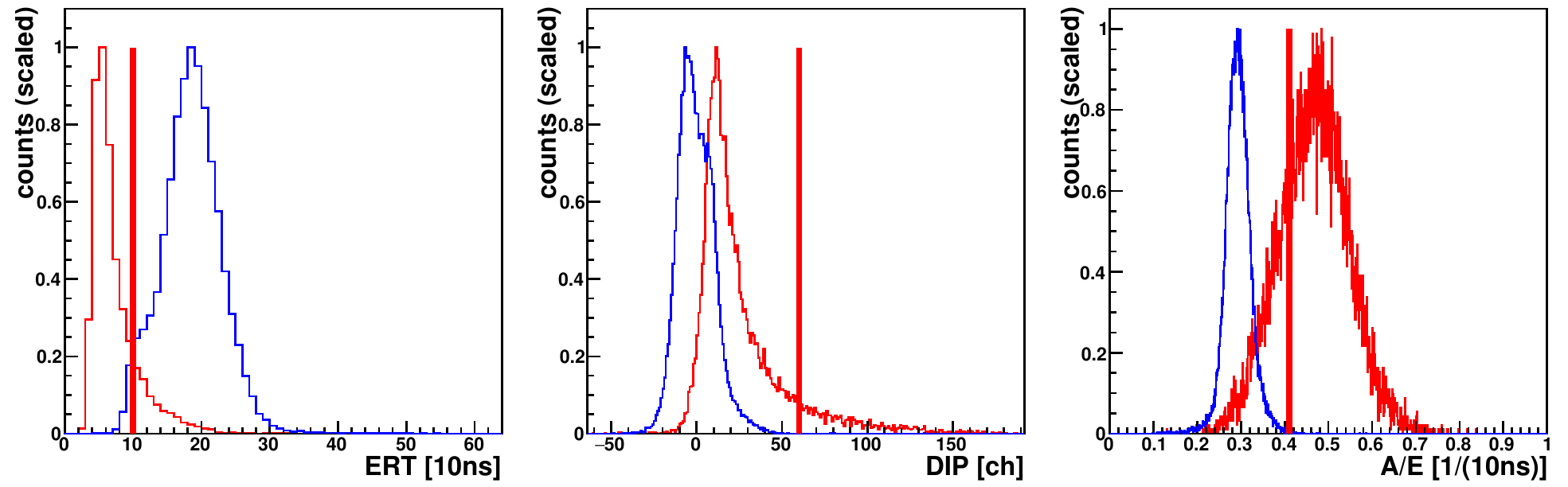}
  \captionsetup{width=0.9\textwidth} 
  \caption[Method to compare the ratio of surviving events of LSE and A/E]
          {Method to compare the ratio of surviving events of LSE and A/E. 
          For each of the quantities ERT, DIP and A/E, a threshold (red line) is determined to keep a certain ratio of gamma events (red). The same threshold is then applied to the distribution of alpha events (blue).}
  \label{fig_comparison_method_LSE_A_over_E}
\end{figure}

Events that did not survive the data cleaning or that are flagged as MSE are excluded from this study.\\
The distributions and calculated thresholds are shown in \autoref{fig_comparison_method_LSE_A_over_E} for a typical example, from which the ratio of surviving gamma and alpha events is calculated.

\begin{SCfigure}
 \centering
  \includegraphics[width=0.8\textwidth]{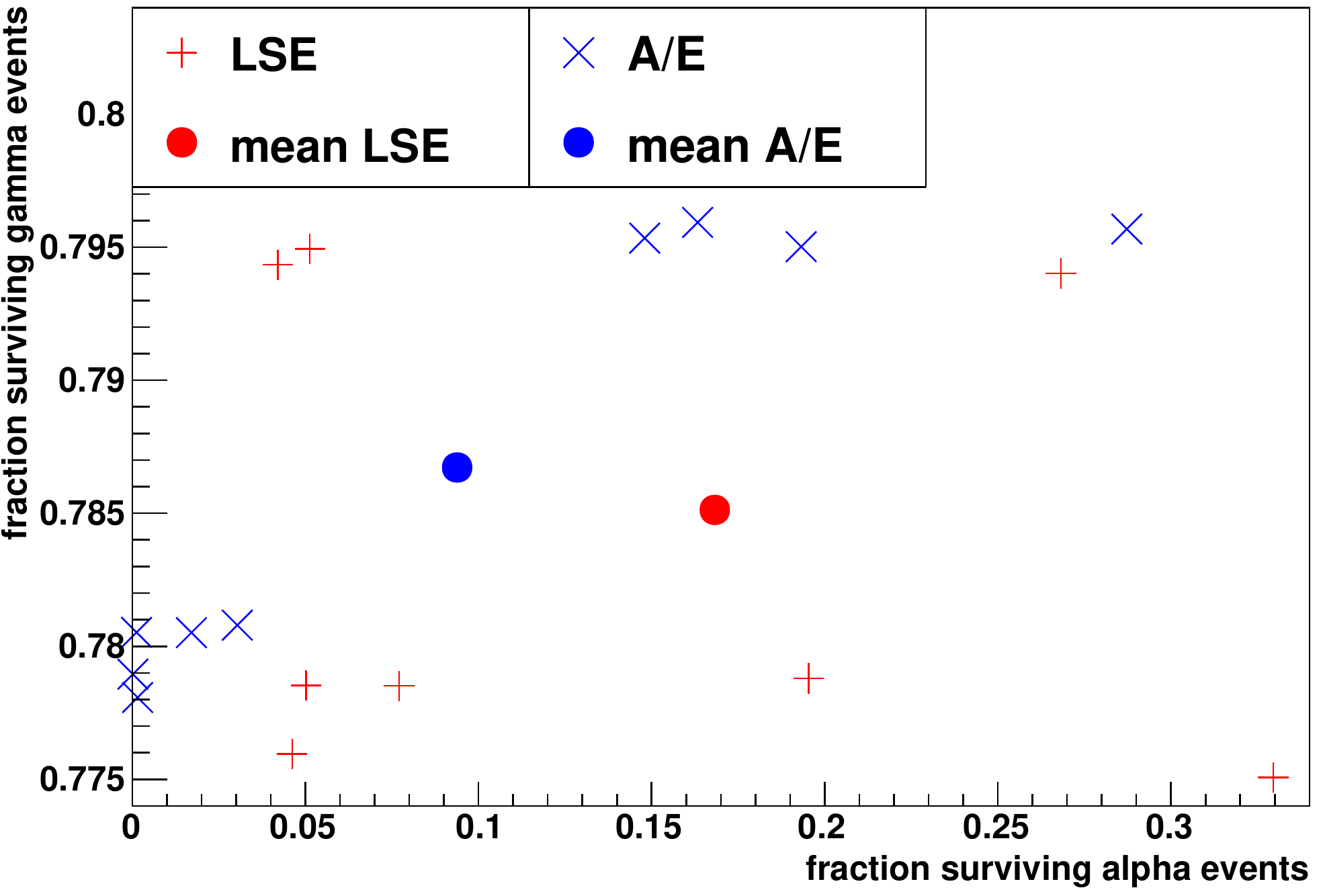}
  \caption[Comparison of LSE and A/E efficiency]
          {Comparison of LSE and A/E efficiency.}
  \label{fig_comparison_result_LSE_A_over_E}
\end{SCfigure}

The result is shown in \autoref{fig_comparison_result_LSE_A_over_E}: The ratio of surviving gamma events are plotted versus the ratio of the surviving alpha events.
The resulting ratios do not reach the desired value of \SI{80}{\%} exactly:
The thresholds are varied in discrete values, and for the LSE ratio, two of those thresholds have to be applied independently.
However, this does not alter the conclusion of this comparison.

For these detectors under test, A/E is the better PSA discrimination tool. It removes approximately \SI{50}{\%} more surface events while keeping the same fraction of central events. 
The mean values and standard deviations are given in the results compilation \autoref{tab_comparison_A/E_LSE}.
The relatively large values for the standard deviations for the surviving alpha events resemble the fact, that this quantity has a huge spread from about \SI{30}{\%} to the per mill level.\\
The initial histograms of the PSA quantities have several ten thousand entries, so the statistical uncertainty is ignored here.
This analysis is used to give a first hint on how the different PSA mechanisms perform. 
It is not intended to give a complete efficiency calculation. Consequently, no systematic uncertainties are considered here.

\begin{SCtable} 
 \begin{tabular}{|c|c|c|}
  \hline 
            & \multicolumn{2}{c|}{surviving events [\%]}   \\
            & gamma                          & alpha \\ \hline
  A/E & \num{0.787(8)}                 & \num{0.094(106)}  \\
  LSE & \num{0.785(9)}                 & \num{0.168(151)} \\ \hline
 \end{tabular}
 \caption[Ratio of surviving alpha  and gamma events for the A/E and LSE PSA]
         {Ratio of surviving alpha  and gamma events for the A/E and LSE PSA. Values taken from \autoref{fig_comparison_result_LSE_A_over_E}.}
 \label{tab_comparison_A/E_LSE}
\end{SCtable}

%%%%%%%%%%%%%%%%%%%%%%%%%%%%%%%%%%%%%%%%%%%%%%%%%%%%%%%%%%%%%%%%%%%%%%%%%%%%%%%%%%%%%%%%%%%%%%%%%%%%%%%%%%%%%%%%%%%%%%%%%%%%
\FloatBarrier
\subsection{Conclusion to LSE and A/E comparison}
\label{sec_conclusion_LSE_A/E_comparison}
Both PSA methods are based on certain characteristics of the weighting potential. 
According to the Shockley-Ramo theorem (\autoref{eq_Shockly-Ramo}), the measured pulses are directly proportional to the weighting potential.
The LSE method relies on the fact, that the weighting potential shows distortions towards the detector surfaces compared to the center, while A/E is based on a weighting potential with a larger gradient in the detector center compared to the surfaces.\\
One advantage of A/E is its simplicity:
It is one cut, consequently one threshold is needed to discriminate surface events from central events.
At the Germanium experiments GERDA and MAJORANA, it is even used to discriminate SSE from MSE, which has not been tested here.
In contrast to this, the LSE discrimination method consists of two independent cuts, and a third one to veto MSE.\\
The cut threshold at the A/E criterion seems to be more stable. It is based on the largest slope of the pulse, which is less susceptible to distortions than comparative ratios of the pulse heights.\\
Furthermore, the efficiency to discriminate surface events was about \SI{50}{\%} higher for A/E than LSE.
However, this surface events discrimination power was vice versa for one of the detectors under study here.\\

Concluding, the A/E method should be investigated further, also using the physics data from the COBRA demonstrator.

%%%%%%%%%%%%%%%%%%%%%%%%%%%%%%%%%%%%%%%%%%%%%%%%%%%%%%%%%%%%%%%%%%%%%%%%%%%%%%%%%%%%%%%%%%%%%%%%%%%%%%%%%%%%%%%%%%%%%%%%%%%%
%%%%%%%%%%%%%%%%%%%%%%%%%%%%%%%%%%%%%%%%%%%%%%%%%%%%%%%%%%%%%%%%%%%%%%%%%%%%%%%%%%%%%%%%%%%%%%%%%%%%%%%%%%%%%%%%%%%%%%%%%%%%
\FloatBarrier
\section{Alpha scanning measurements}
\label{sec_alpha_scanning}
To study the behavior of surface events position-dependent, alpha scanning measurements of the detectors were done.
Some of the open questions are the validity of the interaction depth calculation for surface events, and the investigation of the areas at the edges, where the LSE-tagging might not work (mentioned in \autoref{sec_LSE_recognition}).

%%%%%%%%%%%%%%%%%%%%%%%%%%%%%%%%%%%%%%%%%%%%%%%%%%%%%%%%%%%%%%%%%%%%%%%%%%%%%%%%%%%%%%%%%%%%%%%%%%%%%%%%%%%%%%%%%%%%%%%%%%%%
\FloatBarrier
\subsection{Scanning setup}
\label{sec_scanning_setup}
\begin{figure}
 \centering
  \includegraphics[height=0.45\textwidth]{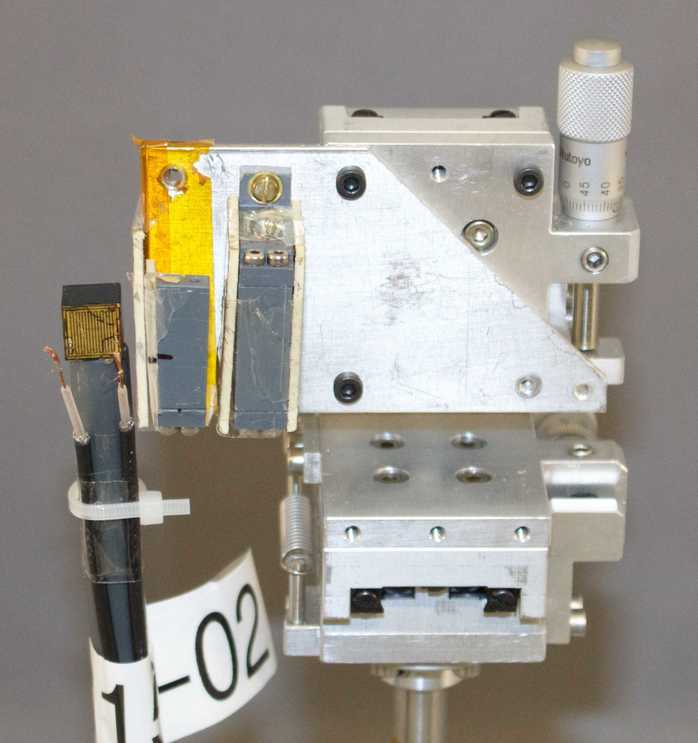}
   \includegraphics[height=0.45\textwidth]{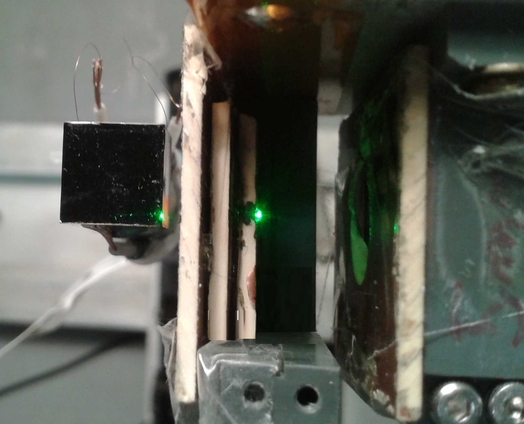}
   \captionsetup{width=0.9\textwidth} 
   \caption[Photograph of scanning setup]
           {Photograph of the scanning setup consisting of a detector mounted on a stick and the scanning device, no radioactive source is inserted. Left: side view. One micrometer screw can be seen in the top right corner. 
           The top view (right) shows the absolute calibration of the collimator and detector. A laser pointer (not visible) is inserted instead of a radioactive source. The laser spot (green) can be seen hitting the collimator (center of picture), and detector surface (left).}
   \label{fig_scanning_setup}
\end{figure}

To be able to do alpha scanning measurements, the setup has to guarantee precise positions of source, collimator and detector.
Especially for alpha radiation, the source has to be very close to the detector.
To satisfy this requirement, a new scanning device was designed.
It is built from a source holder and a collimator. Two micrometer-screws allow very precise (relative) position variations of \SI{1/100}{mm} in horizontal and vertical axis. 
The collimator is made of cardboard, \SI{6}{mm} thick, with \SI{0.3}{mm} drill diameter.\\
A \isotope[241]Am alpha source is used, with an alpha energy of \SI{5.5}{MeV} and an activity of \SI{330}{kBq}. 
The source can be inserted into the scanning device until it is directly in front of the collimator.
To verify the absolute calibration between collimator and detector position, a laser-pointer can beam through the collimator onto the detector surface if the source is removed.\\ 
The detectors are glued with their side four (defined in \autoref{fig_detector_sides_axis_conventions}) on a plastic stick with a diameter of \SI{10}{mm}.
Hence, this particular side is not accessible to measurements, but the remaining five sides are.
To guarantee stable conditions for such precision measurements, the detector and the scanning device are placed on an optical bench.
The standard COBRA EMI-shielding is used, as well as the readout electronics and DAQ system. 
\autoref{fig_scanning_setup} shows the scanning device and exemplarily the absolute calibration method.
Other sources can be added for calibration purpose if needed.
To change the scanning position, the BV and the GB have to be switched off and the EMI shielding has to be opened. 
By biasing the detector again, the energy calibration may change.
Consequently the energy calibration typically varies by a few \%, if only an alpha source is used, which cannot be used as calibration source. 
The position of the scanning points can be chosen using the micrometer screws.\\

For measurements of side 2, the detector is tilted by \ang{90}. 
But the detectors are known to be piezoelectric \cite{PhysRevB.42.11392}, and mechanical tensions can influence the detector performance.
Especially a detector which is glued to one of its sides and then tilted, could suffer from these effects. 
Test measurements were done to see if this is an important effect to consider.
The detector and source were arranged firmly in a fixed position. This whole device is now tilted and rotated that each detector side can face any direction. 
The measurements were done with a \isotope[137]Cs gamma and a \isotope[241]Am alpha source. 
These tests showed, that no significant differences appeared.

%%%%%%%%%%%%%%%%%%%%%%%%%%%%%%%%%%%%%%%%%%%%%%%%%%%%%%%%%%%%%%%%%%%%%%%%%%%%%%%%%%%%%%%%%%%%%%%%%%%%%%%%%%%%%%%%%%%%%%%%%%%%
%%%%%%%%%%%%%%%%%%%%%%%%%%%%%%%%%%%%%%%%%%%%%%%%%%%%%%%%%%%%%%%%%%%%%%%%%%%%%%%%%%%%%%%%%%%%%%%%%%%%%%%%%%%%%%%%%%%%%%%%%%%%
\FloatBarrier
\subsection{Analysis and results}
\label{sec_analysis_principle}
The collimator reduces the flux of the alpha particles by approximately four orders of magnitude.
As a consequence, the measurement duration was \SI{20}{min} to \SI{30}{min} to capture enough data. 
This is enough time to measure a significant amount of laboratory background events.
Consequently, a discrimination of the alpha events versus background is needed. 
This is done according to the peak-finding at the gamma scanning measurements (\autoref{sec_gamma_scanning}).
No information of the position of the collimator is included as input, because the interaction depth calculation is not always reliable for surface events (see e.g. \autoref{fig_alpha_surface_sensitivity}), and is investigated here as well.
Furthermore, the occasionally occurring charge-sharing distortions discussed in \autoref{sec_hyperbolic} lead also to a wrongly calculated interaction depth.

Due to the collimator, the events of the alpha radiation build a peak at a certain position in the interaction depth distribution. 
From the maximum value of the distribution, the interval borders to both sides are searched for until an empirically determined background level is reached. 
This background level is found as follows:
The largest background contribution is at the cathode (interaction depth between 0.96 and 1.1), because the BV is supplied there, which results in electrostatic attraction of contaminations.
The distribution of the background level at the cathode of all measurements per detector and measurement campaign is taken, where the scanning position does not reach into the cathode area.
The threshold is set above this maximal background level.
As the threshold is based on the background level at the cathode (which is not surface-dependent), it is a global threshold for each detector.

\begin{SCfigure}
  \centering
   \includegraphics[width=0.7\textwidth]{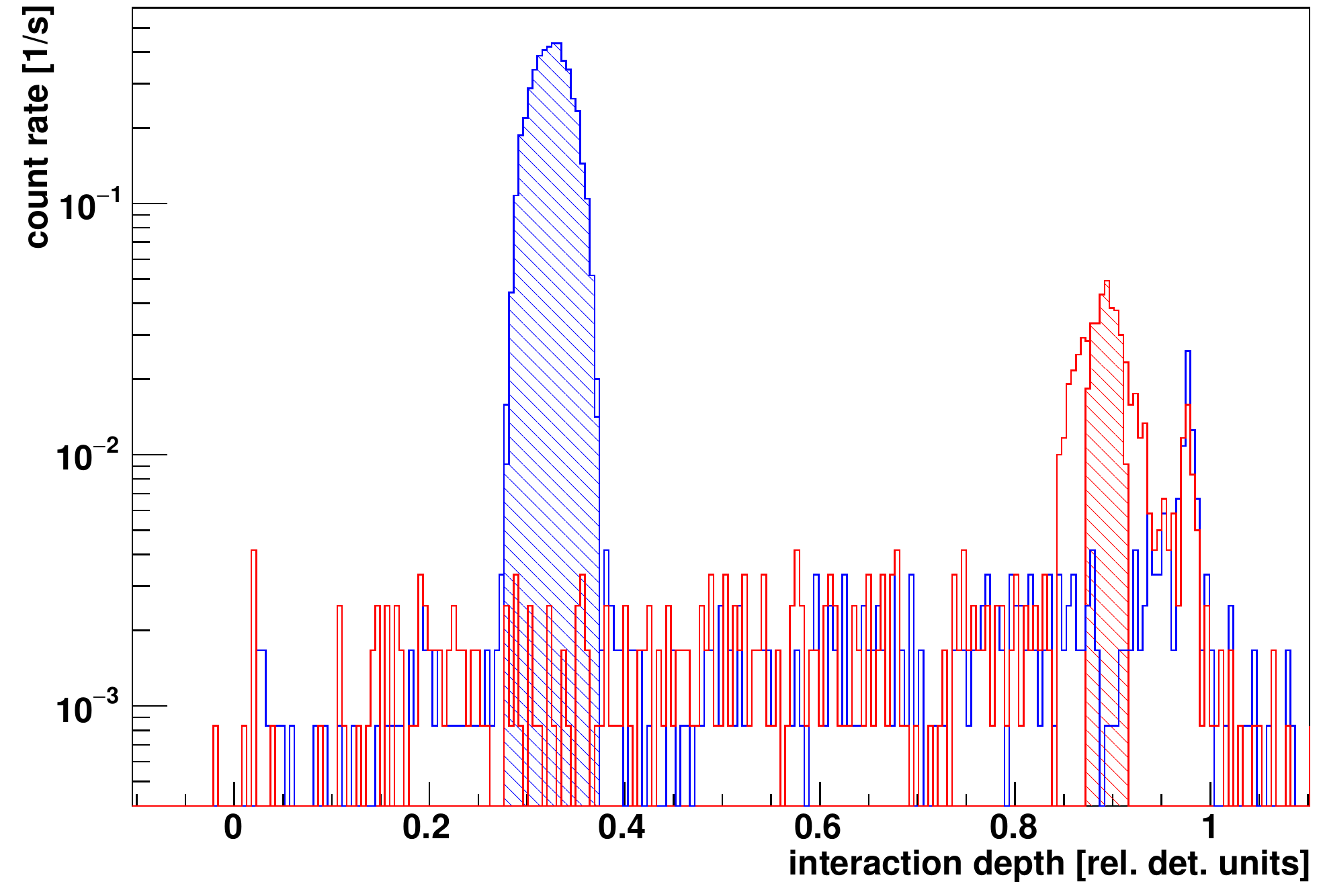}
   \caption[Example of alpha peaks in scanning measurements]
           {Two examples of alpha peaks in scanning measurements. Blue: alpha peak clearly above the background. Red: low surface sensitivity, peak is less distinctive above the background. Filled areas: Events chosen as peak content.}
   \label{fig_analysis_principle_ipos_ztc}   
\end{SCfigure}

The interval found like this contains the selected alpha events.
Most measurements show an alpha peak clearly above the background. At other positions where the surface is not fully sensitive, less alpha radiation is measured. 
The alpha peaks lie sometimes hardly above the maximal background level.
\autoref{fig_analysis_principle_ipos_ztc} shows two typical examples of alpha peaks and backgrounds, plotted into the same histogram.
The first is an alpha peak at a position of full sensitivity, the peak is clearly above the background.
In contrast, the second peak is less distinctive, but still significant above the background.
The diameter of the alpha beam can be estimated to \SI{1}{mm} in these interaction depth distributions.

\begin{figure}[h!]
 \centering
  \includegraphics[width=0.49\textwidth]{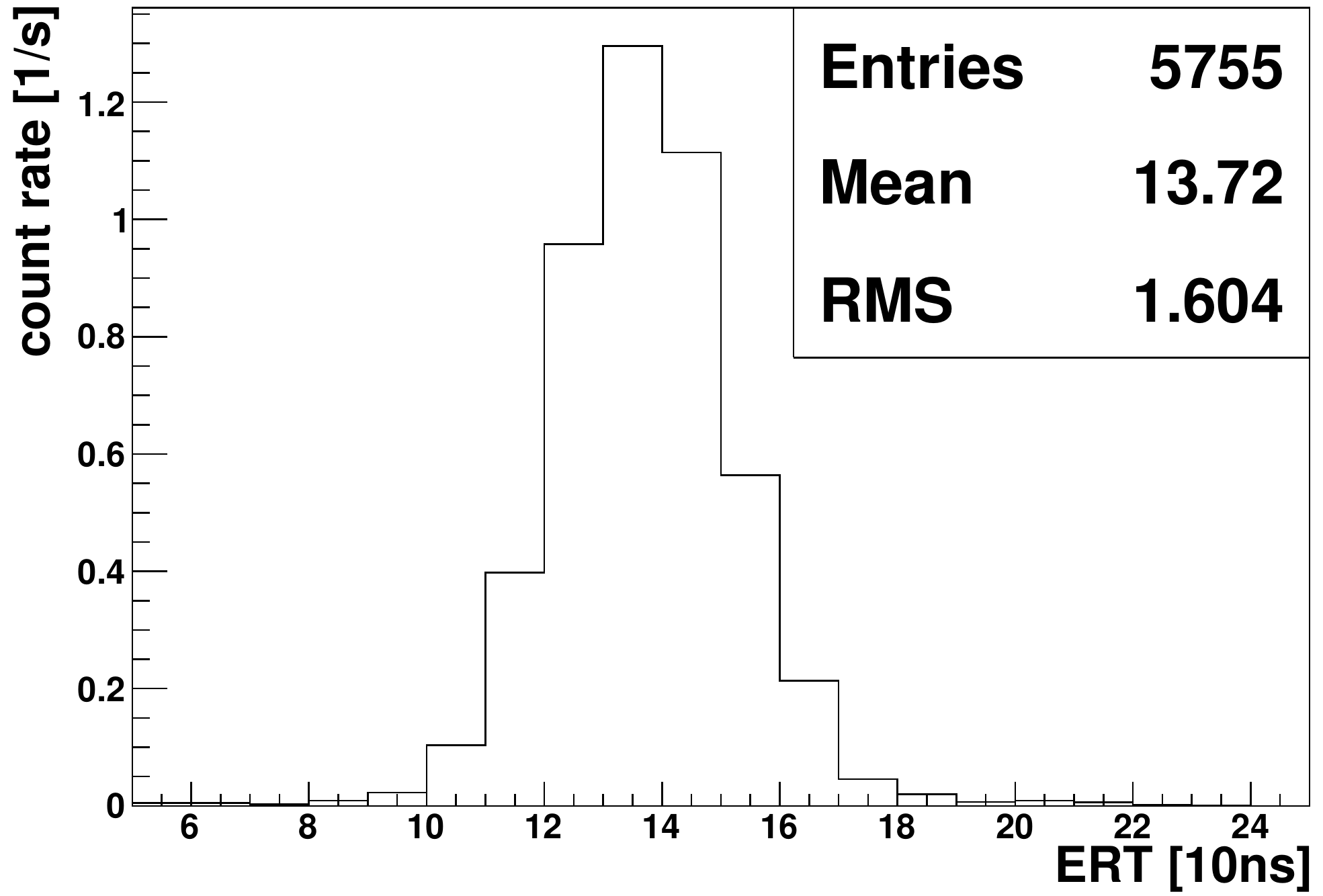}
  \includegraphics[width=0.49\textwidth]{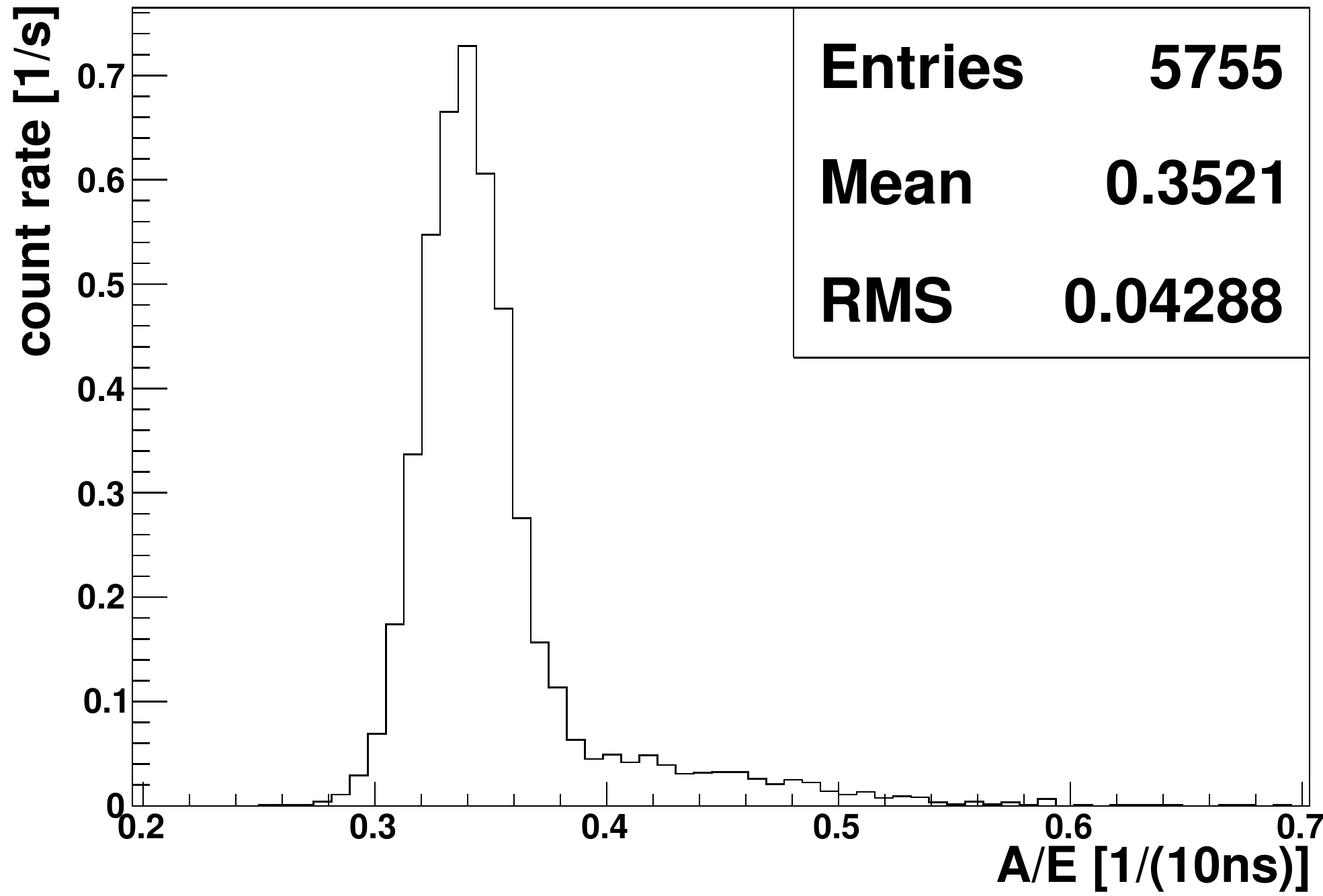}
  \captionsetup{width=0.9\textwidth} 
  \caption[ERT and A/E distributions of the selected alpha events]
          {ERT and A/E distributions of the selected alpha events. The mean values and standard deviation (spuriously titled as ``RMS'') are taken for the following graphical representations.}
  \label{fig_analysis_principle_ert_A_over_E}
\end{figure}

\autoref{fig_analysis_principle_ert_A_over_E} shows exemplarily the ERT and A/E distributions of the selected alpha events of the blue curve in \autoref{fig_analysis_principle_ipos_ztc}. Their mean values and standard deviations are used for the following analyses. 

%%%%%%%%%%%%%%%%%%%%%%%%%%%%%%%%%%%%%%%%%%%%%%%%%%%%%%%%%%%%%%%%%%%%%%%%%%%%%%%%%%%%%%%%%%%%%%%%%%%%%%%%%%%%%%%%%%%%%%%%%%%%
\FloatBarrier
\subsubsection{Surface events recognition}

\begin{figure}[t!] 
 \centering
  \includegraphics[width=0.49\textwidth]{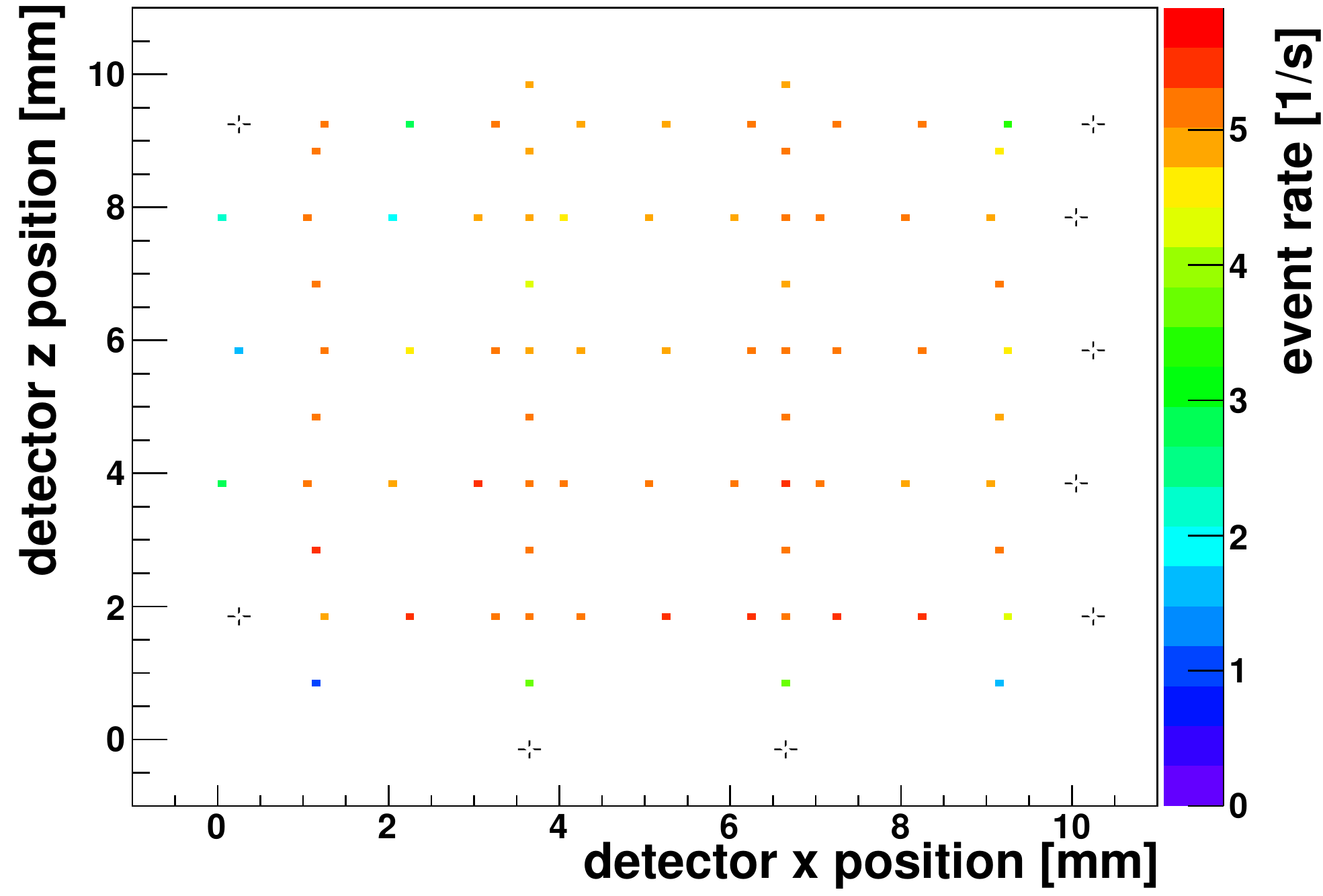}
  \includegraphics[width=0.49\textwidth]{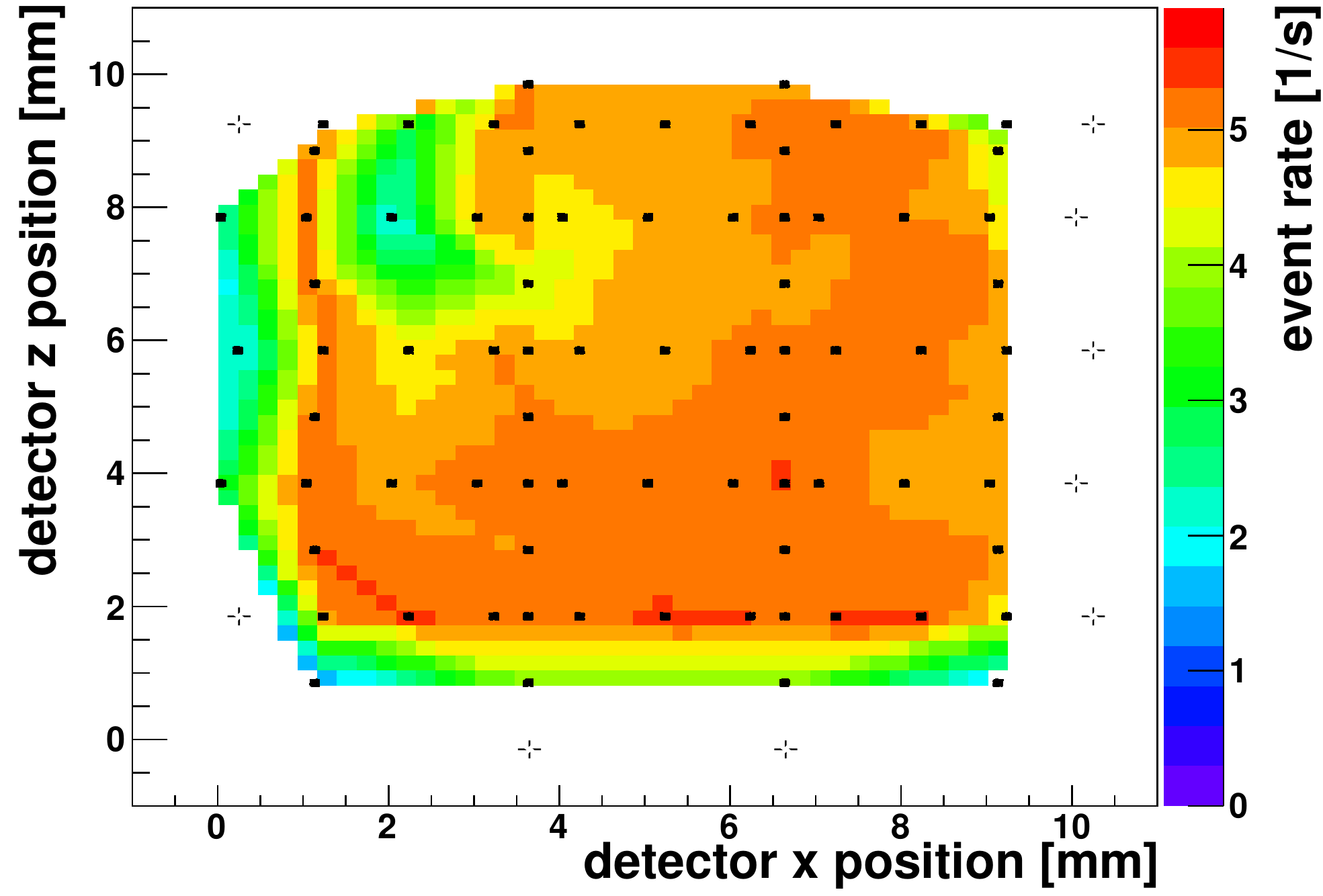}
  \captionsetup{width=0.9\textwidth} 
  \caption[Event rate shows non-fully sensitive areas]
          {Detector 667615-02 side 1 CA. 
           Left: Measured event rate as a function of the position of its scanning point. The cross-hair(``\includegraphics[height=0.35cm]{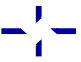}'')
           indicates measurements, where no alpha peak could be found. These are rejected for the following analyses.
           Right: same measurement, but interpolated using the Delaunay-method. The black rectangles indicate the position of the scanning points.}
  \label{fig_alpha_scanning_667615_02_side1_CA_event_rate}  
\end{figure}

\paragraph{Event rate}
In \autoref{fig_alpha_scanning_667615_02_side1_CA_event_rate}, the event rate of each scanning measurement of detector 667615-02 side 1 CA is shown. 
The event rate is the number of entries without any surface events cuts, divided by the measurement time.
As the same distance from detector to source is used, the event rate is expected to be constant.
It can be seen that the surface sensitivity does not vary much over a wide range, but changes locally. 
In this particular case, an area near the cathode is less sensitive.
Furthermore, the detector edges are not fully sensitive. Probably, this could only be partially explained by the fact that the collimated beam does not hit the detector fully at some outer positions.

\begin{SCfigure}%[b!]
  \centering  
    \includegraphics[width=0.69\textwidth]{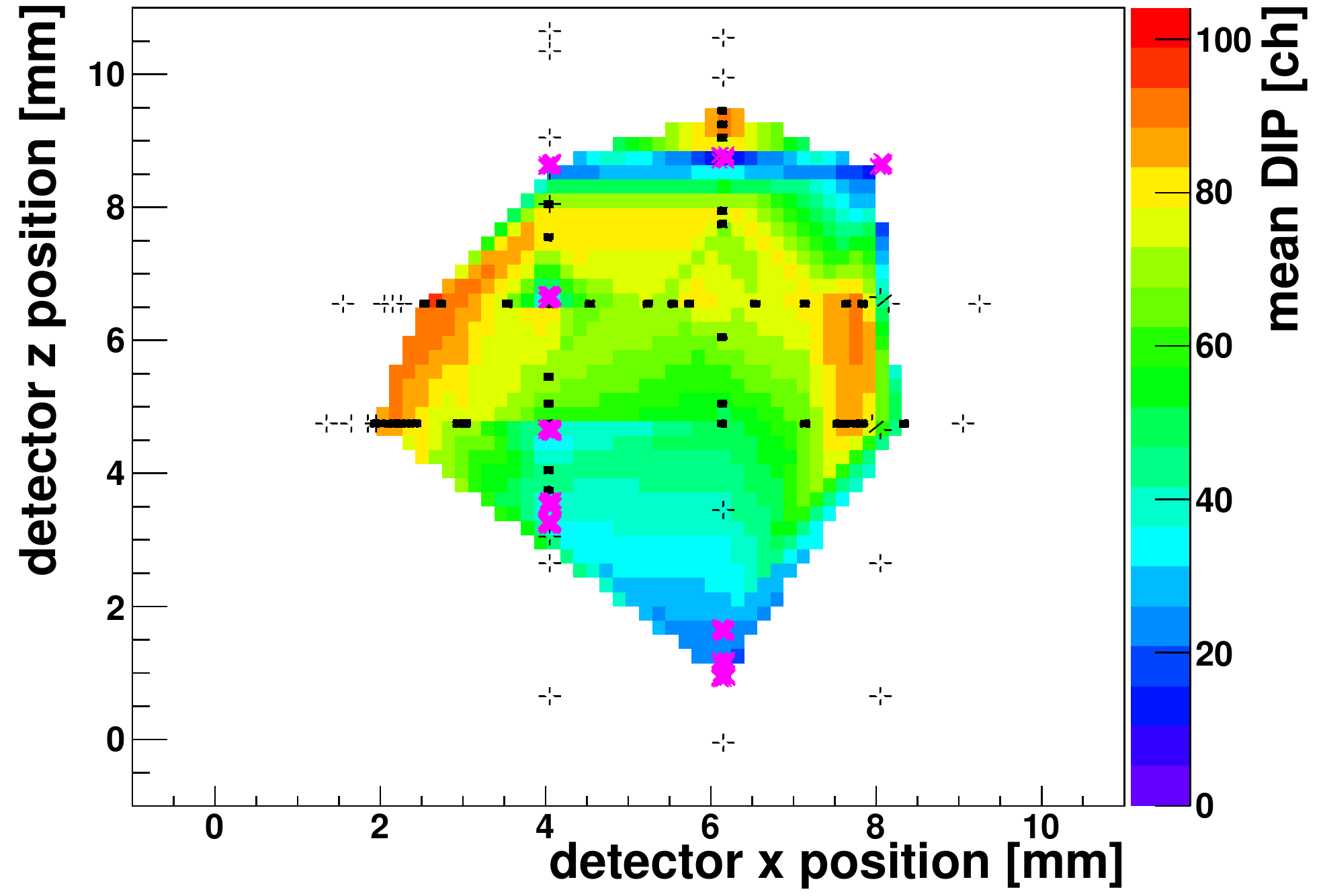}
    \caption[DIP values for detector 631972-06 side 3 NCA]
            {Mean DIP values as a function of their position for detector 631972-06 side 3 NCA. 
            The cross-hair indicates scanning points, where no alpha peak could be identified, the magenta X shows measurements that were not tagged as surface events.}
    \label{fig_631972_06_side3_NCA_rdip_3D}
\end{SCfigure}

The left representation shows the calculated values as a function of their scanning positions. 
The size of the colored rectangles is smaller than the irradiated areas on the detector surface. 
Scanning positions, where no alpha peak could be found using the method described above, are marked with the cross-hair ``\includegraphics[height=0.35cm]{pictures/surface_events/root_crosshair.png}'' and discarded for the following analyses. 
The right representation shows the interpolation of the measured points using the Delaunay triangulation method, mentioned on page \pageref{text_delaunay}.
In the interpolated representation, the position of the scanning points are marked with the black rectangles.
In the following, only the interpolated representations are shown mostly.
The measurements that were not tagged by the surface cuts (if applicable), are marked with the magenta X.\\
This detector surface shows a very high surface sensitivity. 
For a different behavior see \autoref{fig_631972_06_side3_NCA_rdip_3D} for example, where no alpha peaks could be identified at many scanning positions.

\paragraph{DIP}
\autoref{fig_631972_06_side3_NCA_rdip_3D} shows a result of the scanning measurements for the quantity DIP of detector 631972-06 side 3 NCA. 
It shows the position and the mean DIP values of the scanning-points on the detector surface. 
The DIP values tend to decrease slightly with interaction depth z. 
In x-direction, the values seem to be lower in the center, and higher towards the edges.
This is plausible as the LSE discrimination is based on a distorted weighting potential at the surfaces in comparison to the center.
The DIP values at the two edges might show a different behavior: At one edge, the neighboring surface has the same bias, while at the other the ERT bias is adjacent.
A complete comparison of the two edges is hardly possible here due to two facts:
The number of scanning points is low, and additionally, this NCA surface suffers from a lot of surface insensitivity.
However, one can state that the area of the highest DIP values above \SI{90}{ch} is much larger at low x-values. This is the edge where the ERT surface is adjacent, and the GR contact-pad is placed. 
The weighting potential seems to be most distorted there.\\
As the NCA surfaces of the detectors that were investigated using alpha scanning measurements suffered from insensitivities a lot, no additional surface was scanned completely. 
Consequently, no other result can be shown here for comparison.

\begin{SCfigure}[][b!]
 \centering
    \includegraphics[width=0.69\textwidth]{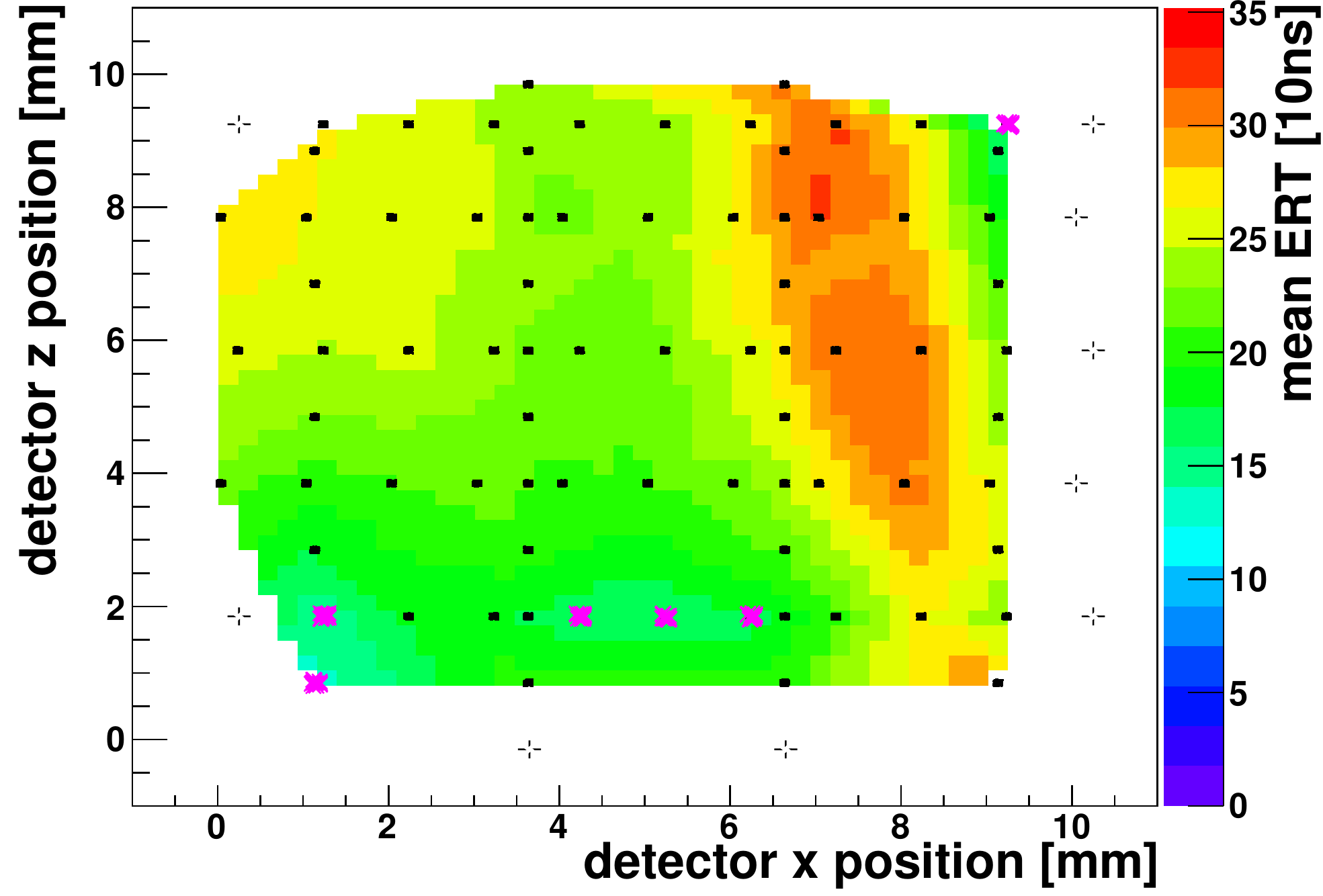}
        \caption[Map of ERT distribution of detector 667615-02 side 1 CA]
                {Detector 667615-02 side 1 CA.
                Interpolation indicating the position of the scanning points.
                The cross-hair indicates scanning points, where no alpha peak could be identified, the magenta X shows measurements that were not tagged as surface events.}
        \label{fig_alpha_scanning_667615_02_side1_CA_3D}
\end{SCfigure}

\paragraph{ERT}
An example for a result of the alpha scanning for the quantity ERT is shown in \autoref{fig_alpha_scanning_667615_02_side1_CA_3D}. 
At some positions, which are not at the edges, no alpha peaks could be found (marked with the magenta X). 
It cannot be determined, if in these areas the detector itself is insensitive, or the lacquer is too thick.
Like in \autoref{fig_631972_06_side3_NCA_rdip_3D} for the quantity DIP, the values decrease in z-direction toward the anodes, and show a variation in detector x-direction: 
Higher values at the sides, and lower in the center.
Towards the edge at high x-values, the ERT distribution shows the highest levels. 
There, the NCA side is adjacent. 
This could hint to a larger distortion in the weighting potential there, than at the edges where the same biases are applied.
Interestingly, the ERT values increase towards the edge, but decrease directly at the last measurable points at the edge.
This could hint to a cancellation of the ERT and DIP characteristics, which was discussed in \autoref{sec_LSE_recognition}.
Furthermore, not the whole \SI{10x10}{mm} surface is sensitive.

\begin{SCfigure}[][t!]
  \centering
   \includegraphics[width=0.69\textwidth]{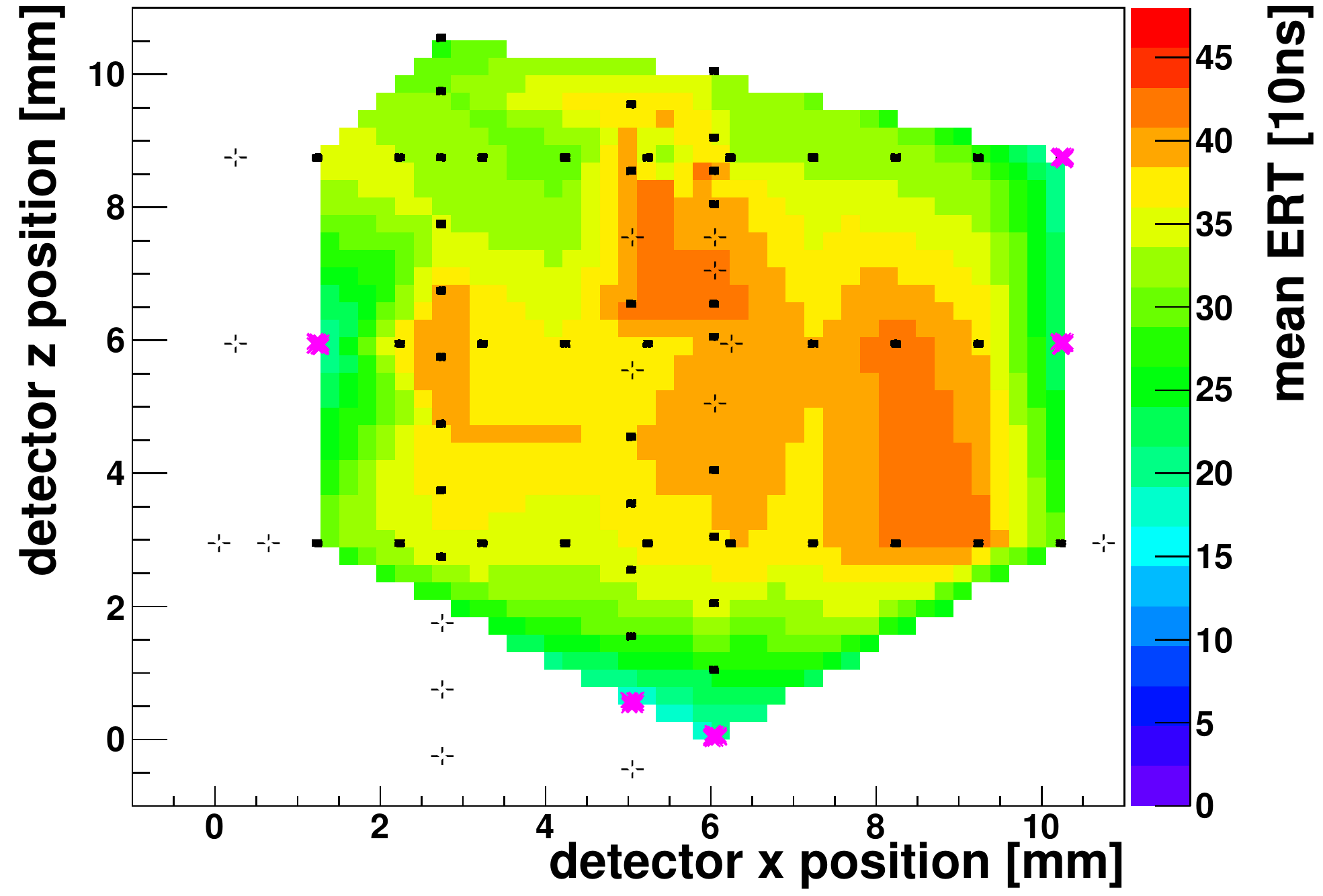}
       \caption[Map of ERT distribution of detector 631972-06 side 1 CA]
               {Detector 631972-06 side 1 CA. 
	       Interpolation indicating the position of the scanning points.
               The cross-hair indicates scanning points, where no alpha peak could be identified, the magenta X shows measurements that were not tagged as surface events.}
        \label{fig_alpha_scanning_631972_06_side1_CA_3D}
\end{SCfigure}

However, at detector 631972-06, the ERT distribution does not show the same characteristics (qualitatively):
The values vary over the detector surface, but no higher values in x-direction towards the edges and lower in the center can be seen.
Additionally, no decreasing values in z-direction are measured.
At several scanning points, also in the center, no alpha peaks could be identified, as shown in \autoref{fig_alpha_scanning_631972_06_side1_CA_3D}.

\begin{figure}[t!]
  \includegraphics[width=0.49\textwidth]{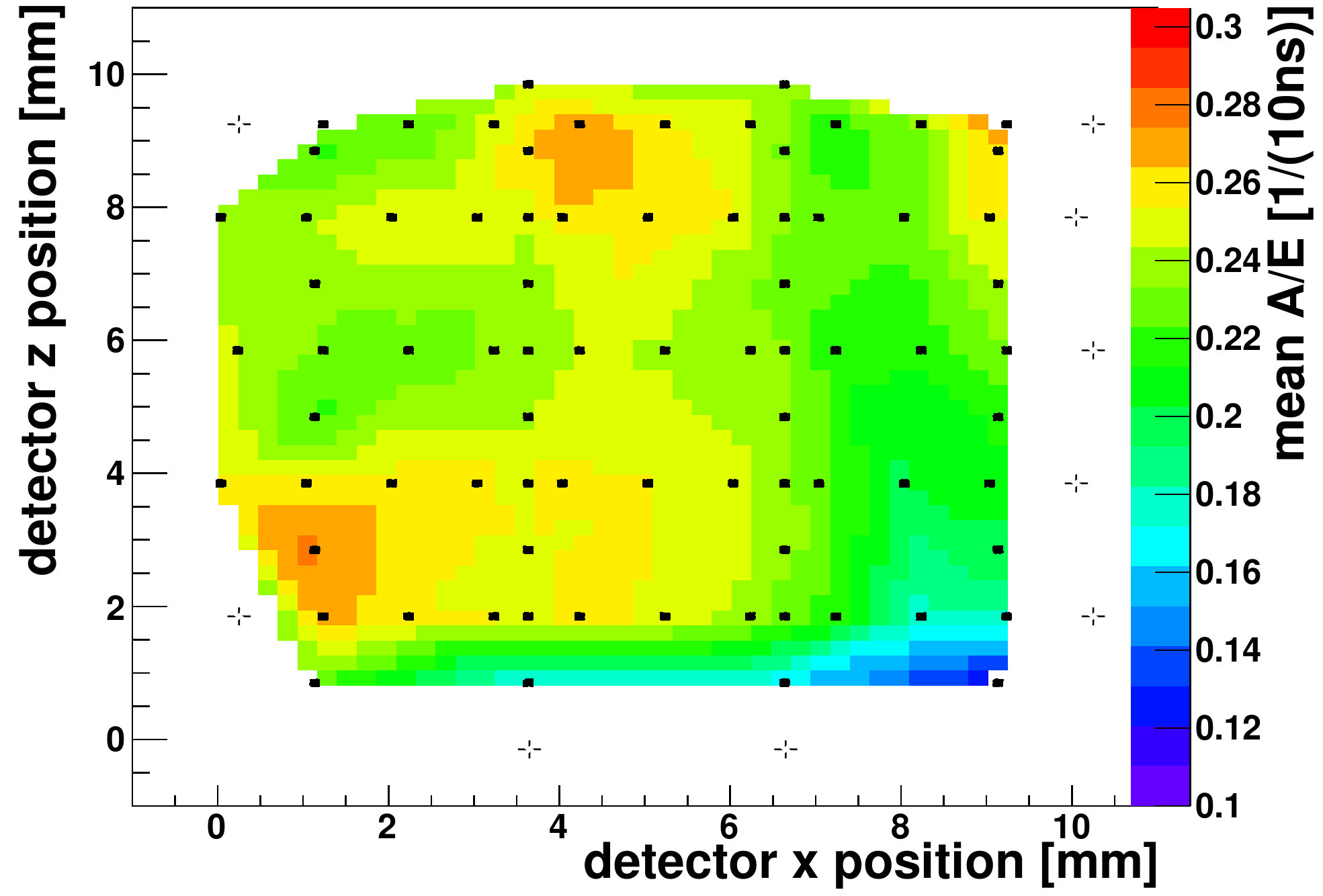}
   \includegraphics[width=0.49\textwidth]{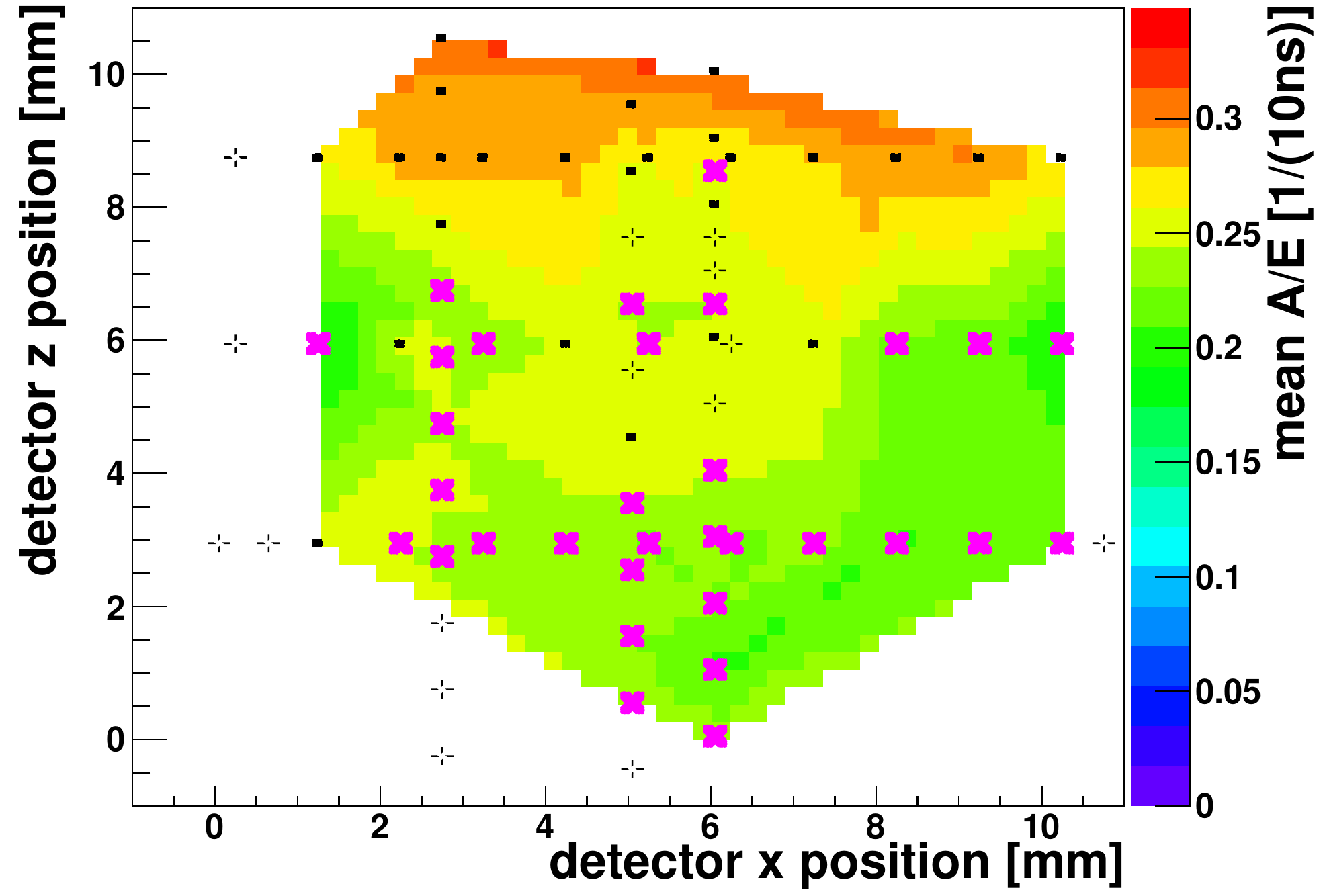}
    \begin{minipage}{0.45\textwidth} 
        \caption[Map of A/E distributions]
                {Map of the A/E distributions. Same detector sides as shown for DIP and ERT before. 
		Top left: Detector 667615-02 side 1 CA. Top right: 631972-06 side 1 CA. Bottom: 631972-06 side 3 NCA. Top left and bottom showed the characteristic peak-and-valley structure in x-direction for ERT and DIP.}
        \label{fig_alpha_scanning_A_over_E}
    \end{minipage}
       \hfill
    \begin{minipage}{0.49\textwidth}
    \includegraphics[width=\textwidth]{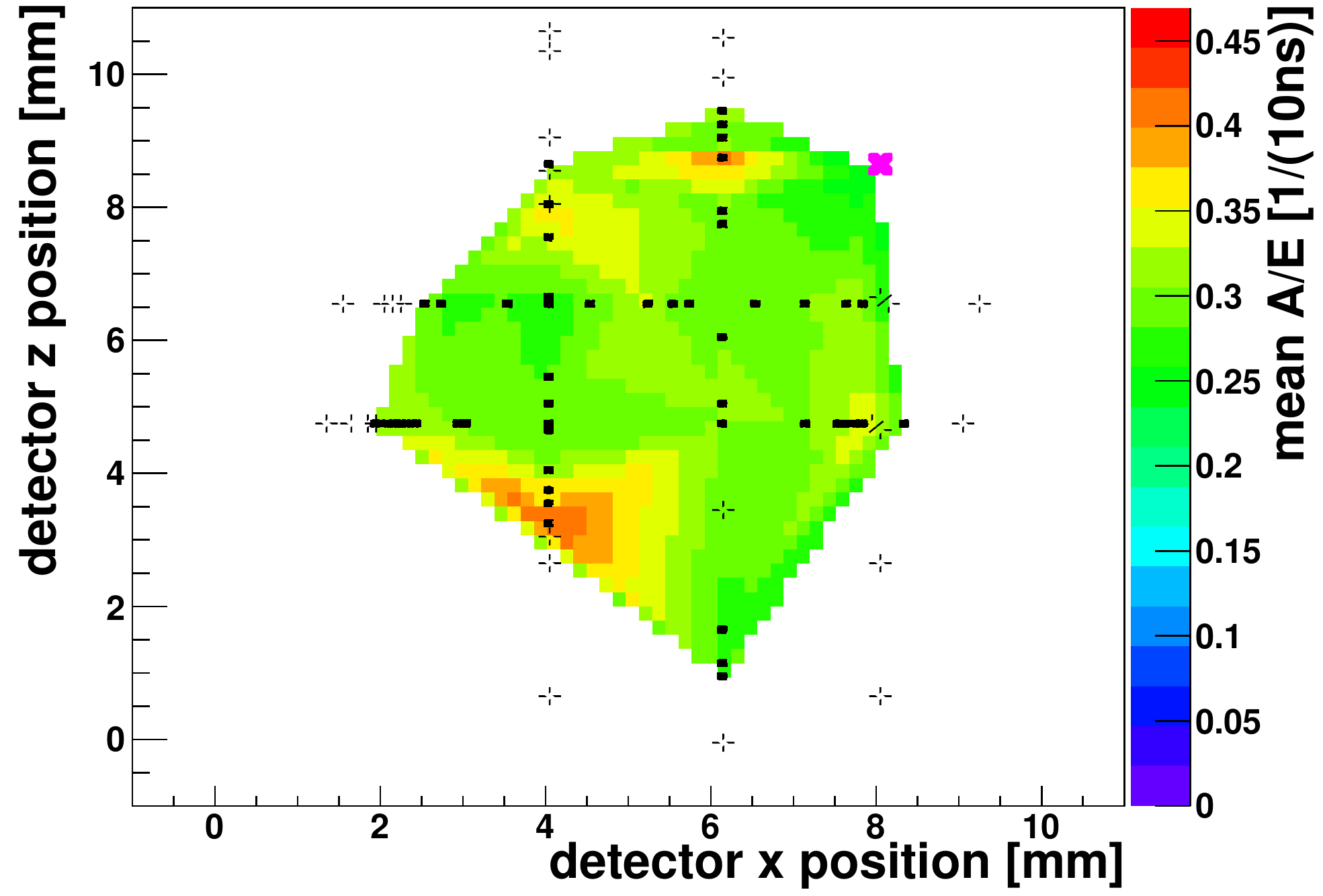}
    \end{minipage}
\end{figure}

\paragraph{A/E}    
For the same measurements that have been shown for DIP and ERT, the analog distributions for the quantity A/E are shown in \autoref{fig_alpha_scanning_A_over_E}.
Note, that A/E shows a behavior complementary to LSE: central events have high values, surface events low values. 
As a consequence, the distribution of the A/E values can be somehow inverted to the ones shown before.
Although some scanning results seem to show a characteristic variation, 
no general behavior in the distributions can be identified concerning the variations in z- or x-directions, as seen partly for the LSE distributions.
The variations differ for each detector surface configuration.

%%%%%%%%%%%%%%%%%%%%%%%%%%%%%%%%%%%%%%%%%%%%%%%%%%%%%%%%%%%%%%%%%%%%%%%%%%%%%%%%%%%%%%%%%%%%%%%%%%%%%%%%%%%%%%%%%%%%%%%%%%%%
\FloatBarrier
\subsubsection{Cathode-scanning}
\label{sec_cathode_scanning}

\begin{SCfigure}[][b!]
 \centering
  \includegraphics[width=0.69\textwidth]{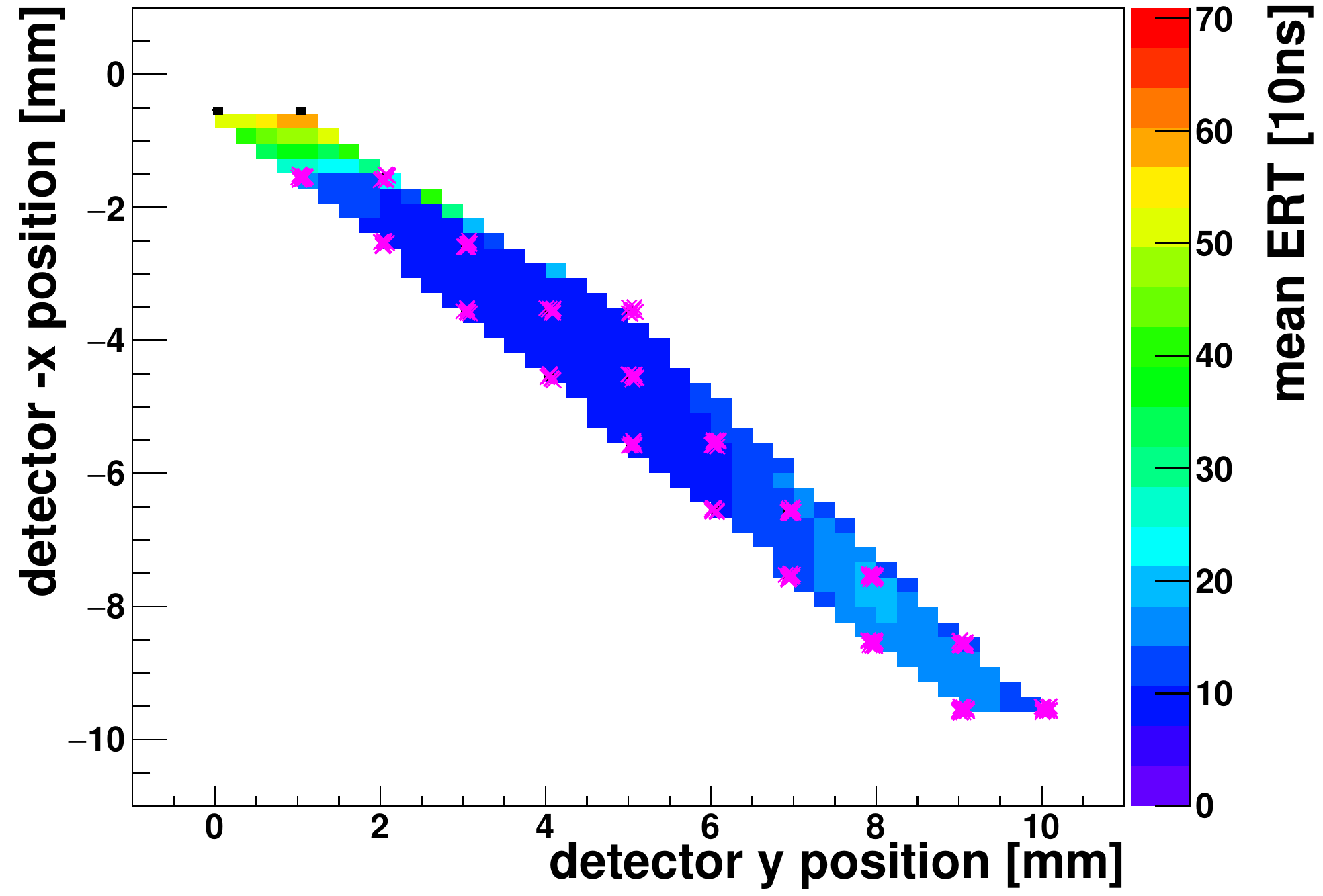}
  \caption[Cathode scanning at detector 631972-06]
          {Cathode scanning at detector 631972-06. Interpolated mean values as a function of their position. All points except two are marked as being below surface events thresholds, which is desired for such central events.}
  \label{fig_cathode_scanning_631972_06}
\end{SCfigure}

Surface events at the cathode are not supposed to be lateral surface events, but central events.
Hence, no lateral surface tagging is desired.
Dedicated scanning measurements diagonally across the whole cathode side of detector 631972-06 were performed to investigate the homogeneity of these central events, and hence of the weighting potential.
The results in \autoref{fig_cathode_scanning_631972_06} exemplarily for the quantity ERT show central events character over a wide range, except for the outermost area.
Here, the values indicate a lateral surface event character. The distribution for DIP and A/E show qualitatively a similar behavior.
This indicates, that the weighting potential has a very homogeneous level across most of the whole surface.\\  
    
\begin{figure}[]
 \centering
  \includegraphics[width=.29\textwidth]{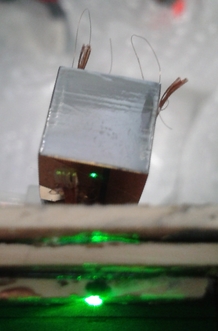}
  \includegraphics[width=.69\textwidth]{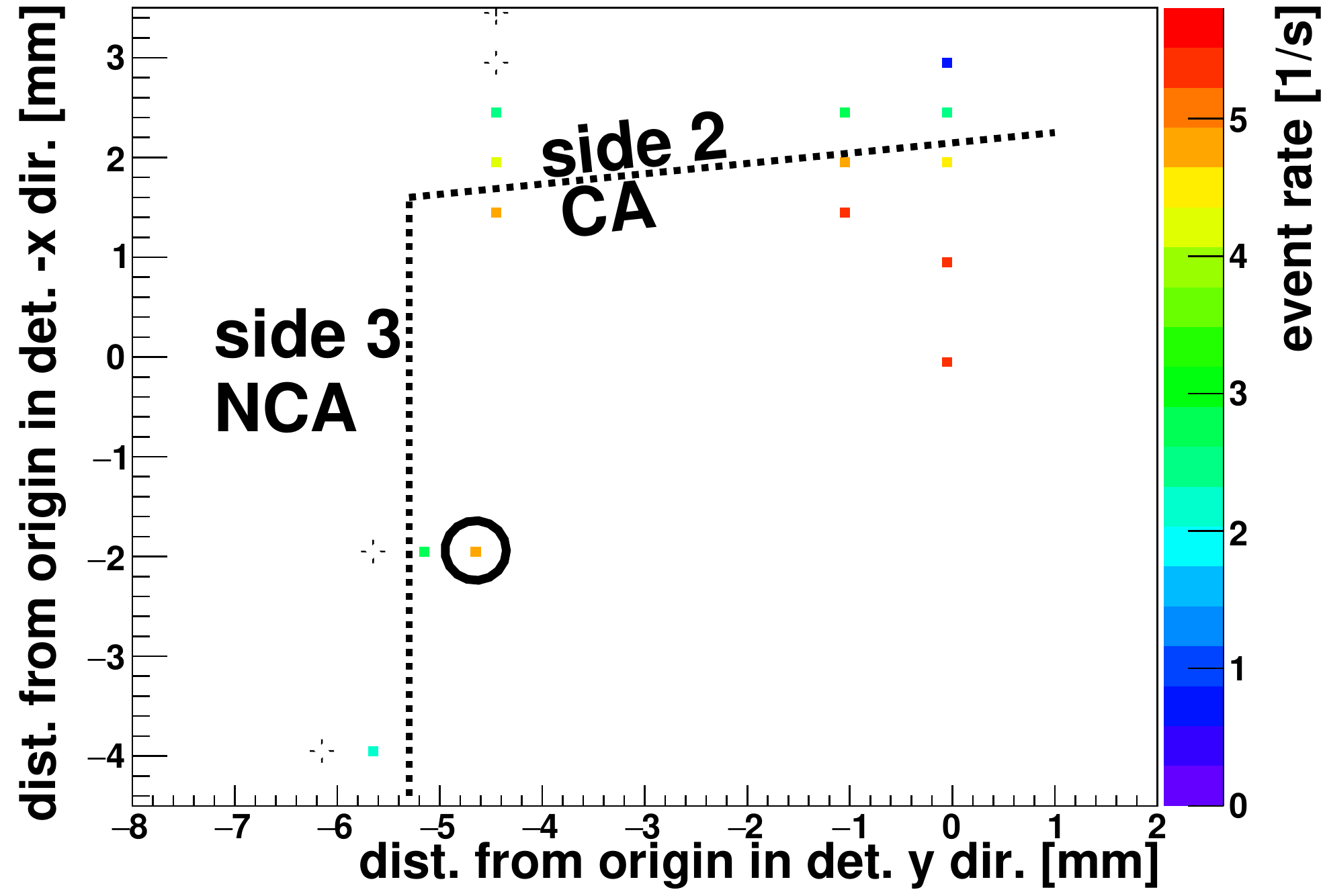}
  \captionsetup{width=0.9\textwidth} 
  \caption[Setup of cathode-scanning]
          {Left: Setup for cathode-scanning. The green laser point shows the position of the origin of coordinates used here, the collimator is at the bottom of the picture. 
	  Right: Positions of the center of the scanning points. The black lines indicate the detector boundaries. Note that the detector is tilted and rotated, and that the beam has a diameter of approximately \SI{1}{mm}. The point marked with a black circle suffered from the charge-sharing distortions and is discussed in \autoref{sec_hyperbolic}. }
  \label{fig_cathode_scan_setup}
\end{figure}

To test the central and potential lateral surface events character at the cathode especially towards the lateral sides, detector 667615-02 was investigated at the cathode towards side 2 and side 3.
In that particular setting, at the surfaces one and two, the CA was the outermost anode rail, for surface three the NCA.
The detector is tilted and rotated by approximately \SI{10}{\degree} each, so that possible overshooting radiation does not hit the lateral sides.  
The beam of alpha radiation has a diameter of about \SI{1}{mm}, which is larger than the colored box in the graphical representation.
Due to this, even when the collimator aims just past to the detector edge, and hence the center of the beam does not hit the detector anymore, 
some parts of the beam still hit the surface.
See \autoref{fig_cathode_scan_setup} for a picture of the setup, where the relative system of coordinates for this particular measurement is shown. Furthermore, the result for the measured event-rate is given, which clearly shows the positions where the beam does not hit the detector completely.
Due to the tilted and rotated detector, the beam is not perpendicular to the detector surface anymore. Additionally, the scales of the collimator and the detector differ. But as this difference is proportional to $\cos(\SI{10}{\degree}) \approx \SI{2}{\%}$, this is ignored here.
Furthermore, the detector edges do not form a right angle in the two-dimensional projection anymore.

\begin{SCfigure}
 \centering
  \includegraphics[width=0.629\textwidth]{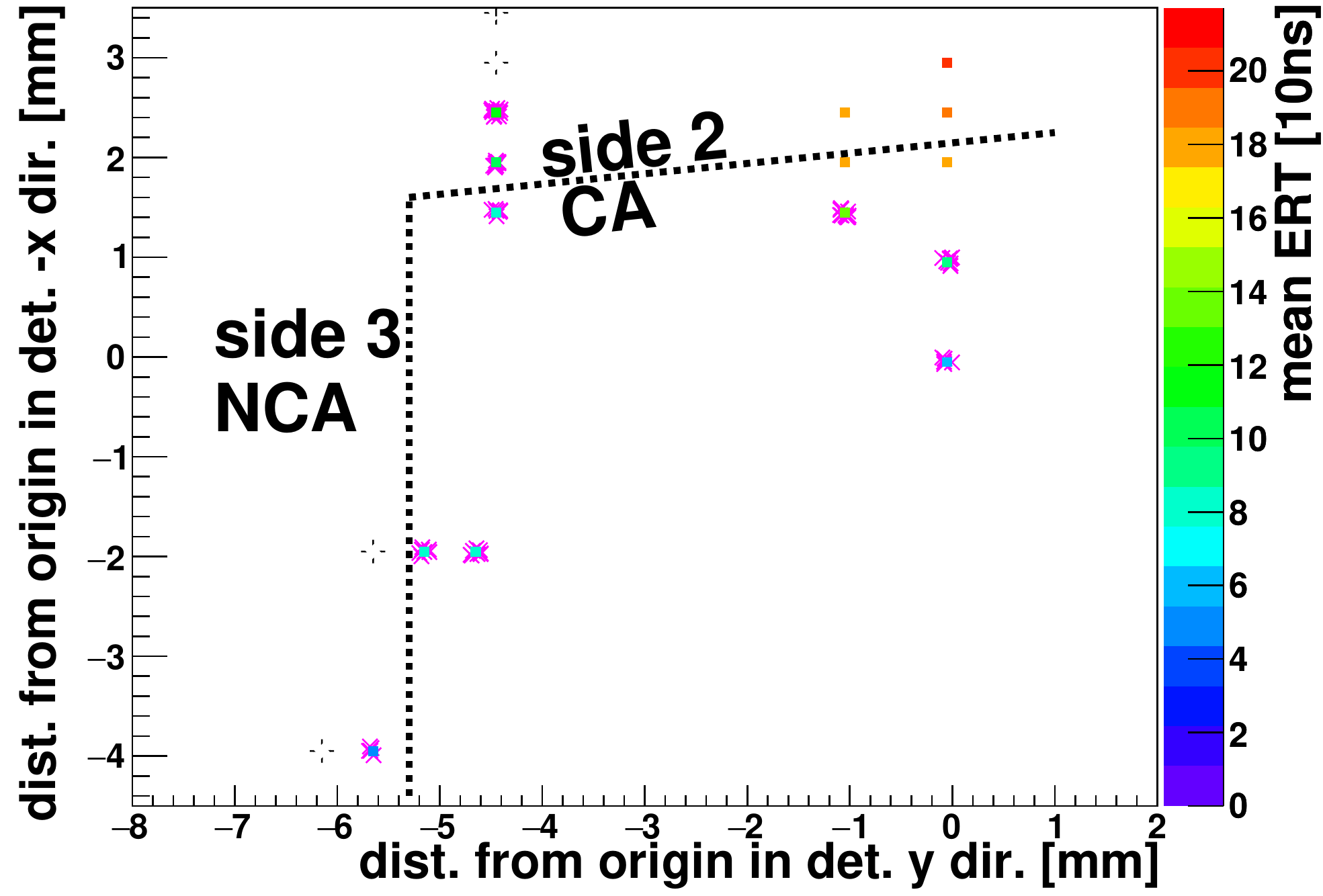}
  \caption[Surface event cuts at cathode]
          {Surface event cuts at the cathode. At the top edge, side two in CA bias is adjacent, at the left edge, side three on NCA bias.}
  \label{fig_cathode_scanning_surface_events}
\end{SCfigure}

The corners of the surfaces two and three were tested for LSE and A/E characteristics, shown exemplarily for the quantity ERT in \autoref{fig_cathode_scanning_surface_events}:
The ERT algorithm tags five events. These are towards the edge where the surface two at CA bias is applied.
The tagged positions are close to the edge, the center of the beam is about \SI{0.7}{mm} and \SI{1.5}{mm} distant from the edge. 
The quantity DIP tags one event. Interestingly, this is also at side two and not at side three, where the NCA bias is applied. 
A/E does not tag any events.
From these measurements one can conclude that the events at the cathode are influenced by a homogeneous weighting potential over a wide range of the surface, which results in a constant behavior typically for central events.
However, at the edges, some events can show lateral surface character, even up to a distance of about a millimeter from the edge.
The quantity A/E performs slightly better here than the LSE mechanism.

%%%%%%%%%%%%%%%%%%%%%%%%%%%%%%%%%%%%%%%%%%%%%%%%%%%%%%%%%%%%%%%%%%%%%%%%%%%%%%%%%%%%%%%%%%%%%%%%%%%%%%%%%%%%%%%%%%%%%%%%%%%%
\FloatBarrier
\subsubsection{Validation of interaction depth calculation}
\begin{SCfigure}
 \centering
  \includegraphics[width=.6\textwidth]{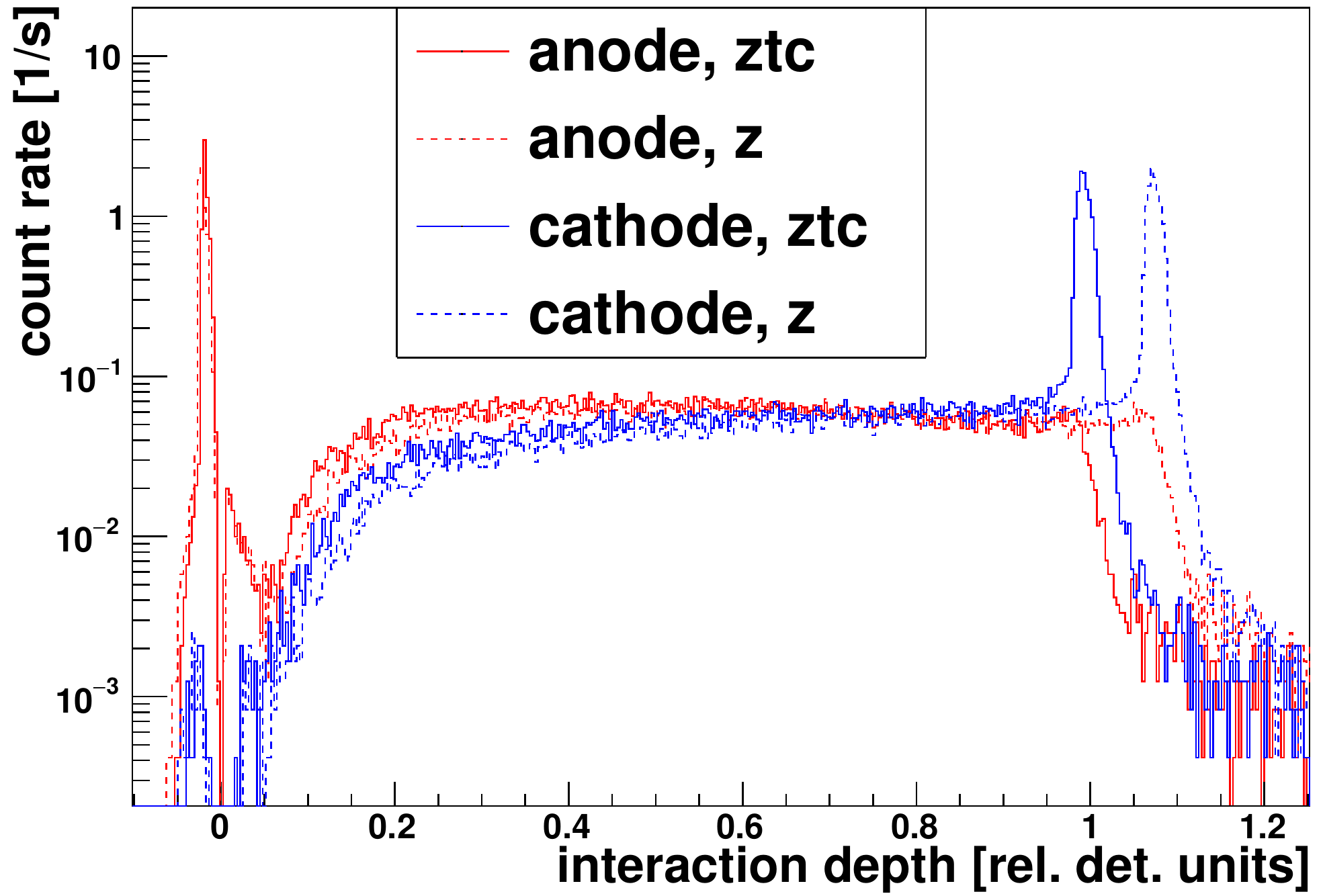}
  \caption[Interaction depth calculation for alpha radiation to anode and cathode]
          {Interaction depth calculation for alpha radiation to anode and cathode surfaces. Red: irradiation of anodes. Blue: irradiation of cathode.}
  \label{fig_interaction_depth_alphas_anode_cathode}
\end{SCfigure}

\begin{figure}[t!]
        \includegraphics[width=0.49\textwidth]{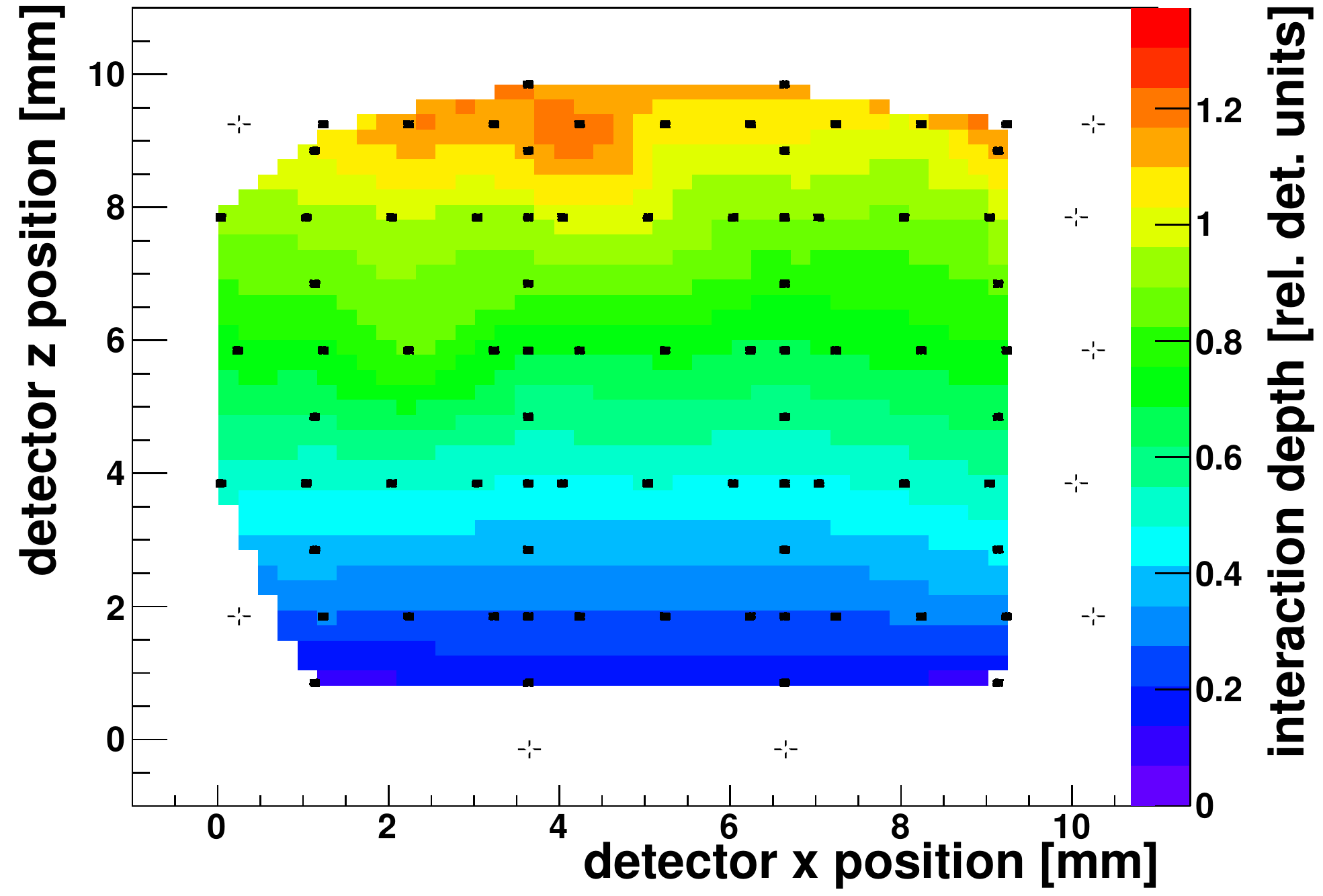}
        \includegraphics[width=0.49\textwidth]{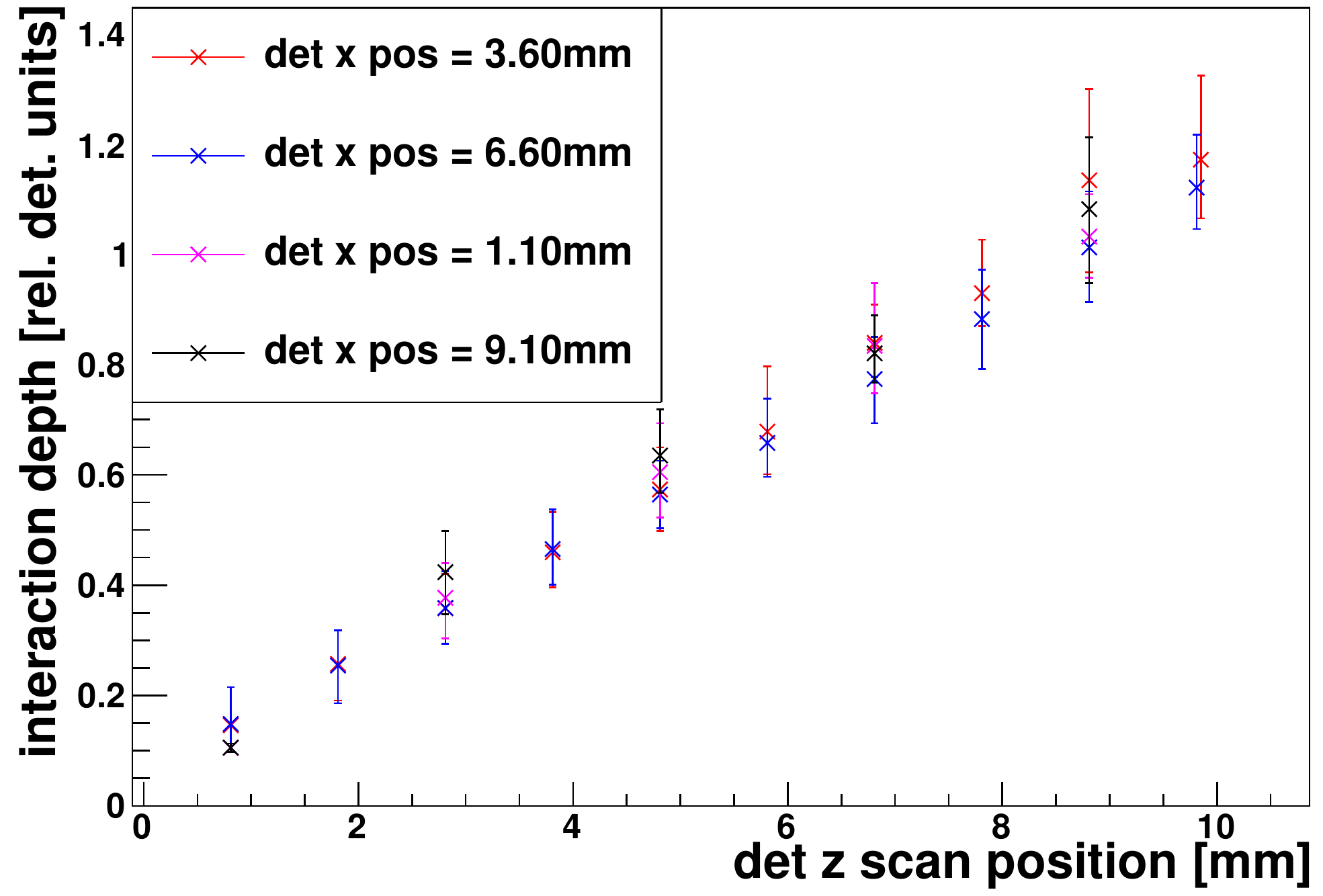}
          \caption[Interaction depth calculation of detector 667615-02]
                  {Detector 667615-02 side 1 CA. 
		  Left: Interpolation of the mean values of the interaction depth calculation.
		  Right: One-dimensional representation of the same measurement.
		  The interaction depth calculation shows a linear behavior in z-direction, and almost constant values in x-direction, but values above 1 occur.}
          \label{fig_667615_02_side1_CA_interaction_depth_3D}
\end{figure}

As a first step in validating the interaction depth calculation for (surface) alpha radiation, the reconstruction for anode- and cathode events is tested.    
Furthermore, a comparison between the two models $z$ (\autoref{eq_cal_ipos_z}) and $z_{tc}$ (\autoref{eq_cal_ipos_ztc}) can be done.
The measurement is done by irradiating the cathode and anode surfaces with \isotope[241]Am alpha radiation. 
Anode- and cathode events are central, not lateral surface events.
The interaction depth reconstruction works fine at the anodes, almost without any differences between the two models.
The energy-doubling effect described in \autoref{fig_60keV_Am241_energy_doubling} does not affect the interaction depth calculation.
For the measurement at the cathode, the model including trapping corrections results in values closer to one, see \autoref{fig_interaction_depth_alphas_anode_cathode}.
This behavior has been identified and published in \cite{Fritts:2012xq} for data from the COBRA demonstrator. 
There, the background, which is mainly alpha radiation, accumulates most prominently at the cathode due to electrostatic attraction. 
Furthermore, these events form a sharper peak for the $z_{tc}$ model. 
As shown with the gamma scanning data in \autoref{fig_gamma_scanning_interaction_depth}, no differences between the two models arise in the bulk of the detectors.
Consequently, the model including trapping corrections is used as the standard method to calculate the interaction depth.\\
The interaction depth calculation at the lateral surfaces is tested with the alpha scanning measurements:\\
At detector 667615-02, the interaction depth calculation has a linear behavior in z-direction, and the variation in x-direction is small, as desired. 
However, the values extend to values larger than one, as shown in \autoref{fig_667615_02_side1_CA_interaction_depth_3D}.

\begin{SCfigure}[][b!]
 \centering
  \includegraphics[width=0.69\textwidth]{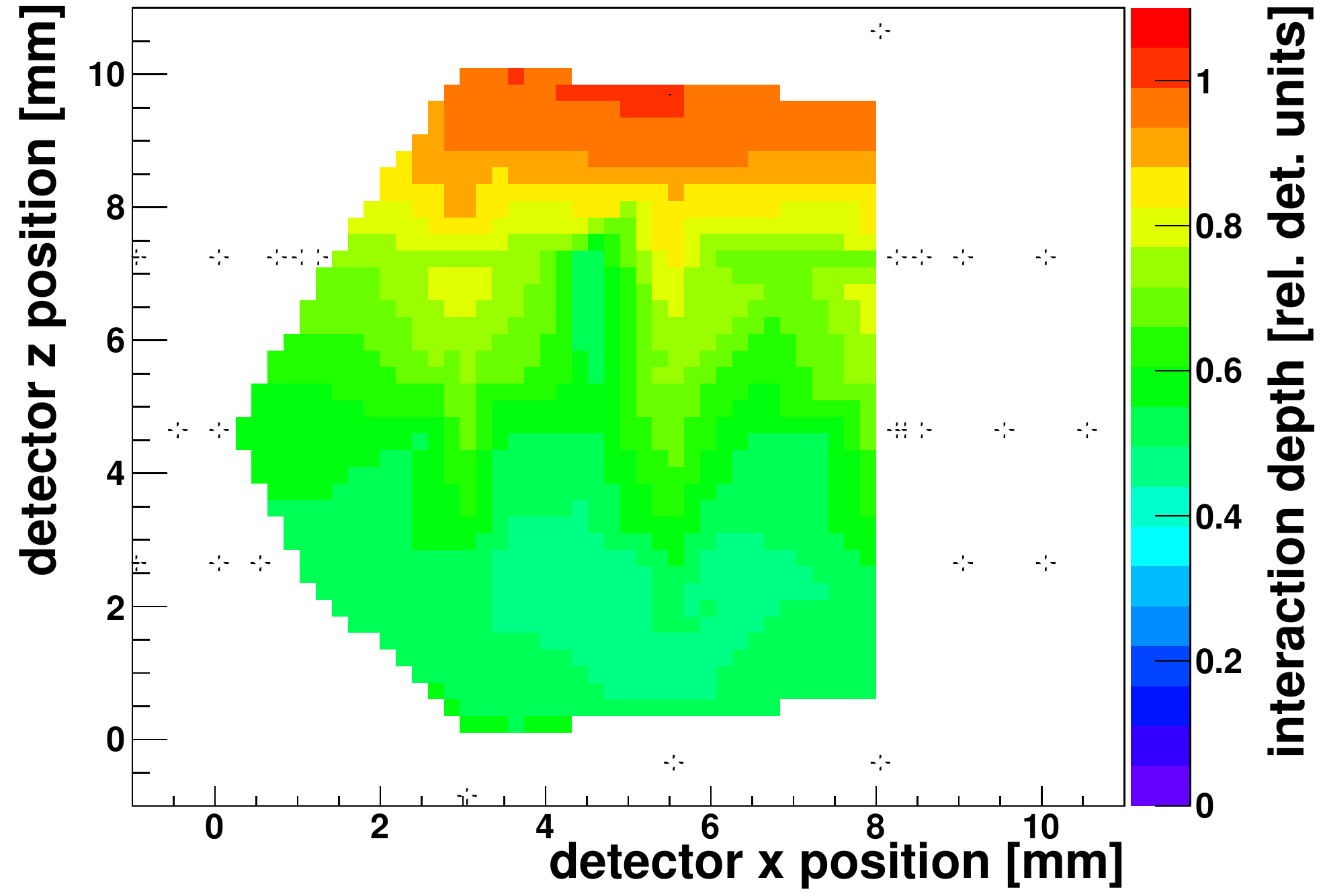}
  \captionsetup{width=0.9\textwidth} 
  \caption[Interaction depth calculation for surface events of detector 631972-06]
          {Detector 631972-06 side 3 CA. The interaction depth calculation is not valid for surface events here. No values below z = 0.4 are reconstructed.}
  \label{fig_631972_06_side3_CA_interaction_depth}
\end{SCfigure}

In contrast, the reconstruction is not working correctly at detector 631972-06. No values below z = 0.4 are calculated, even very close to the anodes, see \autoref{fig_631972_06_side3_CA_interaction_depth}.
This has no severe effects for the analyses, as the events on the lateral surfaces are vetoed by the PSA cuts which are independent of the interaction depth calculation. 
The interaction depth calculation method is used for anode and cathode events where it works fine.

%%%%%%%%%%%%%%%%%%%%%%%%%%%%%%%%%%%%%%%%%%%%%%%%%%%%%%%%%%%%%%%%%%%%%%%%%%%%%%%%%%%%%%%%%%%%%%%%%%%%%%%%%%%%%%%%%%%%%%%%%%%%
\FloatBarrier
\subsection{Conclusion to alpha scanning measurements}    
\label{sec_conclusion_alpha_scanning}    
In general, the surface discrimination methods work well: 
No areas could be identified, where the surface event discrimination showed systematic insensitivities. 
These were assumed to be at the sides, especially where CA and NCA meet. 
No hints were found on how or where the surface discrimination method could be improved.

At all detectors, local variations of the surface events characteristics were measured:
At some detector surfaces, the LSE values show a decline in z-direction.
Furthermore, a peak-and-valley variation in x-direction could also be seen sometimes.
The values are lower in the center, and higher towards the edges, although the outermost measured values decline sometimes. 
This could indicate a cancellation of the opposing ERT and DIP characteristics.
Furthermore, the edges where the transition to the other bias happens (CA to NCA or vice versa), show sometimes even higher values than the edges without bias transition.
This behavior is plausible, as the LSE method relies on a distorted weighting potential at the surfaces. Especially at the edges, the weighting potential can be even more distorted.\\
However, this behavior (qualitatively) could not be identified in general, the effects do not appear at all detectors.\\
The same is true for the A/E distributions: No general position-depending effects could be identified.\\
Additionally, only the z-position of the events can be determined, and even this can suffer from a distorted calculation for surface events. 
As a consequence, the variations of the values at the surfaces cannot be used to improve the surface events discrimination by making it position-dependent.

The scanning measurements at the cathode show constant values typical for central events across the whole cathode surface, as expected. 
Only towards the edges, some scanning points start to show surface events character.

The interaction depth calculation for alpha radiation is valid for (central) events at the anode and cathode. 
At the lateral surfaces, it is not always valid. It can be checked for each detector individually. 
But as the veto of lateral surface events is not based on the interaction depth calculation, this has no impact for the analysis methods of COBRA.

The complete results of more than \num{600} alpha scanning measurements concerning the surface events discrimination methods LSE and A/E are shown in \autoref{tab_surface_events_efficiency_results_compilation}.
For each detector side that was investigated using alpha scanning measurements, it shows the number of scanning points, and the number of alpha peaks that could be identified, as discussed at the beginning of \autoref{sec_analysis_principle}. 
Furthermore, the ratios of the numbers of scanning points, whose values were tagged correctly as surface events, and those where the tagging was not successful, are given absolute and relative for each of the surface events discrimination methods.\\

Concluding, one can state the A/E criterion tags more lateral surface events than LSE for the detectors 667615-02 and F15, but not for detector 631972-06.
The use of the A/E criterion should be investigated further, especially for the detectors of the COBRA demonstrator.\\

\begin{table}[t!]
  \begin{tabular}{|c|c|c|c|c|c|}
    \hline
    \multirow{2}{*}{detector}   & \multirow{2}{*}{side}       & \multirow{2}{*}{\makecell{found\\alpha peak}} & \multirow{2}{*}{\makecell{surface events\\discrimination method}} & \multicolumn{2}{c|}{tagged events} \\ 
                                &                             &                         &   & absolute & [\%] \\ \hline
    \multirow{5}{*}{667615-02}  & \multirow{2}{*}{side 1 CA}  & \multirow{2}{*}{80/90}  &  A/E & 80/80 & 100 \\ %\hline
                                &                             &                         &  ERT & 74/80 & 93  \\  \cline{2-6}
                                & \multirow{3}{*}{cathode}    & \multirow{3}{*}{18/23}  &  A/E & 0/18  & 0   \\ %\hline
                                &                             &                         &  ERT & 5/18  & 28  \\ %\hline
                                &                             &                         &  DIP & 0/18  & 0   \\ \hline
    \multirow{5}{*}{F15}        & \multirow{2}{*}{side 2 CA}  & \multirow{2}{*}{59/67}  &  A/E & 59/59 & 100 \\ %\hline
                                &                             &                         &  ERT & 55/59 & 93  \\ \cline{2-6}
                                & \multirow{3}{*}{side 3 CA}  & \multirow{3}{*}{60/65}  &  A/E & 60/60 & 100 \\ %\hline
                                &                             &                         &  ERT & 57/60 & 95  \\ \hline %\cline{2-6}
    \multirow{14}{*}{631972-06} & \multirow{2}{*}{side 1 CA}  & \multirow{2}{*}{57/75}  &  A/E & 32/57 & 56  \\ %\hline
                                &                             &                         &  ERT & 52/57 & 91  \\ \cline{2-6}
                                & \multirow{2}{*}{side 1 NCA} & \multirow{2}{*}{26/73}  &  A/E & 11/26 & 42  \\ %\hline
                                &                             &                         &  DIP & 18/26 & 69  \\ \cline{2-6}
                                & \multirow{2}{*}{side 3 CA}  & \multirow{2}{*}{66/99}  &  A/E & 26/66 & 39  \\ %\hline
                                &                             &                         &  ERT & 39/66 & 59  \\ \cline{2-6}
                                & \multirow{2}{*}{side 3 NCA} & \multirow{2}{*}{57/90}  &  A/E & 1/57  & 2  \\ %\hline
                                &                             &                         &  DIP & 47/57 & 82   \\ \cline{2-6}
                                & \multirow{3}{*}{cathode}    & \multirow{3}{*}{21/21}  &  A/E & 0/21  & 0   \\ 
                                &                             &                         &  ERT & 2/21  & 10  \\ 
                                &                             &                         &  DIP & 0/21  & 0   \\ \cline{2-6}
                                & \multirow{3}{*}{anode}      & \multirow{3}{*}{6/6}    &  A/E & 0/6   & 0   \\ 
                                &                             &                         &  ERT & 0/6   & 0   \\ 
                                &                             &                         &  DIP & 0/6   & 0   \\ \hline
  \end{tabular}
  \captionsetup{width=0.9\textwidth} 
  \caption[Results of alpha scanning surface events discrimination methods]
          {Results of the alpha scanning surface events discrimination methods. As DIP and ERT are designed specifically for NCA and CA surfaces, their values are only shown for that particular surface. Note that cathode and anode events are central events, so low tagging-ratios are desired. }
  \label{tab_surface_events_efficiency_results_compilation}
\end{table}

%%%%%%%%%%%%%%%%%%%%%%%%%%%%%%%%%%%%%%%%%%%%%%%%%%%%%%%%%%%%%%%%%%%%%%%%%%%%%%%%%%%%%%%%%%%%%%%%%%%%%%%%%%%%%%%%%%%%%%%%%%%%
\FloatBarrier
\subsection{Investigating charge-sharing reconstruction artifacts}    
\label{sec_hyperbolic}
An effect already seen in the author's diploma thesis is a wrong event-reconstruction that can appear more or less distinctive, called charge-sharing.

\begin{figure}[]
 \centering
  \includegraphics[width=0.49\textwidth]{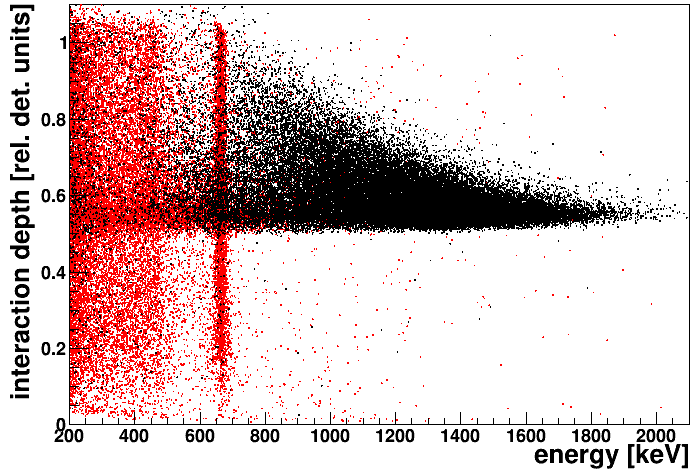}
  \includegraphics[width=0.49\textwidth]{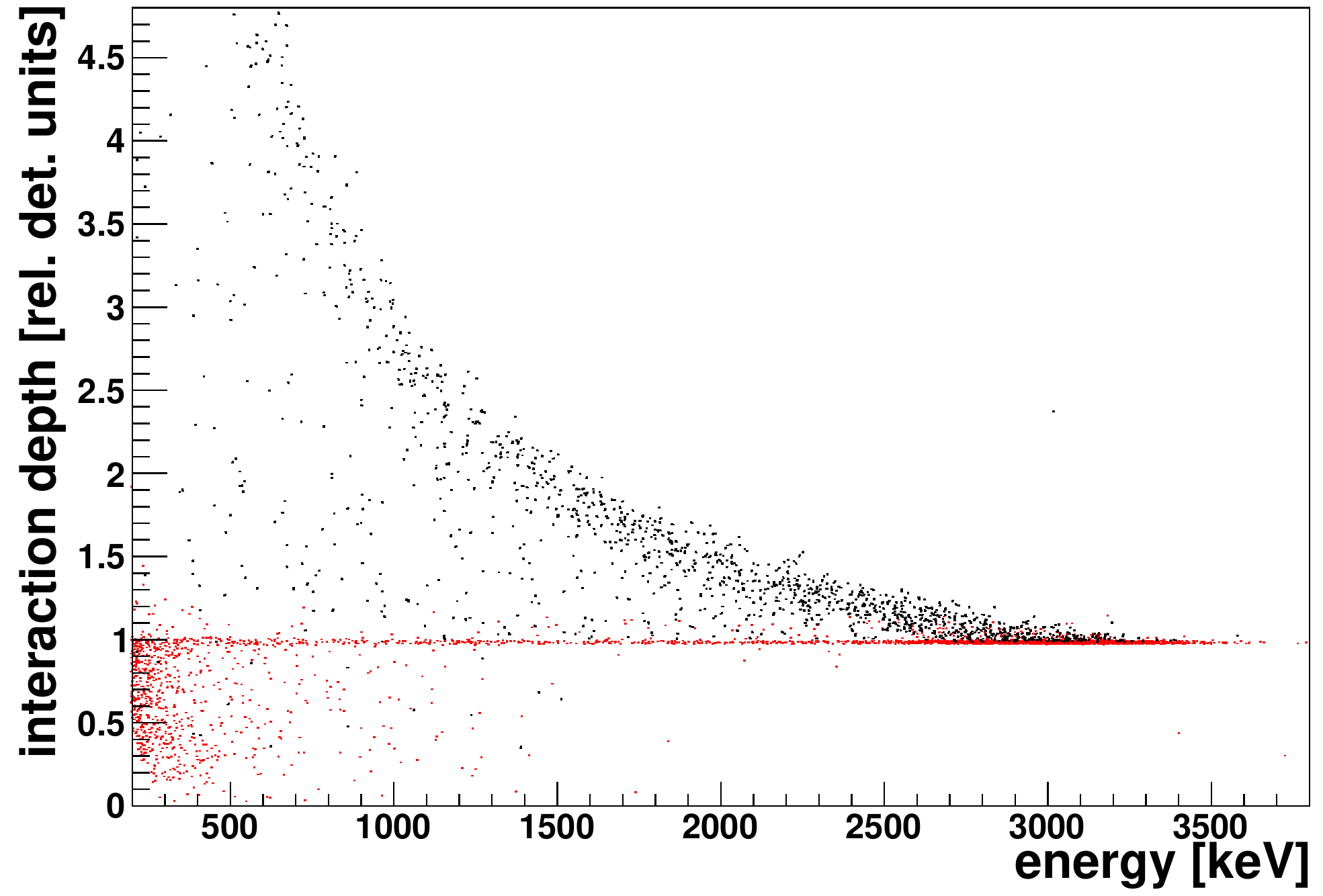}
  \captionsetup{width=0.9\textwidth} 
  \caption[Discrimination of charge-sharing events by DIP]
          {Discrimination of charge-sharing events by DIP. The different colors can be ignored for now, these are discussed on page \pageref{position_black_and_right_colors}. Left: \isotope[137]Cs gamma source and \isotope[241]Am alpha source directed to a lateral surface. Right: Alpha source irradiating the cathode.}
  \label{fig_hyperbolic_rdip_veto}
\end{figure}

The charge-sharing leads to a hyperbolic shape in the distribution of calculated interaction depth. 
Exemplary results are shown here for an alpha scanning measurement of a lateral surface on NCA potential, and one at the cathode see \autoref{fig_hyperbolic_rdip_veto}. 
The different colors can be ignored for now, they are discussed on page \pageref{position_black_and_right_colors}.
The left plot shows events of a \isotope[137]Cs gamma source with \SI{662}{keV}, which are distributed homogeneously within detector volume. These central events do not suffer from charge-sharing.
Additional events of a \isotope[241]Am source (at NCA side) form a peak and a hyperbolic shape. 
It contains the events that suffer from charge-sharing. 
The right plot in \autoref{fig_hyperbolic_rdip_veto} shows an alpha scanning measurement to the cathode. 
This is the measurement marked with a black circle in \autoref{fig_cathode_scan_setup}. 
Interestingly, it has both normal and charge-sharing behavior. 
\SI{70}{\%} of all events are reconstructed correctly at the measurement at the cathode, while \SI{30}{\%} are distorted. 
This could indicate, that the cause for this type of distortion is locally strictly limited. The collimated beam with a diameter of \SI{1}{mm} irradiates the area where charge-sharing occurs, as well as neighboring areas which lead to an undistorted reconstruction as well.

\begin{figure}[b!]
 \centering
  \includegraphics[width=0.49\textwidth]{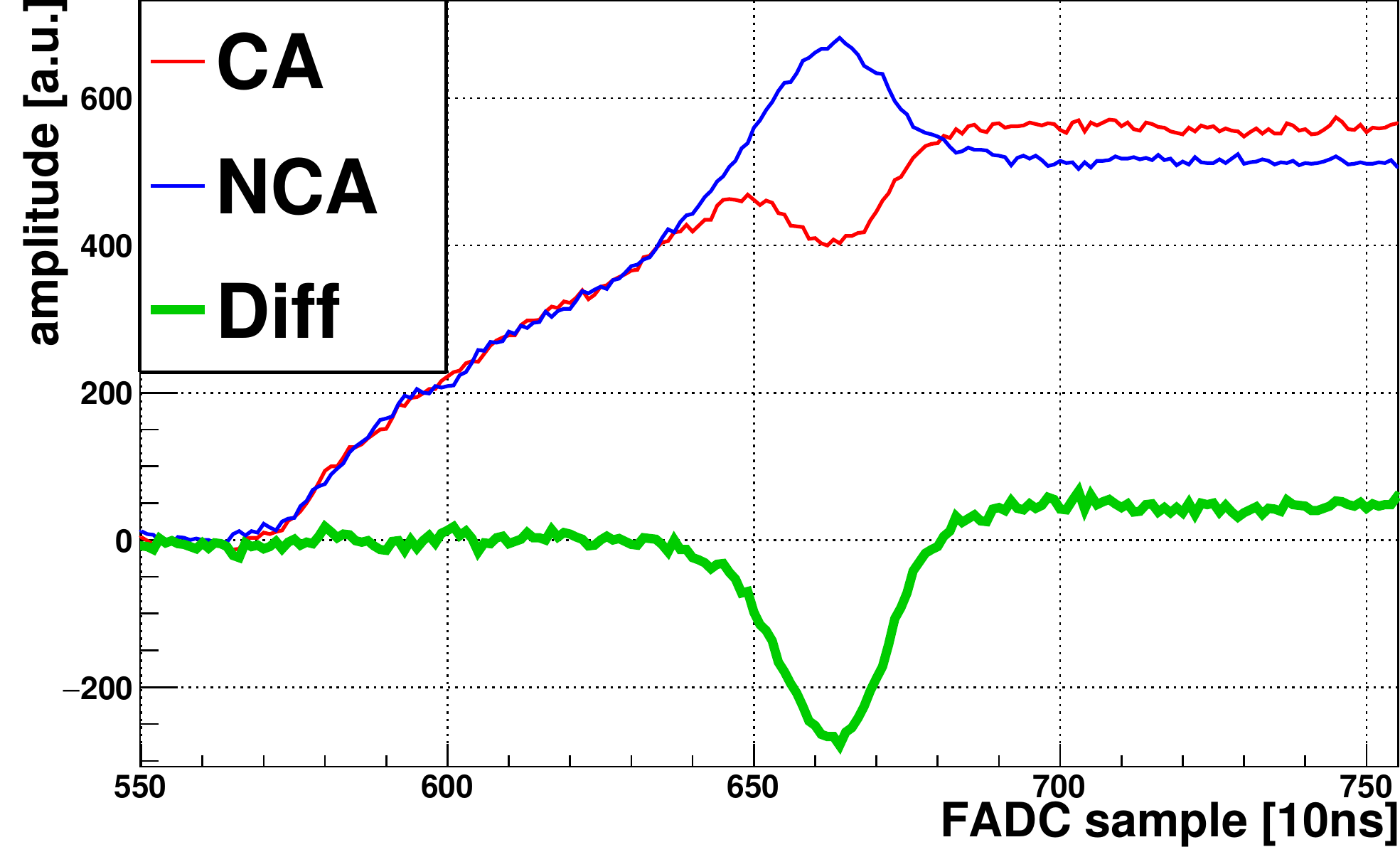}
  \includegraphics[width=0.49\textwidth]{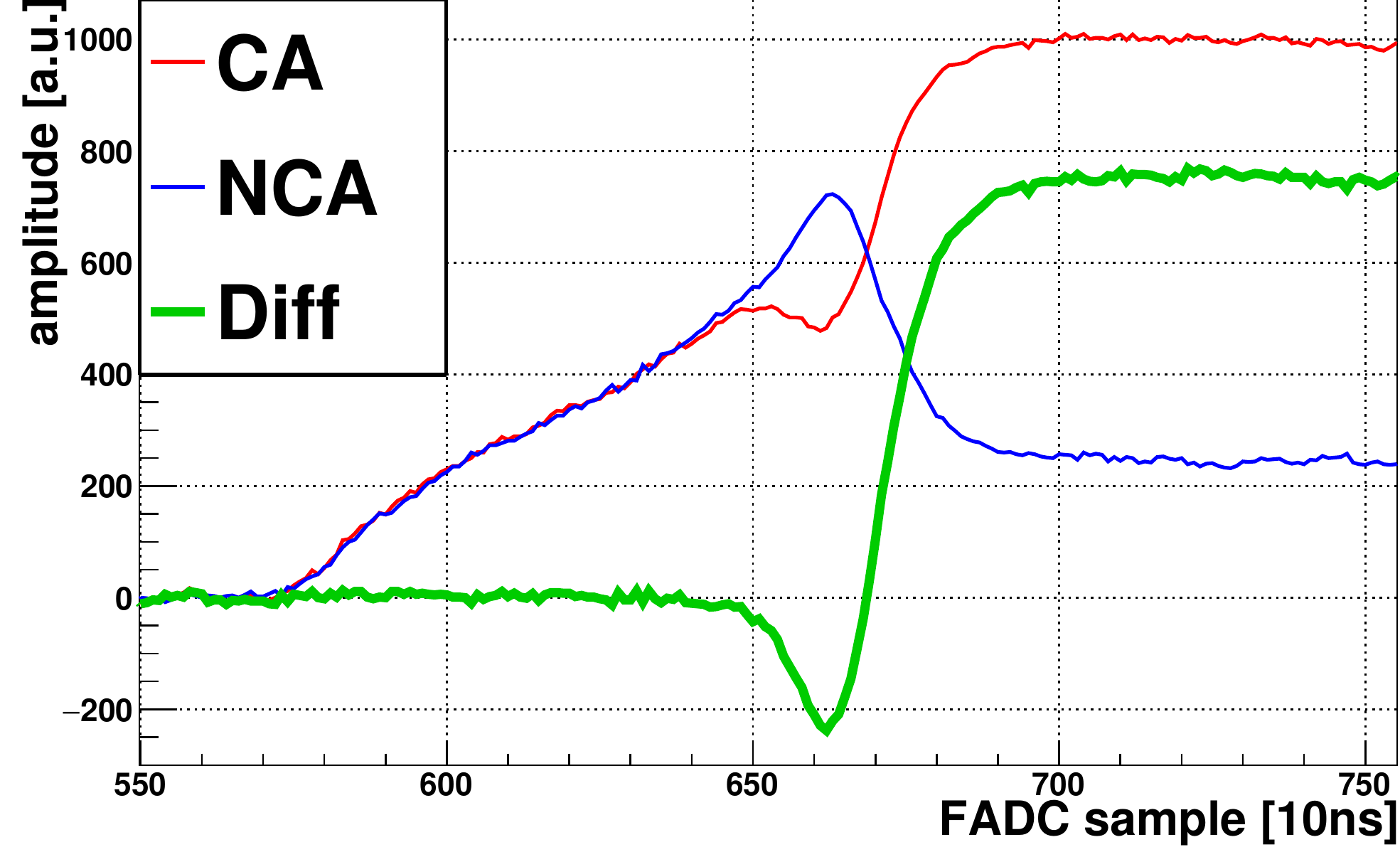}
  \captionsetup{width=0.9\textwidth} 
  \caption[Pulse shapes of charge-sharing events]
          {Pulse shapes of charge-sharing events with a low and high reconstructed energy (left and right).}
  \label{fig_ps_hyperbolic_cath}
\end{figure}

If charge-sharing occurs, the charge cloud of the incident particle is shared between the CA and NCA, so the CPG principle is not fulfilled, and does not work reliably anymore.
Two examples of pulse shapes of events suffering from charge-sharing are shown in \autoref{fig_ps_hyperbolic_cath}.
One event has a low reconstructed energy, the other a higher energy.
However, the energy is probably not reconstructed correctly.
The first event stems from the end of the hyperbolic structure with low energy and high interaction depth, while the second event is situated nearly at the other end of the hyperbola close to the alpha peak.
The signals of CA and NCA have a common rise, but when being influenced by the GB, the distortions occur:
The CA signal shows a little dip, before rising to its full amplitude.
When the dip happens, the NCA signal still rises, before it declines a little. 
The decline is by far not strong enough. In a normal pulse, the NCA signal drops to the baseline or below, depending on the interaction depth of the incident particle, see examples of normal pulse shapes in \autoref{fig_cpg_normal_pulse_shapes}.
Consequently, the difference pulses of charge-sharing events show large DIP values, by which they can easily be discriminated.
This is indicated in \autoref{fig_hyperbolic_rdip_veto} by the black and red colors: 
\label{position_black_and_right_colors}
Events that are tagged by their high DIP value are marked in black, those that are not discriminated in red.
In the left plot, all alpha events are vetoed by the DIP cut, as also the correctly reconstructed events show a high DIP value, as this measurement is done at a NCA surface.
In the right plot, only the charge-sharing events are removed by the DIP cut, as normal events at the cathode do not have a high DIP value.

\begin{figure}[t!]
 \centering
  \includegraphics[width=0.7\textwidth]{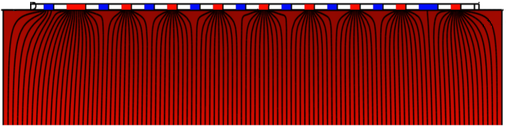}
  \captionsetup{width=0.9\textwidth} 
  \caption[Calculation of electric field lines]
	  {Calculation of the electric field lines at a GB of \SI{-20}{V}. A cross section through the detector center is shown at the near-anode region. The blue and red rectangles indicate the NCA and CA anode rails. Field lines end on the outermost NCA anode, especially on the left where it is the outermost anode at all. Simulation done by J\"urgen Durst, taken from \cite{tebruegge}.}
  \label{fig_field_line_simulation}
\end{figure}

The explanation so far why charge-sharing occurs, is based on calculations of the electric field (done by Jürgen Durst, ECAP Erlangen): A GB is used that is too low, so that there is not enough ''separation power`` between the two anodes. 
Especially when the NCA is the outermost anode rail, some field lines end there. 
This can be seen at a field line calculation of the near-anode region in \autoref{fig_field_line_simulation} for a low GB of \SI{-20}{V}. 
If \SI{-100}{V} is used, no field lines end on an NCA anymore.
Consequently, these distortions occur mostly at the NCA surface, not at the CA.

The hypothesis, that the effect of charge-sharing originates from low GB cannot be confirmed here. 
\begin{SCfigure}[][b!]
 \centering
  \includegraphics[width=0.69\textwidth]{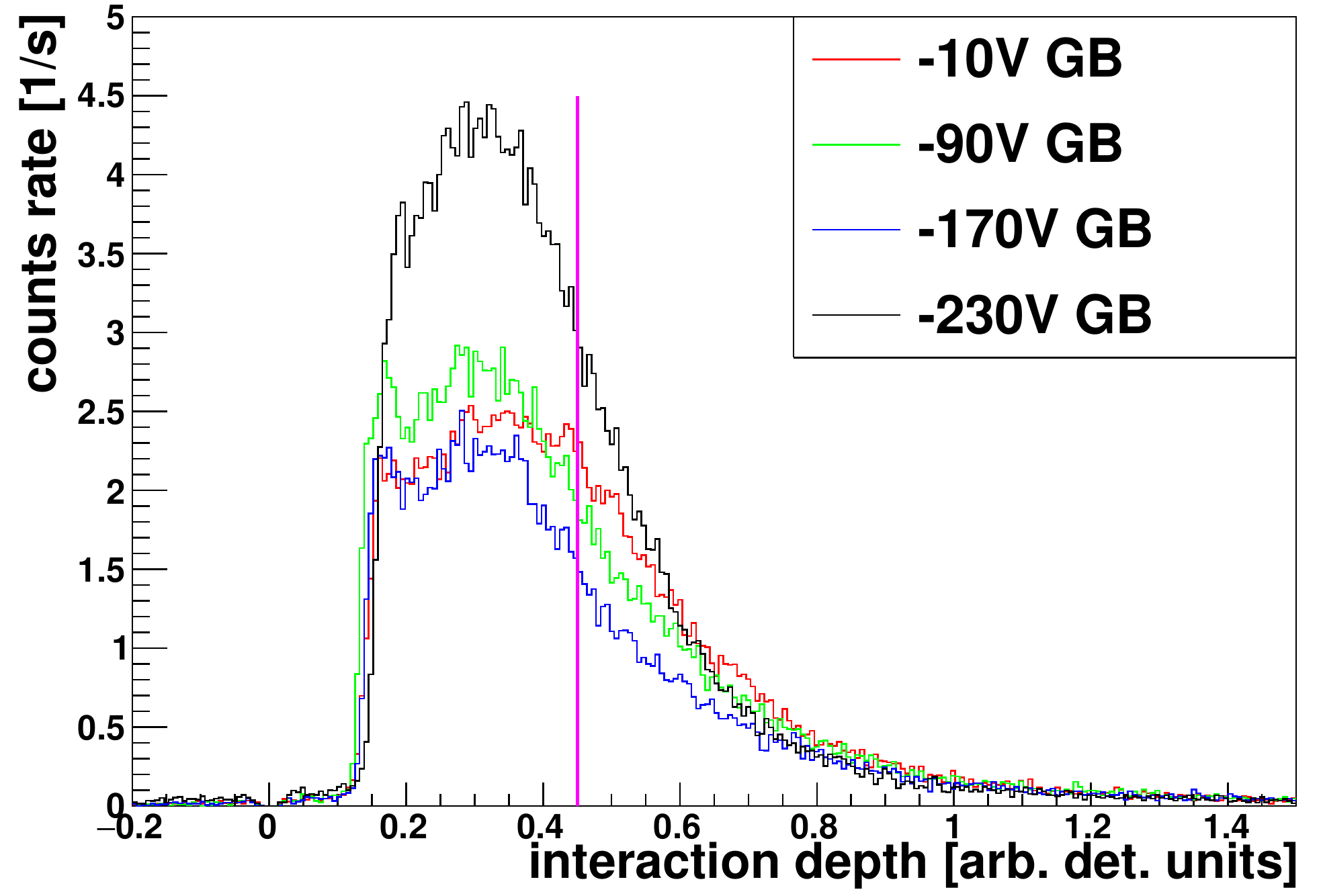}
  \captionsetup{width=0.9\textwidth} 
  \caption[GB dependency of charge-sharing distortions]
          {The charge-sharing distortions are not an effect of too low GB. The magenta line indicates the the region where a non-distorted distribution should decline steeply.}
  \label{fig_hyperbolic_gb_dependency}
\end{SCfigure}
Several measurements with different GB were done, using an alpha source that is slightly collimated to irradiate a spot at the detector surface, see \autoref{fig_hyperbolic_gb_dependency}.
The detector performance varies with different GB, but the charge-sharing remains, independent of the applied GB.
These distortions result in a tailing to a higher interaction depth. 
The magenta line indicates approximately the region where a non-distorted distribution should fall steeply.
This detector is capable of an extremely high GB of up to \SI{-230}{V} before surface leakage currents occur. Other detectors, that are capable of a GB of about \SI{-100}{V} are considered to be good.
At this high GB, the detector seems to have a much better detection efficiency, as the measured count rate is nearly twice as high as at lower GB.\\
    
Concluding, one can state that the charge-sharing effect occurs mostly for surface events if the NCA is the outermost anode rail.
The moving charge cloud is not collected by the CA exclusively, but also by the NCA.
Charge-sharing does not occur because of too low grid bias, at least, it does vanish with reasonable GB values that can be applied to the detector. 
This behavior was discussed in the publication mentioned at the beginning of \autoref{sec_general_surface_sensitivity} (\cite{5075926}) ''\dots the charges generated by incident particles might be driven towards wrong electrodes``.
Interestingly, a cathode scanning point shows normal and charge-sharing behavior at the same time. 
This could be investigated using smaller scanning points.
The main cause for the charge-sharing is still not understood completely. 
But as these distortions happen mostly for alpha events on the detector surfaces, which can be vetoed, this does not have a large impact to the analyses.

%%%%%%%%%%%%%%%%%%%%%%%%%%%%%%%%%%%%%%%%%%%%%%%%%%%%%%%%%%%%%%%%%%%%%%%%%%%%%%%%%%%%%%%%%%%%%%%%%%%%%%%%%%%%%%%%%%%%%%%%%%%%
%%%%%%%%%%%%%%%%%%%%%%%%%%%%%%%%%%%%%%%%%%%%%%%%%%%%%%%%%%%%%%%%%%%%%%%%%%%%%%%%%%%%%%%%%%%%%%%%%%%%%%%%%%%%%%%%%%%%%%%%%%%%
\FloatBarrier
\section[Instrumentation of the guard ring]{Instrumentation of the guard ring}
\label{sec_guard_ring}

\begin{SCfigure}[][b!]
 \centering
  \includegraphics[width=0.4\textwidth]{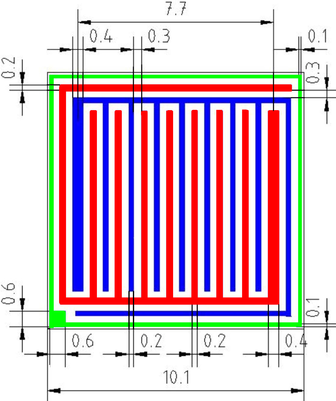}
  \caption[Detailed sketch of anode grid]
          {Detailed sketch of the anode grid, all dimensions in mm. The GR is the green boundary electrode surrounding the NCA and CA (red and blue). The alpha irradiation discussed in this chapter is directed to side 1 (right detector side). Picture taken from \cite{tebruegge}. }
  \label{fig_anode_grid_with_dimensions}
\end{SCfigure}

The COBRA standard detectors have a GR to improve the performance due to a better balanced weighting potential.
At the COBRA demonstrator, the GR is left floating. 
To gain more information on the detected events, the GR could be instrumented, i.e. supplied with a certain bias voltage, and even be read out.\\ 
One idea is to instrument the GR on the potential of the CA, and read out its signals to improve the surface events discrimination power.\\
The GR has a width of \SI{100}{\mu m}, and has some distance to the detector side. The next following anode rail with a thickness of \SI{200}{\mu m} is separated by \SI{300}{\mu m}, see a detailed sketch of the anode surface in \autoref{fig_anode_grid_with_dimensions}. 
As calculated in \autoref{sec_alpha_radiation}, the maximal size of a charge cloud originating from alpha radiation is about \SI{350}{\mu m}. 
Hence, one can clearly state, that the charge clouds of alpha particles from the surface are much closer to the GR and drift towards it, than to the next CA or NCA rail, respectively.
As a consequence, charge clouds of surface events should generate larger signals on the GR than central events.
These induce signals on the GR as well, because of image mirror charges, like they do on NCA and CA.\\
At a standard CPG detector, two outermost anode rails are on CA and two on NCA potential, so there is a difference between the two modes in comparison to the GR. 
A better separation of the charge cloud measurement should be possible in the case that the outermost anode rail is on NCA potential, if the GR is on CA.\\
First laboratory measurements to investigate the GR instrumentation are discussed in this section.

%%%%%%%%%%%%%%%%%%%%%%%%%%%%%%%%%%%%%%%%%%%%%%%%%%%%%%%%%%%%%%%%%%%%%%%%%%%%%%%%%%%%%%%%%%%%%%%%%%%%%%%%%%%%%%%%%%%%%%%%%%%%
\FloatBarrier
\subsection{Measurements with gamma radiation}
\label{sec_guard_ring_gamma}

\begin{figure}   %%   reference detector 
 \centering
  \includegraphics[width=.7\textwidth]{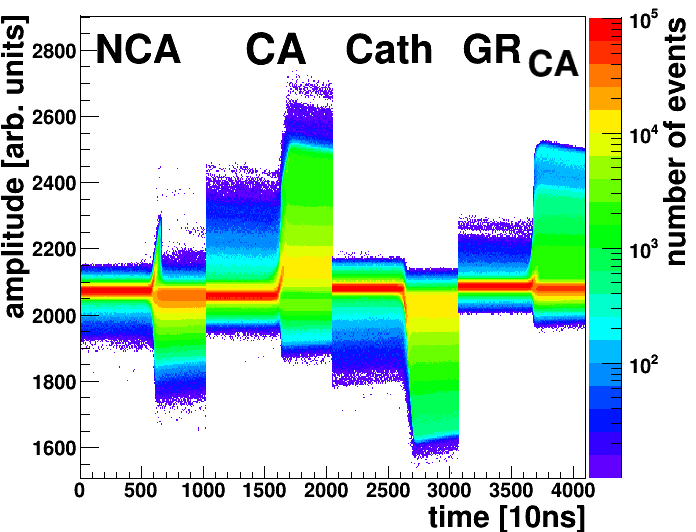}
  \captionsetup{width=0.9\textwidth} 
  \caption[Pulse shapes of gamma radiation with instrumented GR]
	  {Superposition of all pulses of gamma irradiation of the reference detector with GR on CA potential. The four vertical areas of \SI{1024}{bins} each are the four read-out signals, happening at the same time, but plotted successively. So the whole event is \SI{1024}{bins} = \SI{10.24}{\mu s} long, as explained in \autoref{sec_cobra_demonstrator}. The cathode readout in the third area can be ignored here. See \autoref{fig_cpg_normal_pulse_shapes} for a comparison to typical pulse shapes.} 
  \label{fig_guardring_gamma_pulse_shapes}
\end{figure}

At first, the effect of an instrumented GR is tested with \SI{662}{keV} gamma radiation. These central events should not have large GR signals. The pulse shapes of such a measurement are shown in \autoref{fig_guardring_gamma_pulse_shapes}.
The NCA and CA show normal behavior (compare to \autoref{fig_cpg_normal_pulse_shapes}). The GR signals show mostly NCA-like behavior, very seldom CA-like behavior, suppressed by approximately two orders of magnitude.
This can be explained as follows: All electrodes measure induced signals of the drifting charge cloud. Only the electrode, that eventually collects the charge cloud, has a CA-like behavior, all others are NCA-like. As the interactions happen in the whole detector volume, only a small fraction of charge clouds are so close to the surface that they are collected by the GR.

\begin{figure}
 \centering
  \includegraphics[width=0.49\textwidth]{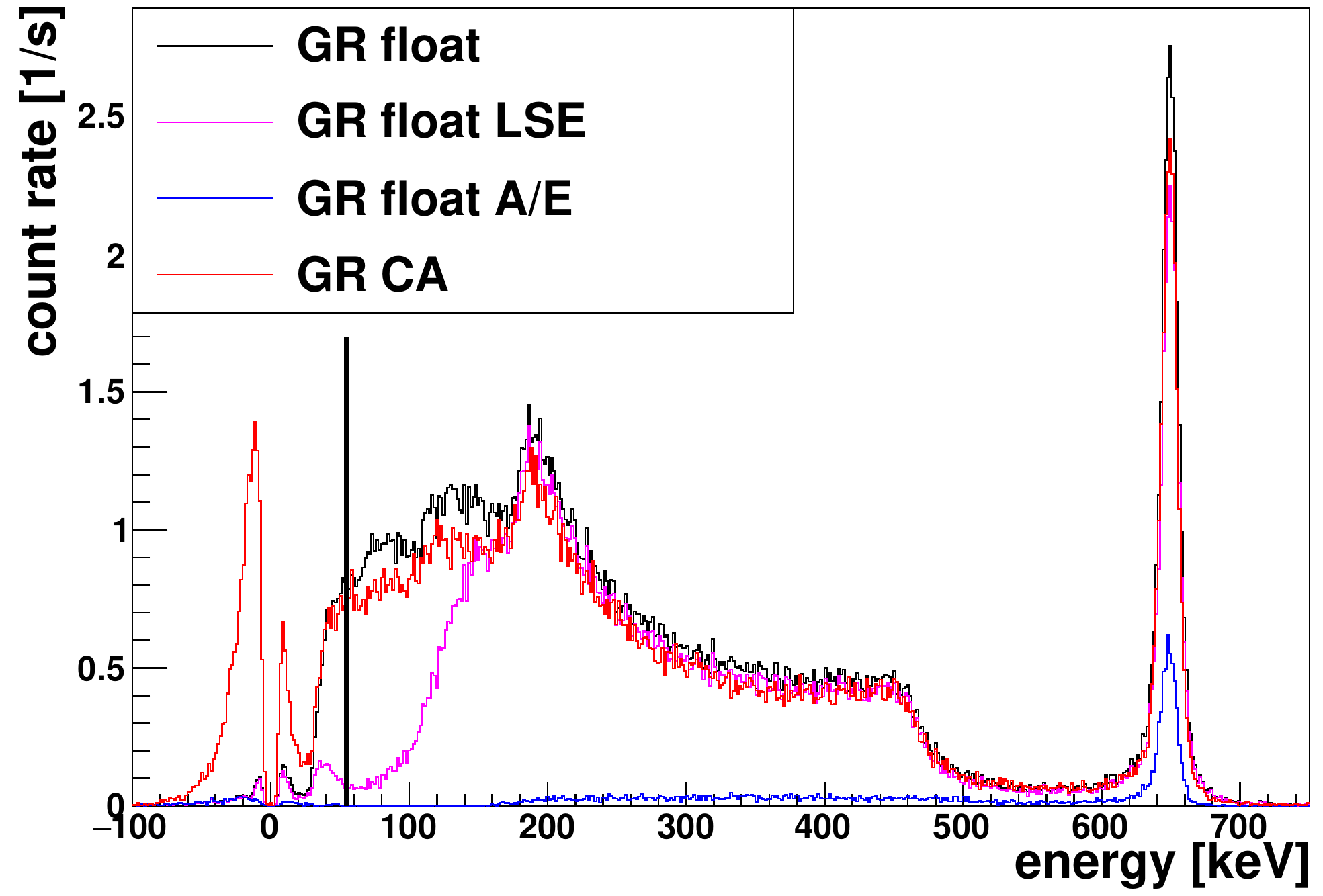}
  \includegraphics[width=0.49\textwidth]{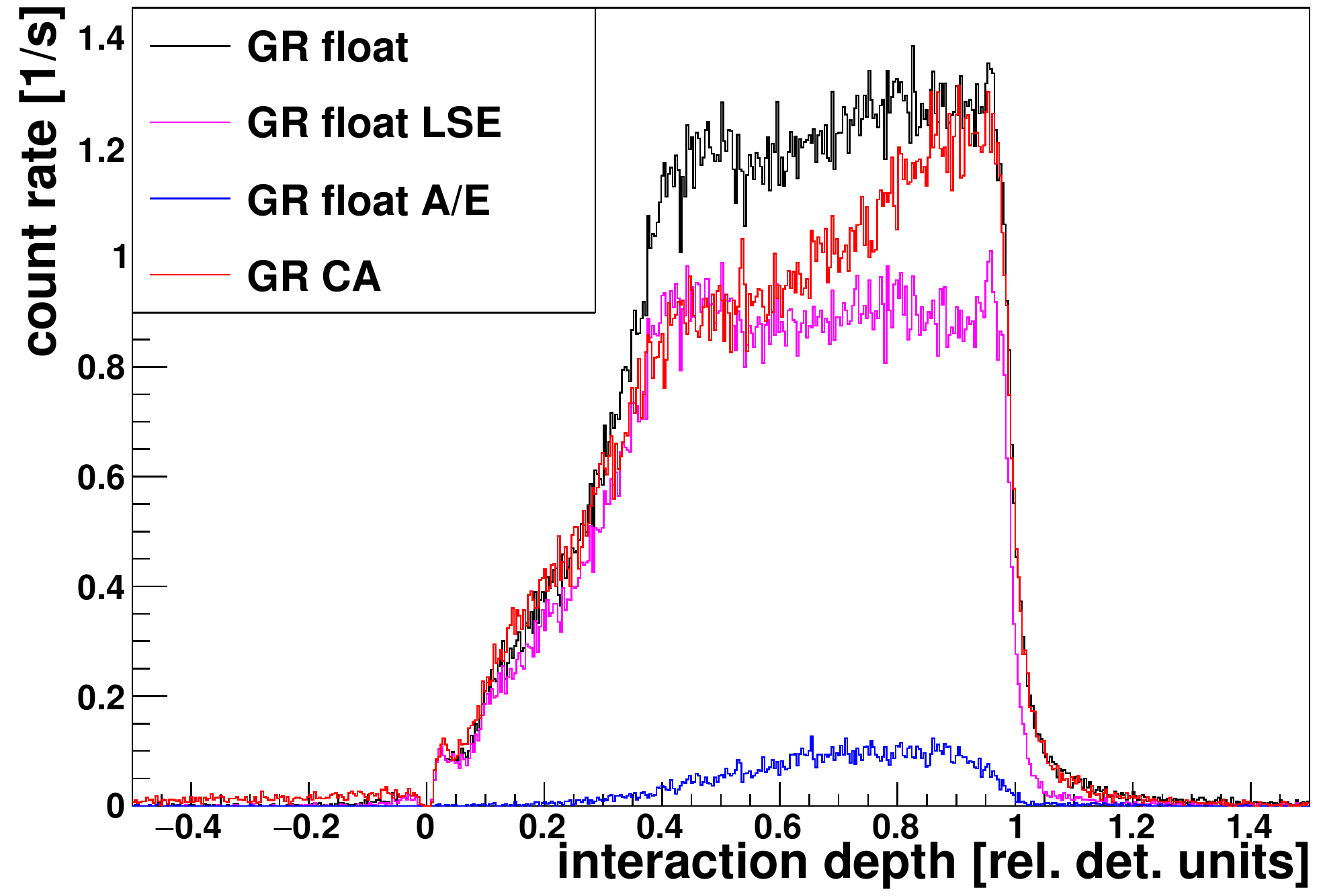}
  \captionsetup{width=0.9\textwidth} 
  \caption[Gamma radiation with instrumented GR]
	  {Comparison between gamma irradiation with GR floating (COBRA-standard) and GR on CA bias.
          Left: energy spectrum. The thick black vertical line around \SI{60}{keV} is the trigger threshold. Right: Calculated interaction depth}
  \label{fig_guardring_gamma_energy}
\end{figure}

The energy spectra shown in \autoref{fig_guardring_gamma_energy} are calculated in the standard way as the weighted difference of CA and NCA according to \autoref{eq_energy_calculation}. The GR signal is ignored for now. An inclusion of the GR signals is discussed later in \autoref{sec_include_gr_signals}.
No data-cleaning cuts are used to show the effects of the reconstructed GR-instrumentation.
For comparison, a measurement with the standard case of a floating GR is shown as well.
The measurement with an instrumented GR shows no deterioration in the spectral shape above threshold, but has events around zero.
The threshold stems from the FADC sampling and indicates the full sensitivity. Events below threshold are ignored in the analysis, they are shown here only to demonstrate the effect of the instrumented GR.
The events reconstructed around zero are those, where the CA has an NCA-like behavior and the GR is collecting the charge cloud. The deposited energy is the difference of CA and NCA, which is close to zero for CA = NCA.
LSE and A/E PSA cuts result in much more spectral distortions than GR-instrumentation. 
However, the PSA methods are not really meaningful at this low energy range.

The calculated interaction depth shown in the same \autoref{fig_guardring_gamma_energy} indicates a distorted distribution for the instrumented GR.
The slow rise for small values is a common feature, but a plateau is expected, which is not the case for the GR-instrumentation. 
This has to be investigated further.

The energy resolution is even slightly better for the measurement with GR on CA potential, but with a relatively large uncertainty: \SI{2.16(63)}{\%} at \SI{662}{keV} versus \SI{2.22(63)}{\%}. 
This behavior was also stated in \cite{He2005291}, because the interactions influenced by the possibly distorted weighting potential at the surface are not included.\\
The measurement with an instrumented GR has \SI{3}{\%} less entries at same detector-source distance and measurement duration. This could hint to a slightly worse detection efficiency. 

In the standard-case of a floating GR, \SI{95}{\%} of all events are above threshold, while this is true for \SI{87}{\%} of all events at GR on CA. 
As nothing was changed between these two measurement, the difference stems from the GR-instrumentation.
Hence, an instrumented GR on CA potential keeps \SI{92}{\%} of events above threshold. 
This is better than the results of the PSA methods (LSE or A/E) with \SI{80}{\%} (for high-energetic events above \SI{1.5}{MeV}), discussed in \autoref{sec_daq_analysis}.

\SI{3}{\%} of the events in this measurement have positive amplitudes (above noise level) for CA and GR, these events are wrongly reconstructed if the GR signals are ignored. 
This is an effect of charge-sharing between the CA and the GR on CA potential: the charge-cloud splits up between the two of them. 
This probably happens for events in the proximity of the surface between the GR and outermost CA anode rail, which is a transition zone between surface and central events.

Summarizing one can say, that an instrumented GR on CA potential does not affect the reconstructed values more than the PSA cuts.

%%%%%%%%%%%%%%%%%%%%%%%%%%%%%%%%%%%%%%%%%%%%%%%%%%%%%%%%%%%%%%%%%%%%%%%%%%%%%%%%%%%%%%%%%%%%%%%%%%%%%%%%%%%%%%%%%%%%%%%%%%%%
\FloatBarrier
\subsection{Measurements with alpha radiation}
\label{sec_guard_ring_alpha}

\begin{figure} 
 \centering
  \includegraphics[width=.49\textwidth]{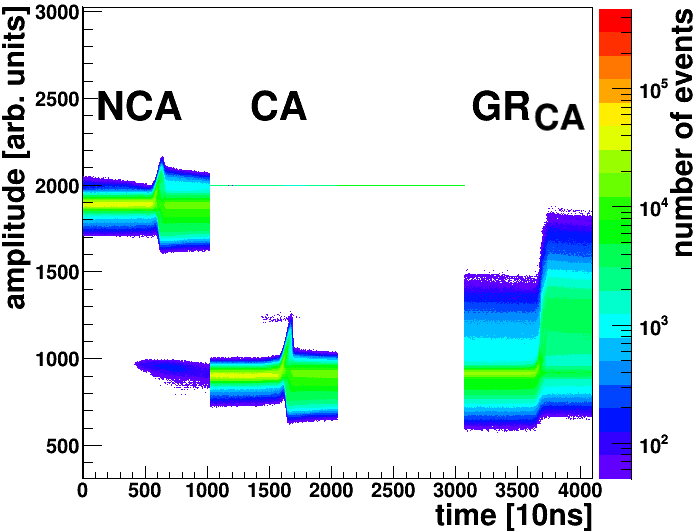}
  \includegraphics[width=.49\textwidth]{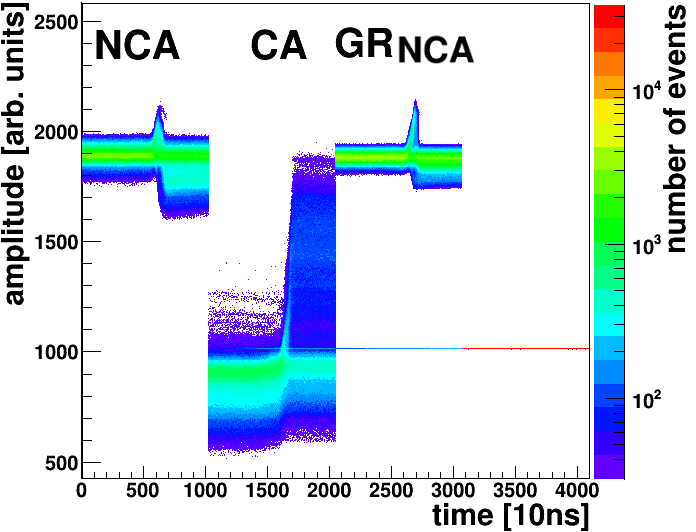}
  \captionsetup{width=0.9\textwidth} 
  \caption[Pulse shapes of alpha radiation with instrumented GR]
	  {Alpha radiation to side 1 CA.
	  Left: GR on CA bias: The CA shows NCA-like behavior, the GR is the only effective CA.
	  Right: GR on NCA bias. The CA shows CA-like, the other two electrodes NCA-like pulse shape.}
  \label{fig_guardring_CA_alpha_pulse_shapes}
\end{figure}

The difficult case of alpha irradiation of detector 631972-06 side 1 on CA, and GR on CA potential is tested here mainly.
In this case, two CA biases are the outermost electrodes, so that charge-sharing between the two could happen.
Nevertheless, the other case of GR on NCA potential is shown here briefly as well.
The superposition of all pulse shapes of two measurements are shown in \autoref{fig_guardring_CA_alpha_pulse_shapes}: One with GR on CA potential, the other with GR on NCA potential. 
Slow baseline oscillations occurred which do not alter the pulses, but result in an offset and hence in a broad signal range (especially at GR on CA). The z-axis does not start at zero to suppress distorted pulses and baseline-oscillations, nevertheless some artifacts in certain regions remain. 
One can see that the CA acts as an effective NCA, if the GR is on CA potential, which is expected when comparing the charge-cloud and detector dimensions.
This can be explained as follows. All electrodes measure the drift of the electron charge cloud through the detector volume. This is why all pulses have a common rise. Only the electrode that collects the charge cloud has a sharp rise and becomes an effective CA, all other signals show a sharp decline. As the charge clouds from alpha surface events are always below the GR and collected by it, this is the only effective CA. The original CA transforms into an effective NCA. 
If the GR is on NCA potential, NCA and CA detect normal signals, the GR NCA-like signals. 
This configuration probably does not result in an improved surface events recognition or energy resolution. 

\begin{figure}
 \centering
  \includegraphics[width=0.49\textwidth]{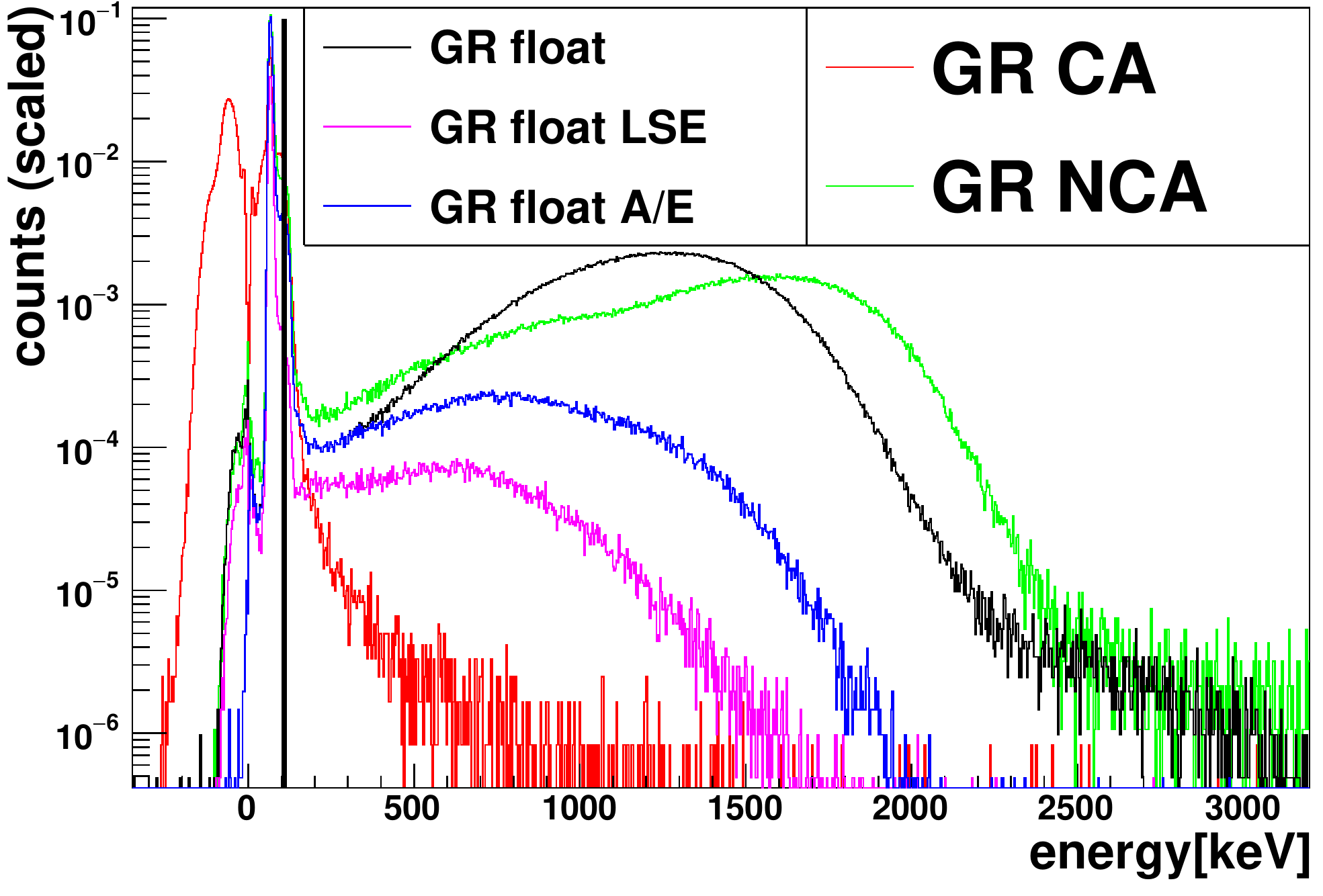}
  \includegraphics[width=0.49\textwidth]{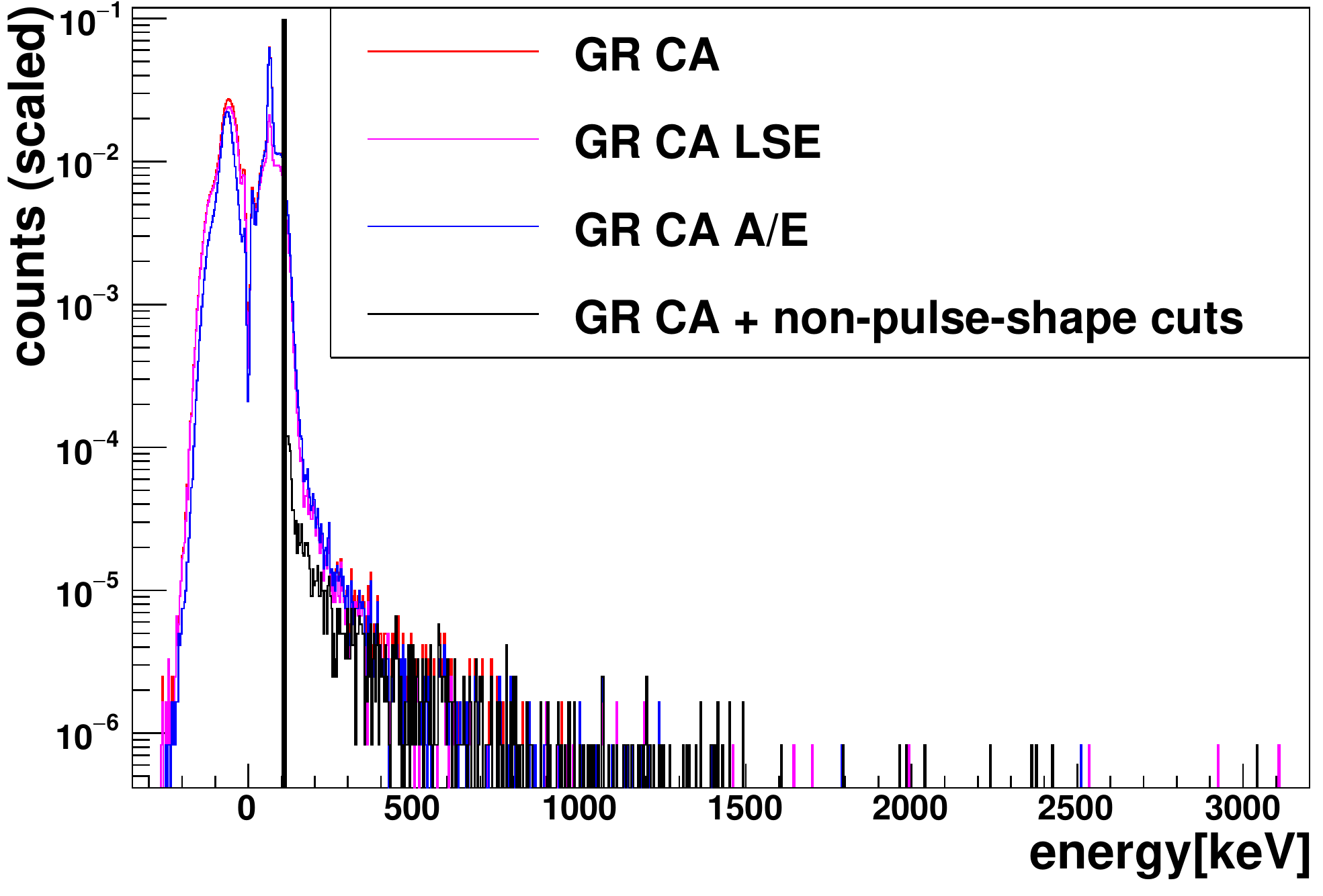}
  \captionsetup{width=0.9\textwidth} 
  \caption[Energy spectra of alpha radiation with instrumented GR]
	  {Three measurements (scaled by their number of entries) of alpha radiation to side 1 CA: GR floating, on CA and on NCA potential.
	  Left: If the GR is on CA bias, most events are reconstructed below trigger threshold.
	  Right: LSE and A/E PSA cuts in addition to the GR on CA ``hardware discrimination'' does not improve the discrimination much.}
  \label{fig_guardring_CA_alpha_spectra}
\end{figure}

The energy reconstruction is a weighted difference of CA and NCA.
Consequently, the energy of surface events can cancel to zero if the GR is the effective CA and hence the CA an effective NCA.
Thus, the surface events do not appear in further analyses. 
\autoref{fig_guardring_CA_alpha_spectra} left shows energy spectra for an instrumented GR.
The use of additional PSA cuts (LSE or A/E) does not improve the discrimination power (right).
This is probably because of the low energy of the remaining events. 
PSA works better for higher energies, which corresponds to a better SNR (see e.g. \autoref{sec_beta_tl-204_measurements} as an example for not-working PSA due to a bad SNR).

The present standard method is a floating GR, and the use of PSA cuts (LSE or A/E) to veto surface events. The fraction of events above thresholds are \SIlist{56; 10; 16}{\%} for the raw-data, LSE and A/E cuts, respectively.
In comparison to these, the instrumentation of the GR on CA potential has a much better discrimination power, only \SI{4}{\%} of events are above threshold. 
Nearly all of those are at low energies of some \SI{100}{keV}, which is far away of COBRA's ROI around \SI{2.8}{MeV}.
If the other standard non-PSA cuts (data-cleaning, cathode and anode surface removal) are applied as well, only \SI{2}{\permil} of the entries remain.\\
This is by far the best surface events discrimination result.

If the GR is on NCA potential and its signals are ignored, a wrong energy reconstruction to higher values results:
The NCA measures less signals, because of charge-sharing to the GR. As the calculated energy is a weighted difference, the energy is higher. 
This GR instrumentation on NCA potential does not improve the analysis results. Consequently, it is not investigated further here.

Due to a different number of entries in the different measurements, the absolute numbers cannot be compared directly.

If the CA turns into an effective NCA due to the GR-instrumentation and those signals are ignored, the interaction depth calculation of surface events does not work reliably anymore. 
But this reconstruction of surface events does not work necessarily under standard conditions, as discussed in \autoref{fig_631972_06_side3_CA_interaction_depth}. 
Furthermore, these events are vetoed mostly, so this can be ignored.\\

The very strong surface events suppression power of an instrumented GR can be compared to tests done with the Polaris CdZnTe pixel detector system. This $11\times11$ pixel detector was operated at LNGS for testing purpose by COBRA in 2009-2010 for 4 months. \SI{4.3}{kg\,d} of low-background physics data were collected. Measurements and simulations in \cite{Schulz} and \cite{koettig_2012_phd} show that discarding the outermost pixel line results in a background suppression of several orders of magnitude.

Furthermore, a similar description can be found in \cite{He2005291}:
\begin{quote}
'' [\dots] the \textbf{c}ollecting \textbf{a}node can be biased to the same potential as the boundary electrode [GR], 
while the \textbf{n}on-\textbf{c}ollecting \textbf{a}node is set to a lower potential. This results in the collection of electrons by the boundary electrode when charges
are generated near the sides of the detector. Only electrons generated in the central region of the
device can be collected by the \textbf{c}ollecting \textbf{a}node, where the difference of weighting potentials of
coplanar anodes is minimum. Therefore, the detector performance is expected to be better in
the later mode of operation since the charge induced from electrons formed in the central
region of the detector will be more uniform.`` 
\end{quote}

%%%%%%%%%%%%%%%%%%%%%%%%%%%%%%%%%%%%%%%%%%%%%%%%%%%%%%%%%%%%%%%%%%%%%%%%%%%%%%%%%%%%%%%%%%%%%%%%%%%%%%%%%%%%%%%%%%%%%%%%%%%%
\FloatBarrier
\subsubsection{Inclusion of GR signals to energy reconstruction}
\label{sec_include_gr_signals}

\begin{SCfigure}
 \centering
  \includegraphics[width=0.6\textwidth]{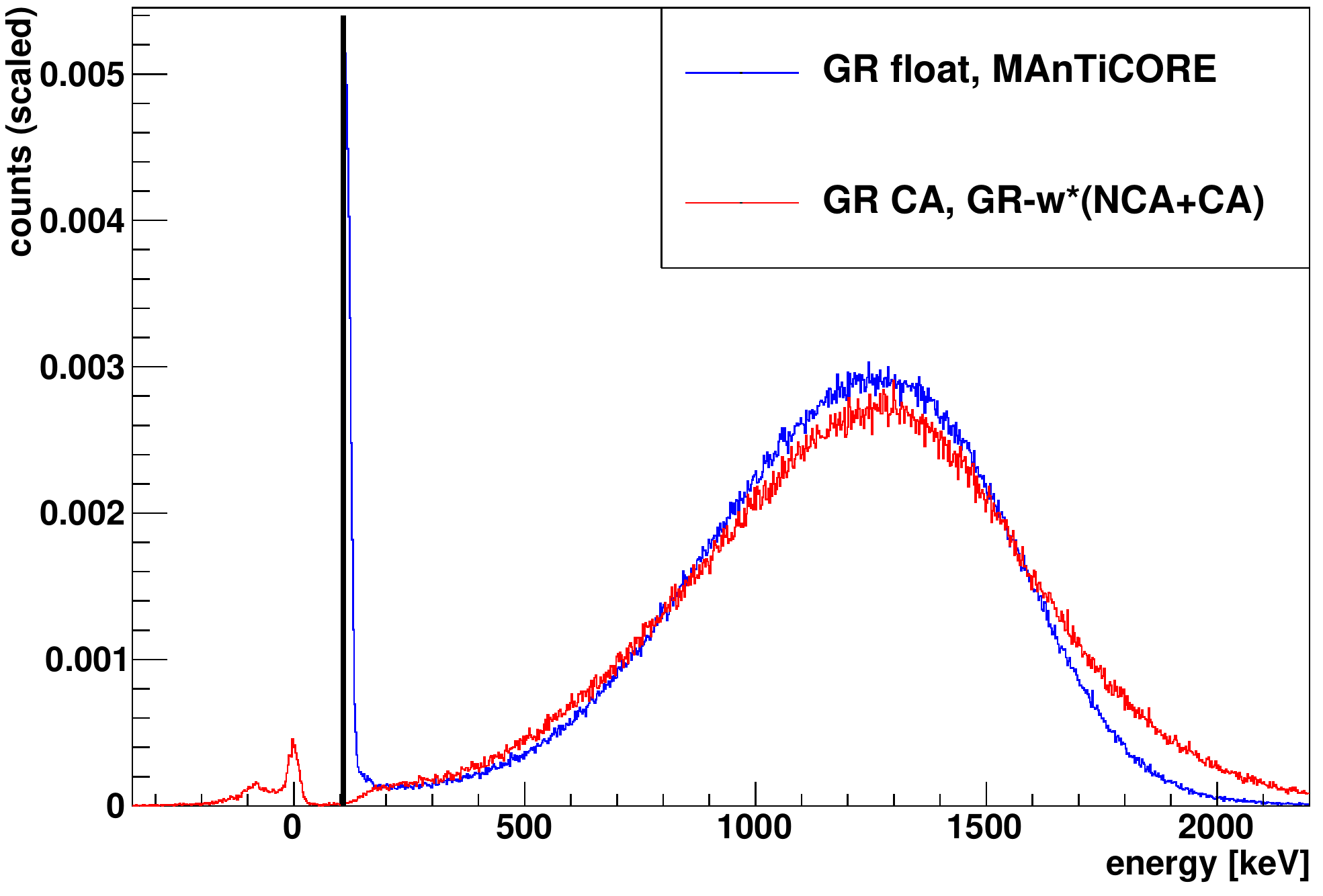}
  \captionsetup{width=0.9\textwidth} 
  \caption[Energy reconstruction including GR signals]
	  {Energy reconstruction including GR signals. Blue: Reconstructed energy of the measurement with floating GR, calculated by MAnTiCORE using all standard reconstruction methods. Red: Reconstructed energy for GR on CA potential using \autoref{eq_energy_calculation_GR_inclusion}, including a threshold not to sum up too much noise.}
  \label{fig_guardring_energy_reconstruction}
\end{SCfigure}

According to the Shockley-Ramo theorem (\autoref{eq_Shockly-Ramo}), a drifting charge-cloud induces charges on all electrodes. If the GR is read out, its signals have to be included to the reconstruction methods. This is also known from the large CPqG (\textbf{c}o\textbf{p}lanar \textbf{q}uad-\textbf{g}rid) detectors, mentioned in \autoref{sec_larg_scale}.\\
A basic idea for the inclusion of the GR signals is the following: Find and sum all CA-like pulses to build an effective amplitude $A_{CA_{eff}}$, sum all others (NCA-like) to form an effective total $A_{NCA_{eff}}$ (for each event separately). 
In most cases there is only one CA-like signal, because charge-sharing is seldom. 
Then calculate the energy analog to \autoref{eq_energy_calculation}. In the case here, this is written as follows: 
\begin{equation}
 \begin{split}
    E_{\text{deposited}} &\propto A_{CA_{eff}} - w \cdot A_{NCA_{eff}}\\
			 &= A_{GR} - w \cdot (A_{NCA} + A_{CA}).
 \label{eq_energy_calculation_GR_inclusion}
 \end{split}
\end{equation}
A certain threshold of the pulse heights at the summation of the amplitudes is used to avoid summing up noise. 
It is determined by the noise level of the baseline of the pulse.
This is demonstrated in \autoref{fig_guardring_energy_reconstruction} where the reconstructed energy according to \autoref{eq_energy_calculation_GR_inclusion} of the measurement with GR on CA potential looks similar to the reconstructed energy of the measurement with floating GR. 
In the latter case, the reconstruction is done by COBRA's analysis tool MAnTiCORE which produces the blue curve in \autoref{fig_guardring_energy_reconstruction}.
Comparing the two results, one can see that the reconstruction according to \autoref{eq_energy_calculation_GR_inclusion} shows a spectral deviation. 
The peak is lower and has a larger width. 
The reconstruction using an effective NCA in the weighted difference is the easiest case, and can probably be enhanced by using more sophisticated methods for noise-exclusion and amplitude calculation, for example. \\

As a proof of principle, this demonstrates that an instrumented GR offers new reconstruction and analyses possibilities.
In particular, using an instrumented GR seems to be the most powerful method of background reduction COBRA has tested so far.%, without losing too much information on the other hand.

%%%%%%%%%%%%%%%%%%%%%%%%%%%%%%%%%%%%%%%%%%%%%%%%%%%%%%%%%%%%%%%%%%%%%%%%%%%%%%%%%%%%%%%%%%%%%%%%%%%%%%%%%%%%%%%%%%%%%%%%%%%%
%%%%%%%%%%%%%%%%%%%%%%%%%%%%%%%%%%%%%%%%%%%%%%%%%%%%%%%%%%%%%%%%%%%%%%%%%%%%%%%%%%%%%%%%%%%%%%%%%%%%%%%%%%%%%%%%%%%%%%%%%%%%
\chapter{Calculation of the electron mobility and lifetime}
\label{sec_drift_time}
As explained in \autoref{sec_cpg_principles}, the signals of the NCA and CA have a common rise due to the drift of the charge cloud towards the anodes.
In a kind of TOF (\textbf{t}ime \textbf{o}f \textbf{f}light)-measurement, this can be used to calculate the electron mobility $\mu_e$, which is basically the drift distance divided by the drift time.
For alpha irradiation of the cathode, the drift distance is defined by the dimension of the detector. Hence, only the drift time has to be measured to calculate $\mu_e$.

A compilation of literature values of $\mu_e$ in CdZnTe shown in \autoref{tab_drift_time_literatur_values} shows a large spread.
\begin{SCtable}%[htbp]
 \centering
  \begin{tabular}{|c|c|c|}
    \hline
    $\mu_e$ [$\text{cm}^2/(\text{Vs})$] & year & author \\ \hline
    1350      & 1993 & Burshtein \\ \hline
    1050      & 1993 & Johnson \\ \hline
    1000      & 1995 & Luke \\ \hline
    1330      & 1997 & Rosaz \\ \hline
    1100      & 1997 & Eisen \\ \hline
    1000      & 1997 & Toussignant \\ \hline
    1000-1300 & 1998 & He \\ \hline
    700-800   & 2000 & James \cite{Schlesinger2001103} \\ \hline
    1000      & 2013 & eV products \\ \hline
  \end{tabular}
  \caption[Literature values for $\mu_e$ of CdZnTe]
          {Literature values for $\mu_e$ of $\text{\textbf{C}a\textbf{d}mium}_{0.45}-\text{\textbf{Z}i\textbf{n}c}_{0.05}-\text{\textbf{Te}lluride}_{0.5}$. 
	  The last value is from the manufacturer of the detectors under study, taken from \autoref{tab_properties_CdZnTe}, the other non-cited values are from \cite{1999NIMPA.428...45J} and references therein.}
  \label{tab_drift_time_literatur_values}
\end{SCtable}

\begin{figure}
 \centering
  \captionsetup{width=0.9\textwidth} 
  \includegraphics[width=0.49\textwidth]{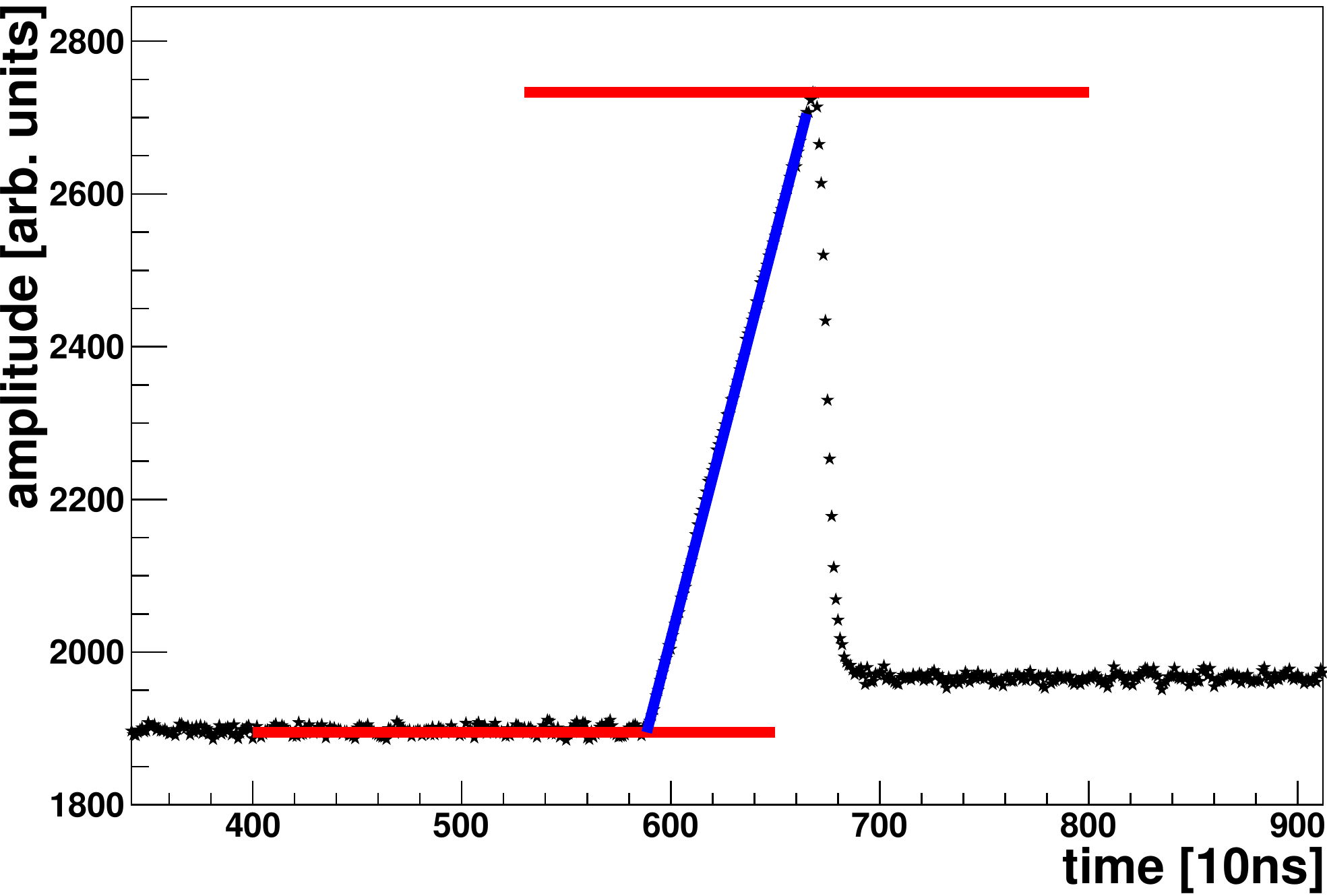}
  \includegraphics[width=0.49\textwidth]{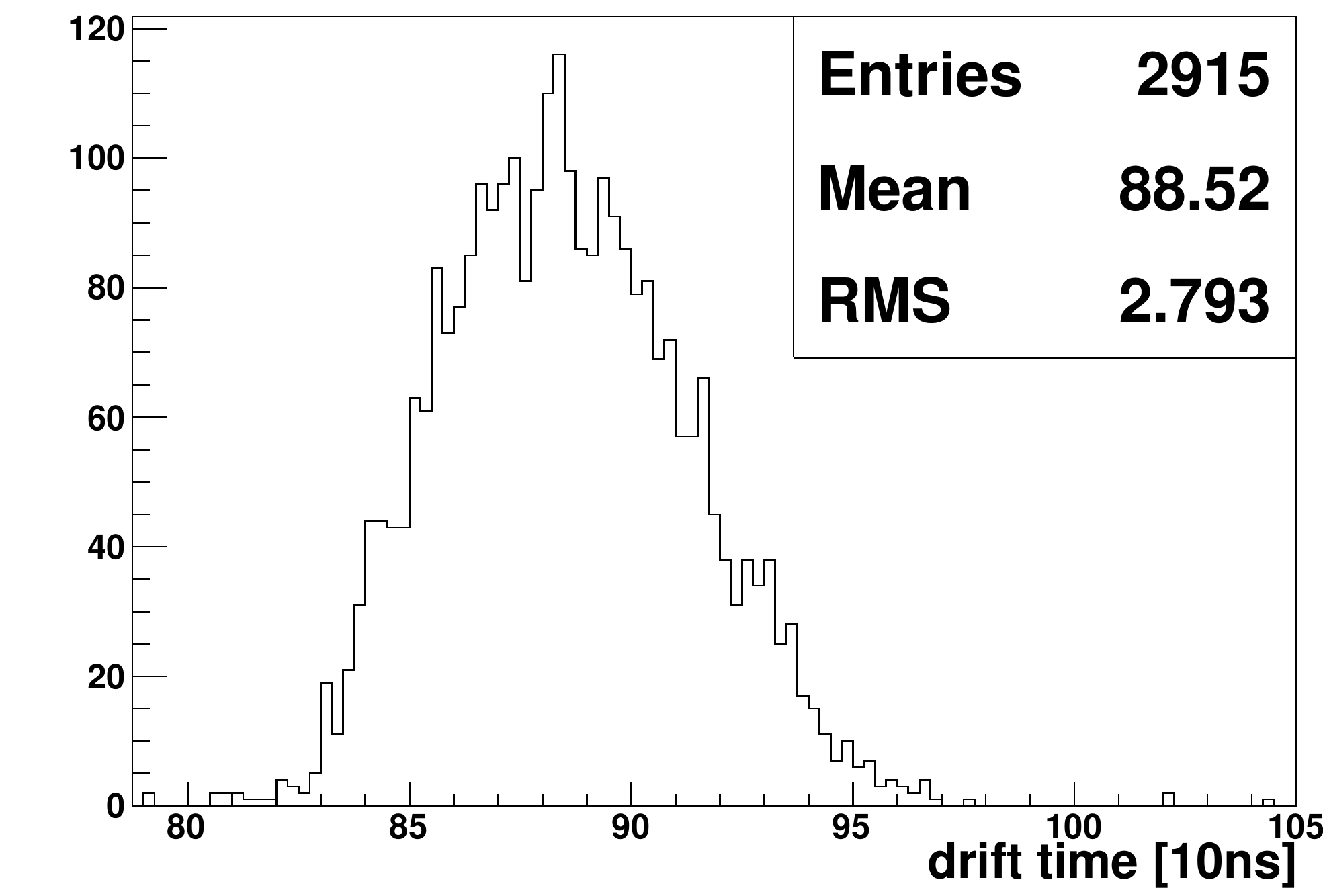}
  \caption[Calculation and distribution of drift time]
          {Left: Calculation of the drift time of one event in a TOF measurement. The red lines are the level of the baseline and maximal amplitude, the blue line shows the slope of the pulse.
           Right: Distribution of all drift times of a measurement of alpha radiation to the cathode.}
  \label{fig_ps_drift_time}
\end{figure}

The drift time can be calculated by measuring the length of the common drift of the NCA and CA signals. 
This is done here using the NCA signals, because their maximal value is the end of the electron drift time due to the BV before entering the near-anode region.
The data-processing software MAnTiCORE calculates the level of the baseline and the maximal value of each pulse, as well as the slope of the drift-part of the pulse. 
The drift time is then calculated according to the gradient triangle as the amplitude of the pulse divided by the slope. 
\autoref{fig_ps_drift_time} shows exemplarily the calculation of the drift time at a measured NCA pulse, and the distribution of the drift times of all events of a measurement of alpha radiation to the cathode.

The quantity $\mu_e$ is then calculated as follows \cite{1589323}:
\begin{equation}
 \mu_{e} = \frac{ d_{drift} }{ t_{drift} \cdot \left( |BV|/d_{drift}\right) } = \frac{ d_{drift}^{\ 2} }{ t_{drift} \cdot |BV| }
\end{equation}

Furthermore, the $\tau_e$ (electron lifetime) of the drifting electrons can be calculated according to \cite{Fritts:2012xq} as
\begin{equation}
  \tau_e = \frac{d_{drift}^{\ 2} \cdot \lambda}{\mu_e \cdot \text{|BV|}} \,.
\end{equation}
$\lambda$ is defined in \autoref{eq_cal_ipos_ztc} as $ \lambda = \dfrac{1+w}{1-w}$, w is the weighting factor to compensate electron trapping.

The maximum value of the pulse in \autoref{fig_ps_drift_time} is just before the pulse is influenced by the near-anode region.
The drift distance $d_{drift}$ is the length of the detector (in z-direction) minus the near-anode region of \SI{440(40)}{\mu m} (calculated in \autoref{sec_60keV_gamma_radiation}).
The dimensions of the detectors vary due to the production processes like cutting and polishing.
For this study, they were determined using a digital optical microscope.
Two edges of each detectors were measured. The mean value was used as detector length. 
The maximal spread of all length was taken as a conservative estimate as uncertainty of the length. 
This uncertainty is summed quadratically with the uncertainty of the near-anode region to the total uncertainty of \SI{110}{\mu m} of the drift distance.

An uncertainty of the BV of $\pm\SI{5}{V}$ is assumed.
This arises from the adjustment knob of the voltage supply device, which displays the applied voltages in steps of \SI{10}{V}.

The uncertainties are calculated using Gaussian error-propagation.

Several measurements of alpha radiation irradiating the cathode are used to calculate $\mu_e$, $\tau_e$ and $(\mu\tau)_e$.
As a result, the weighted mean values (calculated according to \autoref{eq_weighted_mean}) are calculated to 
\begin{equation}
 \begin{split}
  \mu_e &=  \SI{968(28)}{\frac{cm^2}{Vs}}\\
  \tau_e &= \SI{4.39(33)E-6}{\mu s}\\
  (\mu\tau)_e &= \SI{4.22(35)E-3}{\frac{cm^2}{V}}\,.
 \end{split}
\label{eq_result_mu_tau}
\end{equation}
These results are in very good agreement with the values quoted by the manufacturer.
A compilation of all results is shown in \autoref{tab_drift_time_results}.
\begin{table}[t!]
  \begin{tabular}{|c|c|c|c|c|c|c|}
    \hline
    Detector  & \makecell{measure-\\ment} & \makecell{BV\\{[}V]} & \makecell{mean $t_{drift}$\\{[$10^{-7}$}s]} & \makecell{$\mu_e$\\$\left[\frac{\text{cm}^2}{\text{Vs}}\right]$} & \makecell{$\tau_e$\\{[}$\mu s]$} & \makecell{$(\mu\tau)_e$\\$\left[10^{-3}\frac{\text{cm}^2}{\text{V}}\right]$} \\ \hline
    6000-2                                  & 1$\times$S central                        & -1000 & \num{9.29(29)} & \num{984(38)}  & \num{5.14(66)} & \num{5.06(68)} \\ \hline
    F15                                     & F                                         & -1200 & \num{8.00(79)} & \num{926(94)}  & \num{4.64(74)} & \num{4.30(81)} \\ \hline
    \multirow{2}{*}{\makecell{631972-06}}   & F                                         & -1200 & \num{8.16(38)} & \num{952(49)}  & \num{3.95(58)} & \num{3.77(59)} \\ \cline{2-7} 
                                            & \makecell{21$\times$S centr.}             & -1200 & \num{7.99(79)} & \num{972(99)} & \num{3.97(68)} & \num{3.86(77)} \\ \hline
    \makecell{667615-02}                    & \makecell{14$\times$S edges}              & -1300 & \num{9.07(12)} & \num{805(21)}  & \num{10.1(7)}  & \num{8.12(59)} \\ \hline
    \multicolumn{3}{|c|}{weighted mean central positions}                                       & \num{8.75(21)} & \num{968(28)}  & \num{4.39(33)} & \num{4.22(35)} \\ \hline        
    \multicolumn{4}{|c|}{comparison: values of manufacturer}                                                     & \num{1000}     & \num{3}        & 3-5            \\ \hline
  \end{tabular}
  \captionsetup{width=0.9\textwidth} 
  \caption[Results of the calculation of $\mu_e$, $\tau_e$ and $(\mu\tau)_e$]
          {Results of the calculation of $\mu_e$, $\tau_e$ and $(\mu\tau)_e$ of alpha radiation directed to the cathode. ''S`` denotes a scanning measurement, ''F`` a flush measurement. For comparison, the last line shows the values quoted by the manufacturer, taken from \autoref{tab_properties_CdZnTe}. The values of the scanning measurement at the edges are significantly smaller, shown in \autoref{fig_667615_02_side1_CA_drift_time_3D}. These are ignored for the calculation of the total weighted mean value.}
  \label{tab_drift_time_results}
\end{table}

\begin{SCfigure}[][b!]
 \centering
  \includegraphics[width=0.59\textwidth]{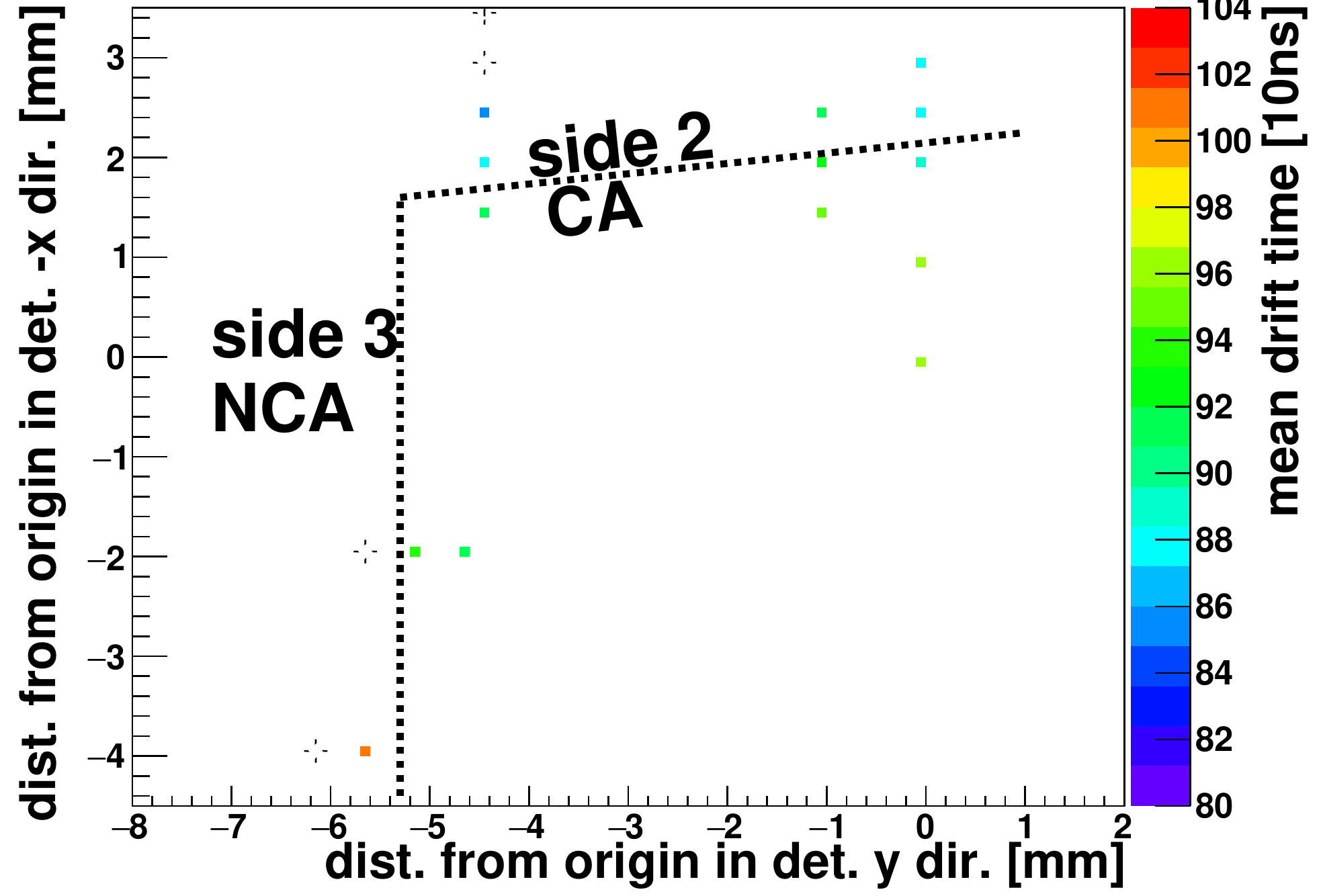}
  \caption[Drift time at edges]
          {The drift time at the top edge (CA) is lower than for central positions, while it seems to be higher at the NCA edge (left). Same scanning measurements as discussed in \autoref{sec_cathode_scanning}.}
  \label{fig_667615_02_side1_CA_drift_time_3D}
\end{SCfigure}

As discussed in \autoref{sec_cathode_scanning}, one measurement series were scanning measurements at the cathode, but at the edges to the lateral surfaces. 
These positions show significantly smaller drift times at the edge to side two, where the CA is the outermost anode rail. 
At the edge to side three (NCA), this is not the case. 
However, it cannot be stated as a systematic difference between the different biases, as just one scanning point shows this behavior at side three, shown in \autoref{fig_667615_02_side1_CA_drift_time_3D}. 
That the drift time is lower at the edge can be interpreted as a weaker electric field there.
Hence, these values are ignored for the calculations of the results in \autoref{eq_result_mu_tau}.

\begin{figure}[b!]
  \centering 
   \includegraphics[width=0.49\textwidth]{pictures/surface_events/2015-02-26_734_667615_02_lngs_return/alpha_scan_ipos_ztc_3D_side1_CA_TH2F__combination.pdf}
   \includegraphics[width=0.49\textwidth]{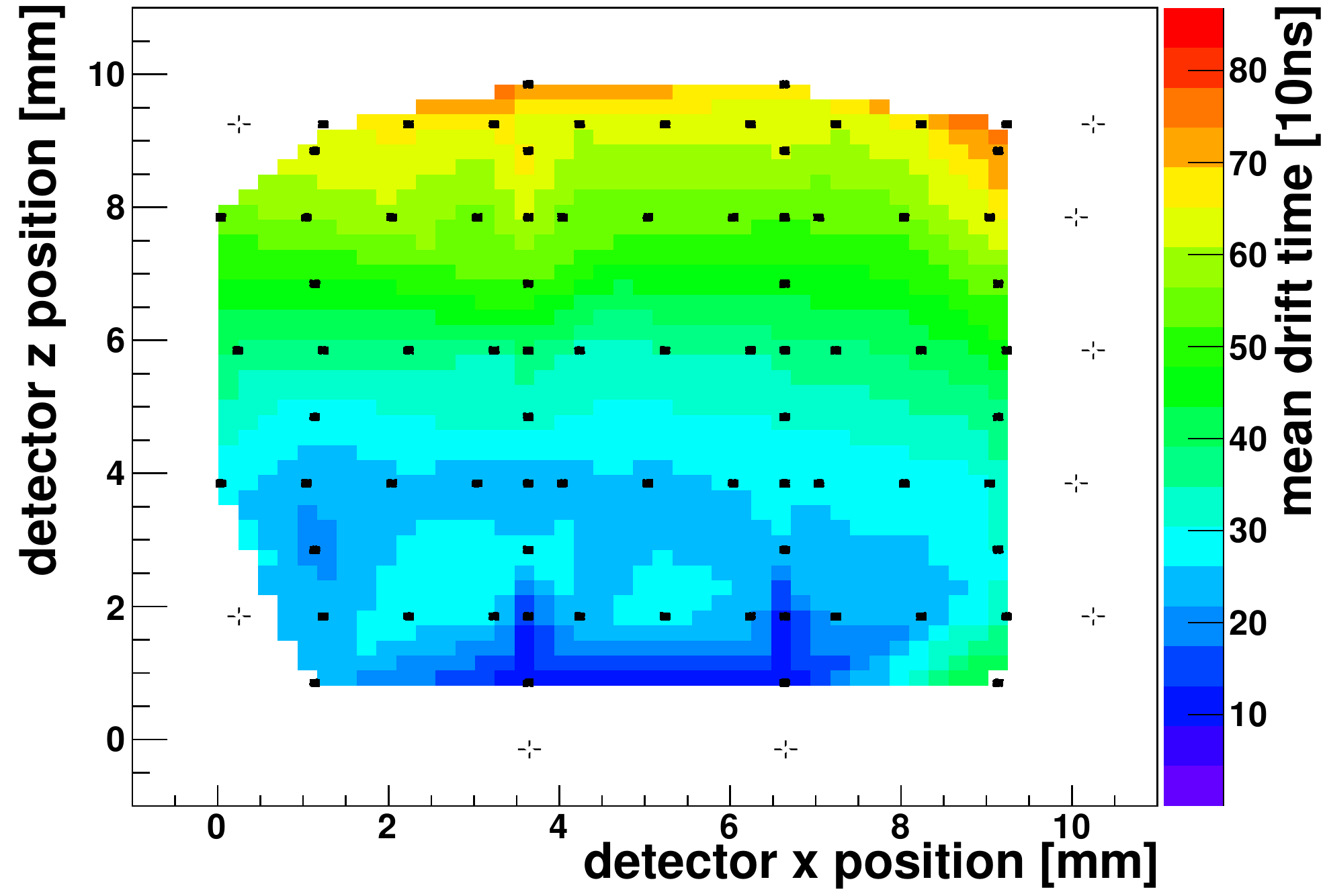}
   \captionsetup{width=0.9\textwidth} 
   \caption[Comparison of calculation of interaction depth and drift time]
           {Comparison of calculation of interaction depth and electron drift time using alpha scanning measurements of detector 667615-02 of side 1 CA (interpolation): Left, copy of \autoref{fig_667615_02_side1_CA_interaction_depth_3D}. Right: Drift time.}
          \label{fig_comparison_drift_time__ipos_tc}
\end{figure}

\begin{SCfigure}[][b!]
 \centering
   \includegraphics[width=0.69\textwidth]{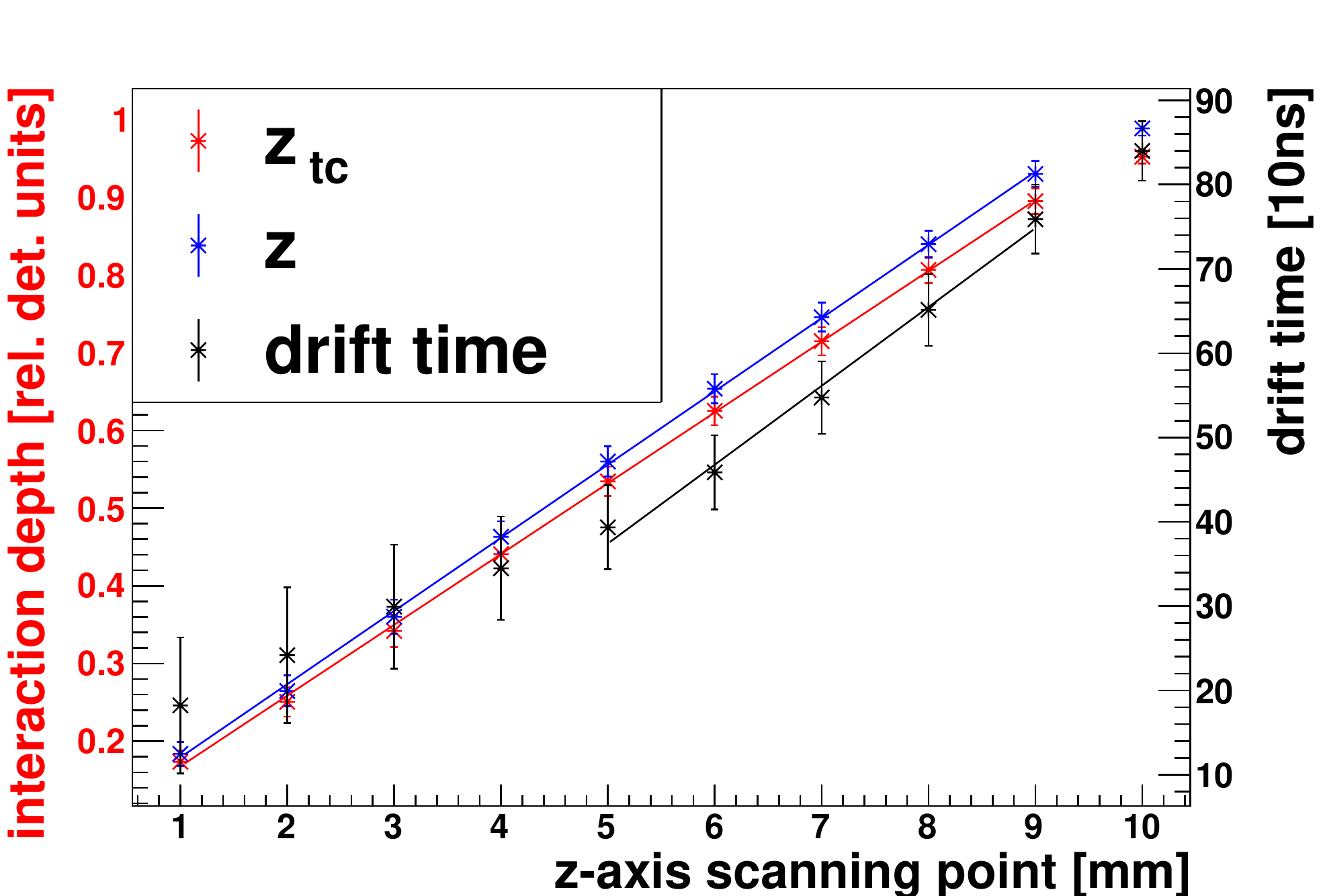}
   \caption[Comparison of calculation of interaction depth and drift time]
           {Comparison of calculation of interaction depth and electron drift time using gamma scanning measurements, adapted from \autoref{fig_gamma_scanning_interaction_depth}. The black curve is based on a restricted fit-range above z=\SI{5}{mm}.}
          \label{fig_comparison_drift_time__ipos_tc_gamma}
\end{SCfigure}

The drift time of the electrons depends on the z-position, which is expected for geometrical reasons. Furthermore, it depends on the lateral position as well.
These two quantities cannot be determined independently reliable, especially not for surface events. 
As a consequence, the drift time cannot be used to determine one of these quantities better than the standard methods described before.
This is shown for the z-dependence of the calculated drift times:

As the drift time is also a measure for the distance the charge cloud has to travel from its origin to the anodes, it could be used to gain information about the point of interaction in z-direction.
A comparison of the complementary methods of calculating the interaction depth using the $z_{tc}$ calculation (\autoref{eq_cal_ipos_ztc}) and the drift time by alpha scanning measurements is shown in \autoref{fig_comparison_drift_time__ipos_tc}. 
The method using the interaction depth calculation $z_{tc}$ yields better results, especially for low z-values.  
The reason for this behavior is, that the calculation using ratios of pulse amplitudes does not suffer from noise and a poor SNR. This is especially true for near-anode or low energy events with a short drift time, where the precision of the fitting of the slope (discussed in \autoref{fig_ps_drift_time}) is low.
This can be seen in a comparison using gamma scanning data, shown in the \autoref{fig_comparison_drift_time__ipos_tc_gamma}, which is adapted from \autoref{fig_gamma_scanning_interaction_depth}. As described there, the error bars are based on the  widths of the distributions.
The values of the drift time show significant deviations from a linear behavior. 
However, the drift times above z=\SI{5}{mm} show a linear behavior, indicated by the linear fit in black. 
Furthermore, the error bars are much larger.
The slope of this curve is very close to those of the interaction depth calculations. 
All fits discard the last scanning point at z=\SI{10}{mm}, as the beam does not hit the detector fully there, as discussed in \autoref{sec_gamma_scanning}.

\chapter{Towards a large-scale COBRA-experiment}
\label{sec_larg_scale}
The COBRA demonstrator with 64 (\num{10 x 10 x 10}){\,mm$^3$} detectors has been installed and commissioned at the LNGS successfully and is taking data reliably.

Plans for future developments go eventually to a large-scale experiment, scaled in detector mass by a factor of \num{1000}.

The half-life sensitivity of the $0\nu\beta\beta$-decay of \isotope[116]Cd of such an experiment shall reach more than \SI{2E26}{yr}.
This corresponds to an effective Majorana neutrino mass of less than \SI{50}{meV}, which would probe the inverted neutrino mass hierarchy, see \autoref{fig_neutrino_hierarchy}.

The formula for the half-life sensitivity (\autoref{eq_half_live_neutrino_mass}) shows, which factors can be used to increase the sensitivity by several orders of magnitude:
\begin{equation*}
  T_{1/2} \sim a\cdot \epsilon \sqrt{\frac{M\cdot t}{\Delta E\cdot B}}
\end{equation*}
The natural isotopic abundance $a$ can be enhanced by using detectors enriched in \isotope[116]Cd, from natural abundance now (7.5\%, see \autoref{tab_double_beta_isotopes}) by at least a factor of ten. 
A recent publication \cite{Danevich:2016eot} reports an enrichment in \isotope[116]Cd to \SI{82}{\%}. 
As the abundance acts linearly in \autoref{eq_half_life_sensitivity}, this is an important factor. Nevertheless, this is not a physics problem to work on, but mainly a question of funding and reliability of the detector production process.\\
Once the detector geometry is fixed, the detection efficiency $\epsilon$ cannot be changed much. 
The measurement time $t$ of the experiment cannot be enlarged by reasonable means.
The energy resolution $\Delta E$ can be improved, but not by orders of magnitude:
The intrinsic energy resolution of a CdZnTe CPG detector is given as \SI{1.26}{\%} FWHM at \SI{662}{keV} in \cite{sturm_diss}.
At the COBRA demonstrator, the energy resolution is about \SI{2.24}{\%} FWHM at \SI{662}{keV} \cite{Ebert:2015rda}. 
From this one can estimate, that the energy resolution can be improved by about a factor of two.
To scale the mass $M$ one needs to use more detectors, which is mainly a question of funding.
So the main item to work on is improving the background rate $B$.

One promising option is to use larger detectors with a better surface-to-volume ratio. 
These offer a better detection efficiency $\epsilon$ and better background rate $B$ due to geometrical reasons:
The particles that shall be measured have a higher probability to be stopped completely within the detector volume. 
Hence, the detection efficiency increases by \num{10} percentage points.
Furthermore, the background level due to surface contaminations is lower, as the larger detectors contain more mass at a (relative) smaller surface.
As a consequence, the surface-to-volume ratio is better by about \SI{50}{\%}.

In order to reach the sensitivity of the half-life of the $0\nu\beta\beta$-decay of \isotope[116]Cd of more than \SI{2E26}{yr},
the background has to be lowered to \SI{E-3}{\cpkky} at least. 
As given in \autoref{tab_results_cobra_demonstrator}, the background rate at the COBRA demonstrator is currently \SI{2.7}{\cpkky}, while the best detectors achieve values below \SI{1}{\cpkky}.\\

A scaled experimental setup includes also shielding, housing, DAQ and electronics.
The custom-made discrete read-out electronics and DAQ used for the COBRA demonstrator, described in \autoref{sec_cobra_demonstrator}, cannot be scaled. The required space, waste heat and amount of cables would be too much.
Furthermore, the energy resolution at the COBRA demonstrator is limited also by the DAQ, mainly by the preamplifier devices.

That is why the transition to a scalable, integrated read-out system with a better energy resolution is desired.

A commercially available ASIC (\textbf{a}pplication \textbf{s}pecific \textbf{i}ntegrated \textbf{c}ircuit)-based read\-out-system for CdZnTe detectors was tested and purchased.\\

The next step for COBRA is to construct a $3\times3$ detector module of larger detectors, using an ASIC-based readout system to build the smallest scalable module for a large-scale setup.
In this chapter, the use of large detectors and mainly the use of the integrated read-out electronics is discussed.

%%%%%%%%%%%%%%%%%%%%%%%%%%%%%%%%%%%%%%%%%%%%%%%%%%%%%%%%%%%%%%%%%%%%%%%%%%%%%%%%%%%%%%%%%%%%%%%%%%%%%%%%%%%%%%%%%%%%%%%%%%%%
\FloatBarrier
\section{Large detectors}
\begin{SCfigure}
 \centering
  \includegraphics[width=0.4\textwidth]{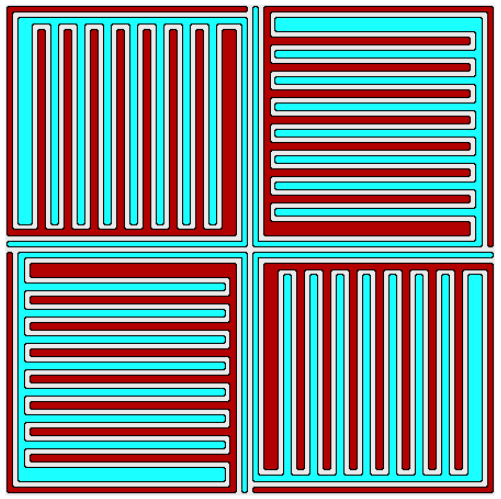}
  \caption[Sketch of the anode of a large CPqG detector]
	  {Sketch of the anode of a large CPqG detector. The GR surrounding the whole CPqG is not shown.}
  \label{fig_sketch_CPqG}
\end{SCfigure}

Latest improvements in CdZnTe crystal production processes allow to build detectors with the volume of (\num{20 x 20 x 15}){\,mm$^3$}, which is six-times larger than the COBRA standard.
Two possible anode configurations were tested: 
The first is one large CPG, extended to (\num{20 x 20}){\,mm$^2$}, the other a so-called CPqG detector with four standard-size CPG anode structures, rotated by \ang{90} against each other, as shown in \autoref{fig_sketch_CPqG}.
The applied voltages of the eight anodes can be chosen individually.
The first anode configuration was discarded as test results were not satisfying \cite{theinert_master}. 

The CPqG detectors were tested extensively, see \cite{Ebert:2015bxa}. Apart from their better surface-to-volume ratio, they offer new techniques like a better event-localization in the x-y-plane and hence a better MSE-recognition. 
The use of the GR shall be investigated thoroughly to study the possibility of background-reduction, as shown in \autoref{sec_guard_ring}.
Especially, it is possible to have all NCA anodes as outermost anodes.
This should be a reasonable measure of background reduction (as discussed in \autoref{sec_results_of_general_surface_sensitivity} for the \SI{1}{cm^3} detectors), if the effect of the GB-dependent surface dead-layer of the standard \SI{1}{cm\cubed} detectors appears here as well.

%%%%%%%%%%%%%%%%%%%%%%%%%%%%%%%%%%%%%%%%%%%%%%%%%%%%%%%%%%%%%%%%%%%%%%%%%%%%%%%%%%%%%%%%%%%%%%%%%%%%%%%%%%%%%%%%%%%%%%%%%%%%
\FloatBarrier
\section{ASIC-based readout system}
\label{sec:asic_based_readout_system}
The Norwegian company IDEAS (\textbf{I}ntegrated \textbf{D}etector \textbf{E}lectronics \textbf{AS}) offers an ASIC-based, commercially available continuously sampling readout system. 
It could be a replacement for the current, custom-designed readout system.
Its key features comprise:
\begin{itemize}
 \item 130 \textbf{c}harge-\textbf{s}ensitive pre\textbf{a}mplifier (CSA): 128 inputs for negative charges from CdZnTe anodes, 2 inputs for positive charges from the cathode
 \item 130 shaper and comparators with \SI{500}{ns} shaping time
 \item full wave form sampling with \SI{100}{MHz} maximum sampling frequency
 \item 160 cell analogue sample-and-hold pipeline at CSA output
 \item digitalization by 14-bit ADC (\textbf{a}nalog to \textbf{d}igital \textbf{c}onverter). The system has 3 ADCs which are external to the ASIC
 \item four programmable gain settings with an energy range in CdZnTe up to \SIlist{0.7; 3; 7; 9}{MeV}
 \item programmable trigger thresholds with readout of the cathode and four, eight or all anodes
 \item \SI{270}{mW} power consumption
\end{itemize}

\begin{SCfigure}
 \centering
  \includegraphics[width=0.7\textwidth]{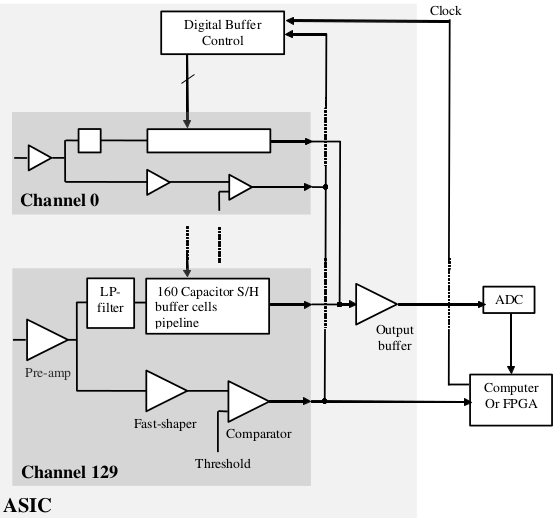}
  \caption[Simplified schematics of the ASIC-based readout system]
          {Simplified schematics of the read-out system with CSA, trigger mechanism and pipeline buffer. Taken from \cite{ideas}.}
  \label{fig_shortened_schematics_ideas}
\end{SCfigure}

The basic functionality is explained in the data-sheet \cite{ideas}:
\begin{quote}
``Each preamplifier output is continuously sampled and the values are stored in an analogue ring buffer (160 cell pipeline). 
In addition, each preamplifier output is also connected to a shaping amplifier and comparator, which generate a trigger when the input charge exceeds a programmable threshold. 
After a programmable delay, the trigger stops sampling the preamplifier outputs. The chip then outputs the pipeline samples from all cathodes and a programmable set of anodes channels.''
\end{quote}
Concerning the noise of the electronics: An energy resolution of better than \SI{2}{keV} FWHM at \SI{3}{MeV} has been measured by IDEAS.
\autoref{fig_shortened_schematics_ideas} shows a simplified schematics of the readout system.

\begin{SCfigure}
 \centering
  \includegraphics[width=0.7\textwidth]{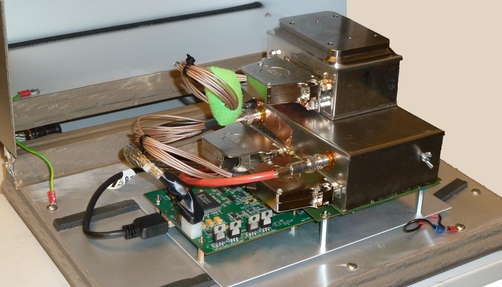}
  \caption[Picture of the ASIC-based measurement setup]
          {Picture of the measurement setup with IDEAS's electronics on the PCB (bottom), and the supply- and connection-adapter and detectors in the metal housings on top of it. Note the extra shielded cable and the edges sealed with conductive metal tape.}
  \label{fig_ideas_setup}
\end{SCfigure}

This system seems to be an ideal system for COBRA. To test it and provide the proof of principle, it was decided to investigate this read-out system at IDEAS' laboratories in Oslo.
The author built an adapter to connect four COBRA detectors to the ASIC-based readout system and supply the detectors with GB and BV.
The setup in Oslo was not shielded like the standard COBRA-setup, because the system is designed to be used for pixelated CdZnTe detectors. These do not have long cables and are therefore not affected by noise that much. 
Consequently, much effort had to be made to reduce problems with noise, like electrically tightly closing the setup, and add an extra shielding-layer to all cables entering the setup, as described in \autoref{sec_cobra_demonstrator}. A picture of the setup is shown in \autoref{fig_ideas_setup}.

\begin{figure}
 \centering
  \includegraphics[width=0.49\textwidth]{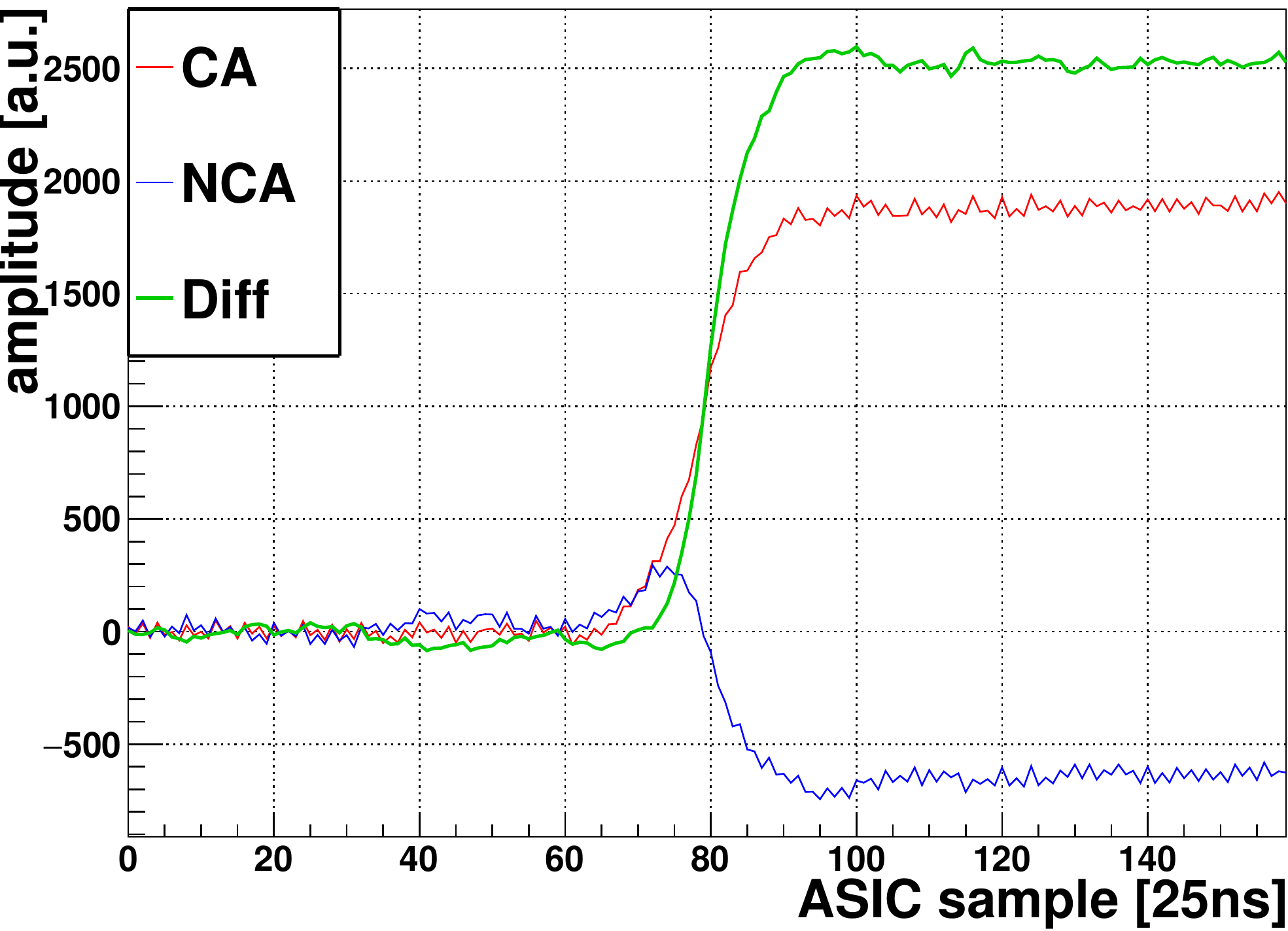}
  \includegraphics[width=0.49\textwidth]{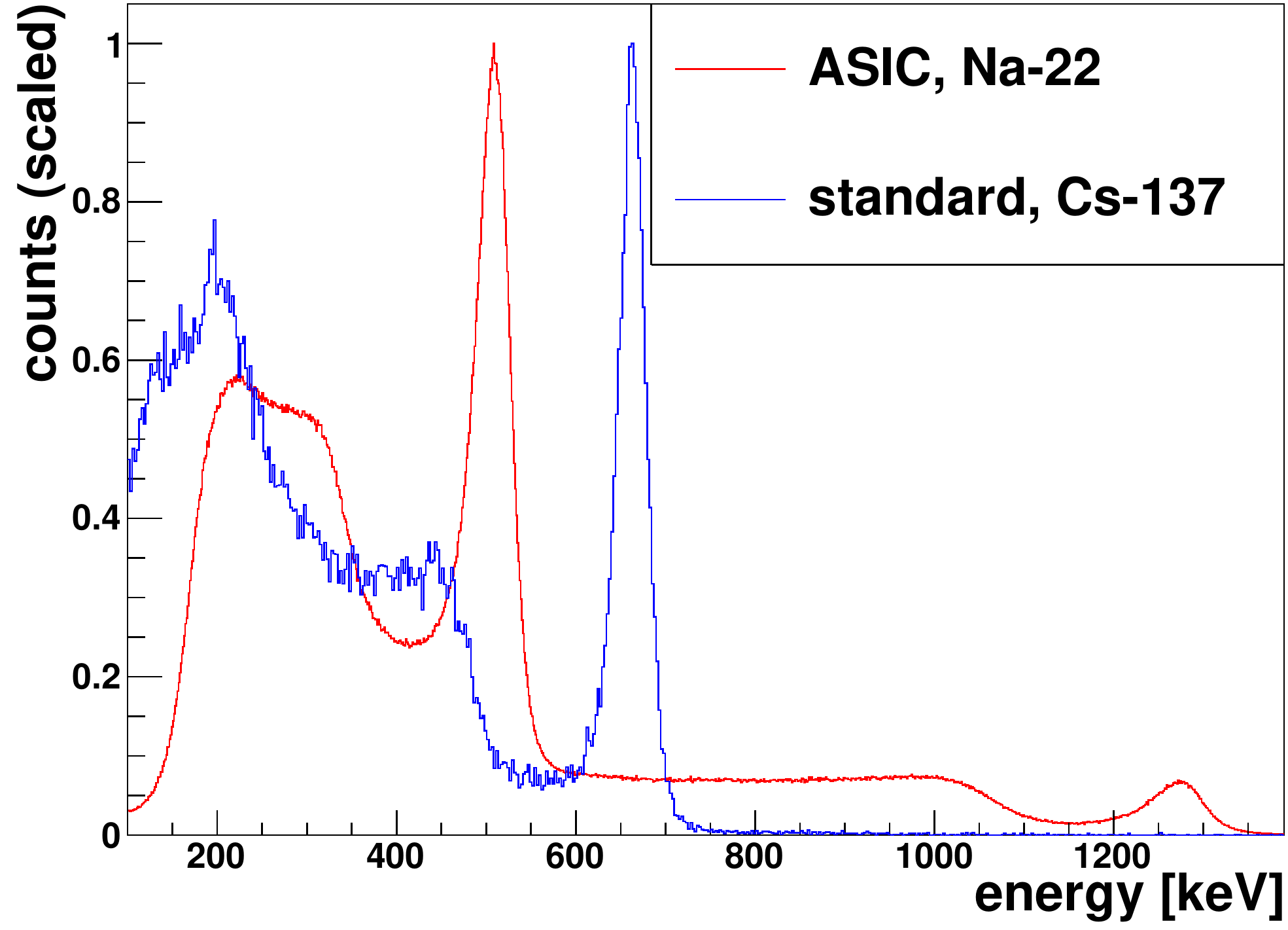}
  \captionsetup{width=0.9\textwidth} 
  \caption[Comparison ASIC-based and COBRA standard readout system]
          {Left: Pulse shapes with \SI{40}{MHz} sampling of an event from a \isotope[22]Na source with \SI{1.3}{MeV}. 
          Right: Comparison of measurements with the ASIC-based readout system at \SI{40}{MHz} at IDEAS' laboratories, and a measurement with the same detectors with the COBRA standard readout system in Dortmund. The energy resolution of the ASIC-based readout system is worse, but due to other properties it cannot be compared directly to the other measurement. }
  \label{fig_comparison_ASIC__standard}
\end{figure}

After improving the shielding analog to the COBRA shielding, several measurements could be done.
An example of pulse shapes of one event and the resulting energy spectrum is shown in \autoref{fig_comparison_ASIC__standard}.

The energy resolution obtained with the ASIC-based readout system is about \SI{8.8}{\%} FWHM at \SI{511}{keV}, compared to \SI{5.5}{\%} at \SI{662}{keV} with the same detectors in Dortmund. 
These values cannot be compared directly, as the setup, settings and noise level were different. Furthermore, due to other properties like pulse lengths or sampling frequencies, COBRA's standard data-cleaning mechanisms could not be used.\\ 

The testing campaign was successful. As a consequence, the readout system was purchased to be used for the detector module of nine large detectors with a volume of (\num{20 x 20 x 15}){\,mm$^3$} each.
Extensive testing has to be done, especially to validate if PSA is still possible despite the reduced event length that is recorded. 
Furthermore, the integration to the planned detector module and the inclusion into COBRA's DAQ-system has to be investigated.

\chapter{Discussion and outlook}
\label{sec_outlook}
In the course of this work, the COBRA demonstrator has been commissioned and completed to its final stage.
A special focus was on reducing noise and improving the signal quality.
Especially the raw signals turned out to be very susceptible to interferences.
Overall, the demonstrator performs well.
Several papers about the demonstrator and the data collected with it were published by the collaboration, see a list on page \pageref{sec_list_of_publications}.
One conclusion of the operation of the demonstrator is that background reduction is a crucial issue.
The major background are events on the detector surfaces, especially alpha radiation.\\
Hence detailed laboratory studies on surface events were done.

The behavior of surface events is mainly dominated by the characteristics of the electric field lines close to the surfaces.
The surface event sensitivity differs for each detector due to individual detector crystal characteristics.
The drifting charge clouds of the incident particles might be trapped in areas with low electric field.
A consequence of this is a dead-layer at the surfaces, where the detector does not collect charges meaningful, or at all.
If the size of the dead-layer is only $\sim \SI{10}{\mu m}$ and affects only alpha radiation, this would be appreciated in terms of background reduction.
However, a larger dead-layer in the size of mm seriously decreases the detection efficiency.\\
On the other hand, the charge clouds might be driven to the wrong anode (NCA (\textbf{n}on-\textbf{c}ollecting \textbf{a}node)). 
This results in charge-sharing distortions. These events are not reconstructed correctly.
Both effects appear more often if the NCA is the outermost anode rail associated to that particular surface.
The CA (\textbf{c}ollecting \textbf{a}node) surfaces are typically about a factor of three more sensitive to alpha events than the NCA surfaces.
The charge-sharing distortions occur also mostly at the NCA surface.
These effects are not bias-dependent, at least with reasonable biases that can be applied to the detectors.

Furthermore, the newly developed PSA (\textbf{p}ulse-\textbf{s}hape \textbf{a}nalysis) method LSE (\textbf{l}ateral \textbf{s}urface \textbf{e}vents) was tested.
It is based on a distorted weighting potential close to the surfaces of the detector compared to the center.
Additionally, the A/E (\textbf{A} divided by \textbf{E}) PSA criterion used by other experiments to discriminate surface events was tested and compared to LSE. 
A/E is based on a weighting potential with a larger gradient in the center compared to the surface.
In the laboratory measurements, A/E has a higher background reduction efficiency, and performs more stably than LSE.
However, some detectors have a lower surface background using the LSE cut than A/E.
Detailed alpha scanning measurements were done for an extensive test of the position-dependence of the surface sensitivity, and the LSE and A/E surface events discrimination power.
Plausible variations of the characteristics on the detector surfaces could be identified sometimes: 
Values indicating a stronger surface events character occurred towards the edges compared to the center. Additionally,
values suggesting less distinctive surface event character were found at the transition edge between NCA and CA bias.
However, no systematic position-dependent characteristics were identified.
No hints were found how to improve the PSA mechanisms.\\

The instrumentation of the GR (\textbf{g}uard \textbf{r}ing) was investigated. 
This offers a very good reduction of surface background events, without losing too much fiducial volume. 
If the GR is on CA potential, it collects the charge-clouds of surface events. 
Consequently, these are very seldom measured by the CA and NCA.
In a laboratory test measurement, only \SI{0.2}{\%} of alpha events remained. 
This is the best surface event reduction achieved so far.

The $\mu_e$ (electron mobility) was determined using the laboratory measurements. The results are in very good agreement with literature values. 
Indications for a lower weighting potential at the lateral surfaces were found.
The drift time of the charge clouds depends on the lateral position (x/y) of the initial interaction, as well as the z-position. 
As none of these quantities can be calculated reliably for surface events independently, the drift time cannot be used to determine one of them.

First test measurements of an ASIC (\textbf{a}pplication \textbf{s}pecific \textbf{i}ntegrated \textbf{c}ircuit)-based readout system provided the proof of principle, that this scalable integrated DAQ (\textbf{d}ata \textbf{a}c\textbf{q}uisition) system could be used by COBRA.

As the COBRA demonstrator is completed, no hardware modifications like GR instrumentation can be done.
However, for the data-analysis it is recommended to investigate the A/E PSA criterion, which was superior in laboratory measurements.
Furthermore, no constant global thresholds for surface events cuts, as done now, should be used.
The thresholds should be determine individually for each detector, maybe even for each physics run, using the calibration data.\\

The following topics should be addressed for the \num{3x3} detector module which shall be installed at LNGS within the next year.
The dead-layer of each detector should be determined. Either using alpha radiation to test for a short-ranged dead-layer, or using a \isotope[207]Bi source for testing the dead-layer in three larger ranges by means of the IC (\textbf{i}nternal \textbf{c}onversion) electrons.
For the large CPqG (\textbf{c}o\textbf{p}lanar \textbf{q}uad-\textbf{g}rid) detectors, all outer surfaces should be biased with NCA potential, which is best in terms of background reduction, if no other objections occur.
The GR should be instrumented to further reduce the surface background.
The ASIC-based readout system has to be investigated in detail to demonstrate its suitability.
Last, but not least, when installing and commissioning the \num{3x3} detector module at LNGS, special care should be taken to the raw signals.
The position and surrounding of their cables is important to reduce noise, enable data taking with low thresholds and to guarantee stable operations.

\begin{figure}[b!]
 \centering
  \includegraphics[width=0.99\textwidth]{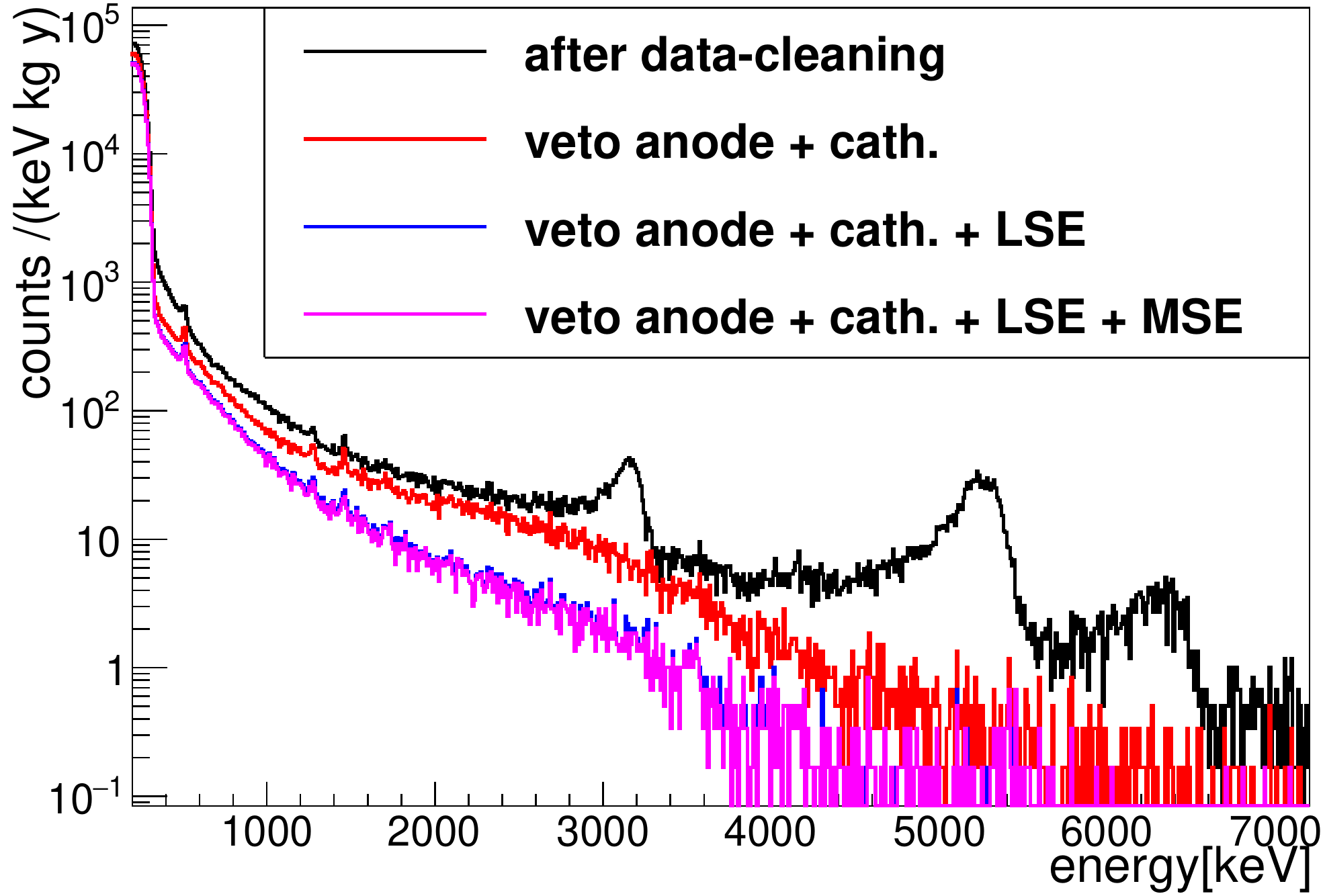} 
  \captionsetup{width=0.9\textwidth} 
  \caption[Background spectrum of the COBRA demonstrator]
          {Background spectrum of the COBRA demonstrator at LNGS.}
  \label{fig_lngs_spectrum_cuts}
\end{figure}

\newpage
Some of the results of this thesis are concluded in \autoref{fig_lngs_spectrum_cuts}.
The data collected with the COBRA (\textbf{C}admium Zinc Telluride \textbf{0}$\nu$ double \textbf{B}eta \textbf{R}esearch \textbf{A}pparatus) demonstrator is used to calculate an energy spectrum using the same dataset as in \cite{Ebert:2015rda}. 
If all events after data-cleaning cuts are used, the black curve results. This shows the effect of all passive measures like shielding and material selection that were done at the COBRA demonstrator.
The distinctive peaks above \SI{3}{MeV} stem from alpha radiation, mainly at the cathode. 
This is because the metalization contains alpha emitting isotopes (\isotope[190]Pt). Furthermore, the BV (\textbf{b}ias \textbf{v}oltage) is applied there, so contaminations gather due to electrostatic attraction (\isotope[210]Pb).
If the anode and cathode areas are vetoed using the interaction depth calculation, the red curve results. 
The alpha peaks are removed by this.
To veto the other four lateral surfaces, the LSE method is used, which lowers the count rate significantly.
As gamma radiation is no major background component, the additional MSE (\textbf{m}ulti-\textbf{s}ite \textbf{e}vent) cut does not improve the result much.
In the ROI (\textbf{r}egion \textbf{o}f \textbf{i}nterest) of \isotope[116]Cd from \SI{2723}{keV} to \SI{2904}{keV} \cite{Ebert:2015rda}, the background level is reduced to \SI{25}{\%} (from \SI{11}{\cpkky} to \SI{2.7}{\cpkky}) using the LSE cut. 
The previous applied cut on the interaction depth removes about \SI{13}{\%} of the lateral surfaces as well.
If this is accounted for, the LSE cut reduces the count rate to \SI{21}{\%}.
This is in good agreement with the number of \SI{17}{\%} which was calculated in \autoref{tab_comparison_A/E_LSE} using laboratory measurements.
Using the A/E cut instead of LSE could offer the potential to reduce the background level further by more than \SI{50}{\%} to about \SI{1.5}{\cpkky}, if the laboratory measurements prove valid again.

If the background consists only of surface events, and the GR instrumentation on CA potential works as good as first laboratory test measurements indicate, the background rate could be reduced by up to another two orders of magnitude.
The count-rate could be around \SI{E-2}{\cpkky} in the ROI of \isotope[116]Cd.
With more restrictive cuts, which remove an assumed partial surface contamination, the best detectors of the COBRA demonstrator concerning the background level (detector layer 1) have a background rate of about \SI{0.5}{\cpkky}.
Additionally, the improvements of the large detectors should provide an extra benefit:
The detection efficiency increases by \num{10} percentage points, while the surface-to-volume ratio is better by about \SI{50}{\%}, which should lead to a lower rate of surface contaminations.
Moreover, the possibility to have only NCA bias at the four outermost anode rails could further improve the background level.

Concluding, reaching a background rate of \SI{E-3}{\cpkky} is not easy, but could be possible for COBRA.

\chapter*{Acknowledgements}
\addcontentsline{toc}{chapter}{Acknowledgements}
I want to thank the people who gave me the opportunity to write this Ph.D. thesis.
First off, this is Mr. Gößling, as my supervisor and first examiner. 
Mr. Zuber kindly agreed to being my second examiner. Furthermore, he is the spokesman of the COBRA collaboration, which provides the context of this thesis.
Mrs Siegmann participated in my examination process, and helped with laboratory infrastructure.

All current and former members of the chair Experimentelle Physik IV provided a good atmosphere. 
I enjoyed working there in the last years.
Special thanks to Robert and Thomas for helping with computer-related problems.

A lot of devices were manufactured by the workshops of the physics department, especially by Jens. Without this many topics described in this thesis would not have been possible.

I appreciated the cooperation with the members of the COBRA collaboration in the daily work, collaboration meetings and in particular at working shift at LNGS. 
Working there was tough, but interesting. Especially Daniel contributed to the completion of the demonstrator.
Perhaps he agrees, that after a hard day full of frustrating work at the COBRA demonstrator, when the causality of the own work was in doubt (see \autoref{fig_lngs_data_rate_after_shift}), everything was fine again with antipastini, arrosticini or bistecca and vino rosso de la casa at Franco's Trattoria Fuore de la Mura \cite{franco}.
Many things at LNGS were only possible because the two Matthias' helped a lot.\\

Most of all, I want thank my family for the support I received in the last years.
Especially in the last busy months writing this thesis, my wife Annabell encouraged and urged me to proceed, and provided so much help.
Coming home after work was so enjoyable, when Clara and Jonathan ran and crawled to me, mostly laughing and chatting, sometimes crying.
Without this, I would not have succeeded.

\chapter*{List of publications}
\label{sec_list_of_publications}
\addcontentsline{toc}{chapter}{List of publications}

\section*{Articles}

\begin{itemize}
    \item J. Ebert, ..., J. Tebrügge, et al.\\
         \textsc{Current Status and Future Perspectives of the COBRA Experiment}, \textit{Adv.High Energy Phys. (2013) 703572}
 
    \item J. Ebert, ..., J. Tebrügge, et al.\\
         \textsc{ Pulse-shape discrimination of surface events in CdZnTe detectors for the COBRA experiment}, \textit{Nucl.Instrum.Meth. A749 (2014) 27-34}

   \item J. Ebert, ..., J. Tebrügge, et al.\\
         \textsc{The COBRA demonstrator at the LNGS underground laboratory}, \textit{Nucl.Instrum.Meth. A807 (2016) 114-120}

   \item J. Ebert, ..., J. Tebrügge, et al.\\
         \textsc{Long-term stability of underground operated CZT detectors based on the analysis of intrinsic $^{113}$Cd $\beta^{-}$-decay}, \textit{Nucl.Instrum.Meth. A821 (2016) 109-115}

   \item J. Ebert, ..., J. Tebrügge, et al.\\
         \textsc{Characterization of a large CdZnTe coplanar quad-grid semiconductor detector}, \textit{Nucl.Instrum.Meth. A806 (2016) 159-168}

   \item J. Ebert, ..., J. Tebrügge, et al.\\
         \textsc{Results of a search for neutrinoless double-beta decay using the COBRA demonstrator}, \textit{ submitted to Phys. Rev. C }

\end{itemize}

\section*{Presentations}

\begin{itemize}
   \item Jan Tebrügge\\
         \textsc{Pulse Shape Analyse for Coplanar Grid CdZnTe Detektoren für das COBRA-Experiment}\\
         \textit{Spring Meeting, German Physical Society, Göttingen 29.02.2012}
   \item Jan Tebrügge on behalf of the COBRA collaboration\\
         \textsc{Status des COBRA-Experiments}\\
         \textit{Spring Meeting, German Physical Society, Dresden 04.03.2013}
   \item Jan Tebrügge on behalf of the COBRA collaboration\\
         \textsc{COBRA - An ultra low background application of CdZnTe}\\
         \textit{International Committee for Future Accelerators School, Poster, \newline Bogot\'a, 03.12.2013}
   \item Jan Tebrügge\\
         \textsc{Pulse-shape discrimination of lateral surface events for the COBRA experiment}\\
         \textit{Spring Meeting, German Physical Society, Mainz 24.03.2014}
   \item Jan Tebrügge on behalf of the COBRA collaboration\\
         \textsc{Status and Perspectives of the COBRA Experiment}\\
         \textit{International Workshop on Gross Properties of Nuclei and Nuclear Excitations, Hirschegg 15.01.2015}
   \item Jan Tebrügge\\
        \textsc{ Detecting surface events at the COBRA Experiment}\\
        \textit{ Spring Meeting, German Physical Society, Wuppertal 09.03.2015}
   \item Jan Tebrügge on behalf of the COBRA collaboration\\
         \textsc{Status and Perspectives of the COBRA Experiment}\\
         \textit{Neutrinos and Dark Matter in Nuclear Physics, Jyväskulä 03.06.2015}
\end{itemize}

\listoftables
\addcontentsline{toc}{chapter}{List of Tables}

\listoffigures
\addcontentsline{toc}{chapter}{List of Figures}

%%%%%%%%%%%%%%%%%%%%%%%%%%%%%%%%%%%%%%%%%%%%%%%%%%%%%%%%%%%% Literaturverzeichnis:

 \bibliography{dissertation_tebruegge}
 \bibliographystyle{alpha}
 \addcontentsline{toc}{chapter}{Bibliography}

\end{document}